\RequirePackage[l2tabu,orthodox]{nag}

\documentclass[a4paper,11pt,twoside,onecolumn,openright,final,titlepage]{book}

\usepackage[T1]{fontenc}
\usepackage[utf8]{inputenc}
\usepackage[inner=2.5cm,outer=1.5cm,top=2.5cm,bottom=2.5cm,includeheadfoot]{geometry}
\usepackage[english]{babel}

\usepackage{fancyhdr}
\usepackage{amsmath} 
\usepackage{esint}
\usepackage{amsfonts} 
\usepackage{dsfont}
\usepackage{bm}
\usepackage{bbm}
\usepackage{upgreek}
\usepackage{slashed} 
\usepackage{amssymb} 
\usepackage{graphicx} 
\usepackage{caption}
\usepackage{multirow}
\usepackage{epstopdf}
\usepackage{feynmp-auto}
\usepackage{textpos}
\usepackage{datetime}
\usepackage{todonotes}
\let\temptodo\todo\renewcommand\todo[1]{\temptodo[inline]{#1}}
\usepackage{footmisc}
\usepackage{tocbibind}
\usepackage{longtable}
\usepackage{url}
\usepackage{float}

\usepackage{cellspace}
\usepackage[protrusion=true,expansion=true]{microtype}

\usepackage[margin=10pt,textfont={small,sl},labelfont=bf]{caption}
\usepackage[margin=15pt,textfont={small,sl},labelfont=bf]{subcaption}
\makeatletter
\providecommand\phantomcaption{\caption@refstepcounter\@captype}
\makeatother

\usepackage{listings}
\lstnewenvironment{Mathematica}{\lstset{basicstyle=\ttfamily,columns=[rl]fixed}}{}
\newcommand{\MathematicaInline}[1]{\lstinline[basicstyle=\ttfamily]{#1}}

\usepackage[pdftex,
pagebackref,
pdftitle={Heavy quarkonium electric dipole transitions in non-relativistic quantum field theory},
pdfauthor={Sebastian Steinbeisser},
pdfsubject={Electric dipole transitions in heavy quarkonia using the non-relativistic quantum field theory called pNRQCD.},
pdfcreator={pdflatex},
pdfproducer={Latex with hyperref and bibtex},
pdfkeywords={QCD, EFT, pNRQCD, heavy quarkonia, electric dipole transition, decay width},
hypertexnames=true,
plainpages=false,
pdfpagelabels=true,
pdfstartview=FitH,
bookmarks=true,
bookmarksnumbered=true,
bookmarksopen=true,
bookmarksopenlevel=\maxdimen,
colorlinks=true,
linkcolor=black,
citecolor=black,
filecolor=black,
urlcolor=black]{hyperref}
\renewcommand*{\backref}[1]{}
\renewcommand*{\backrefalt}[4]{[{\small%
		\ifcase #1 Not cited.%
		\or Cited on page~#2.%
		\else Cited on pages #2.%
		\fi%
	}]}

\renewcommand{\L}{\mathcal{L}}

\renewcommand{\)}{\right)}
\renewcommand{\{}{\left\lbrace}
\renewcommand{\}}{\right\rbrace}

\newcommand{\e}[1]{\mathrm{e}^{#1}}
\renewcommand{\exp}[1]{\mathrm{exp}\left(#1\right)}
\renewcommand{\ln}[1]{\mathrm{ln}\left({#1}\right)}
\newcommand{\lnSquare}[1]{\mathrm{ln}^2\left({#1}\right)}
\renewcommand{\d}{\mathrm{d}}
\newcommand{\1}{\mathds{1}}
\newcommand{\N}{\mathbb{N}}

\newcommand{\R}{\mathbb{R}}
\newcommand{\C}{\mathbb{C}}
\renewcommand{\O}{\mathcal{O}}
\newcommand{\M}{\mathcal{M}}
\renewcommand{\i}{\mathrm{i}}
\newcommand{\mr}{m_{\mathrm{r}}}
\newcommand{\OEone}{\mathcal{O}_{\text{E1}}}
\renewcommand{\l}{\ell}
\renewcommand{\r}[2]{\rho_{#1,#2}}

\renewcommand{\bar}[1]{\overline{#1}}

\allowdisplaybreaks

\begin{document}

\frontmatter

\begin{titlepage}
\begin{center}
\begin{tabular}{c}
\hline
\hline \\
\Huge{\textbf{Heavy quarkonium electric dipole}} \\
\Huge{\textbf{transitions in non-relativistic}} \\
\Huge{\textbf{quantum field theory}} \\
\\
\hline
\hline
\end{tabular}
\newline \\
\vspace*{4cm}
\Large{Master Thesis} \\
\large{by} \\
\Large{Sebastian Steinbeißer} \\
\large{(03616204)} \\
\vspace*{1cm}
\Large{21.02.2017} \\
\vspace*{7cm}
\begin{tabular}{lcr}
\hline
\hline \\
\multirow{3}{*}{\includegraphics[height=1.5cm]{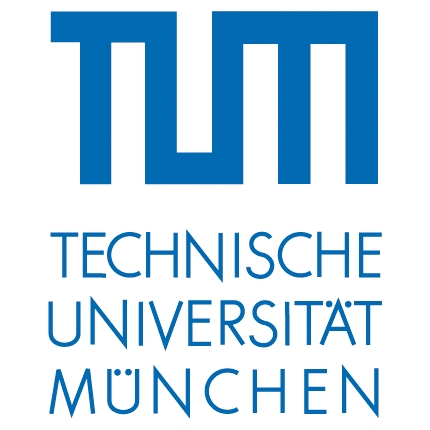}} & Technische Universität München & \multirow{3}{*}{\includegraphics[height=1.5cm]{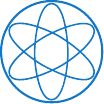}} \\
 & Physics Department, Group T30f & \\
 & N. Brambilla \& A. Vairo & \\
\\
\hline
\hline
\end{tabular}
\end{center}
\end{titlepage}

\clearpage
\thispagestyle{empty}
\clearpage

\tableofcontents

\begin{quotation}
\chapter{Abstract}

Electromagnetic E1 (and M1) multipole transitions have been studied since the early days of hadron spectroscopy because they allow to access heavy quarkonium states which are below open-flavor threshold. Moreover, they are interesting by themselves because they are an important tool to check particular regions of the hadrons' wave function and thus to determine their internal structure and dynamics.\\
From a theoretical point of view, electromagnetic transitions between heavy quarkonium states have been treated for a long time by means of potential models using non-relativistic reductions of phenomenological interactions. However, the progress made in effective field theories (EFTs) for studying heavy quarkonia and the new large set of accurate experimental data taken in the heavy quark sector by B-factories (BaBar, Belle and CLEO), $\tau$-charm facilities (CLEO-c, BESIII) and even proton-(anti)proton colliders (CDF, D0, LHCb, ATLAS, CMS) ask for a systematic and model-independent analysis.\\
In this work we use the low-energy EFT called potential non-relativistic QCD (pNRQCD) to calculate the partial decay width of different $b\bar{b}$-states undergoing an electric dipole transition at next-to-next-to leading order (NNLO). Explicitly, the $\chi_{b J}$, for $J=0,1,2$, and the $h_b$ are investigated by computing the processes $\chi_{b J} \to \Upsilon + \gamma$ and $h_b \to \eta_b + \gamma$. Relativistic corrections of relative order $v^2$ to the leading electric dipole operator are included. The analysis separates those contributions that account for the electromagnetic interaction terms in the pNRQCD Lagrangian, which are $v^2$ suppressed, and those that account for quarkonium state corrections of relative order $v$ and $v^2$. Within the last ones, corrections come from higher order potentials ($\frac{1}{m}$ and $\frac{1}{m^2}$ terms), and from higher order Fock states which account for the coupling of the quark-antiquark state to other low-energy degrees of freedom and thus demand non-perturbative input.\\
Finally, the experimentally known branching fractions are used to predict the total decay with of the respective initial states.

\end{quotation}

\clearpage
\thispagestyle{empty}
\clearpage

\mainmatter

\chapter{Introduction}
\label{chp:Introduction}

Electromagnetic transitions are often significant decay modes for bottomonium states below the $B\bar{B}$ threshold ($10.56$~GeV), making them a suitable experimental tool to access the lowest spectrum of bottomonia. For instance, the first $b\bar{b}$ states not directly produced in $e^{+}e^{-}$ collisions were the six triplet-$P$ states, $\chi_{b}(2P_{J})$ and $\chi_{b}(1P_{J})$ with $J=0,1,2$, discovered in radiative decays of the $\Upsilon(3S)$ and $\Upsilon(2S)$ in $1982$ \cite{Han:1982zk,Eigen:1982zm} and $1983$ \cite{Klopfenstein:1983nx,Pauss:1983pa}, respectively.\\
One important feature of electromagnetic transitions is that they can be 
classified in a series of electric and magnetic multipoles. The most important 
ones are the E1 (electric dipole) and the M1 (magnetic dipole) transitions; 
higher order multipole modes E2, M2, E3, etc. appear in the spectrum, but since 
they are further suppressed one usually does not consider them. Processes 
involving electric dipole (E1) transitions happen more frequently than the ones 
induced by a magnetic dipole (M1). The branching fraction for E1 transitions 
can indeed be significant for some lowest bottomonium states like the ones we 
shall study herein \cite{PDG:2016}: $\mathcal{B}(\chi_{b0}(1P) \to 
\Upsilon(1S) + \gamma) = (1.76 \pm 0.35)\,\%$ (note that it is the largest 
exclusive branching fraction reported by the Particle Data Group (PDG) 
\cite{PDG:2016}), $\mathcal{B}(\chi_{b1}(1P) \to \Upsilon(1S) + \gamma) = (33.9 
\pm 
2.2)\,\%$, $\mathcal{B}(\chi_{b2}(1P) \to \Upsilon(1S) + \gamma) = (19.1 \pm 
1.2)\,\%$ and $\mathcal{B}(h_b(1P) \to \eta_b(1S) + \gamma) = 
(52^{+6}_{-5})\,\%$.\\
The E1 (and M1) electromagnetic transitions have been treated for a long time by means of potential models that basically use non-relativistic reductions of QCD-based quark-antiquark interactions (see, e.g., Ref. \cite{Segovia:2016xqb} for a recent application to the bottomonium system). However, the progress made in effective field theories (EFTs) for studying heavy quarkonia \cite{Brambilla:2004jw} and the new large set of accurate experimental data taken in the heavy quark sector by B-factories (BaBar, Belle and CLEO), $\tau$-charm facilities (CLEO-c, BESIII) and even proton-(anti)proton colliders (CDF, D0, LHCb, ATLAS, CMS) ask for a systematic and model-independent analysis (see, e.g., Refs. \cite{Brambilla:2010cs,Brambilla:2014jmp} for reviews).\\
Formulae and numerical treatment of M1 transitions within the effective field theory named potential non-relativistic QCD (pNRQCD) can be found in Refs.~\cite{Brambilla:2005zw,Pineda:2013lta}. Therein, the relativistic corrections to the leading order (LO) expression were computed in two different expansion schemes: (i) strict weak-coupling regime and (ii) including exactly the static potential in the LO Hamiltonian. Within the same theoretical framework, the corresponding formulae for E1 transitions have been presented in Ref.~\cite{Brambilla:2012be}. In this case, the relativistic corrections to the LO decay width are much more involved, covering not only higher order terms in the E1 transition operator but also corrections to the initial and final state wave functions due to higher order potentials and higher order Fock states.\footnote{It should be mentioned that such corrections in principle also affect M1 transitions, but the color octet contributions vanish analytically.} These facts have avoided numerical computations of the E1 radiative decays within pNRQCD. This work aims to close this gap and to calculate the decay rate of the transitions $\chi_{bJ}(1P) \to \Upsilon(1S) + \gamma$ with $J=0,1,2$ and $h_b(1P) \to \eta_b(1S) + \gamma$.\\
The thesis is organized as follows: In \textbf{Chapter~\ref{chp:BasicConcepts}} we review some basic properties of quantum chromodynamics (QCD), introduce the concept of EFTs in the framework of heavy quarkonium physics, discuss the EFT called non-relativistic QCD (NRQCD) and state the relevant properties of electromagnetic dipole transitions. In \textbf{Chapter~\ref{chp:pNRQCDQuantumMechanicalPerturbationTheory}} we introduce the EFT called potential non-relativistic QCD (pNRQCD) that will be the theoretical framework used throughout this work. We furthermore introduce the concept of quantum mechanical perturbation theory and derive the key concepts and equations in order to compute the mass spectrum and the E1 decay widths we are interested in. In \textbf{Chapter~\ref{chp:QuarkoniumSpectrumPNRQCDWeakCouplingUpToMAlpha4}} we compute the $b\bar{b}$-mass spectrum up to NNLO, $\O(m\alpha_s^4)$, in pNRQCD at weak coupling and confirm results already found in \cite{Brambilla:2001fw,Peset:2015vvi}. We furthermore address the issue of renormalons in the perturbative series and use a particular scheme (the so-called renormalon subtraction scheme) to cure this issue. In \textbf{Chapter~\ref{chp:E1Transitions}} we fulfill the main objective of this work, namely computing the E1 decay width of $2\,^3\!P_J \to 1\,^3\!S_1 + \gamma$, with $J=0,1,2$ and $2\,^1\!P_1 \to 1\,^1\!S_0 + \gamma$ up to NNLO, $\O(m\alpha_s^6)$, in pNRQCD at weak coupling. We include all the relevant corrections at relative order $v$ and $v^2$ which include relativistic corrections to the leading order electric dipole operator and corrections to the initial and final quarkonium states. The latter ones are either induced by higher order corrections to the static potential or by relativistic corrections in the $\frac{1}{m}$-expansion or by higher order Fock states induced by color octet effects. We apply our results to the transitions $\chi_{b J}(1P) \to \Upsilon(1S) + \gamma$, with $J=0,1,2$ and to $h_b(1P) \to \eta_b(1S) + \gamma$. Numerical results are given in \textbf{Chapter~\ref{chp:NumericalResults}} where we also investigate the scale dependence and discuss particular issues arising with this analysis. Our main results will be the prediction of the partial and total widths of the $b\bar{b}$-states, $\chi_{b J}$ and $\eta_b$. Finally, a summary and an outlook is given in \textbf{Chapter~\ref{chp:SummaryOutlook}}.\\
In \textbf{Appendix~\ref{app:chp:NotationsAndConvention}}, we introduce the notation and convention we will be following throughout this work and list important constants as well as the explicit radial and angular expressions of the first few Coulomb wave functions. We furthermore give two examples on the usage of the MATHEMATICA package RunDec, which we use in order to implement the running of the strong fine structure constant $\alpha_s$. In \textbf{Appendix~\ref{app:chp:FunctionsExpectationValues}}, we list several functions, their properties, and useful relations that are used in this work. We furthermore list and derive single and double potential insertion expectations values, needed in order to compute the mass spectrum and show the exact divergence cancellation with respect to the Coulomb Green function approach. Finally, in \textbf{Appendix~\ref{app:chp:AlternativeFirstOrderWaveFunction}}, we derive an alternative method to compute the first order correction to the wave function in perturbation theory.\\
This work is mainly based on the Refs.~\cite{Brambilla:2004jw,Brambilla:2005zw,Pietrulewicz:2011aca,Pineda:2011dg,Brambilla:2012be}.

\clearpage
\thispagestyle{empty}
\clearpage

\chapter{Basic concepts}
\label{chp:BasicConcepts}

In this chapter we explain the basic concepts of quantum chromodynamics (QCD) and effective field theories (EFTs). We then describe the EFT called non-relativistic QCD (NRQCD) that can be obtained from QCD by integrating out the heavy quark mass (hard scale) and finally state the relevant properties of electromagnetic dipole transitions.

\section{Quantum Chromodynamics (QCD)}
\label{sec:QCD}

The framework used to describe the strong interaction, the interaction of the fundamental hadronic degrees of freedom, quarks and gluons, is the quantized local gauge theory called quantum chromodynamics (QCD) with the gauge group being SU(3). From a phenomenological point of view QCD may be characterized by its main properties:
\begin{description}
\item[Asymptotic freedom] which describes the observation that at high energy or, equivalently, low distance quarks and gluons behave as if they were free particles.
\item[Confinement] which is related with the empirical fact that no color-charged particles, e.g., quarks or gluons, have been observed as isolated particles. Only the color-neutral hadrons like mesons (a state of a bound quark-antiquark pair) and baryons (a state of three bound quarks) can be observed.
\item[Dynamical chiral symmetry breaking] which allows for a description of QCD in the low energy regime using, e.g., chiral perturbation theory ($\chi$PT). The breaking of chiral symmetry generates the pions as Nambu-Goldstone bosons and the non-zero masses of the light u- and d-quark, which may be obtained via the Higgs mechanism of the standard model, make them Pseudo-Goldstone bosons with non-zero but small masses.
\end{description}
One can argue that the first two of these features can be connected to the running coupling of QCD that we will explain briefly in Sec.~\ref{subsec:RunningCouplingQCD}. The latter feature originates from the existence of a non-vanishing quark condensate, but it is not relevant for heavy quark physics because chiral symmetry is explicitly broken due to the heavy quark mass. For further reading on chiral symmetry breaking see, e.g., Refs.~\cite{Nambu:1961tp,Nambu:1961fr}.

\subsection{The QCD Lagrangian}
\label{subsec:QCDLagrangian}

The QCD Lagrangian is given by \cite{Griffiths:IntroductionElementaryParticles,Peskin:IntroductionQuantumFieldTheory,Povh:TeilchenKerneEinfuehrungPhysikalischenKonzepte,Skands:2012ts}
\begin{equation}
\label{eqn:QCD-Lagrangian} \L_{\mathrm{QCD}} = \bar{\psi}_q^{\,i} (\i\slashed{D}_{i j} - m_q \delta_{i j}) \psi_q^j - \frac{1}{4} \mathcal{F}_{\mu\nu}^a \mathcal{F}^{\mu\nu\,a} \,,
\end{equation}
where $\psi_q^i$ denotes a quark field with fundamental color index $i \in \lbrace 1,2,3 \rbrace$. $\slashed{D} = \gamma^\mu D_\mu$, where $\gamma^\mu$ is a Dirac matrix\footnote{The Dirac matrices and some of their important properties are listed in Appendix~\ref{app:chp:NotationsAndConvention}.} that makes explicit the vector nature of the strong interaction and $\mu$ is a Lorentz index. $m_q$ are the masses of the different quark flavors that may be generated by Yukawa couplings to the Higgs-sector of the standard model. The gluon field strength tensor
\begin{align}
& \mathcal{F}_{\mu\nu} = t^a \mathcal{F}_{\mu\nu}^a = -\frac{1}{g} [D_\mu,D_\nu] \,, \\
& \mathcal{F}_{\mu\nu}^a = \partial_{\mu} A_{\nu}^a - \partial_{\nu} A_{\mu}^a + g f_{a b c} A_{\mu}^b A_{\nu}^c \,,
\end{align}
with adjoint color indices $a,b,c \in \lbrace 1,\dots,8 \rbrace$ encodes the self-interacting nature of the gluons. This self-interaction arises, because the covariant derivative
\begin{equation}
(D_{\mu})_{i j} = \delta_{i j} \partial_{\mu} - \i g t_{i j}^a A_{\mu}^a \,,
\end{equation}
where $g$ is the QCD coupling, contains the generators
\begin{equation}
t_{i j}^a = \frac{1}{2} \lambda_{i j}^a
\end{equation}
of the non-abelian gauge group SU(3) whose structure constants $f_{a b c} \neq 0$ do not vanish. The hermitian and traceless Gell-Mann matrices, $\lambda_a$, are listed in Appendix~\ref{app:chp:NotationsAndConvention} together with some important properties.

\subsection{The running coupling of QCD}
\label{subsec:RunningCouplingQCD}

In QCD, as in all quantum field theories, the strong coupling $g$ and thus the strong fine structure constant
\begin{equation}
\alpha_s = \frac{g^2}{4\pi}
\end{equation}
is, in contrast to classical theories, not a constant but runs with the energy scale.
\begin{figure}[t]
\centering
{\includegraphics[
clip,trim={7.2cm 23cm 6.4cm 4.3cm},width=0.7\textwidth]{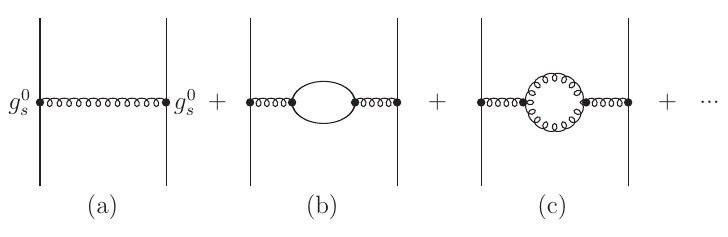}}
\caption[Diagrams which contribute to the running coupling constant.]{Diagrams which contribute to second order in $g^0$ to the running coupling constant $\bar{g}(Q^2)$, taken from \cite{Segovia:2012gka}.}
\label{fig:RunningCoupling}
\end{figure}
Since the gluons also carry color charge, a virtual gluon emitted from a quark does not only see the color charge of another quark (diagram (a) in Fig.~\ref{fig:RunningCoupling}) and of a virtual quark-antiquark pair (diagram (b)), but also the color charge of virtual gluons (diagram (c)). The combined effect of the Feynman diagrams in Fig.~\ref{fig:RunningCoupling}, together with vertex corrections and ghost contributions that are not depicted, yields the effective fine structure constant \cite{Gross:1973ju,Politzer:1973fx,PhysRevLett.30.1343,Gross:1974cs}
\begin{equation}
\label{eq:EffectiveFineStructureConstant} \alpha_s(Q^2) = \frac{g^2(Q^2)}{4\pi} = \alpha_s^{(0)} \left[ 1 - \frac{2}{3}n_f\frac{\alpha_s^{(0)}}{4\pi}\ln{\frac{\Lambda^2}{Q^2}} + 11 \frac{\alpha_s^{(0)}}{4\pi}\ln{\frac{\Lambda^2}{Q^2}} \right] \,,
\end{equation}
where $\alpha_s^{(0)} = \frac{(g^{(0)})^2}{4\pi}$, $g^{(0)}$ is the bare strong coupling, $n_f$ encodes the number of active flavors at the considered energy scale, $Q^2$ is the space-like momentum transfer carried by the virtual gluon, and $\Lambda$ is a cutoff parameter. The second term in the square brackets is the contribution of the virtual quark-antiquark pairs. It is negative and leads to charge screening. The third term comes from the virtual gluon loop. It is positive and gives anti-screening. The appearance of these two competing terms of opposite sign is a peculiarity of non-abelian gauge theories.\\
To incorporate these in a systematic manner, one defines the so-called beta function,
\begin{equation}
Q^2 \frac{\d \alpha_s(Q^2)}{\d Q^2} = \frac{\d \alpha_s(Q^2)}{\d \ln{Q^2}} = \beta(\alpha_s) \,,
\end{equation}
where the beta function, driving the energy dependence, is defined as
\begin{equation}
\beta(\alpha_s) = -\alpha_s \left( \beta_0 \frac{\alpha_s}{4\pi} + \beta_1 \frac{\alpha_s^2}{(4\pi)^2} + \dots \right) \,.
\end{equation}
The coefficients $\beta_i$ can be calculated perturbatively and $\beta_{0,1}$ are explicitly given in Appendix~\ref{app:sec:Constants}. $\beta_0$ and $\beta_1$ are the only coefficients that are renormalization scheme independent. We use the MATHEMATICA package RunDec (\cite{Chetyrkin:2000yt}) to determine the value of $\alpha_s(\nu)$ at a given scale $\nu$. Higher order coefficients, $\beta_{i \geq 2}$, are incorporated in the $\bar{\text{MS}}$-scheme in the RunDec package. The explicit procedure and examples are given in Appendix~\ref{app:sec:RunDec}.

\subsection{Symmetries of the QCD Lagrangian}
\label{subsec:SymmetriesQCDLagrangian}

Besides the invariance under the SU(3)$_{\text{c}}$ color gauge group, with the corresponding transformations
\begin{align}
& \psi_q(x) \mapsto U(x) \psi_q(x) \,, \\
& A_\mu(x) \mapsto U(x) \left[A_\mu(x) - \frac{i}{g} U^{-1}(x) \partial_\mu U(x)\right] U^{-1}(x) \,,
\end{align}
where $U(x) = \exp{-\i\theta_a(x)t^a} \in$ SU(3) is such a gauge transformation, the QCD Lagrangian is Poincaré and CPT invariant and exhibits other global symmetries. The breaking patterns of these global symmetries are summarized in the following diagram for $N_f$ quark flavors, taken from \cite{Hell:2010}:
\begin{center}
{\includegraphics[
clip,trim={3.3cm 17.4cm 2.0cm 2.9cm},width=0.7\textwidth]{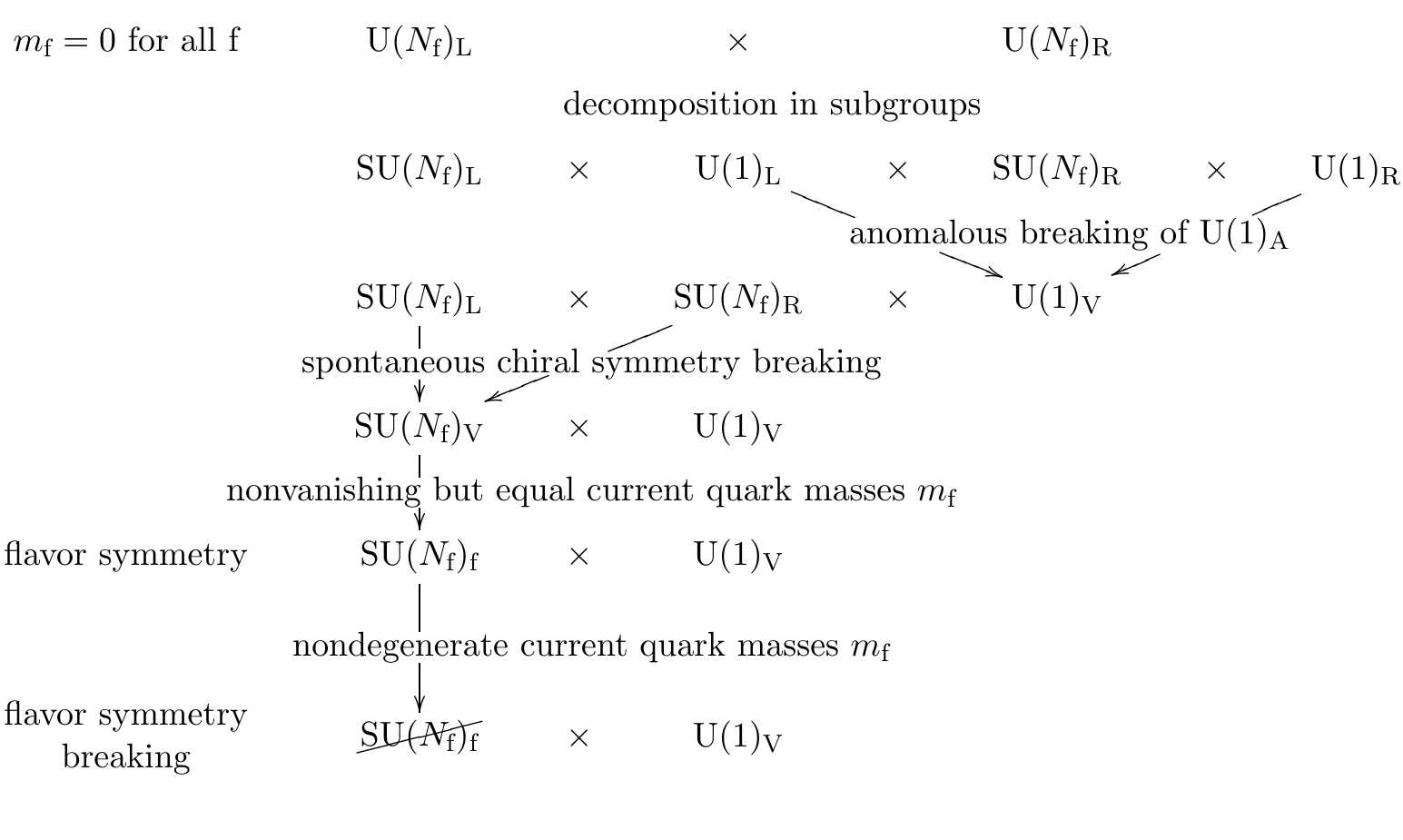}}
\end{center}
For vanishing quark masses, QCD is not sensitive to different flavors and one observes that the Lagrangian is invariant under a chiral U($N_f$)$_L$ $\times$ U($N_f$)$_R$ flavor symmetry group that can be divided in special unitary SU($N_f$) and unitary U(1) groups. The chiral SU($N_f$)$_L$ $\times$ SU($N_f$)$_R$ symmetry is spontaneously broken by the non-trivial QCD vacuum to the vector SU($N_f$)$_V$ with the formation of chiral condensates of quark flavors with current quark masses smaller than the typical QCD scale $\Lambda_{\text{QCD}} \sim 250~\mathrm{MeV}$. The axial symmetry U(1)$_A$ is exact in the classical theory, but broken in the quantum theory by the axial anomaly. Its appearance is strongly related to the non-trivial structure of the QCD vacuum.\footnote{The so-called conformal symmetry is another symmetry that is present in the classical theory but anomalously broken in the quantized version, since dimensional transmutation induces the scale $\Lambda_{\text{QCD}}$ even in massless QCD.} The vector symmetry, U(1)$_V$ corresponds to the baryon number which is an exact symmetry.\\
For non-vanishing current quark masses, the QCD Lagrangian is not invariant under the chiral symmetry SU($N_f$)$_L$ $\times$ SU($N_f$)$_R$ because of the mass term $\delta\L_m = \bar{\psi}\hat{m}\psi$, where $\hat{m} = \mathrm{diag}(m_1,\dots,m_{N_f})$ is the quark mass matrix, that mixes left- and right-handed components of the Dirac field. However, if all masses are assumed to be equal but non-zero, the remnant SU($N_f$)$_V$ symmetry is still a symmetry of the QCD Lagrangian, while axial SU($N_f$)$_A$ is explicitly broken. The remnant SU($N_f$)$_V$ symmetry is often called flavor symmetry and denoted by SU($N_f$)$_f$. For instance, for $N_f=3$ one recovers degenerate meson octets as predicted by the eightfold way even before QCD was established. Finally, if the current quark masses are different, then SU($N_f$)$_f$ is explicitly broken and one obtains non-degenerate meson $N_f^2-1$ multiplets.

\section{Effective field theories and quarkonium physics}
\label{sec:EftQuarkoniumPhysics}

\begin{figure}[t]
\centering
{\includegraphics[
clip,trim={6.8cm 16.6cm 7.2cm 6.9cm},width=0.5\textwidth]{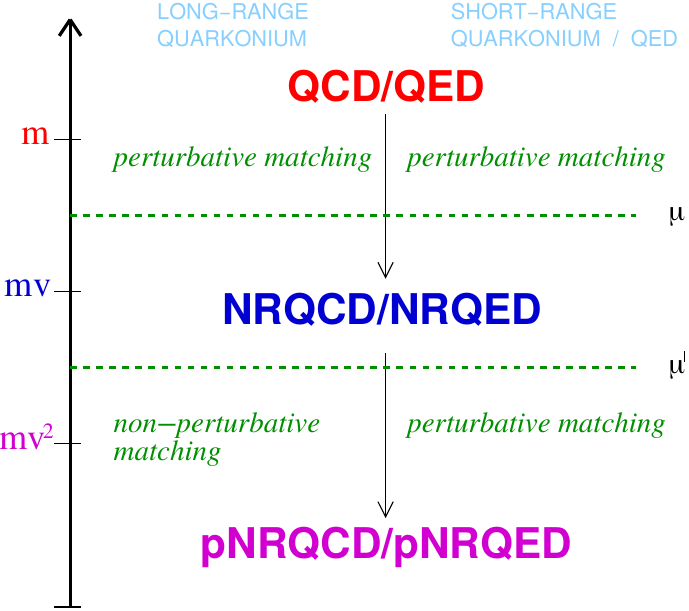}}
\caption[EFTs for bound states in QED and, for heavy quarks, in QCD.]{EFTs for bound states in QED and, for heavy quarks, in QCD, taken from \cite{Vairo:2009rs}.}
\label{fig:EFTsBoundStates}
\end{figure}

Effective field theories may be motivated best by the following quote, taken from \cite{Pich:1998xt}:
\begin{quote}
In order to analyze a particular physical system amid the impressive richness of the surrounding world, it is necessary to isolate the most relevant ingredients from the rest, so that one can obtain a simple description without having to understand everything. The crucial point is to make an appropriate choice of variables, able to capture the physics which is most important for the problem at hand.
\end{quote}
A typical problem consists of a physical system that has multiple and well separated energy scales and one can thus identify a clear hierarchy. If the system under consideration satisfies this condition, an effective field theory may be suited best to study the properties of this system at a given scale of interest. Since the scales are well separated one may neglect smaller or higher scales by putting their parameters to zero or infinity and thereby obtaining a simplified version of the original theory describing the region of interest as an approximation. This approximation may be improved by considering the neglected parts of the original theory as small perturbations.\\
It is common in low energy physics to use effective field theories, where low refers to some energy scale $\Lambda$ of the underlying theory. To construct them one integrates out the states with $M > \Lambda$ from the action and therewith only takes into account the states with $m < \Lambda$ that lie in the region of interest. The price to pay, however, is that the resulting theory is, in general, non-renormalizable but the interaction among the relevant, light, degrees of freedom can be organized in a systematic power expansion energy/$\Lambda$. This allows for an order-by-order renormalization, since at a given order only finitely many couplings exist. These effective couplings encode all the information on the heavy degrees of freedom and can be obtained either by matching the effective field theory to the underlying theory or by fitting them to experimental or lattice data. The latter approach might be unavoidable, for instance, in bottom-up approaches like chiral perturbation theory ($\chi$PT), where the matching coefficients are non-perturbative. The procedure of matching means the equating of the same observables, e.g. Green functions, at the same energy scale, that have independently been calculated in both theories. The matching coefficients, also called Wilson coefficients, allow to determine the effective couplings of the effective field theory. Because, at a given order in the energy/$\Lambda$ expansion, only finitely many effective couplings exist this amounts to calculating finitely many Wilson coefficients order-by-order. This in turn allows for a systematic construction of the effective Lagrangian, which has to be consistent with the underlying symmetries and may be organized in a power series in $\frac{1}{M}$.\\
We now end this very general introduction on effective field theories, there are books and especially lecture notes dealing with the subject in a much deeper way, e.g., Refs~\cite{Pich:1998xt,Kaplan:2005es:MB}, and we refer to them for further reading.\\\\
Quarkonia are a special type of mesons, namely the ones formed by a quark $q_f$ 
with flavor $f$ and the corresponding antiquark $\bar{q}_f$ of the same flavor 
such that the resulting hadron has no net charge or flavor. The ones made up of 
the light quarks (u,d,s) mix quantum mechanically due to the small mass 
difference of their constituents, such that an identification of pure 
$q\bar{q}$-states in experiment is impossible. Therefore the term quarkonia is 
used mainly for the $q\bar{q}$-systems made up by the heavy flavors, namely 
c-quark and b-quark. The resulting hadrons are then called charmonia 
($c\bar{c}$, $J/\psi$ family) and bottomonia ($b\bar{b}$, $\Upsilon$ family, 
main focus of this work). Theoretically, by the means of QCD only, toponia 
($t\bar{t}$) could be possible and have been favored, c.f. 
Refs.~\cite{Leutwyler:1980tn,Voloshin:1979uv}, before the top quark has been 
discovered. However, due to the weak interaction, the t-quark seems to decay 
before a $t\bar{t}$ bound state can be formed \cite{PDG:2016}.\\
The PDG, \cite{PDG:2016}, lists the following quark masses. The u-, d-, and 
s-quark masses are estimates of so-called "current-quark masses", in a 
mass-independent subtraction scheme such as $\bar{\text{MS}}$. The 
$\bar{\text{MS}}$ masses have been normalized at a renormalization scale of 
$\mu = 2$~GeV. The c- and d-quark masses correspond to the "running" masses in 
the $\bar{\text{MS}}$ scheme. The t-quark mass is given (i) as it has been 
determined from $t\bar{t}$ event kinematics and (ii) as the $\bar{\text{MS}}$ 
mass, extracted from $t\bar{t}$ cross-sections using theory calculations.
\begin{align}
\begin{aligned}
& m_{\text{u}} = 2.2^{+0.6}_{-0.4}~\text{eV} \,, && m_{\text{d}} = 4.7^{+0.5}_{-0.4}~\text{eV} \,, & m_{\text{s}} = 96^{+8}_{-4}~\text{MeV} \,, \\
& \bar{m}_{\text{c}}(\bar{m}_{\text{c}}) = (1.27 \pm 0.03)~\text{GeV} \,, && \bar{m}_{\text{b}}(\bar{m}_{\text{b}}) = 4.18^{+0.04}_{-0.03}~\text{GeV} \,, & \\
& m_{\text{t}} = (173.21 \pm 0.51 \pm 0.71)~\text{GeV} \,, && \bar{m}_{\text{t}}(\bar{m}_{\text{t}}) = 160.0^{+4.8}_{-4.3}~\text{GeV} \,.
\end{aligned}
\end{align}
It is useful to distinguish light from heavy quarks due to the obvious mass gap, as well as due to the separation with respect to the dynamically generated scale $\Lambda_{\text{QCD}}$:
\begin{equation}
m_{\text{u,d,s}} \equiv m_{\text{light}} \ll \Lambda_{\text{QCD}} \ll m_{\text{heavy}} \equiv m_{\text{c,b,t}} \,.
\end{equation}
Furthermore, asymptotic freedom, c.f. Refs.~\cite{Gross:1973ju,Politzer:1973fx,PhysRevLett.30.1343,Gross:1974cs}, implies $\alpha_s(m_{\text{heavy}}) \ll 1$.\\
The low energy regime of quarkonium physics then is an ideal system to be described in terms of an effective field theory \cite{Brambilla:2004jw}, because heavy quarkonia can be assumed to be non-relativistic. This assumption of non-relativistic kinematics, $v \ll 1$, where $v$ is the relative velocity in the bound state, generates well separated scales, namely: (i) the hard scale, characterized by the heavy quark mass $m$; (ii) the soft scale, characterized by the relative momentum $p \sim m v$ of the bound state and (iii) the ultra-soft scale, characterized by the binding energy $E \sim m v^2$. For heavy quarkonium produced below threshold we can identify the hierarchy of scales
\begin{equation}
m \gg p \sim m v \gg E \sim m v^2 \quad\quad \text{and} \quad\quad m \gg \Lambda_{\text{QCD}} \,.
\end{equation}
Integrating out the different scales (hard, soft) gives rise to different EFTs. This process is not only possible for QCD, but also has applications in QED, c.f. Refs.~\cite{Pineda:1997bj,Pineda:1998kn}. Figure~\ref{fig:EFTsBoundStates} depicts the fundamental theories QED and QCD and derived effective field theories in their range of applicability.

\section{Non-relativistic QCD (NRQCD)}
\label{sec:NRQCD}

Starting from QCD and integrating out the hard scale $m \gg \Lambda_{\text{QCD}}$, one obtains non-relativistic QCD (NRQCD). This was established by Caswell, Lepage, Bodwin and Braaten in Refs.~\cite{Caswell:1985ui,Bodwin:1994jh}, under the assumptions
\begin{equation}
m \gg m v \,, m v^2 \,, \Lambda_{\text{QCD}} \,.
\end{equation}
Doing so, one obtains the leading order Lagrangian \cite{Brambilla:2004jw} that we present in a form already coupled to electromagnetism\footnote{We already incorporate the electromagnetic terms here in order to shorten the discussion, because the main focus of this work is pNRQCD and not NRQCD.}
\begin{equation}
\L_{\text{NRQCD}} = \L_\psi + \L_\chi + \L_{4f} + \L_{\text{light}} \,.
\end{equation}
The coefficients coming with the operators of the NRQCD Lagrangian can be determined via matching with the non-relativistic limit of QCD order by order in the inverse heavy quark mass $m$ \cite{Manohar:1997qy}. These so-called Wilson coefficients (short distance matching coefficients) are functions of $m$ and the factorization scale $\mu$ (cut-off). The matching can be performed in perturbation theory, since $\alpha_s(m) \ll 1$. The two fermion part is given by $\L_\psi + \L_\chi$, where the relevant degrees of freedom are the heavy quarks and antiquarks that can be described by Pauli spinors $\psi(x)$ and $\chi(x)$ that transform in the fundamental representation of SU(3)$_{\text{c}}$.\footnote{It is sufficient to use Pauli spinors, since the energy scale we are dealing with does not allow the creation of additional heavy degrees of freedom.} They are given, at the relevant order in the $\frac{1}{m}$-expansion, by
\newpage
\begin{align}
\L_\psi &= \psi^\dagger \left(\i D_0 + \frac{1}{2m}\vec{D}^{\,2} + \frac{1}{8 m^3}\vec{D}^{\,4}\right) \psi \\
\nonumber & + g \psi^\dagger \left(\frac{c_F}{2m} \vec{\sigma} \cdot \vec{B} + \i \frac{c_S}{8m^2} \vec{\sigma} \cdot (\vec{D} \times \vec{E} - \vec{E} \times \vec{D}) + \frac{c_D}{8m^2}(\vec{D} \cdot \vec{E} - \vec{E} \cdot \vec{D})\right) \psi \\
\nonumber & + e e_Q \psi^\dagger \left(\frac{c_F^{e/m}}{2m} \vec{\sigma} \cdot \vec{B}^{\,e/m} + \i \frac{c_{\text{s}}^{e/m}}{8m^2} \vec{\sigma} \cdot (\vec{D} \times \vec{E}^{\,e/m} - \vec{E}^{\,e/m} \times \vec{D}) + \frac{c_D^{e/m}}{8m^2} (\vec{D} \cdot \vec{E}^{\,e/m} - \vec{E}^{\,e/m} \cdot \vec{D})\right) \psi \\
\nonumber & + e e_Q \psi^\dagger \left(\frac{c_{W1}^{e/m}}{8m^3} (\vec{D}^{\,2} (\vec{\sigma} \cdot \vec{B}^{\,e/m}) + (\vec{\sigma} \cdot \vec{B}^{\,e/m}) \vec{D}^{\,2}) - \frac{c_{W2}^{e/m}}{4m^3} (\vec{D}^{\,i} (\vec{\sigma} \cdot \vec{B}^{\,e/m}) \vec{D}^{\,i}\right) \psi \\
\nonumber & + e e_Q \psi^\dagger \left(\frac{c_{p' p}^{e/m}}{8m^3} [(\vec{\sigma} \cdot \vec{D})(\vec{B}^{\,e/m} \cdot \vec{D}) + (\vec{D} \cdot \vec{B}^{\,e/m})(\vec{\sigma} \cdot \vec{D})]\right) \psi \\
\nonumber & + e e_Q \psi^\dagger \left(\i \frac{c_M^{e/m}}{8m^3} [\vec{D} \cdot (\vec{D} \times \vec{B}^{\,e/m} + \vec{B}^{\,e/m} \times \vec{D}) + (\vec{D} \times \vec{B}^{\,e/m} + \vec{B}^{\,e/m} \times \vec{D}) \cdot \vec{D}]\right) \psi \,, \\
\L_\chi &= \chi^\dagger \left(\i D_0 - \frac{1}{2m}\vec{D}^{\,2} - \frac{1}{8 m^3}\vec{D}^{\,4}\right) \chi \\
\nonumber & + g \chi^\dagger \left(-\frac{c_F}{2m} \vec{\sigma} \cdot \vec{B} + \i \frac{c_S}{8m^2} \vec{\sigma} \cdot (\vec{D} \times \vec{E} - \vec{E} \times \vec{D}) + \frac{c_D}{8m^2}(\vec{D} \cdot \vec{E} - \vec{E} \cdot \vec{D})\right) \chi \\
\nonumber & + e e_Q \chi^\dagger \left(-\frac{c_F^{e/m}}{2m} \vec{\sigma} \cdot \vec{B}^{\,e/m} + \i \frac{c_{\text{s}}^{e/m}}{8m^2} \vec{\sigma} \cdot (\vec{D} \times \vec{E}^{\,e/m} - \vec{E}^{\,e/m} \times \vec{D}) + \frac{c_D^{e/m}}{8m^2} (\vec{D} \cdot \vec{E}^{\,e/m} - \vec{E}^{\,e/m} \cdot \vec{D})\right) \chi \\
\nonumber & + e e_Q \chi^\dagger \left(-\frac{c_{W1}^{e/m}}{8m^3} (\vec{D}^{\,2} (\vec{\sigma} \cdot \vec{B}^{\,e/m}) + (\vec{\sigma} \cdot \vec{B}^{\,e/m}) \vec{D}^{\,2}) + \frac{c_{W2}^{e/m}}{4m^3} (\vec{D}^{\,i} (\vec{\sigma} \cdot \vec{B}^{\,e/m}) \vec{D}^{\,i}\right) \chi \\
\nonumber & + e e_Q \chi^\dagger \left(-\frac{c_{p' p}^{e/m}}{8m^3} [(\vec{\sigma} \cdot \vec{D})(\vec{B}^{\,e/m} \cdot \vec{D}) + (\vec{D} \cdot \vec{B}^{\,e/m})(\vec{\sigma} \cdot \vec{D})]\right) \chi \\
\nonumber & + e e_Q \chi^\dagger \left(-\i \frac{c_M^{e/m}}{8m^3} [\vec{D} \cdot (\vec{D} \times \vec{B}^{\,e/m} + \vec{B}^{\,e/m} \times \vec{D}) + (\vec{D} \times \vec{B}^{\,e/m} + \vec{B}^{\,e/m} \times \vec{D}) \cdot \vec{D}]\right) \chi \,,
\end{align}
where $\vec{E}$ and $\vec{B}$ are chromo-electric and chromo-magnetic fields, $\vec{E}^{\,e/m}$ and $\vec{B}^{\,e/m}$ are electric and magnetic fields. The covariant derivatives are given by $\i D_0 = \i\partial_0 - g t^a A_0^a - e e_Q A_0^{e/m}$ and $\i\vec{D} = \i\vec{\nabla} + g t^a \vec{A}^{\,a} + e e_Q \vec{A}^{\,e/m}$ and contain the coupling to electromagnetism; $A_\mu$ and $A_\mu^{e/m}$ are the gluon and photon field, respectively.\\
$\L_\psi$ and $\L_\chi$ are related via charge-conjugation, because C-parity is a symmetry of QCD. Thus they are related via
\begin{equation}
\psi \to \i \sigma^2 \chi^* \,, \quad\quad A_\mu \to - A_\mu^T \quad\quad \text{and} \quad\quad A_\mu^{e/m} \to - A_\mu^{e/m} \,.
\end{equation}
The four fermion Lagrangian $\L_{4f}$ contains operators of dimension 6 or higher, and once coupled to electromagnetism the relevant operators are of dimension 8 at least, see Ref.~\cite{Brambilla:2006ph}, and thus $\L_{4f}$ does not contribute to the decay width in pNRQCD at the relative order $v^2$ (NNLO) we are interested in. Finally the light quark Lagrangian accounts for the combined QED-QCD Lagrangian of light quark fields $q_f$ with flavors $f$ and includes the soft Yang-Mills part of QCD. It is given by
\begin{equation}
\L_{\text{light}} = -\frac{1}{4} G_{\mu\nu}^a G^{\mu\nu\,a} -\frac{1}{4} F_{\mu\nu}^{e/m} F^{\mu\nu\,e/m} + \sum\limits_f \bar{q}_f \i\slashed{D} q_f + \O\left(\frac{1}{m^2}\right) \,,
\end{equation}
where higher order terms in the $\frac{1}{m}$-expansion are irrelevant and $G_{\mu\nu}^a$ and $F_{\mu\nu}^{e/m}$ are the field strength tensors of QCD and QED, respectively. The light degrees of freedom, $q_f$, remain unchanged with respect to QCD and are thus represented by Dirac spinors. Further degrees of freedom are soft and ultra-soft gluons appearing in covariant derivatives $D_\mu$ and field strength tensors $G_{\mu\nu}$.\\
The general NRQCD Lagrangian is a power series in $\frac{1}{m}$ and it is crucial to note that the symmetries are the same as in QCD, albeit Lorentz invariance is no longer explicit but must be enforced via the Wilson coefficients. This has first been shown in the Refs.~\cite{Luke:1992cs,Manohar:1997qy} for the bilinear sector. They are explicitly given, e.g., in Ref.~\cite{Brambilla:2004jw}. We do not consider loops of light quarks explicitly, because we can treat the u-quark, d-quark and s-quark as massless, since their energy is orders of magnitude smaller than the typical momentum. Furthermore, if the emitted photon couples to a loop of massless quarks, the sum over the light flavors gives a vanishing contribution to the matrix element, since the sum of the electric charges of the three light flavors is 0. However, effects due to c-quark loops should be taken into account, but they go beyond our accuracy, since hard loops are suppressed by $\alpha_s^2(m_b) \sim v^4$ (N$^4$LO). See Sec.~\ref{subsec:PowerCounting} for the power counting.

\section{Electromagnetic dipole transitions}
\label{sec:ElectromagneticDipoleTransitions}

\begin{figure}[t]
\centering
{\includegraphics[clip,trim={0.0cm 0.0cm 0.0cm 0.0cm},width=0.7\textwidth]{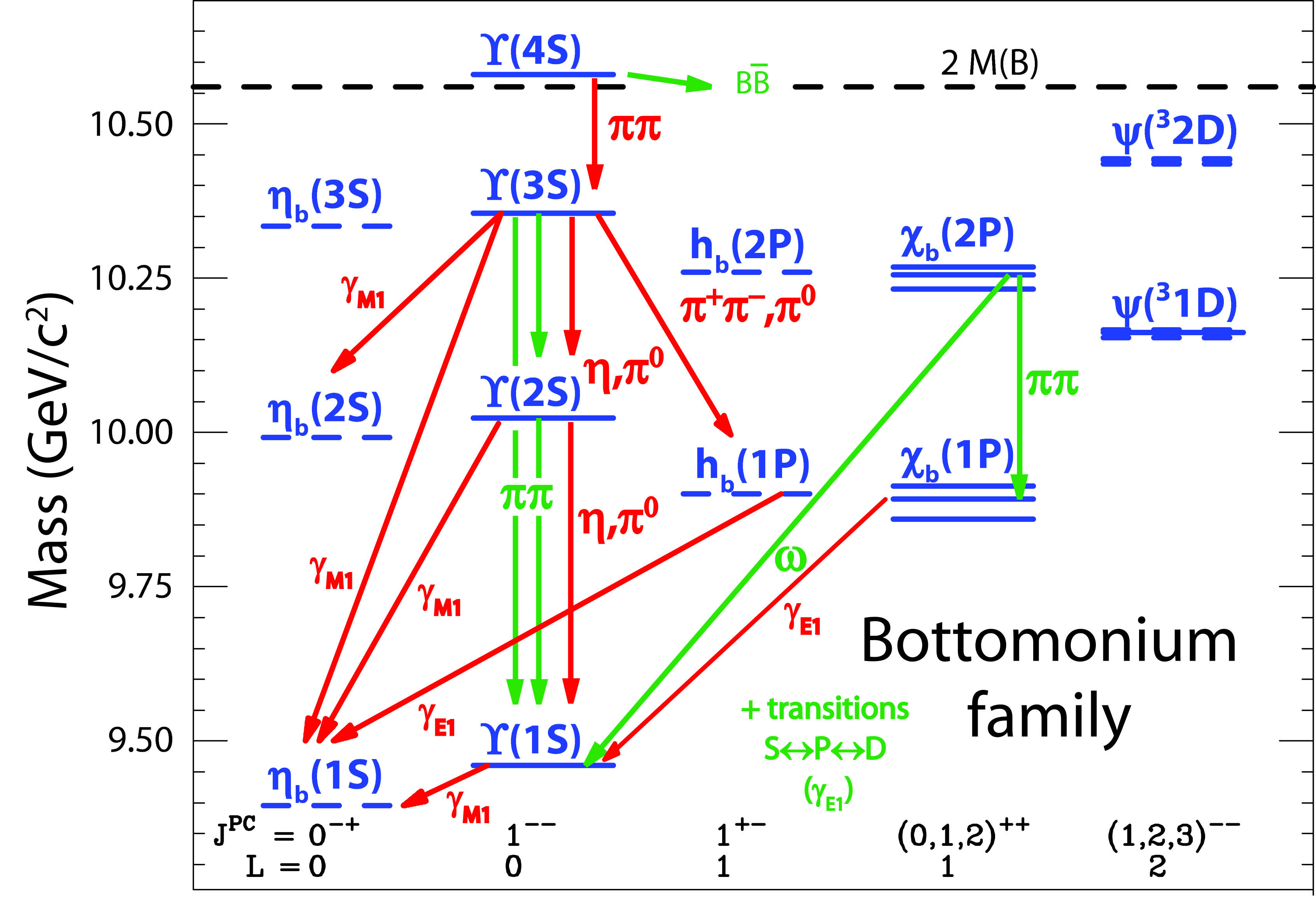}}
\caption[The level scheme of the $b\bar{b}$ states.]{The level scheme of the $b\bar{b}$ states, adopted from \cite{Eichten:2007qx}. The figure shows electric and magnetic dipole transitions and several hadronic transitions. The quantum numbers partially follow the spectroscopic notation, Eq.~\eqref{eq:SpectroscopicNotation}, however in the figure the radial quantum number $n_{r}$ is used instead of the principal quantum number $n$. $J$ denotes the total orbital angular momentum and P and C denote parity and charge parity, respectively. $L=0,1,2$ corresponds to the typical notation $S,P,D$ known from atomic or molecular physics.}
\label{fig:bbSystem}
\end{figure}

Transitions between two different quarkonium states $H(n,\l,s,J)$ and $H'(n',\l',s',J')$, involving the emission or absorption of a photon $\gamma$ are called electromagnetic dipole transitions. These can be subdivided into electric (E1) and magnetic (M1) dipole transitions. Their properties are summarized in Table~\ref{tab:PropertiesE1M1Transitions}.
\begin{table}[t]
\centering
\caption[Properties of E1 and M1 transitions.]{Properties of E1 and M1 transitions, adapted from \cite{Pietrulewicz:2011aca}.}
\label{tab:PropertiesE1M1Transitions}
\begin{tabular}{c|c|c}
property & \quad\quad E1 \, \quad\quad & \quad\quad M1 \, \quad\quad \\
\hline
$|\Delta L|$ & 1 & 0 \\
$|\Delta S|$ & 0 & 1 \\
changes parity & yes & no \\
changes charge parity & yes & yes \\
\end{tabular}
\end{table}
The defining feature of E1 transitions is that they change the orbital angular momentum by one unit. The spin of the states remains unchanged, yielding a change in parity, since $P = (-1)^{\l+1}$, as well as in charge parity, since $C = (-1)^{\l+s}$, with respect to initial and final states. The electric dipole operator $\OEone \propto \vec{r} \cdot \vec{E}$ enters the decay width already at leading order and, since it's expectation value is non-trivial, this allows the usage of E1 transitions to gain insight into the structure of quarkonium states.\\
In contrast thereto, M1 transitions change the spin by one unit and leave the orbital angular momentum unchanged. Thus, parity is conserved and charge parity changes. Furthermore, in contrast to E1 transitions, at leading order the wave functions do not enter the decay width, since the magnetic dipole operator $\mathcal{O}_{\text{M1}} \propto \vec{\sigma} \cdot \vec{B}$ has a trivial expectation value. M1 transitions can be subdivided into allowed ($n = n'$) and hindered ($n \neq n'$) transitions. Allowed transitions are possible at leading order due to quantum mechanical selection rules, while the latter ones can only occur at higher order.\\
The leading order decay widths, see Refs.~\cite{Brambilla:2005zw,Pietrulewicz:2011aca,Brambilla:2012be}, for p-wave ($\l=1$) to s-wave ($\l=0$) transitions in pNRQCD at weak coupling, are given by
\begin{align}
\label{eq:LeadingOrderE1} & \Gamma_{\text{E1}}^{(0)} = \frac{4}{9} \alpha_{e/m} e_Q^2 k_\gamma^3 \left[I_3^{(0)}(n1 \to n'0)\right]^2 \sim \frac{k_\gamma^3}{m^2 \alpha_s^2} \,, \\
\label{eq:LeadingOrderM1} & \Gamma_{\text{M1}}^{(0)} = \frac{4}{3} \alpha_{e/m} e_Q^2 \frac{k_\gamma^3}{m^2} \delta_{n n'} \sim \frac{k_\gamma^3}{m^2} \,,
\end{align}
where $\alpha_{e/m}$ is the electromagnetic fine structure constant, $e_Q$ is the electric charge of the quarks constituting the quarkonium and $k_\gamma$ is the photon energy. For E1 transitions $k_{\gamma}$ is of order of the Energy and counts as $m \alpha_{s}^{2}$, whereas for allowed M1 transitions $k_{\gamma}$ is of the order of hyperfine splitting and counts as $m \alpha_{s}^{4}$ and for hindered M1 transitions $k_{\gamma}$ is of order of the Energy and counts as $m \alpha_{s}^{2}$. It becomes apparent that M1 transitions are suppressed by a factor of $\alpha_s^2$ with respect to E1 transitions, making the latter ones an experimental more feasible observable to measure. The power counting behind this will be explained in detail in Chapter~\ref{chp:pNRQCDQuantumMechanicalPerturbationTheory}, once pNRQCD is established. The dependence of the E1 decay width on the wave function is encoded in the square of the matrix element $I_3^{(0)}$. The generalized form of this matrix element is given by
\begin{equation}
\label{eq:INk} I_N^{(k)}(n\l \to n'\l') = \int\limits_0^\infty \d r \, r^2 r^{N-2} R_{n'\l'}(r) \left( \frac{\d^k}{\d r^k} R_{n\l}(r) \right) \,,
\end{equation}
and will appear in several expressions throughout this work. To illustrate the above, Fig.~\ref{fig:bbSystem} shows the quantum numbers and decay channels of the lowest lying bottomonium states.
The quantum numbers $n$,$\l$, $s$ and $J$ characterize the principle quantum number, the orbital angular momentum, the spin and the total angular momentum of a given quarkonium state using the spectroscopic notation
\begin{equation}
\label{eq:SpectroscopicNotation} n\,^{2s+1}\!\l_J \,,
\end{equation}
where $\l=0,1,2,\dots$ correspond to the typical notation $S,P,D,\dots$ known from atomic or molecular physics. The $c\bar{c}$ and $b\bar{b}$ ground states, $\eta_c$ and $\eta_b$ can thus be identified by $\eta_c(1\,^1\!S_0)$ and $\eta_b(1\,^1\!S_0)$. The corresponding spin-excited states are the famous $J/\psi(1\,^3\!S_1)$ and the $\Upsilon(1\,^3\!S_1)$, respectively. The first orbital angular momentum excited states are the $h_c(1\,^1\!P_1)$ and the $h_b(1\,^1\!P_1)$, respectively. Finally, the respective first spin and orbital angular momentum excited states are the $\chi_{cJ}(1\,^3\!P_0)$ and the $\chi_{bJ}(1\,^3\!P_0)$.\\
The quantum numbers mentioned in Eq.~\eqref{eq:SpectroscopicNotation} originate from the operators $\vec{J} = \vec{L} + \vec{S}$, $\vec{S} = \vec{S}_1 + \vec{S}_2 = \frac{1}{2} (\vec{\sigma}_1 + \vec{\sigma}_2)$ and $\vec{L} = \vec{r} \times \vec{p}$. The square of these operators are physical observables with the following eigenvalues
\begin{align}
\label{eq:ExpectationValuesPhysicalOperators}
\begin{aligned}
\chi_{J^2} &= \langle n\l s J | \vec{J}^{\,2} | n\l s J \rangle = J(J+1) \,, \\
\chi_{L^2} &= \langle n\l s J | \vec{L}^{\,2} | n\l s J \rangle = \l(\l+1) \,, \\
\chi_{S^2} &= \langle n\l s J | \vec{S}^{\,2} | n\l s J \rangle = s(s+1) \,, \\
\chi_{LS} &= \langle n\l s J | \vec{L} \cdot \vec{S} | n\l s J \rangle = \frac{1}{2} [J(J+1) - \l(\l+1) - s(s+1)] \,,
\end{aligned}
\end{align}
where one uses the relation $\vec{J}^{\,2} = \vec{L}^{\,2} + 2 \vec{L} \cdot \vec{S} + \vec{S}^{\,2}$ in order to derive the last expectation value and the term $\vec{L} \cdot \vec{S}$ describes the spin-orbit coupling of the quarkonium under consideration.

\clearpage
\thispagestyle{empty}
\clearpage

\chapter{pNRQCD and quantum mechanical perturbation theory}
\label{chp:pNRQCDQuantumMechanicalPerturbationTheory}

In this chapter we describe the EFT called potential non-relativistic QCD (pNRQCD) that will be the framework for the computations throughout this work. Furthermore, we introduce the concept of quantum mechanical perturbation theory and derive the key equations that will allow us to compute the $b\bar{b}$-mass spectrum and, later on, the E1 decay width.

\section{Potential non-relativistic QCD (pNRQCD)}
\label{sec:pNRQCD}

In the same way as NRQCD arises from QCD by integrating out the hard scale $m$, 
potential non-relativistic QCD (pNRQCD) arises from NRQCD by going one step further and integrating out the soft scale $p \sim mv$. This means integrating out quarks and gluons with energy and momentum of order $m v$ and potential gluons with momentum and energy of order $m v$ and $m v^2$, respectively. This has been demonstrated first by Pineda, Soto, Brambilla and Vairo in Refs.~\cite{Pineda:1997bj,Brambilla:1999xf}. Now the relative size between the scales $m v^2$ and $\Lambda_{\text{QCD}}$ is important and defines the weak ($\Lambda_{\text{QCD}} \le m v^{2}$) and the strong ($\Lambda_{\text{QCD}} \ge m v^{2}$) coupling regimes, respectively. The first one allows for a full perturbative treatment, i.e. the Wilson coefficients can be determined in perturbation theory, and will be used throughout this work. In the latter one the Wilson coefficients have to be determined non-perturbatively.\\
We would like to start this section by pointing out the success of pNRQCD computations, since
\begin{enumerate}
\item The $q\bar{q}$-spectrum at weak coupling has been computed to very high accuracy over the years: NLO ($m\alpha_s^3$) by Billoire (1980), NNLO ($m\alpha_s^4$) by Pineda and Yndurain (1997), NNNLO ($m\alpha_s^5 \, \text{ln} \, \alpha_s$ only) by Brambilla, Pineda, Soto and Vairo (1999), NNLL ($m\alpha_s^{4+n} \, \text{ln}^n \, \alpha_s$) by Pineda (2001), NNNLO ($m\alpha_s^5$, almost complete) by Kniehl, Penin, Smirnov and Steinhauser (2002) and Beneke, Kiyo and Schuller (2005) and finally NNNLO ($m\alpha_s^5$, complete) by Smirnov, Smirnov and Steinhauser (2009). The spectrum at NNNLO ($m\alpha_s^5$) for unequal masses has been computed by Peset, Pineda and Stahlhofen (2016) in \cite{Peset:2015vvi}.
\item The computation of magnetic dipole transitions in heavy quarkonia has been established and performed successfully by Brambilla, Jia, Vairo, Pineda and Segovia in Refs.~\cite{Brambilla:2005zw,Pineda:2013lta} and this work aims to give a contribution to the success of pNRQCD by providing the first numerical determination of the electric dipole transitions, analogously.
\end{enumerate}

\subsection{The Lagrangian}

The weak coupling regime \cite{Brambilla:2004jw,Pineda:2011dg} is determined by the condition $\Lambda_{\text{QCD}} \leq m v^2$ ($p \gg E \gtrsim \Lambda_{\text{QCD}}$). Then $v \sim \alpha_{s}$ and the degrees of freedom are quark-antiquark pairs, ultra-soft gluons and light quarks. The quark-antiquark pair can be cast either as separate fields $\psi$ and $\chi$ representing the quark and the antiquark, respectively; or as a single field $\Psi$ for the pair of them. The first choice allows for a smooth connection with NRQCD, while the latter one allows for the decomposition of the quark-antiquark pair into color singlet, $S$, and color octet fields, $O$. The pNRQCD Lagrangian as given by \cite{Brambilla:1999xf,Brambilla:2004jw,Pineda:2011dg}, where
\begin{equation}
\Psi(\vec{x}_1,\vec{x}_2)_{\alpha\beta} \sim \psi_\alpha(\vec{x}_1) \chi_\beta^\dagger(\vec{x}_2) \,,
\end{equation}
describes aforesaid wave function field consisting of a quark antiquark pair, takes the form
\begin{align}
\L_{\text{pNRQCD}} &= \int \d^3 x_1 \d^3 x_2 \, \text{tr}\left\lbrace \Psi^\dagger(\vec{x}_1,\vec{x}_2) \left( \i D_0 + \frac{\vec{D}_{x_1}^{\,2}}{2m_1} + \frac{\vec{D}_{x_2}^{\,2}}{2m_2} + \dots \right) \Psi(\vec{x}_1,\vec{x}_2) \right\rbrace \\
\nonumber & - \int \d^3 x \, \frac{1}{4} G_{\mu\nu}^a(x) G^{\mu\nu\,a}(x) + \int \d^3 x \, \sum\limits_f \bar{q}_f(x) \i\slashed{D} q_f(x) + \dots \\
\nonumber & + \int \d^3 x_1 \d^3 x_2 \, \text{tr}\left\lbrace \Psi^\dagger(\vec{x}_1,\vec{x}_2) V(\vec{r},\vec{p}_1,\vec{p}_2,\vec{S}_1,\vec{S}_2) \times (\text{US gluon fields}) \Psi(\vec{x}_1,\vec{x}_2) \right\rbrace \,,
\end{align}
where $\i D_0 \Psi(\vec{x}_1,\vec{x}_2) = \i\partial_0 \Psi(\vec{x}_1,\vec{x}_2) - g A_0(\vec{x}_1) \Psi(\vec{x}_1,\vec{x}_2) + g \Psi(\vec{x}_1,\vec{x}_2) A_0(\vec{x}_2)$, $\vec{r} = \vec{x}_1 - \vec{x}_2$, $\vec{p} = -\i\vec{\nabla}_r$, $\vec{S}_i = \frac{\vec{\sigma_i}}{2}$ and the dots stand for higher order terms in the $\frac{1}{m}$-expansion. One can enforce the gluons to be ultra-soft by multipole expanding them in $\vec{r}$. This spoils the manifest gauge invariance, which may be restored by introducing aforementioned color singlet $S$ and color octet fields $O$. They have the following normalizations and transformation properties with respect to homogeneous gauge transformations $g(\vec{R},t)$, with respect to the center of mass coordinate $\vec{R}$,
\begin{align}
& S = S \frac{\1_{\text{c}}}{\sqrt{N_{\text{c}}}} \,, && O = O^a \frac{t^a}{\sqrt{T_F}} \,, \\
& S \mapsto S \,, && O \mapsto g(\vec{R},t) O g^{-1}(\vec{R},t) \,.
\end{align}
This explicitly establish gauge invariance at the level of the Lagrangian and allows for a multipole expansion in the relative coordinate $r$, since it is explicit and much smaller than the typical length of the light degrees of freedom. As in NRQCD the discrete symmetries C, P and T remain unbroken but Poincaré symmetry is realized in an non-linear manner. We briefly discuss pNRQCD symmetries below in Sec.~\ref{subsec:Symmetries}.\\
After multipole expanding one may organize the pNRQCD Lagrangian as an expansion in $\frac{1}{m}$ and $r$ \cite{Caswell:1985ui,Brambilla:1999xf,Pineda:2011dg}
\begin{align}
\label{eq:pNRQCDLagrangian} \L_{\text{pNRQCD}} &= \int \d^3 r \, \text{tr}\left\lbrace S^\dagger (\i\partial_0 - H_{\text{s}}(\vec{r},\vec{p},\vec{P}_r,\vec{S}_1,\vec{S}_2)) S + O^\dagger (\i D_0 - H_{\text{o}}(\vec{r},\vec{p},\vec{P}_r,\vec{S}_1,\vec{S}_2)) O \right\rbrace \\
\nonumber & + V_A(r) \, \text{tr}\lbrace O^\dagger g \vec{r} \cdot \vec{E} S + S^\dagger g \vec{r} \cdot \vec{E} O \rbrace + \frac{V_B(r)}{2} \, \text{tr}\lbrace O^\dagger g \vec{r} \cdot \vec{E} O + O^\dagger O g \vec{r} \cdot \vec{E} \rbrace \\
\nonumber & - \frac{1}{4} G_{\mu\nu}^a G^{\mu\nu\,a} + \sum\limits_f \bar{q}_f \i\slashed{D} q_f \,,
\end{align}
where
\begin{align}
& H_{\text{s}}(\vec{r},\vec{p},\vec{P}_r,\vec{S}_1,\vec{S}_2) = \frac{1}{2} \left\lbrace c_{\text{s}}^{(1,-2)}(r) , \frac{\vec{p}^{\,2}}{2\mr} \right\rbrace + c_{\text{s}}^{(1,0)}(r) \frac{\vec{p}_R^{\,2}}{2(m_1 + m_2)} + \frac{\vec{p}^{\,4}}{32 \mr} +  V_{\text{s}}(\vec{r},\vec{p},\vec{P}_r,\vec{S}_1,\vec{S}_2) \,, \\
& H_{\text{o}}(\vec{r},\vec{p},\vec{P}_r,\vec{S}_1,\vec{S}_2) = \frac{1}{2} \left\lbrace c_{\text{o}}^{(1,-2)} (r) , \frac{\vec{p}^{\,2}}{2\mr} \right\rbrace + c_{\text{o}}^{(1,0)}(r) \frac{\vec{p}_R^{\,2}}{2(m_1 + m_2)} + \frac{\vec{p}^{\,4}}{32 \mr} + V_{\text{o}}(\vec{r},\vec{p},\vec{P}_r,\vec{S}_1,\vec{S}_2) \,, \\
& V_{\text{s}} = V^{(0)}_{\text{s}} + \frac{V^{(1,0)}_s}{m_1} + \frac{V^{(0,1)}_s}{m_2} + \frac{V^{(2,0)}_s}{m_1^2} + \frac{V^{(0,2)}_s}{m_2^2} + \frac{V^{(1,1)}_s}{m_1m_2} \,, \\
& V_{\text{o}} = V^{(0)}_{\text{o}} + \frac{V^{(1,0)}_{\text{o}}}{m_1} + \frac{V^{(0,1)}_{\text{o}}}{m_2} + \frac{V^{(2,0)}_{\text{o}}}{m_1^2} + \frac{V^{(0,2)}_{\text{o}}}{m_2^2} + \frac{V^{(1,1)}_{\text{o}}}{m_1m_2} \,,
\end{align}
where $\i D_0 O = \i \partial_0 O - g (A_0(\vec{R}) O + O A_0(\vec{R})$, $\vec{P}_R = -\i \vec{D}_R$, $\vec{p} = -\i \vec{\nabla}_r$ and $\mr = \frac{m_1 m_2}{m_1 + m_2}$. The expansion can be performed either for equal masses or for unequal masses. We are interested in the equal mass case, $m_1 = m_2 = m$. We drop the labels s and o for the singlet and octet and organize the potentials in powers of $\frac{1}{m}$ \cite{Brambilla:2012be}, where we distinguish between spin independent (SI) and spin dependent (SD) contributions. Furthermore, we concentrate on the singlet case. The scale dependent potentials can be calculated in perturbation theory and the terms relevant for our further computations are given by
\begin{align}
\label{eq:GeneralPotential} & V(r) = V^{(0)}(r) + \frac{V^{(1)}(r)}{m} + \frac{V^{(2)}}{m^2} + \dots \,, \\
& V^{(2)} = V^{(2)}_{\text{SI}} + V^{(2)}_{\text{SD}} \,, \\
& V_{\text{SI}}^{(2)} = V_r^{(2)} + \frac{1}{2} \lbrace V_{p^2}^{(2)},-\nabla_r^2 \rbrace + V_{L^2}^{(2)} \vec{L}^{\,2} \,, \\
& V_{\text{SD}}^{(2)} = V_{LS}^{(2)} \vec{L} \cdot \vec{S} + V_{S^2}^{(2)} \vec{S}^{\,2} + V_{S_{12}}^{(2)} S_{12} \,,
\end{align}
where $\vec{S} = \vec{S}_1 + \vec{S}_2$, $\vec{L} = \vec{r} \times \vec{p}$ and $S_{12}(\hat{r}) = 3 \hat{r} \cdot \vec{\sigma}_1 \, \hat{r} \cdot \vec{\sigma}_2 - \vec{\sigma}_1 \cdot \vec{\sigma}_2$.\\
The functions $V_s$, $V_{\text{o}}$, $c_{\text{s}}^{(1,-2)}$, $c_{\text{o}}^{(1,-2)}$, $c_{\text{s}}^{(1,0)}$, $c_{\text{o}}^{(1,0)}$, $V_A$ and $V_B$ are the matching coefficients of the effective theory. At leading order one has $V_A = V_B = 1$, $c_{\text{s}}^{(1,-2)} = c_{\text{o}}^{(1,-2)} = c_{\text{s}}^{(1,0)} = c_{\text{o}}^{(1,0)} =1$, and $V_{\text{s}}^{(0)} = -C_F \frac{\alpha_s}{r}$ and $V_{\text{o}}^{(0)} = \frac{1}{2 N_{\text{c}}} \frac{\alpha_s}{r}$.\footnote{Note that in contrast to the leading order singlet static potential, the leading order octet static potential is not attractive but repulsive.}\\
The involved potentials explicitly read
\begin{align}
\label{eq:ExplicitPotentials}
\begin{aligned}
& V^{(1)} = -\frac{C_F C_A \alpha_s^2}{2 r^2} \,, \quad && V_r^{(2)} = \pi C_F \alpha_s \delta^{(3)}(\vec{r}\,) \,, \\
& V_{p^2}^{(2)} = -\frac{C_F \alpha_s}{r} \,, \quad && V_{L^2}^{(2)} = \frac{C_F \alpha_s}{2 r^3} \,, \\
& V_{LS}^{(2)} = \frac{3 C_F \alpha_s}{2 r^3} \,, \quad && V_{S^2}^{(2)} = \frac{4\pi C_F \alpha_s}{3} \delta^{(3)}(\vec{r}\,) \,, \\
& V_{S_{12}}^{(2)} = \frac{C_F \alpha_s}{4 r^3} \,. &&
\end{aligned}
\end{align}
Furthermore, at $\O(\frac{1}{m^2})$ there are also three operators that act on the center of mass of the system \cite{Brambilla:2004jw}. They are given by
\begin{align}
& V_{\text{SI}}^{(2)} \ni \frac{1}{8} \left\lbrace \vec{P}^{\,2} , V^{(2)}_{\vec{p}^{\,2} \,, \text{CM}} \right\rbrace \sim m^3 \alpha_s^6 \,, \\
& V_{\text{SI}}^{(2)} \ni \frac{(\vec{r} \times \vec{P})}{4 r^2} V^{(2)}_{\vec{L}^{\,2} \,, \text{CM}} \sim m^3 \alpha_s^6 \,, \\
& V_{\text{SD}}^{(2)} \ni \frac{(\vec{r} \times \vec{P}) \cdot (\vec{S}_1 - \vec{S}_2)}{2} V^{(2)}_{LS \,, \text{CM}} \sim m^3 \alpha_s^5 \,,
\end{align}
but are suppressed by additional factors of $\alpha_s$ and are thus beyond our accuracy of NNLO. In the case of M1 transitions the situation is different. Because the leading order E1 operator is enhanced by $\frac{1}{\alpha_{s}}$ with respect to the leading order M1 operator, spin dependent corrections due to $V_{\text{SD}}^{(2)}$ enter at NLO in hindered M1 transitions. This is relevant for the M1 transition $n\,^{3}\!S_{1} \to n'\,^{1}\!S_{0} + \gamma$, where the p-wave spin-triplet final state component can be reached from the initial $^{3}\!S_{1}$ state through an E1 transition. Nevertheless, as an interesting fact the involved potentials are not independent from one another, but linked by Poincaré invariance (see the following Sec.~\ref{subsec:Symmetries}).\\
The logarithmic corrections to the static potential, induced by hard and soft gluons, counting as $m$ and $m v$, respectively, read
\begin{equation}
\label{eq:SingletStaticPotential} V_{\text{s}}^{(0)} = -C_F \frac{\alpha_s}{r} \left[ 1 + \sum\limits_{k=1}^{\infty} \left(\frac{\alpha_s}{4\pi}\right)^k a_k(r) \right] = -C_F \frac{\alpha_s}{r} + \delta V_{\text{s}}^{(0)} \,.
\end{equation}
The coefficients $a_k(r)$ are known up to $k=3$, \cite{Brambilla:2004jw,Pineda:2011dg,Peset:2015vvi}, and they are, up to $k=2$, given by
\begin{align}
\label{eq:a1} a_1(\nu,r) &= a_1 + 2\beta_0 \ln{\nu \e{\gamma_E} r} \,, \\
\label{eq:a2} a_2(\nu,r) &= a_2 + \frac{\pi^2}{3} \beta_0^2 + (4a_1 \beta_0 + 2\beta_1) \ln{\nu \e{\gamma_E} r} + 4\beta_0^2 \lnSquare{\nu \e{\gamma_E} r} \,,
\end{align}
where the constant $a_1$ was computed in Ref.~\cite{Fischler:1977yf} and $a_2$ 
was computed in Refs.~\cite{Peter:1996ig,Schroder:1998vy}. They are both given 
in Appendix~\ref{app:sec:Constants}.\\
A convenient way in order to distinguish between contributions that are treated exactly and contributions that are treated perturbatively is to split the Hamiltonian as follows:
\begin{align}
& H_{\text{s}} = \frac{\vec{p}^{\,2}}{2 m_r} + V_{\text{s}} \equiv H_{\text{s}}^C + \delta H_{\text{s}} \,, \\
& H_{\text{s}}^C = \frac{\vec{p}^{\,2}}{2 m_r} + V_{\text{s}}^C \,, \\
\label{eq:LeadingOrderStaticPotential} & V_{\text{s}}^C = -C_F \frac{\alpha_s}{r} \,, \\
\label{eq:DeltaHs} & \delta H_{\text{s}} = -\frac{\nabla_r^4}{4 m^3} + \delta V_{\text{s}} \,,
\end{align}
where $H_{\text{s}}^C$ is the leading order Hamiltonian with a Coulombic potential $V_{\text{s}}^C$ that can and will be treated exactly, and $\delta H_s$, c.f. Ref.~\cite{Brambilla:2012be}, encodes next to leading order at least. These feature quartic corrections to the kinetic energy and higher order corrections that are either due to radiative corrections, $\delta V_{\text{s}}^{(0)}$, in the static potential $V_{\text{s}}^{(0)}$ (hard and soft gluons), or due to relativistic corrections ($\frac{1}{m}$-expansion) due to the transition QCD $\to$ NRQCD. Thus
\begin{equation}
\delta V_{\text{s}} = \delta V_{\text{s}}^{(0)} + \frac{V^{(1)}}{m} + \frac{V_{\text{SI}}^{(2)}}{m^2} + \frac{V_{\text{SD}}^{(2)}}{m^2} \,,
\end{equation}
with the individual contributions as above. Furthermore, corrections due to higher order Fock states may be of relevance. These will be discussed once we consider non-perturbative corrections due to color octet effects in Sec.~\ref{sec:NonPerturbativeContributions}.\\
So far, in contrast to the section of NRQCD where we directly incorporated the electromagnetic interaction via the covariant derivatives, we did not couple the photon field to pNRQCD. This can be done by adding the suitable Lagrangian containing all the terms that do cause a change in parity, c.f. Ref.~\cite{Pietrulewicz:2011aca}, which is necessary for E1 transitions
\begin{align}
\label{eq:GammapNRQCDLagrangian} \L_{\gamma\text{pNRQCD}} &= e e_Q \int \d^3 r \, \text{tr}\Big\lbrace V^{r \cdot E} S^\dagger \vec{r} \cdot \vec{E}^{\,e/m} S + V_{\text{o}}^{r \cdot E} O^\dagger \vec{r} \cdot \vec{E}^{\,e/m} O \\
\nonumber & + \frac{1}{24} V^{(r \nabla)^2 r \cdot E} S^\dagger \vec{r} \cdot \left[(\vec{r} \, \vec{\nabla})^2 \vec{E}^{\,e/m} \right] S \\
\nonumber & + \i \frac{1}{4m} V^{\nabla \cdot (r \times B)} S^\dagger \left[ \vec{\nabla} \cdot (\vec{r} \times \vec{B}^{\,e/m}) + (\vec{r} \times \vec{B}^{\,e/m}) \cdot \vec{\nabla} \right] S \\
\nonumber & + \i \frac{1}{12m} V^{(r \nabla) \nabla_r \cdot (r \times B)} S^\dagger \left\lbrace \left[ \vec{\nabla}_r \cdot \left(\vec{r} \times ((\vec{r} \, \vec{\nabla}) \vec{B}^{\,e/m})\right) \right] + \left[ \left(\vec{r} \times ((\vec{r} \, \vec{\nabla}) \vec{B}^{\,e/m})\right) \cdot \vec{\nabla}_r \right] \right\rbrace S \\
\nonumber & + \frac{1}{4m} V^{(r \nabla) \sigma \cdot B} \left( S^\dagger \vec{\sigma} - \vec{\sigma} S^\dagger \right) \cdot \left[ (\vec{r} \, \vec{\nabla}) \vec{B}^{\,e/m} \right] S \\
\nonumber & + \frac{1}{m r} V^{r \cdot E/r} S^\dagger \vec{r} \cdot \vec{E}^{\,e/m} S \\
\nonumber & - \i \frac{1}{4m^2} V^{\sigma \cdot (E \times \nabla_r)} \left( S^\dagger \vec{\sigma} - \vec{\sigma} S^\dagger \right) \cdot \left( \vec{E}^{\,e/m} \times \vec{\nabla}_r \right) S \Big\rbrace \,.
\end{align}
Matching at tree level yields $V^{r \cdot E} = V_{\text{o}}^{r \cdot E} = V^{(r \nabla)^2 r \cdot E} = 1$, $V^{\nabla \cdot (r \times B)} = V^{(r \nabla) \nabla_r \cdot (r \times B)} = 1$, $V^{(r \nabla) \sigma \cdot B} = c_F^{e/m}$, $V^{r \cdot E/r} = 0$ and $V^{\sigma \cdot (E \times \nabla_r)} = c_{\text{s}}^{e/m}$.\\
Beyond tree level, at $\O(\frac{1}{m})$ one finds $V^{(r \nabla) \sigma \cdot B} = c_F^{e/m}$, meaning that there are no soft contributions to the anomalous magnetic moment of the quarkonium. Furthermore, $V^{\nabla \cdot (r \times B)} = V^{(r \nabla) \nabla_r \cdot (r \times B)} = 1$ and $V^{r \cdot E/r} = 0$.\\
At $\O(\frac{1}{m^2})$ one finds $V^{\sigma \cdot (E \times \nabla_r)} = c_{\text{s}}^{e/m}$ and $\O(\frac{1}{m^3})$ only contributes to M1 transitions. The two non-trivial Wilson coefficients are constrained by Poincaré invariance \cite{Brambilla:2003nt} and are related via
\begin{equation}
2 c_F^{e/m} - c_{\text{s}}^{e/m} - 1 = 0 \,.
\end{equation}
Furthermore, they are related to the anomalous magnetic moment
\begin{equation}
\label{eq:AnomalousMagneticMoment} \kappa_Q^{e/m} = c_F^{e/m} - 1 = \frac{c_{\text{s}}^{e/m} - 1}{2} = C_F \frac{\alpha_s}{2\pi} + \O(\alpha_s^2) \,,
\end{equation}
which thus exceeds our accuracy goal, because it comes with an additional factor of $\alpha_s$ at least.\\
Finally, for completeness, we mention the strong coupling regime that is determined by the condition $\Lambda_{\text{QCD}} \leq m v$ ($p \gg \Lambda_{\text{QCD}} \gg E$ or $p \gtrsim \Lambda_{\text{QCD}} \gg E$). This corresponds to the Lagrangian
\begin{equation}
\L_{\text{pNRQCD}}^{\text{strong coupling}} = \int \d^3 x_1 \d^3 x_2 \, S^\dagger (\i\partial_0 - H_s) S \,,
\end{equation}
where
\begin{equation}
H_{\text{s}} = \frac{\vec{p}_1^{\,2}}{2m} + \frac{\vec{p}_2^{\,2}}{2m} + V_{\text{s}} \,,
\end{equation}
and $V_{\text{s}} = V_{\text{s}}^{(0)} + \frac{V_{\text{s}}^{(1)}}{m} + \frac{V_{\text{s}}^{(2)}}{m^2} + \dots$ is a series in the inverse heavy quark mass $m$. Each $V_s$ has to be determined non-perturbatively, since now also the hadronic scale has to be integrated out, but the dynamics of the system in the non-perturbative regime still reduces to a quantum mechanical problem, c.f. Ref.~\cite{Brambilla:2004jw}. The fact that these potentials should be calculated non-perturbatively requires, for instance, lattice QCD. Finally, the degrees of freedom in pNRQCD at strong coupling are color singlets and pseudo Goldstone bosons.

\subsection{Symmetries}
\label{subsec:Symmetries}

We give a short overview on the symmetries of pNRQCD based on Ref.~\cite{Brambilla:2004jw}. The pNRQCD Lagrangian has to fulfill the same symmetries as the QCD one, i.e. charge conjugation, time reversal, parity and Poincaré invariance. See, e.g., Ref.~\cite{Peskin:IntroductionQuantumFieldTheory} on how to derive the exact transformation properties.

\subsubsection{Charge conjugation}

With the relative distance $r$ being explicit in pNRQCD, the QCD charge conjugation translates into pure C-parity supplemented by the exchange of the position of quark and antiquark, i.e. $\vec{r} \to -\vec{r}$. From the transformation properties of Pauli spinors in pNRQCD, which are in turn derived from the Dirac spinors in QCD, it follows that
\begin{align}
\begin{aligned}
& S(\vec{r},\vec{R},t) \to \sigma^{2} S(-\vec{r},\vec{R},t)^T \sigma^{2} \,, \quad\quad && O(\vec{r},\vec{R},t) \to \sigma^{2} O(-\vec{r},\vec{R},t)^T \sigma^{2} \,, \\
& A_{\mu}(\vec{R},t) \to -A_{\mu}(\vec{R},t)^{T} \,, && A_{\mu}^{e/m}(\vec{R},t) \to -A_{\mu}^{e/m}(\vec{R},t) \,, \\
& \vec{E}^{\,e/m}(\vec{R},t) \to -\vec{E}^{\,e/m}(\vec{R},t) \,, && \vec{B}^{\,e/m}(\vec{R},t) \to -\vec{B}^{\,e/m}(\vec{R},t) \,,
\end{aligned}
\end{align}
such that, e.g., the spin dependent operators may only appear in terms of a commutator $[S^\dagger,\vec{\sigma}]$ or anticommutator $\lbrace S^\dagger,\vec{\sigma} \rbrace$.

\subsubsection{Parity}

The parity transformations in pNRQCD can be read almost immediately from the ones in QCD. They are given by
\begin{align}
\begin{aligned}
& S(\vec{r},\vec{R},t) \to -S(-\vec{r},-\vec{R},t) \,, && O(\vec{r},\vec{R},t) \to -O(-\vec{r},-\vec{R},t) \,, \\
& A_{\mu}(\vec{R},t) \to A^{\mu}(-\vec{R},t)^T \,, && A_{\mu}^{e/m}(\vec{R},t) \to A^{\mu\,,e/m}(-\vec{R},t) \,, \\
& \vec{E}^{\,e/m}(\vec{R},t) \to -\vec{E}^{\,e/m}(-\vec{R},t) \,, \quad\quad && \vec{B}^{\,e/m}(\vec{R},t) \to +\vec{B}^{\,e/m}(-\vec{R},t) \,,
\end{aligned}
\end{align}
such that operators like
\begin{equation}
e e_{Q} \frac{1}{m} S^{\dagger}(\vec{r},\vec{R},t) \, \vec{r} \cdot \vec{B}^{\,e/m}(\vec{R},t) \, S(\vec{r},\vec{R},t)
\end{equation}
cannot appear in the Lagrangian. In other words, $\vec{r}$ is a vector, while $\vec{B}^{\,e/m}(\vec{R},t)$ is an axial-vector and thus they do not transform in the same manner under parity and we need a scalar transformation for the Lagrangian to be invariant.

\subsubsection{Time reversal}

The pNRQCD time reversal transformations read
\begin{align}
\begin{aligned}
& S(\vec{r},\vec{R},t) \to \sigma^{2} S(\vec{r},\vec{R},-t) \sigma^{2} \,, \quad\quad && O(\vec{r},\vec{R},t) \to \sigma^{2} O(\vec{r},\vec{R},-t) \sigma^{2} \,, \\
& A_{\mu}(\vec{R},t) \to A^{\mu}(\vec{R},-t) \,, && A_{\mu}^{e/m}(\vec{R},t) \to A^{\mu\,,e/m}(\vec{R},-t) \,, \\
& \vec{E}^{\,e/m}(\vec{R},t) \to +\vec{E}^{\,e/m}(\vec{R},-t) \,, && \vec{B}^{\,e/m}(\vec{R},t) \to -\vec{B}^{\,e/m}(\vec{R},-t) \,,
\end{aligned}
\end{align}
such that operators like
\begin{equation}
e e_{Q} \frac{\i r}{m} S^{\dagger}(\vec{r},\vec{R},t) \, \vec{E}^{\,e/m}(\vec{R},t) \cdot \vec{\nabla}_{r} \, S(\vec{r},\vec{R},t)
\end{equation}
do not appear in the Lagrangian.

\subsubsection{Poincaré invariance}

Poincaré invariance is not explicitly fulfilled for the operators in the pNRQCD Lagrangian, therefore one has to impose constrains on the matching coefficients \cite{Brambilla:2003nt}. Explicitly, this fixes the kinetic terms and the coefficients for some potentials, for instance, for the center of mass spin-orbit potential, the center of mass orbital angular momentum potential and the center of mass kinetic energy, we have the following relations:
\begin{equation}
V_{LS \,, \text{CM}}^{(2)} = -\frac{1}{2r} \frac{\d V^{(0)}}{\d r} \,, \quad V_{\vec{L}^{\,2} \,, \text{CM}} + \frac{r}{2} \frac{\d V^{(0)}}{\d r} = 0 \,, \quad V_{\vec{p}^{\,2} \,, \text{CM}} + V_{\vec{L}^{\,2} \,, \text{CM}} + \frac{V^{(0)}}{2} = 0 \,.
\end{equation}

\subsection{Power counting}
\label{subsec:PowerCounting}

The corrections to the leading Hamiltonian due to \eqref{eq:DeltaHs} contribute to the spectrum and to the decay width. They enter as corrections to the initial and final state wave functions and, in order to obtain consistent results, it is crucial to establish a power counting scheme. The standard one is given, e.g., in Refs.~\cite{Brambilla:1999xf,Brambilla:2004jw}. In the weak coupling regime, $\Lambda_{\text{QCD}} \leq m v^2$, there is not relevant physical scale between $m v$ and $m v^2$ and the pNRQCD Lagrangian~\eqref{eq:pNRQCDLagrangian} only describes ultra-soft degrees of freedom. In this regime we have
\begin{equation}
\alpha_s(m) \ll 1 \,, \quad\quad \alpha_s(p = m v) < 1 \quad\quad \text{and} \quad\quad \alpha_s(E = m v^2) \sim 1 \,.
\end{equation}
The resulting power counting is
\begin{align}
\begin{aligned}
& p \sim \frac{1}{r} \sim m \alpha_s \,, && E \sim \frac{1}{R} \sim m \alpha_s^2 \,, && \nabla_r \sim m \alpha_s \,, \\
& \nabla \equiv \nabla_R \sim m \alpha_s^2 \,, \quad && S \sim m^3 \alpha_s^{9/2} \,, && O \sim m^3 \alpha_s^{9/2} \,, \\
& E,B \sim (m \alpha_s^2)^2 \,, \quad\quad && E^{e/m},B^{e/m} \sim k_\gamma^2 \,, \quad && k_\gamma \sim m \alpha_s^2
\end{aligned}
\end{align}
where $R$ is the center of mass coordinate, $E$ and $B$ denote chromo-electric and chromo-magnetic fields, respectively, $E^{e/m}$ and $B^{e/m}$ denote electric and magnetic fields, respectively, and $k_\gamma$ is the photon energy. The inverse center of mass coordinate $\frac{1}{R}$, and thus the energy $E \sim \frac{1}{R}$, scale like $m \alpha_s^2$ since all gluonic degrees of freedom, that could change the system, have been integrated out except for the ultra-soft ones whose momentum scales like $m \alpha_s^2$. Note that the general expansion is performed in $v$, but, because we are dealing with a Coulombic problem, we have $v \sim \alpha_s$. The resulting power counting for the potentials is then
\begin{align}
H_{\text{s}}^C \sim m\alpha_s^2 \quad \text{(LO)} \,, \quad\quad V_{s,a_1}^{(0)} \sim m\alpha_s^3 \quad \text{(NLO)} \quad\quad \text{and} \quad\quad V_{s,a_2}^{(0)} \,, \delta H_{\text{s}} \sim m\alpha_s^4 \quad \text{(NNLO)} \,.
\end{align}
This power counting has an impact on observables like, e.g., the spectrum or the decay width. Let us therefore consider the following generic potentials with their respective power counting
\begin{equation}
V_\text{LO} \,, \quad\quad V_\text{NLO} \quad\quad \text{and} \quad\quad V_\text{NNLO} \,.
\end{equation}
The respective matrix elements accordingly count as follows (the mathematical details behind this will be derived below in Sec.~\ref{sec:AnalyticSolutionsQuantumMechanicalPerturbationTheory})
\begin{equation}
\M_\text{LO} \sim \,^{(0)}\!\langle | V_\text{LO} | \rangle^{(0)} \,, \quad \M_\text{NLO} \sim \,^{(0)}\!\langle | V_\text{NLO} | \rangle^{(0)} \,, \quad \dots \,,
\end{equation}
where $\,^{(0)}\!\langle |$ and $| \rangle^{(0)}$ represent arbitrary zeroth order initial and final states, respectively. Quantum mechanical perturbation theory, that we introduce and discuss below in Sec.~\ref{sec:AnalyticSolutionsQuantumMechanicalPerturbationTheory}, allows for corrections to these states, denoted by $\,^{(1)}\!\langle |$ and $| \rangle^{(1)}$, due to $\delta H_{\text{s}}$. However, because our desired precision goal is NNLO, only the following matrix elements can contribute to the spectrum:
\begin{align}
\begin{aligned}
& \M_\text{LO} \sim \,^{(0)}\!\langle | V_\text{LO} | \rangle^{(0)} \,, \quad\quad \M_\text{NLO} \sim \,^{(0)}\!\langle | V_\text{NLO} | \rangle^{(0)} \,, \\
& \M_\text{NNLO} \sim \,^{(0)}\!\langle | V_\text{NNLO} | \rangle^{(0)} \,, \,^{(0)}\!\langle | V_\text{NLO} | \rangle^{(1)} \,, \,^{(1)}\!\langle | V_\text{NLO} | \rangle^{(0)} \,.
\end{aligned}
\end{align}
Note that at NNLO, the two corrections due to the NLO potential come with first order corrected states. These corrections can only be induced by the NLO potential, since anything else would exceed our desired precision.\\
The matrix elements entering the decay width are proportional to the expectation value of $r$ (the electric dipole operator is proportional to $\vec{r} \cdot \vec{E}$) and thus are at their respective orders we have
\begin{align}
\begin{aligned}
& \M_\text{LO} \sim \,^{(0)}\!\langle | r | \rangle^{(0)} \,, \quad\quad \M_\text{NLO} \sim \,^{(0)}\!\langle | r | \rangle^{(1)} \,, \,^{(1)}\!\langle | r | \rangle^{(0)} \,, \\
& \M_\text{NNLO} \sim \,^{(0)}\!\langle | r | \rangle^{(1)} \,, \,^{(1)}\!\langle | r | \rangle^{(0)} \,, \,^{(0)}\!\langle | r | \rangle^{(2)} \,, \,^{(2)}\!\langle | r | \rangle^{(0)} \,, \,^{(1)}\!\langle | r | \rangle^{(1)} \,.
\end{aligned}
\end{align}
Here, the first order corrections to the states for the NLO matrix elements are mediated by the NLO potential and the first order corrections to the states for the NNLO matrix elements are mediated by the NNLO potentials. However, the three additional matrix elements including a second order correction to the states or two first order corrections to each state, respectively, can only be mediated by the NLO potential, since anything else would, again, exceed our desired goal.\\
Having this result allows us to discuss its implications on the actual decay width $\Gamma \propto |\M|^2$, that, in contrast to the spectrum, does not depend linearly on the discussed matrix elements, but quadratically (the derivation of this fact is given in Chapter~\ref{chp:E1Transitions}). Let us therefore assume that we have the following set of matrix elements contributing to the total decay width, including their relative power counting:
\begin{align}
\M_{LO} \sim 1 \,, \quad\quad \M_{NLO} \sim \alpha_s \,, \quad\quad \M_{NNLO} \sim \alpha_s^2 \,, \quad\quad \M_{\text{rel.}} \sim \alpha_s^2 \,, \quad\quad \M_{\text{non-pert.}} \sim \alpha_s^2 \,,
\end{align}
where the NLO and NNLO matrix elements account for corrections to the initial and final state wave functions and the relativistic and non-perturbative matrix elements steam from corrections to the Lagrangian and higher order Fock states, respectively (the reason behind this splitting will become apparent in Chapter~\ref{chp:E1Transitions}). The total matrix element is then given by the sum of the partial ones, and the decay with is proportional to its absolute value squared, hence
\begin{align}
\Gamma &\propto |\M_{LO} + \M_{NLO} + \M_{NNLO} + \M_{\text{rel.}} + \M_{\text{non-pert.}}|^2 \\
&= \M_{LO}^2 + \M_{NLO}^2 + \M_{NNLO}^2 + \M_{\text{rel.}}^2 + \M_{\text{non-pert.}}^2 \\
\nonumber & \quad + 2 \M_{LO} \M_{NLO} + 2 \M_{LO} \M_{NNLO} + 2 \M_{LO} \M_{\text{rel.}} + 2 \M_{LO} \M_{\text{non-pert.}} \\
\nonumber & \quad + 2 \M_{NLO} \M_{NNLO} + 2 \M_{NLO} \M_{\text{rel.}} + 2 \M_{NLO} \M_{\text{non-pert.}} \\
\nonumber & \quad + 2 \M_{NNLO} \M_{\text{rel.}} + 2 \M_{NNLO} \M_{\text{non-pert.}} \\
\nonumber & \quad + 2 \M_{\text{rel.}} \M_{\text{non-pert.}} \\
&= \M_{LO}^2 + \M_{NLO}^2 + 2 \M_{LO} \M_{NLO} + 2 \M_{LO} \M_{NNLO} \\
\nonumber & \quad + 2 \M_{LO} \M_{\text{rel.}} + 2 \M_{LO} \M_{\text{non-pert.}} + \O(\alpha_s^3) \\
&= \M_{LO}^2 \left(1 + \frac{\M_{NLO}^2}{\M_{LO}^2} + 2 \frac{\M_{NLO}}{\M_{LO}} + 2 \frac{\M_{NNLO}}{\M_{LO}} + 2 \frac{\M_{\text{rel.}}}{\M_{LO}} + 2 \frac{\M_{\text{non-pert.}}}{\M_{LO}} + \O(\alpha_s^3)\right) \\
&\equiv \M_{LO}^2 \left(1 + R + 2 \bar{\M}_{\text{rel.}} + \O(\alpha_s^3)\right) \,,
\end{align}
where we defined
\begin{equation}
\bar{\M} \equiv \frac{\M}{\M_{LO}} \,.
\end{equation}
All the wave function corrections are now encoded in the function
\begin{equation}
\label{eq:R} R = \bar{\M}_{NLO}^2 + 2 \bar{\M}_{NLO} + 2 \bar{\M}_{NNLO} + 2 \bar{\M}_{\text{non-pert.}} \,.
\end{equation}
We therefore see that, in order to have a consistent power counting, the matrix element entering the decay width at NLO is given by $2 \bar{\M}_{NLO}$; and the matrix element at NNLO is given by $|\bar{\M}_{NLO}|^2 + 2 \bar{\M}_{NNLO} + 2 \bar{\M}_{\text{non-pert.}}$.

\section{Analytic solutions and quantum mechanical perturbation theory}
\label{sec:AnalyticSolutionsQuantumMechanicalPerturbationTheory}

\subsection{Analytic solution of the leading order singlet Schrödinger equation}

The Schrödinger equation induced by the leading order singlet Hamiltonian, $H_{\text{s}}^C$, can be solved exactly due to its similarity to the hydrogen atom that is well known from quantum mechanics. To do so, we make only small adjustments to respect the changes from the electromagnetic interaction to the strong interaction by redefining the Coulomb potential and the Bohr radius accordingly as follows:
\begin{equation}
- \frac{\alpha_{e/m}}{r} \longrightarrow -C_F \frac{\alpha_s}{r} \quad\quad \text{and} \quad\quad a = \frac{1}{\mr \alpha_{e/m}} \longrightarrow \frac{1}{\mr C_F \alpha_s} \quad\quad \text{with} \quad\quad \mr = \frac{m}{2} \,.
\end{equation}
The Schrödinger equation of the heavy $q\bar{q}$-system
\begin{equation}
\label{eq:ZerothOrderSchroedingerEquation} \left( \frac{-1}{2\mr}\nabla_r^2 -C_F \frac{\alpha_s}{r} \right) \psi_{n\l m}^{(0)}(\vec{r}\,) = E_n^{(0)} \psi_{n\l m}^{(0)}(\vec{r}\,)
\end{equation}
has the well known, normalized solution
\begin{equation}
\label{eq:ZerothOrderWaveFunction} \psi_{n\l m}^{(0)}(\vec{r}\,) = R_{n\l}(r) Y_{\l m}(\Omega_r) \,.
\end{equation}
Here $n \in \N$ is the principal quantum number satisfying $n = n_r + \l + 1$ with $n_r \in \N_0$ being the radial quantum number, $\l \in \lbrace 0,1,\dots,n-1 \rbrace$ is the angular momentum quantum number and $m \in \lbrace -\l,\dots,\l \rbrace$ is the third component of the angular momentum. The spherical harmonics $Y_{\l m}(\Omega_r)$ are the eigenfunctions of the angular part of the Laplace operator in spherical coordinates and form an orthonormal basis on the unit sphere, hence they satisfy
\begin{equation}
\int \d\Omega \, Y_{\l m}^*(\Omega) Y_{\l'm'}(\Omega) = \int\limits_0^\pi \d\theta \int\limits_0^{2\pi} \d\phi \, \sin^2(\theta) Y_{\l m}^*(\theta,\phi) Y_{\l'm'}(\theta,\phi) = \delta_{\l\l'} \, \delta_{mm'} \,.
\end{equation}
The corresponding eigenvalues are given by $\l(\l+1)$ and further properties are listed in Appendix~\ref{app:subsec:SphericalHarmonics}.\\
The radial solution is given by
\begin{equation}
\label{eq:RadialWaveFunctionLaguerre} R_{n\l}(r) = N_{n\l} \, \e{-\frac{\rho_n}{2}} \rho_n^\l \, L_{n-\l-1}^{2\l+1}(\rho_n) \,,
\end{equation}
where the normalization reads
\begin{equation}
N_{n\l} = \sqrt{\left( \frac{2}{n a} \right)^3 \frac{(n-\l-1)!}{2n[(n+\l)!]}} \,,
\end{equation}
and we introduced the dimensionless quantity $\rho_n = \frac{2r}{n a}$ and the $L_{n-\l-1}^{2\l+1}(\rho_n)$ are the associated Laguerre polynomials, of which we list several important properties and relations in Appendix~\ref{app:subsec:AssociatedLaguerrePolynomials}. An equivalent form, consistent with the one given, e.g., in Ref.~\cite{BransdenJoachain:QuantumMechanics}, is
\begin{equation}
\label{eq:RadialWaveFunctionHypergeometric} R_{n\l}(r)  = \frac{1}{(2\l+1)!} \sqrt{\left( \frac{2}{n a} \right)^3 \frac{(n+\l)!}{2n[(n-\l-1)!]}} \e{-\frac{\rho_n}{2}} \rho_n^\l \, _1F_1(\l+1-n;2(\l+2);\rho_n) \,,
\end{equation}
where the $_1F_1(\l+1-n;2(\l+2);\rho_n)$ is the Kummer confluent hypergeometric function. The radial wave function is properly normalized and thus satisfies
\begin{equation}
\int\limits_0^\infty \d r \, r^2 R_{n\l}(r) R_{n'\l'}(r) = \delta_{n n'} \, \delta_{\l\l'} \,,
\end{equation}
in such a way that we have
\begin{equation}
\int \d^3 r \, \psi_{n\l m}^{(0)\,*}(\vec{r}\,) \psi_{n'\l'm'}^{(0)}(\vec{r}\,) = \delta_{n n'} \, \delta_{\l\l'} \, \delta_{mm'} \,.
\end{equation}
Explicit expressions for the first few radial and angular wave functions are given in Appendix~\ref{app:sec:WaveFunctions}.\\
In order to avoid confusion and to be as precise as possible, we want to adopt the bra-ket-notation known from quantum mechanics and in order to keep notation short, we may cast the states as
\begin{equation}
| n\l \rangle^{(0)} \,, \dots \quad\quad \text{with} \quad\quad \langle \vec{r} \, | n\l \rangle^{(0)} = \psi_{n\l m}^{(0)}(\vec{r}\,) \,, \dots
\end{equation}
Finally, the leading order energy solution, corresponding to \eqref{eq:ZerothOrderSchroedingerEquation}, is given by the Coulomb energy
\begin{equation}
\label{eq:ZerothOrderEnergy} E_n^{(0)} = -\frac{\mr C_F^2 \alpha_s^2}{2n^2} \,.
\end{equation}

\subsection{Physical quarkonium states}

The complete physical state of a given quarkonium is not fully covered by the solution $| n\l \rangle^{(0)}$ of the Schrödinger equation~\eqref{eq:ZerothOrderSchroedingerEquation}, since its solution lacks the information about spin $s$, total angular momentum $J$ and polarization $\lambda$. One therefore introduces the full quarkonium wave function
\begin{equation}
\Phi_{n\l m s m_{s} J m_{J}}^{(0)}(\vec{r},\lambda) = \psi_{n\l m}^{(0)}(\vec{r}\,) \cdot \chi_{s m_{s} J m_{J}}^{(0)}(\lambda) \,,
\end{equation}
where $\chi_{s m_{s} J m_{J}}^{(0)}(\lambda)$ encodes $s$, $J$ and $\lambda$. This part of the wave function can be derived by decomposing the product of spin $s$ and orbital angular momentum $\l$ into irreducible subspaces of total angular momentum $J$. The general procedure is described, e.g., in Ref.~\cite{GalindoPascual:QuantumMechanicsI} and the application to pNRQCD is shown in Appendix~C of Ref.~\cite{Pietrulewicz:2011aca}.\\
Physically the full wave functions $\Phi_{n\l m s m_{s} J m_{J}}^{(0)}(\vec{r},\lambda)$ are the subset of eigenstates made up by a quark-antiquark pair in a singlet representation. Following the notation from Refs.~\cite{Pietrulewicz:2011aca,Brambilla:2012be}, using bra-ket-notation and Fourier transforming into momentum space, these states can be cast as
\begin{equation}
\label{ketHP00} | H(\vec{P},\lambda) \rangle^{(0)} = \int \d^3 R \int \d^3 r \; \e{\i\vec{P}\cdot\vec{R}} \, \text{tr } \left\lbrace \Phi^{(0)}_{H(\lambda)}(\vec{r}\,) \, S^\dagger(\vec{r},\vec{R}) | 0 \rangle \right\rbrace \,,
\end{equation}
where $| 0 \rangle$ is a state that belongs to the Fock subspace containing no heavy quarks, but an arbitrary number of ultra-soft gluons, photons and light quarks. The state $| 0 \rangle$ is normalized in such a way that
\begin{equation}
\label{eq:StateNormalization} \langle H(\vec{P}',\lambda') | H(\vec{P},\lambda) \rangle = (2\pi)^3 \delta^{(3)}(\vec{P} - \vec{P}') \delta_{\lambda\lambda'}
\end{equation}
is fulfilled. The function $\Phi^{(0)}_{H(\lambda)}(\vec{r}\,) = \langle 0 | S(\vec{r},\vec{R}) | H(\vec{0},\lambda)\rangle^{(0)}$ is an eigenstate of the spin and orbital angular momentum of the quarkonium and satisfies the Schrödinger equation
\begin{equation}
H_{\text{s}}^{(0)}\,\Phi^{(0)}_{H(\lambda)}(\vec{r}\,) = E^{(0)}_H \Phi^{(0)}_{H(\lambda)}(\vec{r}\,) \,,
\end{equation}
that is equivalent to the Schrödinger equation~\eqref{eq:ZerothOrderSchroedingerEquation}, and thus $E^{(0)}_H$ is the leading-order binding energy of the quarkonium $H$, yielding
\begin{equation}
M_H = 2m + E^{(0)}_H \,.
\end{equation}
For $\l = 0$ the states $\Phi_{H(\lambda)}^{(0)}$ can be written in the form
\begin{align}
\label{eq:BeginWaveFunctionDefinition} & \Phi_{n\,^3\!S_1(\lambda)}^{(0)}(\vec{r}\,) = \frac{1}{\sqrt{4\pi}} \, R_{n0}(r) \, \frac{\vec{\sigma} \cdot \hat{e}_{n\,^3\!S_1}(\lambda)}{\sqrt{2}} \,, \\
& \Phi_{n\,^1\!S_0}^{(0)}(\vec{r}\,) = \frac{1}{\sqrt{4\pi}} \, R_{n0}(r) \, \frac{1}{\sqrt{2}} \,,
\end{align}
where $\hat{e}_{n\,^3\!S_1}(\lambda)$ is the polarization vector of the state $n\,^3\!S_1$, normalized as $\hat{e}_{n\,^3\!S_1}^*(\lambda) \cdot \hat{e}_{n\,^3\!S_1}(\lambda') = \delta_{\lambda\lambda'}$.\\
For $\l = 1$ the states $\Phi_{H(\lambda)}^{(0)}$ can be written in the form
\begin{align}
& \Phi_{n\,^1\!P_1(\lambda)}^{(0)}(\vec{r}\,) = \sqrt{\frac{3}{4\pi}} \, R_{n1}(r) \, \frac{\hat{e}_{n\,^1\!P_1}(\lambda) \cdot \hat{r}}{\sqrt{2}} \,, \\
& \Phi_{n\,^3\!P_0}^{(0)}(\vec{r}\,) = \sqrt{\frac{1}{4\pi}} \, R_{n1}(r) \, \frac{\vec{\sigma} \cdot \hat{r}}{\sqrt{2}} \,, \\
& \Phi_{n\,^3\!P_1(\lambda)}^{(0)}(\vec{r}\,) = \sqrt{\frac{3}{8\pi}} \, R_{n1}(r) \, \frac{\vec{\sigma} \cdot (\hat{r} \times \hat{e}_{n\,^3\!P_1}(\lambda))}{\sqrt{2}} \,, \\
\label{eq:EndWaveFunctionDefinition} & \Phi_{n\,^3\!P_2(\lambda)}^{(0)}(\vec{r}\,) = \sqrt{\frac{3}{4\pi}} \, R_{n1}(r) \, \frac{\vec{\sigma}^i \, h^{i j}_{n\,^3\!P_2}(\lambda) \, \hat{r}^j}{\sqrt{2}} \,,
\end{align}
where $\hat{e}_{n\,^1\!P_1}(\lambda)$ and $\hat{e}_{n\,^3\!P_1}(\lambda)$ are polarization vectors satisfying $\hat{e}_{n\,^1\!P_1}^*(\lambda) \cdot \hat{e}_{n\,^1\!P_1}(\lambda') = \hat{e}_{n\,^3\!P_1}^*(\lambda) \cdot \hat{e}_{n\,^3\!P_1}(\lambda') = \delta_{\lambda\lambda'}$, whereas the polarization of the $n\,^3\!P_2$ state is represented by the symmetric and traceless rank-2 tensor $h_{n\,^3\!P_2}^{i j}(\lambda)$, normalized according to $h_{n\,^3\!P_2}^{i j\,*}(\lambda) \, h_{n\,^3\!P_2}^{j i}(\lambda') = \delta_{\lambda\lambda'}$.

\subsection{Quantum mechanical perturbation theory}
\label{subsec:QuantumMechanicalPerturbationTheory}

A general introduction to quantum mechanical perturbation theory can be found in standard literature, c.f. Refs.~\cite{GalindoPascual:QuantumMechanicsI,BransdenJoachain:QuantumMechanics,Schwabl:QuantenmechanikI,Sakurai:ModernQuantumMechanics}. As we have argued in Sec.~\ref{sec:pNRQCD} when introducing the power counting, we need up to second order corrections in the eigenstates and eigenenergies to reach our goal of NNLO, relative order $\alpha_s^2$, accuracy. We therefore now introduce the needed tools\footnote{Note that we are giving expressions mainly for initial states (ket-vectors). However, corrections also affect final states (bra-vectors) and the derivation holds for them as well and can be obtained in a analogous manner. Furthermore, we reduce the discussion to the wave function $\psi_{n\l m}(\vec{r}\,)$ for the moment, since the explicit form of $\chi_{s m_{s} J m_{J}}(\lambda)$ is irrelevant for the discussion to follow. Note also that we drop summation indices $\l$ in sums over intermediate states in order to keep notation simple.}:\\
A generic matrix element of an operator $\O$ is defined as
\begin{equation}
\label{eq:GeneralMatrixElement} \langle n'\l' | \O | n\l \rangle = \int \d^3 r \, \psi_{n'\l'm'}^{(0)\,*}(\vec{r}\,) \, \O \, \psi_{n\l m}^{(0)}(\vec{r}\,) \,,
\end{equation}
where the notation $| n\l \rangle \equiv | n\l \rangle^{(0)}$ is introduced in order to shorten equations. This immediately gives rise to the first order correction of the energy induced by a potential $V$:
\begin{equation}
\label{eq:FirstOrderEnergy} E_n^{(1)} = \langle n\l | V | n\l \rangle \,.
\end{equation}
Integrals of this type can be solved analytically, considering that $V$ takes the form of the potentials listed in the Eqs.~\eqref{eq:SingletStaticPotential} and \eqref{eq:ExplicitPotentials}. The according formulas and expressions of these so-called single potential insertions are listed and derived in Appendix~\ref{app:sec:ExpectationValuesSinglePotentialInsertions}.\\
The first order correction to the wave function involves off-diagonal matrix elements and is given by
\begin{equation}
\label{eq:FirstOrderWaveFunction} | n\l \rangle^{(1)} = \sum\limits_{n' \neq n} \frac{\langle n'\l' | V | n\l \rangle}{E_n^{(0)} - E_{n'}^{(0)}} | n'\l' \rangle = \left( \sum\limits_{n' \neq n} \frac{| n'\l' \rangle \langle n'\l' |}{E_n^{(0)} - E_{n'}^{(0)}} \right) V | n\l \rangle \,,
\end{equation}
and enters the second order correction to the energy, induced by a potential $V$, that is then given by
\begin{equation}
\label{eq:SecondOrderEnergy} E_n^{(2)} = \langle n\l | V | n\l \rangle^{(1)} = \langle n\l | V \cdot \left( \sum\limits_{n' \neq n} \frac{| n'\l' \rangle \langle n'\l' |}{E_n^{(0)} - E_{n'}^{(0)}} \right) \cdot V | n\l \rangle \,.
\end{equation}
Again, with the potentials entering our computations, these integrals can be solved analytically. We list these so-called double potential insertions in Appendix~\ref{app:sec:ExpectationValuesDoublePotentialInsertions} and derive a procedure to compute them as well.\\
Finally, the second order correction to the wave function reads
\begin{align}
\label{eq:SecondOrderWaveFunction}
\begin{aligned}
| n\l \rangle^{(2)} & = \sum\limits_{k_{1} \neq n} \left[ \sum\limits_{k_{2} \neq n} \frac{\langle k_{1}\l | V | k_{2}\l \rangle \langle k_{2}\l | V | n\l \rangle}{(E_{n} - E_{k_{1}}) (E_{n} - E_{k_{2}})} - \frac{\langle k_{1}\l | V | n\l \rangle \langle n\l | V | n\l \rangle}{(E_{n} - E_{k_{1}})^{2}} \right] |k_{1}\l\rangle \\
& -\frac{1}{2} \sum\limits_{k_{2} \neq n} \frac{|\langle k_{2}\l | V | n\l\rangle|^2}{(E_{n} - E_{k_{2}})^{2}} | n\l \rangle \,.
\end{aligned}
\end{align}
The following is known from standard quantum mechanics:
\begin{align}
\label{eq:QuantumMechanicalIdentities}
\begin{aligned}
& \langle n\l | n\l \rangle = 1 \,, && \langle n\l | \1 | n'\l \rangle = \delta_{nn'} \,, \quad\quad && \sum_{n} E_{n}^{(0)} | n\l \rangle\langle n\l | = H^{(0)} \,, \\
& \sum_{n} | n\l \rangle \langle n\l | = \1_{n \times n} \,, \quad\quad && \sum\limits_{\l} | n\l \rangle \langle n\l | = \mathcal{P}(n) \,,
\end{aligned}
\end{align}
where $H^{(0)} \equiv H$ is the unperturbed leading order Hamiltonian and $\mathcal{P}(n')$ is a projection operator that may be represented by a $n \times n$-matrix that has entries $0$ everywhere except for the one entry equal to $1$ in the crossing point of the $n'$-th row and $n'$-th column, while $n \geq n'$ is the dimension of the underlying Hilbert space.\\
Several of the above formulae~\eqref{eq:FirstOrderWaveFunction}, \eqref{eq:SecondOrderEnergy} and \eqref{eq:SecondOrderWaveFunction} involve a sum over all intermediate states $n' \neq n$. Analytically, this is a non-trivial task and a first attempt to overcome this problem is sketched in Appendix~\ref{app:chp:AlternativeFirstOrderWaveFunction}. It turns out though, that it is not applicable for our kinds of potentials and we proceed by decomposing the sum over intermediate states as
\begin{equation}
\label{eq:Decomposition} \sum\limits_{n' \neq n} \frac{| n'\l' \rangle \langle n'\l' |}{E_n^{(0)} - E_{n'}^{(0)}} = \sum\limits_{n'} \frac{| n'\l' \rangle \langle n'\l' |}{E_n^{(0)} - E_{n'}^{(0)}} - \sum\limits_{n'=n} \frac{| n'\l' \rangle \langle n'\l' |}{E_n^{(0)} - E_{n'}^{(0)}} \,,
\end{equation}
by using an equality that is formally correct, albeit both terms on the right hand side of Eq.~\eqref{eq:Decomposition} now diverge. We use above quantum mechanical identities~\eqref{eq:QuantumMechanicalIdentities}, in order to manipulate this expression further, yielding
\begin{equation}
\label{eq:Limit} \sum\limits_{n' \neq n} \frac{| n'\l' \rangle \langle n'\l' |}{E_n^{(0)} - E_{n'}^{(0)}} = \frac{\1}{E_n^{(0)} - H} - \frac{\mathcal{P}(n)}{E_n^{(0)} - E_{n'=n}^{(0)}} = \lim_{E \to E_n^{(0)}} \left( \frac{\1}{E - H} - \frac{\mathcal{P}(n)}{E - E_n^{(0)}} \right) \,,
\end{equation}
and the expression on the right hand side may be cast as
\begin{equation}
\label{eq:CoulombGreenFunction} \lim_{E \to E_n^{(0)}} \left( \frac{\1}{E - H} - \frac{\mathcal{P}(n)}{E - E_n^{(0)}} \right) \equiv \frac{1}{(E_n - H)'} \,,
\end{equation}
in agreement with \cite{Peset:2015vvi} (Eq.~(7.22) therein).\\
The formal results of the Eqs.~\eqref{eq:Decomposition}, \eqref{eq:Limit} and \eqref{eq:CoulombGreenFunction} allow us to recast the second order correction to the energy, Eq.~\eqref{eq:SecondOrderEnergy}, and the generic expectation values of an operator $\O$ as
\begin{align}
\label{eq:SecondOrderEnergyGreenFunction} & E_n^{(2)} = \langle n\l | V \frac{1}{(E_n - H)'} V | n\l \rangle \,, \\
\label{eq:FirstOrderMatrixElement} & \langle n'\l' | \O | n\l \rangle^{(1)} = \langle n'\l' | \O \frac{1}{(E_n - H)'} V | n\l \rangle \,, \\
\label{eq:SecondOrderMatrixElement} & \langle n'\l' | \O | n\l \rangle^{(2)} = \langle n'\l' | \O \frac{1}{(E_n - H)'} V \frac{1}{(E_n - H)'} V | n\l \rangle \\
\nonumber & \quad - \langle n\l | V | n\l \rangle \langle n'\l' | \O \frac{1}{(E_n - H)'} \1 \frac{1}{(E_n - H)'} V | n\l \rangle \\
\nonumber & \quad - \frac{1}{2} \langle n'\l' | \O | n\l \rangle \langle n\l | V \frac{1}{(E_n - H)'} \1 \frac{1}{(E_n - H)'} V | n\l \rangle \,,
\end{align}
where in the last equation we have manipulated the expectation value as
\begin{align}
& \langle n'\l' | \O \sum\limits_{k_{1} \neq n} \sum\limits_{k_{2} \neq n} \frac{\langle k_{1}\l | V | k_{2}\l \rangle \langle k_{2}\l | V | n\l \rangle}{(E_{n} - E_{k_{1}}) (E_{n} - E_{k_{2}})} |k_{1}\l \rangle  \\
\nonumber & \quad\quad - \langle n'\l' | \O \sum\limits_{k_{1} \neq n} \frac{\langle k_{1}\l | V | n\l \rangle \langle n\l | V | n\l \rangle}{(E_{n} - E_{k_{1}})^{2}} | k_{1}\l \rangle \\
\nonumber & \quad\quad - \frac{1}{2} \langle n'\l' | \O \sum\limits_{k_{2} \neq n} \frac{\langle n\l | V | k_{2}\l \rangle \langle k_{2}\l | V | n\l \rangle}{(E_{n} - E_{k_{2}})^{2}} | n\l \rangle \\
& \quad = \langle n'\l' | \O \sum\limits_{k_{1} \neq n} \frac{| k_{1}\l \rangle \langle k_{1}\l |}{(E_{n} - E_{k_{1}})} V \sum\limits_{k_{2} \neq n} \frac{| k_{2}\l \rangle \langle k_{2}\l |}{(E_{n} - E_{k_{2}})} V | n\l \rangle \\
\nonumber & \quad\quad - \langle n\l | V | n\l \rangle \langle n'\l' | \O \sum\limits_{k_{1} \neq n} \frac{| k_{1}\l \rangle \langle k_{1}\l |}{(E_{n} - E_{k_{1}})^{2}} V | n\l \rangle \\
\nonumber & \quad\quad - \frac{1}{2} \langle n'\l' | \O | n\l \rangle \langle n\l | V \sum\limits_{k_{2} \neq n} \frac{| k_{2}\l \rangle \langle k_{2}\l |}{(E_{n} - E_{k_{2}})^{2}} V | n\l \rangle \\
& \quad = \langle n'\l' | \O \sum\limits_{k_{1} \neq n} \frac{| k_{1}\l \rangle \langle k_{1}\l |}{(E_{n} - E_{k_{1}})} V \sum\limits_{k_{2} \neq n} \frac{| k_{2}\l \rangle \langle k_{2}\l |}{(E_{n} - E_{k_{2}})} V | n\l \rangle \\
\nonumber & \quad\quad - \langle n\l | V | n\l \rangle \langle n'\l' | \O \sum\limits_{k_{1} \neq n} \frac{| k_{1}\l \rangle \langle k_{1}\l |}{(E_{n} - E_{k_{1}})} \1 \sum\limits_{k_{2} \neq n} \frac{| k_{2}\l \rangle \langle k_{2}\l |}{(E_{n} - E_{k_{2}})} V | n\l \rangle \\
\nonumber & \quad\quad - \frac{1}{2} \langle n'\l' | \O | n\l \rangle \langle n\l | V \sum\limits_{k_{1} \neq n} \frac{| k_{1}\l \rangle \langle k_{1}\l |}{(E_{n} - E_{k_{1}})} \1 \sum\limits_{k_{2} \neq n} \frac{| k_{2}\l \rangle \langle k_{2}\l |}{(E_{n} - E_{k_{2}})} V | n\l \rangle \,.
\end{align}
We have now a compact set of equations in order to describe corrections to the energy (\eqref{eq:FirstOrderEnergy} and \eqref{eq:SecondOrderEnergyGreenFunction}), and matrix elements of general operators (\eqref{eq:GeneralMatrixElement}, \eqref{eq:FirstOrderMatrixElement} and \eqref{eq:SecondOrderMatrixElement}). However, we still need a procedure to get explicit results. In order to do so, we make use of the Coulomb Green function technique that, later on, can be identified with the expression $\frac{1}{(E_n - H)'}$ in the aforementioned equations.

\subsection{The non-relativistic Coulomb Green function}
\label{subsec:CoulombGreenFunction}

The Green function $G(\vec{r}_{1},\vec{r}_{2};E)$ is defined as the solution of the differential equation \cite{Meixner:1933, Hostler:1964}
\begin{equation}
\label{eq:GreenFunctionDifferentialEquation} \left[\vec{\nabla}_{r_1}^{\,2} + \frac{2k\nu}{r_1} + k^2\right] G(\vec{r}_1,\vec{r}_2;E) = \delta^{(3)}(\vec{r}_1-\vec{r}_2) \,,
\end{equation}
where
\begin{align}
\begin{aligned}
& a = \frac{1}{m Z \alpha} \,, \quad\quad k = \sqrt{2 m E} \,, \quad \mathrm{Im}(k)>0 \,, \\
& \nu = \frac{1}{k a} = \frac{m Z \alpha}{\sqrt{2 m E}} \,, \quad\quad \lambda = \i\nu = \i \frac{m Z \alpha}{\sqrt{2 m E}} \quad \Rightarrow \quad E = -\frac{m Z^{2} \alpha^{2}}{2\lambda^{2}} \,,
\end{aligned}
\end{align}
and it satisfies the following boundary conditions
\begin{align}
\label{eq:GreenFunctionLimit}
\begin{aligned}
& \text{as} \quad \vec{r}_{1} \to 0 \quad
\begin{cases}
r_{1}^{\frac{1}{2}} \, G(\vec{r}_{1},\vec{r}_{2};E) \to 0 \,, & \\
r_{1}^{\frac{1}{2}} \, \vec{r}_{1} \cdot \vec{\nabla}_{r_{1}} \, G(\vec{r}_{1},\vec{r}_{2};E) \to 0 \,, &
\end{cases} \\
& \text{as} \quad \vec{r}_{1} \to \infty \quad
\begin{cases}
r_{1} \, G(\vec{r}_{1},\vec{r}_{2};E) \to 0 \,, & \\
\vec{r}_{1} \cdot \vec{\nabla}_{r_{1}} \, G(\vec{r}_{1},\vec{r}_{2};E) \to 0 \,, &
\end{cases}
\end{aligned}
\end{align}
Here $E$ is a complex, discrete or continuous number, not in the eigenvalue spectrum of the Hamiltonian $H$ of the system. The Green function, as defined by the Eqs.~\eqref{eq:GreenFunctionDifferentialEquation} and \eqref{eq:GreenFunctionLimit}, is unique and symmetric, hence $G(\vec{r}_{1},\vec{r}_{2};E) = G(\vec{r}_{2},\vec{r}_{1};E)$. The Green function is an analytic function of $E$ on the complex $E$ plane, with a cut along the positive real axis (being the continuous spectrum $0 < E < +\infty$), except for simple poles at the distinct bound-state energy eigenvalues of the system.\\
The retarded (advanced) physical Green function, defined for real $E$, is obtained from $G(\vec{r}_{1},\vec{r}_{2};E)$ by taking the limit as $E$ approaches the real axis from above (below). For $E > 0$, the physical Green function has an oscillatory behavior as $r_{1} \to \infty$. At large distances, the retarded Green function only consists of outgoing spherical waves and the advanced Green function only consists of incoming spherical waves. For $E < 0$, the retarded and advanced Green functions coincide and both agree with the general Green function as defined by the Eqs.~\eqref{eq:GreenFunctionDifferentialEquation} and \eqref{eq:GreenFunctionLimit}. These values for $E$ are non-propagating in the sense that the Green functions decay exponentially as $r_{1} \to \infty$.\\
The solution of the Eqs.~\eqref{eq:GreenFunctionDifferentialEquation} and \eqref{eq:GreenFunctionLimit} can be written in the form of an eigenfunction expansion
\begin{align}
\label{eq:GreenFunctionExpansion1}
\begin{aligned}
G(\vec{r}_{1},\vec{r}_{2};E) = & - \frac{1}{2m} \sum\limits_{\l=0}^{\infty} \sum\limits_{m=-\l}^{\l} \int\limits_{0}^{\infty} \d k \, \frac{\psi_{\l m}(k;r_{1}) \psi_{\l m}^*(k;r_{2})}{\frac{k^{2}}{2m} - E} \\
& - \frac{1}{2m} \sum\limits_{n=1}^{\infty} \sum\limits_{\l=0}^{\infty} \sum\limits_{m=-\l}^{\l} \frac{\psi_{n\l m}(r_{1}) \psi_{n\l m}^*(r_{2})}{E_{n\l} - E} \,.
\end{aligned}
\end{align}
The eigenfunctions $\psi$ here are the simultaneous eigenfunctions of the Hamiltonian $H$ and of $\vec{L}^{\,2}$ and $L_{z}$, where $\vec{L}$ is the orbital angular momentum operator. In the first term of Eq.~\eqref{eq:GreenFunctionExpansion1}, we have a summation and integration over the continuous spectrum of $H$, whereas in the second term we have a summation over the discrete spectrum only. Inserting the explicit expressions for the wave functions and performing the integral over the continuous spectrum \cite{Mapleton:1961}\footnote{See also Sec.~II of Ref.~\cite{Hostler:1964} for the analogous calculation for the Klein-Gordon problem.}, we obtain
\begin{equation}
\label{eq:GreenFunctionExpansion2} G(\vec{r}_{1},\vec{r}_{2};E) = \frac{1}{8\pi \i k r_{1}r_{2}} \sum\limits_{\l=0}^{\infty} (2\l+1) P_{\l}(\hat{r}_{1} \cdot \hat{r}_{2}) \Gamma(1+\l-\i\nu) \mathcal{W}_{\i\nu;\l+\frac{1}{2}}(-2\i k r_{1}) \mathcal{M}_{\i\nu;\l+\frac{1}{2}}(-2\i k r_{2}) \,,
\end{equation}
with $r_{1}>r_{2}$. Herein $\Gamma(z)$ and $P_{\l}(\hat{r}_{1}\cdot\hat{r}_{2})$ denote the Gamma function and the Legendre polynomial, respectively (we give some of the important properties and relations in the respective Appendices~\ref{app:subsec:GammaDigammaPolygamma} and \ref{app:subsec:Legendre}), and the Whittaker functions, as defined in Ref.~\cite{Buchholz:KonfluenteHypergeometrischeFunktion}, are given by
\begin{align}
\begin{aligned}
L_{n}^{\mu}(z) &= \frac{\Gamma(n+\mu+1)}{n!} z^{-\frac{1}{2}(1+\mu)} \e{+\frac{1}{2}z} \mathcal{M}_{n+\frac{1}{2}(1+\mu);\frac{1}{2}\mu}(z) \\
& = \frac{(-1)^{n}}{n!} z^{-\frac{1}{2}(1+\mu)} \e{+\frac{1}{2}z} \mathcal{W}_{n+\frac{1}{2}(1+\mu);\frac{1}{2}\mu}(z) \,.
\end{aligned}
\end{align}
Both expressions, \eqref{eq:GreenFunctionExpansion1} and \eqref{eq:GreenFunctionExpansion2}, are quite standard but they are not yet our final result. We introduce, in place of $G$, the Coulomb Green function $\mathcal{G}(\vec{r}_{1},\vec{r}_{2};E)$ of the squared Schrödinger equation, which is related to $G$ by \cite{Zapryagaev:1981}
\begin{equation}
G(\vec{r}_{1},\vec{r}_{2};E) = \left[\vec{\nabla}_{r_{1}}^{\,2} + \frac{2k\nu}{r_{1}} + k^{2}\right] \mathcal{G}(\vec{r}_{1},\vec{r}_{2};E) \,.
\end{equation}
$\mathcal{G}$ can be written in the form
\begin{align}
\begin{aligned}
\mathcal{G}(\vec{r}_{1},\vec{r}_{2};E) &= \sum\limits_{\l=0}^{\infty} \sum\limits_{m=-\l}^{\l} g_{\l}(r_{1},r_{2};E) Y_{\l m}(\Omega_{r_{1}}) Y_{\l m}(\Omega_{r_{2}}) \\
& = \sum\limits_{\l=0}^{\infty} \frac{(2\l+1)}{4\pi} P_{\l}(\hat{r}_{1} \cdot \hat{r}_{2}) g_{\l}(r_{1},r_{2};E) \,.
\end{aligned}
\end{align}
Different representations have been obtained for the radial Green function $g_{\l}(r_{1},r_{2};E)$, in particular, in the form of a parametric integral \cite{Hostler:1964, Granovskii:1974yf} or in the form of an expansion with respect to associated Laguerre polynomials $L_{s}^{\alpha}$ (Sturm expansion) \cite{Zon:1972, Khristenko:1975}:
\begin{equation}
\label{eq:RadialGreenFunction} g_{\l}(r_{1},r_{2};E) = \frac{4m}{a\lambda} \sum\limits_{s=0}^{\infty} \frac{\mathcal{S}_{s}\left(\frac{2r_{1}}{a\lambda}\right) \mathcal{S}_{s}\left(\frac{2r_{2}}{a\lambda}\right)}{s+\l+1-\lambda} \,,
\end{equation}
where
\begin{equation}
\mathcal{S}_{s}(x) = \left[\frac{s!}{\Gamma(s+2\l+2)}\right]^{\frac{1}{2}} 
x^{\l} \e{-\frac{x}{2}} L_{s}^{2\l+1}(x) \,.
\end{equation}
The series~\eqref{eq:RadialGreenFunction} converges point wise for all $r_{1},r_{2}$, except for the point $r_{1} = r_{2} = 0$, as long as the condition $|E| < m$ holds. The final expression is given by
\begin{equation}
\label{eq:GreenFunctionOldFinal} g_{\l}(r_{1},r_{2};E) = \frac{4m}{a\lambda} \, \left(\frac{2r_{1}}{\lambda a}\right)^{\l} \left(\frac{2r_{2}}{\lambda a}\right)^{\l} \e{-\frac{1}{\lambda a}(r_{1} + r_{2})} \sum\limits_{s=0}^{\infty} \frac{L_{s}^{2\l+1}\left(\frac{2r_{1}}{\lambda a}\right) L_{s}^{2\l+1}\left(\frac{2r_{2}}{\lambda a}\right) \, s!}{(s+\l+1-\lambda)(s+2\l+1)!} \,.
\end{equation}
We mention that the Fourier transform of the Sturm expansion of $G(r_{1},r_{2};E)$, with respect to the variables $r_{1},r_{2}$, gives the expansion of $G$ in the momentum representation with respect to four-dimensional spherical harmonics, obtained for the first time by Schwinger in Ref.~\cite{Schwinger:1964}, using group-theoretical considerations.\\
We now introduce a shift in notation, see, e.g., Ref~\cite{Kiyo:2014uca}, in order to become compatible with current literature. Note, however, that this relabeling is equivalent to the final result given in Eq.~\eqref{eq:GreenFunctionOldFinal}. We cast the Coulomb Green function as
\begin{equation}
\label{eq:CoulombGreenFunctionDefinition} G(\vec{r}_1,\vec{r}_2;E) = \sum\limits_{\l=0}^\infty \frac{(2\l+1)}{4\pi} P_\l(\hat{r}_1 \cdot \hat{r}_2) G_\l(r_1,r_2;E) \,.
\end{equation}
The Green function for the partial wave $\l$ is given by an infinite sum\footnote{This representation of $G$ includes contributions of not only the bound states at $E < 0$ but also those of the continuum states at $E > 0$, as introduced above, although this may not be obvious from the summation formula.}:
\begin{equation}
G_\l(r_1,r_2;E) = \sum\limits_{\nu=\l+1}^\infty \mr a^2 \left(\frac{\nu^4}{\lambda}\right) \frac{R_{\nu\l}(\r{\lambda}{1}) R_{\nu\l}(\r{\lambda}{2})}{\nu-\lambda} \,,
\end{equation}
where the $R_{\nu\l}(\r{\lambda}{i})$ is the radial wave function~\eqref{eq:RadialWaveFunctionLaguerre}, evaluated at the dimensionless radial coordinate $\r{\lambda}{i} = \frac{2r_i}{\lambda a}$. The above expression~\eqref{eq:CoulombGreenFunctionDefinition} can be shown to be identical to its formulation in Ref.~\cite{Peset:2015zga} (Eq.~(4.8) therein) since
\begin{align*}
& G(\vec{r}_1,\vec{r}_2;E) \\
& \quad\quad = \sum\limits_{\l=0}^\infty \frac{2\l+1}{4\pi} P_\l(\hat{r}_1 \cdot \hat{r}_2) \sum\limits_{\nu=\l+1}^\infty \mr a^2 \left(\frac{\nu^4}{\lambda}\right) N_{\nu\l}^2 \, \r{\lambda}{1}^\l \r{\lambda}{2}^\l \e{-\frac{1}{2}(\r{\lambda}{1}+\r{\lambda}{2})} \frac{L_{\nu-\l-1}^{2\l+1}(\r{\lambda}{1}) L_{\nu-\l-1}^{2\l+1}(\r{\lambda}{2})}{\nu-\lambda} \\
& \quad\quad = \sum\limits_{\l=0}^\infty \frac{2\l+1}{4\pi} P_\l(\hat{r}_1 \cdot \hat{r}_2) \sum\limits_{\nu=\l+1}^\infty \frac{4 \mr}{\lambda a} \r{\lambda}{1}^\l \r{\lambda}{2}^\l \e{-\frac{1}{2}(\r{\lambda}{1}+\r{\lambda}{2})} \frac{L_{\nu-\l-1}^{2\l+1}(\r{\lambda}{1}) L_{\nu-\l-1}^{2\l+1}(\r{\lambda}{2}) (\nu-\l-1)!}{(\nu-\lambda)(\nu+\l)!} \\
& \quad\quad = \sum\limits_{\l=0}^\infty \frac{2\l+1}{4\pi} P_\l(\hat{r}_1 \cdot \hat{r}_2) \sum\limits_{s=0}^\infty \frac{4 \mr}{\lambda a} \r{\lambda}{1}^\l \r{\lambda}{2}^\l \e{-\frac{1}{2}(\r{\lambda}{1}+\r{\lambda}{2})} \frac{L_{s}^{2\l+1}(\r{\lambda}{1}) L_{s}^{2\l+1}(\r{\lambda}{2}) s!}{(s+\l+1-\lambda)(s+2\l+1)!} \\
& \quad\quad = \frac{\mr^2 C_F \alpha_s}{\lambda\pi} \sum\limits_{\l=0}^\infty (2\l+1) P_\l(\hat{r}_1 \cdot \hat{r}_2) \left(\frac{2\mr C_F \alpha_s}{\lambda} r_1\right)^\l \left(\frac{2\mr C_F \alpha_s}{\lambda} r_2\right)^\ell \e{-\frac{\mr C_F \alpha_s}{\lambda} (r_1+r_2)} \\
& \quad\quad\quad\times \sum\limits_{s=0}^\infty \frac{L_{s}^{2\l+1}\left(\frac{2\mr C_F \alpha_s}{\lambda} r_1\right) L_{s}^{2\l+1}\left(\frac{2\mr C_F \alpha_s}{\lambda} r_2\right) s!}{(s+\l+1-\lambda)(s+2\l+1)!} \,,
\end{align*}
where we defined $s = \nu-\l-1$ and used the explicit definition of $a = \frac{1}{\mr C_F \alpha_s}$. The principal quantum number $\lambda$ is defined as $\lambda = n + \delta\lambda$, such that in the limit $\delta\lambda \to 0$ it becomes the usual principal quantum number $n$. This splitting is done in order to allow for an expansion in $\delta\lambda$. This expansion becomes apparent, once we expand the continuous energy $E$, which appears in the Coulomb Green function and is defined as
\begin{equation}
E = -\frac{\mr C_F^2 \alpha_s^2}{2\lambda^2} \,,
\end{equation}
around $E_{n}$, which is the Coulomb energy~\eqref{eq:ZerothOrderEnergy}:
\begin{equation}
E = -\frac{m_{r} C_{F}^{2} \alpha_s^{2}}{2\lambda^{2}} = E_{n} (1 - \epsilon) = -\frac{m_{r} C_{F}^{2} \alpha_s^{2}}{2 n^{2}} (1 - \epsilon) \,,
\end{equation}
and thus\footnote{The choice of the minus sign is established convention.}
\begin{equation}
\frac{1}{\lambda^{2}} = \frac{1}{n^{2}} (1 - \epsilon) \quad \Rightarrow \lambda = \frac{n}{\sqrt{1 - \epsilon}} \,.
\end{equation}
Thus an expansion in $\delta\lambda$ is sort of interchangeable with the expansion in $\epsilon$. It is to be noted that each of the different, yet equivalent, formulations of the Green function inherently comes with a pole - either in $s$ or in $\nu$, depending on the formulation. This pole arises naturally and can not be avoided, but cured in a particular manner.

\subsection{Conclusion and final expressions}

The connection between the introduced formalism of the Green function and the expression $\frac{1}{(E_n - H)'}$ is given by \cite{Voloshin:1978hc,Voloshin:1979uv}
\begin{equation}
\label{eq:Connection1} \langle n'\l' | \O \frac{1}{(E - H)'} V | n\l \rangle = \int \d^3 r_1 \, \d^3 r_2 \, \psi_{n'\l'}^*(\vec{r}_2) \, \O(\vec{r}_2) \, G'(\vec{r}_2,\vec{r}_1) \, V(\vec{r}_1) \, \psi_{n\l}(\vec{r}_1) \,,
\end{equation}
where
\begin{align}
\label{eq:Connection2}
\begin{aligned}
G'(\vec{r}_1,\vec{r}_2) & \equiv \langle \vec{r}_1 | \frac{1}{(E_n - H)'} | \vec{r}_{2} \rangle = - \langle \vec{r}_1 | \frac{1}{(H - E_n)'} | \vec{r}_2 \rangle \\
& \equiv (-1) \times \lim_{E \to E_n} \left(G(\vec{r}_1,\vec{r}_2,E) - \frac{|\psi_{n\l}|^{2}}{E - E_n}\right) \,,
\end{aligned}
\end{align}
and the global sign corrects a typo in \cite{Peset:2015zga} and makes it compatible with \cite{Pineda:2011dg}.\\
We compare the two different schemes: the one obtained in quantum mechanical perturbation theory (Sec.~\ref{subsec:QuantumMechanicalPerturbationTheory}) and the one using the Green function (Sec.~\ref{subsec:CoulombGreenFunction}). We notice immediately that the Eqs.~\eqref{eq:SecondOrderEnergy}, \eqref{eq:SecondOrderEnergyGreenFunction} and \eqref{eq:FirstOrderMatrixElement} resemble Eq.~\eqref{eq:Connection1}. Furthermore, we notice that Eq.~\eqref{eq:CoulombGreenFunction} is identical to Eq.~\eqref{eq:Connection2}, up to a minus sign. We therefore conclude that
\begin{equation}
\label{eq:Connection3} \frac{1}{(E_n - H)'} = G'(\vec{r}_1,\vec{r}_2) \,,
\end{equation}
and explicitly, in the limit $E \to E_n$,
\begin{align}
& \frac{\1}{E_n - H} = -G(\vec{r}_1,\vec{r}_2;E) \,, \\
& \frac{\mathcal{P}(n)}{E_n - E_{n'=n}} = \frac{|\psi_{n\l}|^{2}}{E - E_n} \,.
\end{align}
Therefore, as explained in \cite{Escobedo:2008sy} and followed in computations of, for instance, Refs.~\cite{Escobedo:2008sy,Kiyo:2014uca,Peset:2015zga}, one may explicitly substitute $G'(\vec{r}_1,\vec{r}_2)$ by $\lim\limits_{E \to E_n} - G(\vec{r}_1,\vec{r}_2;E)$ in Eq.~\eqref{eq:Connection1} and, subsequently one calculates with $\lambda=n$ the finite contribution of this expression, which means performing the sum without the pole. The divergent term of that sum is computed by using $\lambda=\frac{n}{\sqrt{1-\epsilon}}$, expanding in $\epsilon$ and finally picking up the finite term ($\O(\epsilon^{0})$ when $\epsilon \to 0$) only. This procedure heavily simplifies computations, but amounts to explicitly dropping a $\frac{1}{\epsilon}$ divergent term. However, since the $\mathcal{P}(n)$ part does not have any finite terms, this is allowed, since this term exactly cancels aforementioned divergence, as we demonstrate explicitly in Appendix~\ref{app:sec:DemonstrationDivergenceCancellation}. In terms of formulae one arrives at the simplified (with respect to computational effort) version of Eq.~\eqref{eq:Connection3}
\begin{equation}
\frac{1}{(E_n - H)'} = G'(\vec{r}_1,\vec{r}_2) = \left. - G(\vec{r}_1,\vec{r}_2;E) \right|_{\begin{array}{l} E \to E_n \\ \epsilon \to 0 , \O(\epsilon^0) \end{array}} .
\end{equation}
The method will be used to calculate the expectation values for double potential insertions in Appendix~\ref{app:sec:ExpectationValuesDoublePotentialInsertions} which are needed in order to compute the spectrum in Chapter~\ref{chp:QuarkoniumSpectrumPNRQCDWeakCouplingUpToMAlpha4}. Furthermore, we will generalize it in order to calculate matrix elements of general operators with different initial and final states as they appear in the calculations for the decay width in Chapter~\ref{chp:E1Transitions}.

\clearpage
\thispagestyle{empty}
\clearpage

\chapter{\texorpdfstring{The quarkonium spectrum in pNRQCD at weak coupling up to NNLO - {\boldmath$\O(m \alpha_s^4)$}}{The quarkonium spectrum in pNRQCD at weak coupling up to NNLO - O(m alphas**4)}}
\chaptermark{The quarkonium spectrum in pNRQCD at weak coupling up to NNLO - $\O(m \alpha_s^4)$}
\label{chp:QuarkoniumSpectrumPNRQCDWeakCouplingUpToMAlpha4}

Bottomonium was discovered as spin-triplet states called $\Upsilon(1S)$, $\Upsilon(2S)$ and $\Upsilon(3S)$ at Fermilab in 1977 in proton Cu and proton Pb fixed target scatterings \cite{Herb:1977ek,Innes:1977ae}. Later, the six triplet-$P$ states, $\chi_{bJ}(2P)$ and $\chi_{bJ}(1P)$ with $J=0,1,2$, have been discovered in radiative decays of the $\Upsilon(3S)$ and $\Upsilon(2S)$ in $1982$ \cite{Han:1982zk,Eigen:1982zm} and $1983$ \cite{Klopfenstein:1983nx,Pauss:1983pa}, respectively. They were the first $b\bar{b}$ states not directly produced in $e^{+}e^{-}$ collisions.\\
The description of hadrons containing two heavy quarks is a rather challenging problem from the point of view of QCD. A proper relativistic quantum field theoretical treatment of the heavy quarkonium system based on the Bethe-Salpeter equation has proved to be difficult and the two most promising approaches to the bottomonium bound state problem are Effective Field Theories (EFTs) and lattice gauge theories, see, e.g., Ref.~\cite{Vairo:2009rs}. Because of the heavy mass of the b-quark a very fine lattice discretization is needed such that this approach remains challenging and EFTs can be considered the most straight forward approach. Results for the $b\bar{b}$-spectrum have been obtained in perturbative QCD, e.g., in Refs.~\cite{Titard:1993nn,Titard:1994id,Pineda:1997hz} and a very recent and quite comprehensive computation within pNRQCD is given in Ref.~\cite{Kiyo:2014uca}.\\
In this chapter we compute the $b\bar{b}$-spectrum up to NNLO, $\O(m\alpha_s^4)$, in pNRQCD at weak coupling using the tools of perturbation theory derived in Chapter~\ref{chp:pNRQCDQuantumMechanicalPerturbationTheory}.

\section[\texorpdfstring{The $q\bar{q}$ mass spectrum up to NNLO - $\O(m \alpha_s^4)$}{The qqbar mass spectrum up to NNLO - O(m alphas**4)}]{\texorpdfstring{The {\boldmath$q\bar{q}$} mass spectrum up to NNLO - {\boldmath$\O(m \alpha_s^4)$}}{The qqbar mass spectrum up to NNLO - O(m alphas**4)}}
\label{sec:QbarQMassSpectrumUpToOrderMAlpha4}

To calculate the mass spectrum of the $q\bar{q}$-system we assume that the mass $M_{q\bar{q}}$ of the quarkonium under study can be calculated from the quark mass $m$, the leading order binding energy, being the exact solution of the Schrödinger equation~\eqref{eq:ZerothOrderSchroedingerEquation}, and from the corrections to the binding energy $\delta E$, c.f. Ref.~\cite{Kiyo:2014uca}, via
\begin{equation}
M_{q\bar{q}} = 2m + E_n^{LO} + \delta E^{NLO} + \dots \,,
\end{equation}
where the $\delta E$ are computable in perturbation theory. This approach, i.e. computing in perturbation theory, is valid only in the weak coupling regime of pNRQCD, since otherwise a perturbative approach would not be justified and one would need to make use of the Wilson loop approach, see Refs.~\cite{Brambilla:2000gk,Pineda:2000sz}.\\
As mentioned above, the leading order contribution, being the binding energy, is the trivial solution of the Schrödinger equation. Nevertheless, we recall that in quantum mechanics the virial theorem holds and thus the leading order binding energy is equivalent to $\frac{1}{2}$ times the expectation value of the leading order static potential, Eq.~\eqref{eq:LeadingOrderStaticPotential}, and we have
\begin{equation}
E_n^{LO} = E_n^{(0)} = -\frac{\mr C_F^2 \alpha_s^2}{2n^2} = \frac{1}{2} \langle n\l | -C_F \frac{\alpha_s}{r} | n\l \rangle \,.
\end{equation}
The NLO correction to the binding energy is given by the expectation value of the NLO term, $V_{a1}$, in the static potential~\eqref{eq:SingletStaticPotential}. There are no more contributions at this order in the power counting. We thus have, using the identity
\begin{align}
& \ln{\nu \e{\gamma_E} r} = \ln{\nu r} + \gamma_E \,, \\
&
\begin{aligned}
\Rightarrow & \langle n\l | \frac{\ln{\nu \e{\gamma_E} r}}{r} | n\l \rangle = \langle n\l | \frac{\ln{\nu r}}{r} | n\l \rangle + \langle n\l | \frac{\gamma_E}{r} | n\l \rangle \\
& \quad = \frac{1}{an^2} \left(\ln{\frac{an \nu}{2}} + S_1(n+\l)\right) \,,
\end{aligned}
\end{align}
and the expectation values given in Appendix~\ref{app:sec:ExpectationValuesSinglePotentialInsertions},
\begin{align}
\delta E_n^{NLO} &= (\delta E_{a1})^{(1)} = \langle n\l | -C_F \frac{\alpha_s}{r} \frac{\alpha_s}{4\pi} \left[ a_1 + 2\beta_0 \ln{\nu \e{\gamma_E} r} \right] | n\l \rangle \\
\nonumber &= -C_F \frac{\alpha_s^2}{4\pi} \left[ a_1 \langle n\l | \frac{1}{r} | n\l \rangle + 2\beta_0 \langle n\l | \frac{\ln{\nu \e{\gamma_E} r}}{r} | n\l \rangle \right] \\
\nonumber &= -C_F \frac{\alpha_s^2}{4\pi} \left[ \frac{a_1}{an^2} + \frac{2\beta_0}{an^2} \left(\ln{\frac{an \nu}{2}} + S_1(n+\l)\right) \right] \\
\nonumber &= -C_F \frac{\alpha_s^2}{4\pi an^2} \left[ a_1 + 2\beta_0 \left(\ln{\frac{an \nu}{2}} + S_1(n+\l)\right) \right] \,.
\end{align}
At NNLO we have three different types of contributions:\\
First, the single insertion of the potential $V_{a2}$, coming from the NNLO term in the static potential. We may now use the identity
\begin{align}
& \lnSquare{\nu \e{\gamma_E} r} = \lnSquare{\nu r} + 2\gamma_E \ln{\nu r} + \gamma_E^2 \,, \\
&
\begin{aligned}
\Rightarrow & \langle n\l | \frac{\lnSquare{\nu \e{\gamma_E} r}}{r} | n\l \rangle = \langle n\l | \frac{\lnSquare{\nu r}}{r} | n\l \rangle + \langle n\l | \frac{2\gamma_E \ln{\nu r}}{r} | n\l \rangle + \langle n\l | \frac{\gamma_E^2}{r} | n\l \rangle \\
& \quad = \frac{1}{an^2} \left[\lnSquare{\frac{an \nu}{2}} + 2 \psi(n+\l+1) \ln{\frac{an \nu}{2}} + \psi^2(n+\l+1) \right. \\
& \quad\quad + \psi'(n+\l+1) + \theta(n-\l-2) \frac{2\Gamma(n-\l)}{\Gamma(n+\l+1)} \sum\limits_{j=0}^{n-\l-2} \frac{\Gamma(2\l+2+j)}{j!(n-\l-1-j)^2} \\
& \quad\quad \left. + 2\gamma_E \left(\ln{\frac{an \nu}{2}} - \gamma_E + S_1(n+\l)\right) + \gamma_E^2 \right] \,,
\end{aligned}
\end{align}
where $\psi(z)$ and $\psi'(z) = \psi_1(z)$ are the Digamma function and the first Polygamma function, respectively (we give some of the important properties and relations in Appendix~\ref{app:subsec:GammaDigammaPolygamma}). The expectation values yielding above result are given in Appendix~\ref{app:sec:ExpectationValuesSinglePotentialInsertions} and the resulting correction to the energy is then given by
\begin{align}
(\delta E_{a2})^{(1)} & = \langle n\l | -C_F \frac{\alpha_s}{r} \left(\frac{\alpha_s}{4\pi}\right)^2 \left[ a_2 + \frac{\pi^2}{3} \beta_0^2 + (4a_1 \beta_0 + 2\beta_1) \ln{\nu \e{\gamma_E} r} + 4\beta_0^2 \lnSquare{\nu \e{\gamma_E} r} \right] | n\l \rangle \\
\nonumber & = -C_F \frac{\alpha_s^3}{(4\pi)^2} \left[ \left(a_2 + \frac{\pi^2}{3} \beta_0^2\right) \langle n\l | \frac{1}{r} | n\l \rangle + (4a_1 \beta_0 + 2\beta_1) \langle n\l | \frac{\ln{\nu \e{\gamma_E} r}}{r} | n\l \rangle \right. \\
\nonumber & \quad \left. + 4\beta_0^2 \langle n\l | \frac{\lnSquare{\nu \e{\gamma_E} r}}{r} | n\l \rangle \right] \\
\nonumber & = -C_F \frac{\alpha_s^3}{(4\pi)^2} \frac{1}{an^2} \left\lbrace \left(a_2 + \frac{\pi^2}{3} \beta_0^2\right) + (4a_1 \beta_0 + 2\beta_1) \left(\ln{\frac{an \nu}{2}} + S_1(n+\l)\right) \right. \\
\nonumber & \quad + 4\beta_0^2 \left[\lnSquare{\frac{an \nu}{2}} + 2 \psi(n+\l+1) \ln{\frac{an \nu}{2}} + \psi^2(n+\l+1) \right. \\
\nonumber & \quad + \psi'(n+\l+1) + \theta(n-\l-2) \frac{2\Gamma(n-\l)}{\Gamma(n+\l+1)} \sum\limits_{j=0}^{n-\l-2} \frac{\Gamma(2\l+2+j)}{j!(n-\l-1-j)^2} \\
\nonumber & \quad \left. \left. + 2\gamma_E \left(\ln{\frac{an \nu}{2}} - \gamma_E + S_1(n+\l)\right) + \gamma_E^2 \right] \right\rbrace \,.
\end{align}
Second, the double insertion of the potential $V_{a1}$, yielding
\begin{align}
(\delta E_{a_1})^{(2)} &= \langle n\l | V_{a1} \frac{1}{(E_n - H)'} V_{a1} | n\l \rangle \\
\nonumber & = \frac{C_F^2 \alpha_s^4}{(4\pi)^2} \left( a_1^2 \langle n\l | \frac{1}{r} \frac{1}{(E_n - H)'} \frac{1}{r} | n\l \rangle + 4a_1 \beta_0 \langle n\l | \frac{1}{r} \frac{1}{(E_n - H)'} \frac{\ln{\nu \e{\gamma_E} r}}{r} | n\l \rangle \right. \\
\nonumber & \quad \left. + 4\beta_0^2 \langle n\l | \frac{\ln{\nu \e{\gamma_E} r}}{r} \frac{1}{(E_n - H)'} \frac{\ln{\nu \e{\gamma_E} r}}{r} | n\l \rangle \right) \,.
\end{align}
We do not give the full expression for $(\delta E_{a_1})^{(2)}$ due to its length. However, we want to stress that all of the above corrections to the binding energy due to the radiative corrections to the static potential up to NNLO, $\O(m \alpha_s^4)$, may, in full agreement with Refs.~\cite{Kiyo:2014uca,Peset:2015vvi}, be cast as
\begin{equation}
\label{eq:EnergyCorrectionStaticPotentialNNLO} \delta E_{V_{\text{s}}^{(0)}}^{\text{NNLO}} = E_n^{(0)} \left( \frac{\alpha_s}{\pi} P_1(L_\nu) + \left(\frac{\alpha_s}{\pi}\right)^2 P_2^c(L_\nu) \right) \,,
\end{equation}
where
\begin{equation}
\label{eq:P1P2c} P_1(L_\nu) = \beta_0 L_\nu + \frac{a_1}{2} \quad\quad \text{and} \quad\quad P_2^c(L_\nu) = \frac{3}{4} \beta_0^2 L_\nu^2 + \left( -\frac{\beta_0^2}{2} + \frac{\beta_1}{4} + \frac{3\beta_0 a_1}{4} \right) L_\nu + c_2^c \,,
\end{equation}
and
\begin{align}
\label{eq:Lnu} & L_\nu = \ln{\frac{n \nu}{2\mr C_F \alpha_s}} + S_1(n+\l) \,, \\
\label{eq:c2c} & c_2^c = \frac{a_1^2}{16} + \frac{a_2}{8} - \frac{\beta_0 a_2}{4} + \beta_0^2 \left( \frac{n}{2} \zeta(3) + \frac{\pi^2}{8} \left( 1 - \frac{2n}{3} \Delta S_{1a} \right) - \frac{1}{2} S_2(n+\l) + \frac{n}{2} \Sigma_a(n,\l) \right) \,,
\end{align}
and the functions $\Delta S_{1a}$, $S_p$ and $\Sigma_a$ are defined in Appendix~\ref{app:sec:Functions}.\\
The third type of contribution entering at NNLO is due to the relativistic corrections encoded in $\delta H$, Eq.~\eqref{eq:DeltaHs}. These corrections yield, again using the expectation values for single potential insertions listed in Appendix~\ref{app:sec:ExpectationValuesSinglePotentialInsertions}, for the quartic correction to the kinetic energy
\begin{equation}
(\delta E_{\nabla_r^4})^{(1)} = \langle n\l | -\frac{\nabla_r^4}{4 \mr^3} | n\l \rangle = -\frac{1}{4 \mr^3} \frac{1}{(an)^4} \left(\frac{8n}{2\l+1} - 3\right) \,,
\end{equation}
for the $\frac{1}{m}$ correction
\begin{equation}
(\delta E^{(1)})^{(1)} = \langle n\l | -\frac{C_F C_A \alpha_s^2}{2 m r^2} | n\l \rangle = -\frac{C_F C_A \alpha_s^2}{2 m} \frac{2}{(2\l+1) a^2 n^3} \,,
\end{equation}
for the spin independent $\frac{1}{m^2}$ correction
\begin{align}
& (\delta E_r^{(2)})^{(1)} = \langle n\l | \frac{\pi C_F \alpha_s}{m^2} \delta^{(3)}(\vec{r}\,) | n\l \rangle = \frac{\pi \alpha_s C_F}{m^2} \frac{1}{\pi (an)^3} \delta_{\l 0} \,, \\
& (\delta E_{\lbrace V_{p^2}^{(2)},-\nabla_r^2 \rbrace}^{(2)})^{(1)} = \langle n\l | \frac{1}{2 m^2} \left\lbrace -\frac{C_F \alpha_s}{r},-\nabla_r^2 \right\rbrace | n\l \rangle = \frac{C_F \alpha_s}{m^2} \frac{1}{a^3} \left(\frac{1}{n^4} - \frac{4}{(2\l+1) n^3}\right) \,, \\
& (\delta E_{L^2}^{(2)})^{(1)} = \langle n\l | \frac{C_F \alpha_s}{2 m^2 r^3} \vec{L}^{\,2} | n\l \rangle = \frac{C_F \alpha_s}{2 m^2} \frac{2 (1 - \delta_{\l 0})}{\l (\l+1) (2\l+1) (an)^3} \chi_{L^2} \,,
\end{align}
and for the spin dependent $\frac{1}{m^2}$ correction
\begin{align}
& (\delta E_{LS}^{(2)})^{(1)} = \langle n\l | \frac{3 C_F \alpha_s}{2 m^2 r^3} \vec{L} \cdot \vec{S} | n\l \rangle = \frac{3 C_F \alpha_s}{2 m^2} \frac{2 (1 - \delta_{\l 0})}{\l (\l+1) (2\l+1) (an)^3} \chi_{LS} \,, \\
& (\delta E_{S^2}^{(2)})^{(1)} = \langle n\l | \frac{4\pi C_F \alpha_s}{3 m^2} \delta^{(3)}(\vec{r}\,) \vec{S}^{\,2} | n\l \rangle = \frac{4\pi C_F \alpha_s}{3 m^2} \frac{1}{\pi (an)^3} \delta_{\l 0} \chi_{S^2} \,, \\
& (\delta E_{S_{12}}^{(2)})^{(1)} = \langle n\l | \frac{C_F \alpha_s}{4 m^2 r^3} S_{12} | n\l \rangle = \frac{C_F \alpha_s}{4 m^2} \frac{2 (1 - \delta_{\l 0})}{\l (\l+1) (2\l+1) (an)^3} \chi_{S_{12}} \,.
\end{align}
Adding up these results and evaluating them straight forward gives very badly converging results, as presented in Table~\ref{tab:MassesWithoutRenormalonCorrection} below. This is due to the very bad convergence of the underlying perturbative series. The reason for this bad behavior is the existence of the large, $r$-independent contribution proportional to $\beta_{0}$ that deteriorates the convergence of the perturbative series. However, this issue can be cured by subtracting the first renormalon, as we will discuss in the following section.

\section{Renormalon effects and improved convergence of the perturbative series}

In perturbative calculations one usually expresses a physical observable $R$ as a series in a small parameter, e.g., the coupling $\alpha$, such that
\begin{equation}
R \sim \sum\limits_n r_n \alpha^n \,.
\end{equation}
However, if one encounters large contributions $r_{n}$, as it the case here where we find the large contribution proportional to $\beta_{0}$, this series expansion is divergent. The question arising is how to sum, hence how to assign a numerical value to, the series? The following closely follows Ref.~\cite{Beneke:1998ui}, where a detailed discussion of the whole subject is provided. As it is commonly done, we assume that the divergent series is a useful approximation to the quantity $R$ we want to study. A convenient way to sum the divergent series $R$ is Borel summation, where the Borel transformed $B[R]$ is defined as
\begin{equation}
R \sim \sum\limits_{n=0}^\infty r_n \alpha^{n+1} \quad\Rightarrow\quad B[R](t) = \sum\limits_{n=0}^\infty r_n \frac{t^n}{n!} \,,
\end{equation}
and, given that $B[R](t)$ has not singularities for $t \in \R_+$, and does not increase too rapidly at positive infinity, we can, for $\alpha > 0$, define the Borel integral as
\begin{equation}
\tilde{R} = \int\limits_0^\infty \d t \, \e{-\frac{t}{\alpha}} B[R](t) \,,
\end{equation}
that has the same series expansion as $R$. Thus $\tilde{R}$, if the integral exists, gives the Borel sum of the initially divergent series.\\
As an example consider that
\begin{equation}
r_n = K a^n \Gamma(n+1+b) \,.
\end{equation}
For $b \notin -\N$, not a negative integer, we have the Borel transform
\begin{equation}
B[R](t) = \frac{K \Gamma(1+b)}{(1-at)^{1+b}} \,,
\end{equation}
however, for $b = -m$, $m \in \N$, a negative integer the Borel transform
\begin{equation}
B[R](t) = \frac{(-1)^m}{\Gamma(m)} (1-at)^{m-1} \ln{1-at} + \text{ polynomial in } t
\end{equation}
has singularities for positive $t$ (assuming that the series is non-sign-alternating, hence $a > 0$) and the Borel integral does not exist. These singularities in the complex Borel plane are commonly referred to as renormalons. Getting rid of them, or at least of the first, in general significantly improves the convergence of the series under consideration. One way of achieving this goal is the usage of the so-called renormalon subtraction scheme (RS), developed in Ref.~\cite{Pineda:2001zq}.\\
We now return to the $q\bar{q}$ mass spectrum up to NNLO, $\O(m\alpha_s^4)$, and make use of the RS scheme. It was initially derived, confirmed and extended in Refs.~\cite{Pineda:1997hz,Pineda:1998ja,Melnikov:1998ug,Titard:1993nn,Titard:1994id}. Numerical results can be found in Ref.~\cite{Brambilla:2001fw} and we will follow their choice of parameters.\\
We use the abbreviations $m = m_{\text{pole}}$ for the pole mass, satisfying $\mr = \frac{m}{2}$, $\alpha_s = \alpha_s(\nu)$ for the strong fine structure constant in the $\bar{\text{MS}}$-scheme, evaluated at the scale $\nu$, and $\bar{m} = m_{\bar{\text{MS}}}(m_{\bar{\text{MS}}})$ for the $\bar{\text{MS}}$-mass. The mass of a heavy quarkonium state, identified by the quantum numbers $n$, $\l$, $s$ and $J$, is given by
\begin{equation}
\label{eq:Mqq} M_{q\bar{q}}(\nu,\alpha_s,m) = 2m + E_{\text{bin}}(\nu,\alpha_s,m) \,,
\end{equation}
with the binding energy given by
\begin{equation}
\label{eq:Ebin} E_{\text{bin}}(\nu,\alpha_s,m) = -\frac{\mr C_F^2 \alpha_s^2}{2n^2} \sum\limits_{k=0}^2 \epsilon^{k+1} \left(\frac{\alpha_s}{\pi}\right)^k P_{k}(L_\nu) \,,
\end{equation}
where $\epsilon=1$ is the parameter that will be used in order to properly organize the perturbative expansion in view of the $\O(\Lambda_{\text{QCD}})$ renormalon cancellation. In Eq.~\eqref{eq:Mqq} the pole mass $m$ as well as the static potential that enters the binding energy $E_{\text{bin}}$ suffer from renormalons. However, the heavy quark mass $M_{q\bar{q}}$ is a physical observable and thus free of renormalons. The renormalon cancellation therefore takes place between the pole mass and the static potential (see also the discussion, e.g., in Refs.~\cite{Brambilla:2001fw,Pineda:2013lta}). $P_{k}(L_\nu)$ is a $k$-th degree polynomial of the function $L_\nu$, defined in Eq.~\eqref{eq:Lnu}. It is convenient to decompose the polynomials into renormalization-group invariant subsets $P_0 = 1$ and $P_{1,2}$ as defined in Eq.~\eqref{eq:P1P2c} with the substitution $c_2^c \to c_2^c + c_2^{n c}$, where
\begin{equation}
c_2^{n c} = \pi^2 C_F^2 \left[\frac{2}{n (2\l + 1)} - \frac{11}{16 n^2} - \frac{\frac{1}{2} \chi_{S_{12}} + 3 \chi_{LS}}{2n\l (\l + 1)(2\l + 1)} \delta_{\l \geq 1} - \frac{2}{3n} \chi_{S^2} \delta_{\l0} \right] + \frac{\pi^2 C_F C_A}{n (2\l + 1)} \,,
\end{equation}
with $c_2^c$ as defined in Eq.~\eqref{eq:c2c} and the expectation values $\chi$ as defined in Eq.~\eqref{eq:ExpectationValuesPhysicalOperators}.\\
Next, we rewrite the series expansion of $M_{q\bar{q}}$ in terms of the $\bar{\text{MS}}$-mass. This is done by expressing the pole masses $m$ in terms of the renormalization--group-invariant $\bar{\text{MS}}$-mass $\bar{m}$ as
\begin{equation}
\label{eq:mpole} m = \bar{m} \left[ 1 + \frac{4}{3} \epsilon \frac{\alpha_s(\bar{m})}{\pi} + \epsilon^2 \left(\frac{\alpha_s(\bar{m})}{\pi}\right)^2 d_1 + \epsilon^3 \left(\frac{\alpha_s(\bar{m})}{\pi}\right)^3 d_2 \right] \,,
\end{equation}
where the coefficients $d_{1,2}$ were derived in Ref.~\cite{Brambilla:2001qk} and are given in Appendix~\ref{app:sec:Constants}.\\
We want to stress that the counting in $\epsilon$ in Eq.~\eqref{eq:Ebin} and Eq.~\eqref{eq:mpole} does not reflect the order in $\alpha_{s}$ but the wanted renormalon cancellation. One way to understand this is to consider that in the sum of the pole-quark masses and the static QCD potential, $2m + V_{\text{QCD}}(r)$, the renormalon cancellation takes place without reordering of the power counting in $\alpha_{s}$. Moreover, in order to realize the renormalon cancellation at each order of the expansion, it is necessary to expand $m$ and $E_{\text{bin}}$ in the same coupling. Therefore we express $\alpha_s(\bar{m})$ in Eq.~\eqref{eq:mpole} in terms of $\alpha_s$ as
\begin{equation}
\label{eq:alphasmbar} \alpha_s(\bar{m}) = \alpha_s \left\lbrace 1 + \epsilon \frac{\alpha_s}{\pi} \frac{\beta_0}{2} \ln{\frac{\nu}{\bar{m}}} + \epsilon^2 \left(\frac{\alpha_s}{\pi}\right)^2 \left[ \frac{\beta_0^2}{4} \ln{2}{\frac{\nu}{\bar{m}}} + \frac{\beta_1}{8} \ln{\frac{\nu}{\bar{m}}} \right] \right\rbrace \,. 
\end{equation}
Substituting the Eqs.~\eqref{eq:alphasmbar} and \eqref{eq:mpole} into the Eqs.~\eqref{eq:Ebin} and \eqref{eq:Mqq}, we obtain an expression for the mass of the heavy quarkonium states, which depends on $\nu$, $\alpha_s$ and $\bar{m}$, that we can organize as an expansion in $\epsilon$ up to $\O(\epsilon^3)$:
\begin{equation}
\label{eq:binqqEps} M_{q\bar{q}}(\nu,\alpha_s,\bar{m}) = 2\bar{m} + M_{q\bar{q}}^{(1)}(\nu,\alpha_s,\bar{m}) \epsilon + M_{q\bar{q}}^{(2)}(\nu,\alpha_s,\bar{m}) \epsilon^2 +
M_{q\bar{q}}^{(3)}(\nu,\alpha_s,\bar{m}) \epsilon^{3} \,.
\end{equation}
Since the counting in $\epsilon$ explicitly realizes the order $\Lambda_{\text{QCD}}$ renormalon cancellation, and since $\alpha_s$ and $\bar{m}$ are short-range quantities, the obtained perturbative expansion~\eqref{eq:binqqEps} is expected to show a better convergence with respect to the original expansion~\eqref{eq:Ebin}. Due to our incomplete knowledge of the perturbative series, the obtained quarkonium mass $M_{q\bar{q}}$ depends on the scale $\nu$. The scale $\nu$ is normally fixed by demanding stability against variation of the scale:
\begin{equation}
\label{eq:ScaleFixing} \left.\frac{\d}{\d\nu} M_{q\bar{q}}(\nu,\alpha_s,\bar{m})\right|_{\nu = \nu_{q\bar{q}}} = 0 \,,
\end{equation}
which then determines the scale $\nu_{q\bar{q}}$ of minimal sensitivity. When this is done, one expects that the convergence properties of the series become optimal, and the scale should becomes close to the inverse of the physical size of the $q\bar{q}$ bound state. However, if the scale fixed by Eq.~\eqref{eq:ScaleFixing} evidently does not fulfill these expectations, the theoretical predictions obtained in this way should be considered unreliable. This typically happens when the coupling constant becomes bigger than one and can be seen, together with the other results, in Table~\ref{tab:MassesIncludingRenormalonCorrection}. We furthermore, as already mentioned above, display the results without the renormalon correction, in a way one would obtain them following the previous section, in Table~\ref{tab:MassesWithoutRenormalonCorrection}.\\
Since perturbation theory is considered to work best for low lying states, we take the $b\bar{b}$ ground state to fix $\bar{m} = \bar{m}_b = \bar{m}_{\bar{b}}$ by demanding
\begin{equation}
M_{\Upsilon(1S)}(\nu,\alpha_s,\bar{m}) = M_{\Upsilon(1S)}^{\text{exp.}} = 9.460~\text{GeV} \,,
\end{equation}
where the experimental mass of the $\Upsilon(1S)$ has been taken from 
Ref.~\cite{PDG:2016} (PDG). This allows to determine the $b$-quark 
$\bar{\text{MS}}$-mass to be
\begin{equation}
\bar{m} = m_b^{\bar{\text{MS}}}(m_b^{\bar{\text{MS}}}) = 4.203~\text{GeV} \,.
\end{equation}

\begin{table}[t]
\centering
\caption[Numerical results for the $b\bar{b}$ mass spectrum including the 
renormalon correction.]{Numerical results for the $b\bar{b}$ mass spectrum up 
to $\O(m\alpha_s^4)$ including the renormalon correction, adapted from 
\cite{Brambilla:2001fw}. All masses and scales are given in GeV. The 
experimental values $M_{q\bar{q}}^{\text{exp}}$ are taken from \cite{PDG:2014}, 
$M_{q\bar{q}}$ is the theoretical value and $\Delta M_{q\bar{q}}$ denotes the 
difference between them. The $M_{q\bar{q}}^{(i)}$, $i=1,2,3$, denote the 
contributions according to Eq.~\eqref{eq:binqqEps} and the scale $\nu$ is fixed 
according to Eq.~\eqref{eq:ScaleFixing}.}
\label{tab:MassesIncludingRenormalonCorrection}
\begin{tabular}{c|c|c|c|c|c|c|c|c|c}
$q\bar{q}$-state & $n^{2s+1}\l_J$ & $M_{q\bar{q}}^{\text{exp}}$ & $\Delta M_{q\bar{q}}$ & $M_{q\bar{q}}$ & $M_{q\bar{q}}^{(1)}$ & $M_{q\bar{q}}^{(2)}$ & $M_{q\bar{q}}^{(3)}$ & $\nu$ & $\alpha_s(\nu)$ \\
\hline
$\Upsilon(1^3S_1)$ & $1^3S_1$ & 9.460 & 0 & 9.460 & 0.837 & 0.204 & 0.013 & 2.49 & 0.274 \\
$\chi_{b0}(1^3P_0)$ & $2^3P_0$ & 9.859 & -0.046 & 9.905 & 1.381 & 0.115 & 0.003 & 1.18 & 0.409 \\
$\chi_{b1}(1^3P_1)$ & $2^3P_1$ & 9.892 & -0.017 & 9.909 & 1.403 & 0.098 & 0.002 & 1.15 & 0.416 \\
$\chi_{b2}(1^3P_2)$ & $2^3P_2$ & 9.912 & -0.004 & 9.916 & 1.422 & 0.086 & 0.002 & 1.13 & 0.422 \\
$\Upsilon(2^3S_1)$ & $2^3S_1$ & 10.023 & 0.058 & 9.965 & 1.457 & 0.093 & 0.009 & 1.09 & 0.433 \\
$\chi_{b0}(2^3P_0)$ & $3^3P_0$ & 10.232 & -0.030 & 10.262 & 2.366 & -0.658 & 0.154 & 0.693 & 0.691 \\
$\chi_{b1}(2^3P_1)$ & $3^3P_1$ & 10.255 & -0.053 & 10.308 & 3.982 & -3.590 & 1.516 & 0.552 & 1.20 \\
$\chi_{b2}(2^3P_2)$ & $3^3P_2$ & 10.268 & -0.181 & 10.449 & 4.558 & -5.052 & 2.543 & 0.537 & 1.39 \\
$\Upsilon(3^3S_1)$ & $3^3S_1$ & 10.355 & 0.033 & 10.322 & 2.343 & -0.584 & 0.163 & 0.698 & 0.684 \\
$\Upsilon(4^3S_1)$ & $4^3S_1$ & 10.579 & -1.176 & 11.755 & 5.441 & -6.460 & 4.374 & 0.527 & 1.61
\end{tabular} 
\end{table}
\begin{table}[t]
\centering
\caption[Numerical results for the $b\bar{b}$ mass spectrum without the renormalon correction.]{Numerical results for the $b\bar{b}$ mass spectrum up to $\O(m\alpha_s^4)$ without the renormalon correction. All masses and scales are given in GeV. The scale $\nu$ is chosen to be the same as in Table~\ref{tab:MassesIncludingRenormalonCorrection}.}
\label{tab:MassesWithoutRenormalonCorrection}
\begin{tabular}{c|c|c|c|c|c|c|c|c|c}
$q\bar{q}$-state & $n^{2s+1}\l_J$ & $M_{q\bar{q}}^{\text{exp}}$ & $\Delta M_{q\bar{q}}$ & $M_{q\bar{q}}$ & $M_{q\bar{q}}^{(1)}$ & $M_{q\bar{q}}^{(2)}$ & $M_{q\bar{q}}^{(3)}$ & $\nu$ & $\alpha_s(\nu)$ \\
\hline
$\Upsilon(1^3S_1)$ & $1^3S_1$ & 9.460 & 1.618 & 7.842 & -0.140 & -0.220 & -0.204 & 2.49 & 0.274 \\
$\chi_{b0}(1^3P_0)$ & $2^3P_0$ & 9.859 & 2.173 & 7.686 & -0.078 & -0.223 & -0.418 & 1.18 & 0.409 \\
$\chi_{b1}(1^3P_1)$ & $2^3P_1$ & 9.892 & 2.222 & 7.670 & -0.081 & -0.230 & -0.425 & 1.15 & 0.416 \\
$\chi_{b2}(1^3P_2)$ & $2^3P_2$ & 9.912 & 2.257 & 7.655 & -0.083 & -0.236 & -0.432 & 1.13 & 0.422 \\
$\Upsilon(2^3S_1)$ & $2^3S_1$ & 10.023 & 2.353 & 7.670 & -0.088 & -0.206 & -0.443 & 1.09 & 0.433 \\
$\chi_{b0}(2^3P_0)$ & $3^3P_0$ & 10.232 & 3.621 & 6.611 & -0.099 & -0.387 & -1.308 & 0.693 & 0.691 \\
$\chi_{b1}(2^3P_1)$ & $3^3P_1$ & 10.255 & 10.098 & 0.157 & -0.299 & -1.099 & -6.850 & 0.552 & 1.20 \\
$\chi_{b2}(2^3P_2)$ & $3^3P_2$ & 10.268 & 14.512 & -4.244 & -0.401 & -1.386 & -10.860 & 0.537 & 1.39 \\
$\Upsilon(3^3S_1)$ & $3^3S_1$ & 10.355 & 3.634 & 6.721 & -0.097 & -0.324 & -1.264 & 0.698 & 0.684 \\
$\Upsilon(4^3S_1)$ & $4^3S_1$ & 10.579 & 20.779 & -10.200 & -0.303 & -1.409 & -16.890 & 0.527 & 1.61
\end{tabular} 
\end{table}
The difference of the numerical results is remarkable. Subtracting the first renormalon (Tab.~\ref{tab:MassesIncludingRenormalonCorrection}) allows for a fairly accurate determination of the low lying spectrum. The deviation from the experimental values are around several MeV at NNLO such that the relative error is of the order of several percent or even at the per-mill level. The scale of minimal sensitivity with respect to Eq.~\eqref{eq:ScaleFixing} decreases with higher radial excitations in the spectrum and correspondingly $\alpha_s(\nu)$ gets closer or even exceeds one. These are the predictions one has to consider unreliable and also at the same time the difference between the theoretical prediction and the experimental value increases drastically, yielding a relative error of up to 10\%.\\
In the case where the first renormalon is not subtracted (Tab.~\ref{tab:MassesWithoutRenormalonCorrection}), the ground state prediction already has an error of about 10\%. Note, however, that we did not re-fix the scale according to Eq.~\eqref{eq:ScaleFixing}, but for comparison used the same values that have been used for the renormalon subtracted case. Re-fixing the scale might provide a slight improvement, however, we do not expect result even close to the ones including the renormalon subtraction. It is also worth mentioning that in the case without the renormalon subtraction, the theoretical prediction for $M_{q\bar{q}}$ has a very strong dependence on the scale $\nu$ and on $\alpha_s(\nu)$, yielding a zero crossing and subsequently negative masses for $\nu \lesssim 0.55$ and thus $\alpha_s(\nu) \gtrsim 1.2$, which is completely unphysical. We expect that even a redetermination according to Eq.~\eqref{eq:ScaleFixing} would not cure this behavior and at best only milden it. We therefore conclude that subtracting at least the leading renormalon is crucial in order to obtain reliable results.

\clearpage
\thispagestyle{empty}
\clearpage

\chapter{\texorpdfstring{E1 transitions in pNRQCD at weak coupling up to NNLO - {\boldmath$\O(m \alpha_s^6)$}}{E1 transitions in pNRQCD at weak coupling up to NNLO - O(m alphas**6)}}
\chaptermark{E1 transitions in pNRQCD at weak coupling up to NNLO - $\O(m \alpha_s^6)$}
\label{chp:E1Transitions}

In this chapter we derive the general framework and the expressions in order to compute electric dipole (E1) transitions in pNRQCD at weak coupling. Therefore we first introduce the relevant objects, which are the decay width and the electric dipole operator, clarify the notation, and compute the leading order decay width. After that, we incorporate higher order operators and extend the formalism of quantum mechanical perturbation theory, derived in Chapter~\ref{chp:pNRQCDQuantumMechanicalPerturbationTheory}, to account for different initial and final states and subsequently compute first and second order corrections to them, which are induced by higher order potentials. We finally discuss color octet contributions and explicitly compute the impact on the renormalization of the wave function, assuming a local two gluon condensate.

\section{Essentials}
\label{sec:Essentials}

\begin{figure}[t]
\centering
{\includegraphics[
clip,trim={6.4cm 19.2cm 3.2cm 3.2cm},width=0.6\textwidth]{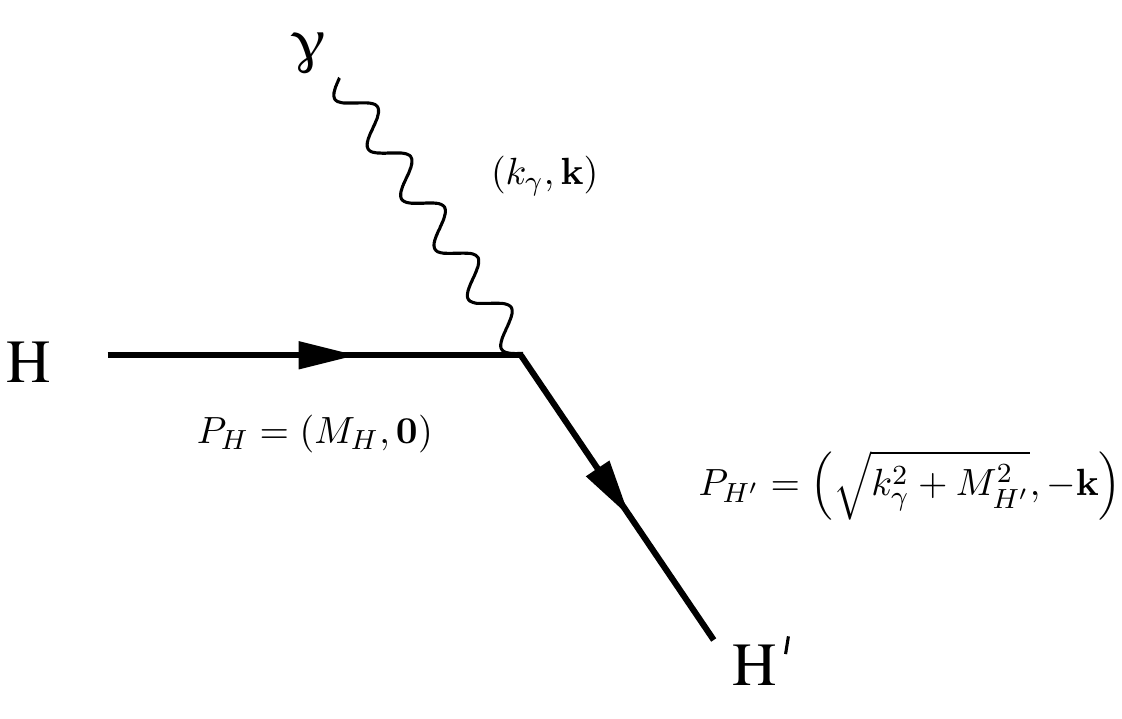}}
\caption[Kinematics of the decay of the quarkonium $H$ into the quarkonium $H'$ and a photon.]{Kinematics of the decay of the excited quarkonium $H$ in its rest frame into the quarkonium $H'$ and a photon $\gamma$, taken from \cite{Brambilla:2005zw}. Energy and spatial momentum of the decay products are determined by conservation of energy and momentum.}
\label{fig:kinematics}
\end{figure}

The kinematics of the decay process of an excited quarkonium $H$ into the quarkonium $H'$ under the emission of a photon $\gamma$ is depicted in Fig.~\ref{fig:kinematics} and the corresponding differential decay width is given by \cite{Peskin:IntroductionQuantumFieldTheory}:
\begin{equation}
\d\Gamma = \frac{(2\pi)^4 \delta^{(4)}(p_f + p_\gamma - p_i)}{2M_i} \bar{|\M_{fi}|^2} \frac{\d^3 p_f}{(2\pi)^3 (2E_f)} \frac{\d^3 p_\gamma}{(2\pi)^3 (2E_\gamma)} \,.
\end{equation}
Integration yields
\begin{align}
& \Gamma = \frac{1}{(2\pi)^2} \frac{1}{2M_i} \bar{|\M_{fi}|^2} \int \delta^{(4)}(p_f + p_\gamma - p_i) \frac{\d^3 p_f}{(2E_f)} \frac{\d^3 p_\gamma}{(2E_\gamma)} \\
\nonumber & = \frac{1}{(2\pi)^2} \frac{1}{2M_i} \frac{1}{2M_f} \left(1-\frac{1}{2}\frac{k_\gamma^2}{M_f^2}\right) \bar{|\M_{fi}|^2} \int \delta^{(1)}(p_\gamma - k_\gamma) \frac{\d p_\gamma}{2p_\gamma} p_\gamma^2 \d\Omega_{p_\gamma} \\
\nonumber & = \frac{1}{(2\pi)^2} \frac{1}{2M_i} \frac{1}{2M_f} \left(1-\frac{1}{2}\frac{k_\gamma^2}{M_f^2}\right) \bar{|\M_{fi}|^2} 4\pi \frac{k_\gamma}{2} \\
\nonumber & \simeq \frac{1}{(2\pi)} \frac{1}{2M_i} \frac{1}{2M_f} \bar{|\M_{fi}|^2} k_\gamma \,,
\end{align}
where we used the kinematic relations
\begin{equation}
\vec{k} = \vec{p}_\gamma = - \vec{p}_f \quad \text{ and } \quad \vec{p}_i = \vec{0} \,.
\end{equation}
The averaged invariant matrix element
\begin{equation}
\bar{|\M_{fi}|^2} = \frac{1}{N_\lambda} \sum\limits_{\lambda,\lambda',\sigma} |\M_{fi}|^2
\end{equation}
is given as the polarization average over the initial state quarkonium (polarization $\lambda$) and the polarization sum over the final state quarkonium and photon (polarizations $\lambda'$ and $\sigma$, respectively) of the invariant amplitude $|\M_{fi}|^2$. The term $-\frac{1}{2}\frac{k_\gamma^2}{M_f^2}$ is a negligible relativistic correction and can be disregarded, since it is suppressed. Furthermore, using non-relativistic normalizations for the bound states, Eq.~\eqref{eq:StateNormalization}, we can get rid of the factors $\frac{1}{2M_i}$ and $\frac{1}{2M_f}$. The photon obeys the Lorentz invariant normalization
\begin{equation}
\langle \gamma(\vec{k}',\sigma') | \gamma(\vec{k},\sigma) \rangle = 2k (2\pi)^3 \delta^{(3)}(\vec{k} - \vec{k}') \delta_{\sigma\sigma'} \,,
\end{equation}
and we arrive at the decay rate formula
\begin{equation}
\Gamma = \frac{k_\gamma}{(2\pi)} \bar{|\M_{fi}|^2} \,.
\end{equation}
From now on, we shall extend the notation of initial, $| n\l \rangle$, and final, $| n'\l' \rangle$, states, respectively, whenever it is necessary in order to make all relevant quantum numbers explicit, such that
\begin{align}
& | n\l \rangle \to | n ; \l , m_\l ; s , m_s ; J , m_J ; 0 \rangle = | n ; \l , m_\l ; s , m_s ; J , m_J \rangle \otimes | 0 \rangle \,, \\
& | n'\l' \rangle \to | n' ; \l' , m_{\l'} ; s' , m_{s'} ; J' , m_{J'} ; \gamma \rangle = | n' ; \l' , m_{\l'} ; s' , m_{s'} ; J' , m_{J'} \rangle \otimes | \gamma \rangle \,.
\end{align}
The respective initial and final Fock states may be decomposed into one particle Hilbert states containing only a quarkonium and a photon (or no photon in case of the initial state), respectively. However, as indicated by the direct product, $\otimes$, this decomposition is not arbitrary, but must follow the standard Clebsch-Gordan decomposition, since total angular momentum is conserved. Decomposing the electric field into creation and annihilation operators yields the important relation
\begin{equation}
\langle \gamma(\vec{k},\sigma) | \vec{E}^{\,e/m} | 0 \rangle = -\i k \hat{\epsilon}^*(\sigma) \e{-\i \vec{k} \cdot \vec{R}} \,.
\end{equation}
The electric dipole operator may be extracted from the dipole term of the pNRQCD Lagrangian~\eqref{eq:GammapNRQCDLagrangian},
\begin{equation}
\label{eq:LeadingOrderGammapNRQCDLagrangian} \L_{\text{E1}} = e e_Q \int \d^3r \, \text{tr}\lbrace S^\dagger \vec{r} \cdot \vec{E}^{\,e/m} S \rbrace \,,
\end{equation}
and is thus given by
\begin{equation}
\OEone = e e_Q (\vec{r} \cdot \vec{E}^{\,e/m}) \,.
\end{equation}
It acts as the identity operator in spin-space, since it does not produce spin flips, and thus
\begin{equation}
\langle s' , m_{s'} | \OEone | s , m_s \rangle = \delta_{s s'} \delta_{m_s m_{s'}} \,.
\end{equation}
The matrix element involving the electric dipole operator may be decomposed into
\begin{equation}
\label{eq:DecompositionMatrixElement} \langle n' ; \l' ; s' ; J' ; \gamma | \OEone | n ; \l ; s ; J ; 0 \rangle = e e_Q \langle n' ; \l' ; s' ; J' | \vec{r} | n ; \l ; s ; J \rangle \cdot \langle \gamma | \vec{E} | 0 \rangle \,,
\end{equation}
and one can, following \cite{Varshalovich:QuantumTheoryAngularMomentum}, represent $\vec{r}$ in spherical harmonics
\begin{equation}
\label{eq:DecompositionVecR} (\vec{r}\,)_\mu = \sqrt{\frac{4\pi}{3}} \, r \, Y_1^{\mu*}(\Omega_r) \,,
\end{equation}
where $\mu = 0,\pm 1$. Note that this decompositions~\eqref{eq:DecompositionMatrixElement} and \eqref{eq:DecompositionVecR} implicitly take care of the needed Clebsch-Gordan decomposition. Since we have
\begin{equation}
\langle \vec{r} | n ; \l , m ; s , m_s ; J , m_J \rangle = \Phi_{n\l m s m_{s} J m_{J}}(\vec{r},\lambda) = \psi_{n\l m}(\vec{r}\,) \cdot \chi_{s m_{s} J m_{J}}(\lambda) \,,
\end{equation}
with the physical quarkonium state $\Phi$, the leading order matrix element encoding the E1 transition may be cast as
\begin{align}
\M_{\text{E1}}^{(0)} &= \langle n' ; \l' ; s' ; J' ; \gamma | \OEone | n ; \l ; s ; J ; 0 \rangle \\
&= e e_Q \sqrt{\frac{4\pi}{3}} \sum\limits_{\mu = -1}^1 \int \d^3 r \, \psi_{n'\l'm_{\l'}}(\vec{r}\,) \, r \, Y_1^{\mu*}(\Omega_r) \psi_{n\l m_{\l}}(\vec{r}\,) \\
\nonumber & \quad \times \chi_{s' m_{s'} J' m_{J'}}(\lambda')^* \chi_{s m_{s} J m_{J}}(\lambda) \times \hat{e}_r \cdot \langle \gamma | \vec{E} | 0 \rangle \\
\label{eq:SplittingLeadingOrderE1} & = e e_Q \sqrt{\frac{4\pi}{3}} I_3^{(0)}(n\l \to n'\l') \\
\nonumber & \quad \times \sum\limits_{m_{\l'} = -\l'}^{\l'} \sum\limits_{\mu = -1}^1 \sum\limits_{m_{\l} = -\l}^\l \int \d\Omega_r Y_{\l'}^{m_{\l'}*}(\Omega_r) Y_1^{\mu*}(\Omega_r) Y_\l^{m_{\l}}(\Omega_r) \\
\nonumber & \quad \times \chi_{s' m_{s'} J' m_{J'}}(\lambda')^* \chi_{s m_{s} J m_{J}}(\lambda) \times \hat{e}_r \cdot \langle \gamma | \vec{E} | 0 \rangle \,,
\end{align}
which is the most general form of this matrix element. Note that the orbital angular momentum integral can be computed using the relations for spherical harmonics, Appendix~\ref{app:subsec:SphericalHarmonics}, yielding
\begin{equation}
\int \d\Omega \, Y_{\l'}^{m_{\l'}*}(\Omega) Y_1^{\mu*}(\Omega) Y_\l^{m_\l}(\Omega) = (-1)^{m_{\l'}+\mu} \sqrt{\frac{(2\l'+1)(2\l+1) 3}{4\pi}} \begin{pmatrix} \l' & \l & 1 \\ 0 & 0 & 0 \end{pmatrix} \begin{pmatrix} \l' & \l & 1 \\ -m_{\l'} & m_{\l} & -\mu \end{pmatrix} \,.
\end{equation}
The first of the two Wigner $3j$-symbols (see Appendix~\ref{app:subsec:WignerSymbol} on how to compute them) can be evaluated easily
\begin{equation}
\begin{pmatrix} \l' & \l & 1 \\ 0 & 0 & 0 \end{pmatrix} = \left\lbrace
\begin{array}{cl}
(-1)^{-\l'} \sqrt{\frac{(\l'+\l'^2-\l-\l^2)^2}{(\l'+\l)(1+\l'+\l)(2+\l'+\l)(1+\l'-\l)!(1-\l'+\l)!}} & \text{ for }
\begin{array}{l}
\l' \leq 1 + \l \\
\& \, \l \leq 1 + \l' \\
\& \, \l' + \l \geq 1 \\
\& \, \l,\l' \geq 0
\end{array} \\
0 & \text{ else}
\end{array} \right.
\end{equation}
yielding the very important result
\begin{equation}
\l' = \l \pm 1 \quad \Leftrightarrow \quad \l = \l' \pm 1 \,,
\end{equation}
which is the defining relation for initial and final state orbital angular momenta with respect to electric dipole transitions. In the next section we derive the leading order decay width in two different ways and therewith show their equivalence. Furthermore, a comparison between these two methods will allow us to define the radial matrix element and therewith significantly simplify the computational effort once we consider higher order corrections to initial and final states.

\section{The leading order decay width}
\label{sec:LeadingOrderDecayWidth}

The leading order E1 decay width is given by
\begin{equation}
\Gamma_{\text{E1}}^{(0)} = \frac{k_\gamma}{(2\pi)} \bar{|\M_{\text{E1}}^{(0)}|^2} \,,
\end{equation}
and one might use the definitions of the different wave functions, Eqs.~\eqref{eq:BeginWaveFunctionDefinition}~-~\eqref{eq:EndWaveFunctionDefinition}, together with
\begin{align}
& \sum\limits_\lambda \hat{e}^{i\,*}_{n\,^{2s+1}\!\l_J}(\lambda) \cdot \hat{e}^{j}_{n\,^{2s+1}\!\l_J}(\lambda) = \delta^{i j} \,, \\
& \sum\limits_\lambda h^{ij\,*}_{n\,^{2s+1}\!\l_J}(\lambda) \cdot h^{kl}_{n\,^{2s+1}\!\l_J}(\lambda) = \frac{1}{2} (\delta^{i k} \delta^{j l} + \delta^{i l} \delta^{j k}) - \frac{1}{3} \delta^{i j} \delta^{kl} \,, \\
& \sum\limits_\sigma \hat{\epsilon}^{i\,*}(\sigma) \cdot \hat{\epsilon}^{j}(\sigma) = \delta^{i j} - \hat{k}^i \hat{k}^j \,,
\end{align}
and the polarization sums witch can be derived from these relations, yielding
\begin{align}
& \sum\limits_{\lambda',\sigma} |\hat{\epsilon}^*(\sigma) \cdot \hat{e}_{n'\,^3\!S_1}(\lambda')|^2 = 2 \,, \\
& \sum\limits_{\lambda,\lambda',\sigma} |\hat{\epsilon}^*(\sigma) \cdot (\hat{e}_{n\,^3\!P_1}(\lambda) \times \hat{e}_{n'\,^3\!S_1}(\lambda'))|^2 = 4 \,, \\
& \sum\limits_{\lambda,\lambda',\sigma} |\hat{e}^i_{n'\,^3\!S_1}(\lambda') h^{i j}_{n\,^3\!P_2}(\lambda) \hat{\epsilon}^{j\,*}(\sigma)|^2 = \frac{10}{3} \,,
\end{align}
in order to arrive at the $J$ independent result
\begin{equation}
\label{eq:LeadingOrderE1_2} \Gamma_{\text{E1}}^{(0)} = \frac{4}{9} \alpha_{e/m} e_Q^2 k_\gamma^3 \left[I_3^{(0)}(n1 \to n'0)\right]^2 \,.
\end{equation}
The same result can also be obtained by following an approach that does not depend on a certain choice of the functions $\chi$, but makes use of the common technique of projecting the $z$-component $m_J$ of the total angular momentum $J$ such that it is maximal. Doing so for all initial and final states and taking into account that $J_{\text{total}}$ is conserved, one can fix the spin wave function. In order to obtain results, we go back to the expression for the matrix element that, now explicitly omitting the functions $\chi$, may be cast as
\begin{align}
\M_{\text{E1}}^{(0)} &= \langle n' ; \l' ; s' ; J' ; \gamma | \OEone | n ; \l ; s ; J ; 0 \rangle = e e_Q \sqrt{\frac{4\pi}{3}} I_3^{(0)}(n\l \to n'\l') \\
\nonumber & \quad \times \sum\limits_{m_{\l'} = -\l'}^{\l'} \sum\limits_{\mu = -1}^1 \sum\limits_{m_{\l} = -\l}^\l (-1)^{m_{\l'}+\mu} \sqrt{\frac{(2\l'+1)(2\l+1) 3}{4\pi}} \begin{pmatrix} \l' & \l & 1 \\ 0 & 0 & 0 \end{pmatrix} \begin{pmatrix} \l' & \l & 1 \\ -m_{\l'} & m_{\l} & -\mu \end{pmatrix} \\
\nonumber & \quad \times \hat{e}_r \cdot \langle \gamma | \vec{E} | 0 \rangle \,.
\end{align}
We will now explicitly compute the leading order decay width for each of the three processes for $n\,^3\!P_{J=0,1,2} \to n'\,^3\!S_1 + \gamma$:
\begin{itemize}
\item We start with the process $n\,^3\!P_2 \to n'\,^3\!S_1 + \gamma$. The quantum numbers of the initial state quarkonium are given by
\begin{equation}
\l = 1 \,, \quad s = 1 \,, \quad  J = 2 \,.
\end{equation}
Choosing the maximal projection with respect to $J$, yields the projections of the quantum numbers
\begin{equation}
m_J = 2 \quad \Rightarrow \quad m_\l = 1 \,, \quad m_s = 1 \,,
\end{equation}
and the corresponding wave function is thus given by
\begin{equation}
| n\,^3\!P_2 \rangle = | n ; 1 , 1 ; 1 , 1 ; 2 , 2 \rangle = R_{n1}(r) Y_1^1(\Omega_r) | \uparrow \uparrow \rangle \,.
\end{equation}
Taking into account the quantum numbers of the photon,
\begin{equation}
J_\gamma = 1 \,, \quad m_{J_\gamma} = 1 \,,
\end{equation}
and the quantum numbers of the final state quarkonium,
\begin{equation}
\l' = 0 \,, \quad s' = 1 \,, \quad J' = 1 \,,
\end{equation}
we consequently have the projections of the final state quarkonium quantum numbers\footnote{Remember that the electric dipole operator acts as the identity in spin-space and thus $s$ and $m_s$ are conserved. Furthermore Lorentz invariance guarantees the conservation of $J$ and $m_J$.}
\begin{equation}
m_{\l'} = 0 \,, \quad m_{J'} = 1 \quad \Rightarrow \quad m_{s'} = 1 \,.
\end{equation}
Thus, the wave function of the final state quarkonium is given by
\begin{equation}
| n'\,^3\!S_1 \rangle = | n' ; 0 , 0 ; 1 , 1 ; 1 , 1 \rangle = R_{n'0}(r) Y_0^0(\Omega_r) | \uparrow \uparrow \rangle \,.
\end{equation}
The matrix element is then given by
\begin{align}
& \M_{\text{E1}}^{(0)}(n\,^3\!P_2 \to n'\,^3\!S_1 + \gamma) = e e_Q \sqrt{\frac{4\pi}{3}} I_3^{(0)}(n1 \to n'0) \\
\nonumber & \quad\quad \times \sum\limits_{\mu = -1}^1 (-1)^{\mu} \sqrt{\frac{9}{4\pi}} \begin{pmatrix} 0 & 1 & 1 \\ 0 & 0 & 0 \end{pmatrix} \begin{pmatrix} 0 & 1 & 1 \\ 0 & 1 & -\mu \end{pmatrix} \times \hat{e}_r \cdot \langle \gamma | \vec{E} | 0 \rangle \\
\nonumber & \quad = e e_Q I_3^{(0)}(n1 \to n'0) \frac{1}{\sqrt{3}} \, \hat{e}_r \cdot \langle \gamma | \vec{E} | 0 \rangle \,,
\end{align}
such that
\begin{equation}
\bar{|\M_{\text{E1}}^{(0)}(n\,^3\!P_2 \to n'\,^3\!S_1 + \gamma)|^2} = e^2 e_Q^2 \left[I_3^{(0)}(n1 \to n'0)\right]^2 \frac{1}{3} \frac{2}{3} k_\gamma^2 \,,
\end{equation}
where summing over the photon polarizations yields the factor $\frac{2}{3} k_\gamma^2$, and thus
\begin{align}
\begin{aligned}
\Gamma_{\text{E1}}^{(0)}(n\,^3\!P_2 \to n'\,^3\!S_1 + \gamma) &= \frac{k_\gamma}{2\pi} \bar{|\M_{\text{E1}}^{(0)}(n\,^3\!P_2 \to n'\,^3\!S_1 + \gamma)|^2} \\
&= \frac{4}{9} \alpha_{e/m} e_Q^2 k_\gamma^3 \left[I_3^{(0)}(n1 \to n'0)\right]^2 \,.
\end{aligned}
\end{align}
\item In the case of $n\,^3\!P_1 \to n'\,^3\!S_1 + \gamma$ the quantum numbers of the initial state quarkonium are given, again choosing maximal projection with respect to $J$, by
\begin{align}
\begin{aligned}
& \l = 1 \,, \quad s = 1 \,, \quad J = 1 \,, \\
& m_J = 1 \quad \Rightarrow \quad m_\l = 0,1 \,, \quad m_s = 1,0 \,,
\end{aligned}
\end{align}
from where we can see, that the wave function of the initial state quarkonium,
\begin{equation}
| n\,^3\!P_1 \rangle = | n ; 1 , 0/1 ; 1 , 1/0 ; 1 , 1 \rangle = R_{n1}(r) \frac{1}{\sqrt{2}} \left(-Y_1^0(\Omega_r) | \uparrow \uparrow \rangle + Y_1^1(\Omega_r) \frac{1}{\sqrt{2}} \left(| \uparrow \downarrow \rangle + | \downarrow \uparrow \rangle\right)\right) \,,
\end{equation}
is a superposition of two different combinatorial possibilities. Therefore the quarkonium final state quantum numbers are given by
\begin{align}
\begin{aligned}
& \l' = 0 \,, \quad s' = 1 \,, \quad J' = 1 \,, \\
& m_{\l'} = 0 \,, \quad m_{s'} = 0,1  \,, \quad m_{J'} = 0,1 \,,
\end{aligned}
\end{align}
which allows for the two possible wave functions for the final state quarkonium
\begin{align}
& \left. | n'\,^3\!S_1 \rangle \right|_{m_{s'} = 0} = | n' ; 0 , 0 ; 1 , 0 ; 1 , 0 \rangle = R_{n'0}(r) Y_0^0(\Omega_r) \frac{1}{\sqrt{2}} (| \uparrow \downarrow \rangle + | \downarrow \uparrow \rangle) \,, \\
& \left. | n'\,^3\!S_1 \rangle \right|_{m_{s'} = 1} = | n' ; 0 , 0 ; 1 , 1 ; 1 , 1 \rangle = R_{n'0}(r) Y_0^0(\Omega_r) | \uparrow \uparrow \rangle \,.
\end{align}
This leads to the two matrix elements
\begin{align}
& \left. \M_{\text{E1}}^{(0)}(n\,^3\!P_1 \to n'\,^3\!S_1 + \gamma) \right|_{m_{s'} = 0} = e e_Q I_3^{(0)}(n1 \to n'0) \frac{1}{\sqrt{2}} \frac{1}{\sqrt{3}} \, \hat{e}_r \cdot \langle \gamma | \vec{E} | 0 \rangle \,, \\
& \left. \M_{\text{E1}}^{(0)}(n\,^3\!P_1 \to n'\,^3\!S_1 + \gamma) \right|_{m_{s'} = 1} = e e_Q I_3^{(0)}(n1 \to n'0) \frac{1}{\sqrt{2}} \frac{1}{\sqrt{3}} \, \hat{e}_r \cdot \langle \gamma | \vec{E} | 0 \rangle \,,
\end{align}
that need to be summed such that they do not interfere, yielding
\begin{align}
\begin{aligned}
& \bar{|\M_{\text{E1}}^{(0)}(n\,^3\!P_1 \to n'\,^3\!S_1 + \gamma)|^2} \\
& \quad = \bar{|\M_{\text{E1}}^{(0)}(n\,^3\!P_1 \to n'\,^3\!S_1 + \gamma)_{m_{s'} = 0}|^2} + \bar{|\M_{\text{E1}}^{(0)}(n\,^3\!P_1 \to n'\,^3\!S_1 + \gamma)_{m_{s'} = 1}|^2} \\
& \quad = e^2 e_Q^2 \left[I_3^{(0)}(n1 \to n'0)\right]^2 \left[\frac{1}{2} \frac{1}{3} +\frac{1}{2} \frac{1}{3}\right] \frac{2}{3} k_\gamma^2 \,,
\end{aligned}
\end{align}
and the decay width is thus finally given by
\begin{align}
\begin{aligned}
\Gamma_{\text{E1}}^{(0)}(n\,^3\!P_1 \to n'\,^3\!S_1 + \gamma) &= \frac{k_\gamma}{2\pi} \bar{|\M_{\text{E1}}^{(0)}(n\,^3\!P_1 \to n'\,^3\!S_1 + \gamma)|^2} \\
&= \frac{4}{9} \alpha_{e/m} e_Q^2 k_\gamma^3 \left[I_3^{(0)}(n1 \to n'0)\right]^2 \,,
\end{aligned}
\end{align}
which is the same expression than the former case.
\item Finally, for the decay $n\,^3\!P_0 \to n'\,^3\!S_1 + \gamma$, the quantum numbers of the initial state quarkonium are given, again choosing maximal projection with respect to $J$, by
\begin{align}
\begin{aligned}
& \l = 1 \,, \quad s = 1 \,, \quad J = 0 \,, \\
& m_J = 0 \quad \Rightarrow \quad m_\l = 0,\pm 1 \,, \quad m_s = 0,\mp 1 \,.
\end{aligned}
\end{align}
Thus, the wave function of the initial state quarkonium,
\begin{align}
\begin{aligned}
| n\,^3\!P_0 \rangle &= | n ; 1 , 1/0/-1 ; 1 , -1/0/1 ; 0 , 0 \rangle \\
&= R_{n1}(r) \frac{1}{\sqrt{3}} \left(Y_1^1(\Omega_r) | \downarrow \downarrow \rangle - Y_1^0(\Omega_r) \frac{1}{\sqrt{2}} (| \uparrow \downarrow \rangle + | \downarrow \uparrow \rangle) + Y_1^{-1}(\Omega_r) | \uparrow \uparrow \rangle\right) \,,
\end{aligned}
\end{align}
is a superposition with respect to the three different combinatorial possibilities. Therefore the quarkonium final state quantum numbers are
\begin{align}
\begin{aligned}
& \l' = 0 \,, \quad s' = 1 \,, \quad J' = 1 \,, \\
& m_{\l'} = 0 \,, \quad \Rightarrow \quad m_{s'} = 0,\pm 1  \,, \quad m_{J'} = 0,\mp 1 \,,
\end{aligned}
\end{align}
which allows for the three possible wave functions for the final state quarkonium
\begin{align}
& \left. | n'\,^3\!S_1 \rangle \right|_{m_{s'} = -1} = | n' ; 0 , 0 ; 1 , -1 ; 1 , -1 \rangle = R_{n'0}(r) Y_0^0(\Omega_r) | \downarrow \downarrow \rangle \,, \\
& \left. | n'\,^3\!S_1 \rangle \right|_{m_{s'} = 0} = | n' ; 0 , 0 ; 1 , 0 ; 1 , 0 \rangle = R_{n'0}(r) Y_0^0(\Omega_r) \frac{1}{\sqrt{2}} (| \uparrow \downarrow \rangle + | \downarrow \uparrow \rangle) \,, \\
& \left. | n'\,^3\!S_1 \rangle \right|_{m_{s'} = 1} = | n' ; 0 , 0 ; 1 , 1 ; 1 , 1 \rangle = R_{n'0}(r) Y_0^0(\Omega_r) | \uparrow \uparrow \rangle \,.
\end{align}
The three matrix elements,
\begin{align}
& \left. \M_{\text{E1}}^{(0)}(n\,^3\!P_0 \to n'\,^3\!S_1 + \gamma) \right|_{m_{s'} = -1} = e e_Q I_3^{(0)}(n1 \to n'0) \frac{1}{\sqrt{3}} \frac{1}{\sqrt{3}} \, \hat{e}_r \cdot \langle \gamma | \vec{E} | 0 \rangle \,, \\
& \left. \M_{\text{E1}}^{(0)}(n\,^3\!P_0 \to n'\,^3\!S_1 + \gamma) \right|_{m_{s'} = 0} = e e_Q I_3^{(0)}(n1 \to n'0) \frac{1}{\sqrt{3}} \frac{1}{\sqrt{3}} \, \hat{e}_r \cdot \langle \gamma | \vec{E} | 0 \rangle \,, \\
& \left. \M_{\text{E1}}^{(0)}(n\,^3\!P_0 \to n'\,^3\!S_1 + \gamma) \right|_{m_{s'} = 1} = e e_Q I_3^{(0)}(n1 \to n'0) \frac{1}{\sqrt{3}} \frac{1}{\sqrt{3}} \, \hat{e}_r \cdot \langle \gamma | \vec{E} | 0 \rangle \,,
\end{align}
yield, upon summing them non-interfering,
\begin{equation}
\bar{|\M_{\text{E1}}^{(0)}(n\,^3\!P_0 \to n'\,^3\!S_1 + \gamma)|^2} = e^2 e_Q^2 \left[I_3^{(0)}(n1 \to n'0)\right]^2 \left[\frac{1}{3} \frac{1}{3} +\frac{1}{3} \frac{1}{3} + \frac{1}{3} \frac{1}{3}\right] \frac{2}{3} k_\gamma^2 \,,
\end{equation}
and the decay width can be written again as
\begin{align}
\begin{aligned}
\Gamma_{\text{E1}}^{(0)}(n\,^3\!P_0 \to n'\,^3\!S_1 + \gamma) &= \frac{k_\gamma}{2\pi} \bar{|\M_{\text{E1}}^{(0)}(n\,^3\!P_0 \to n'\,^3\!S_1 + \gamma)|^2} \\
& = \frac{4}{9} \alpha_{e/m} e_Q^2 k_\gamma^3 \left[I_3^{(0)}(n1 \to n'0)\right]^2 \,.
\end{aligned}
\end{align}
\end{itemize}
From this explicit computation, we confirmed, with a different approach, the $J$ independence of the leading order decay width for $^3\!P_{J=0,1,2} \to \,^3\!S_1 + \gamma$, being
\begin{equation}
\Gamma_{\text{E1}}^{(0)}(n\,^3\!P_{J=0,1,2} \to n'\,^3\!S_1 + \gamma) = \frac{4}{9} \alpha_{e/m} e_Q^2 k_\gamma^3 \left[I_3^{(0)}(n1 \to n'0)\right]^2 \,,
\end{equation}
and we thus successfully reproduced Eqs.~\eqref{eq:LeadingOrderE1} and \eqref{eq:LeadingOrderE1_2} that are consistent with the general non-relativistic formula \cite{Eichten:1978tg}
\begin{equation}
\Gamma_{\text{E1}}^{(0)}(n\,^{2s+1}\!\l_J \to n'\,^{2s+1}\!\l'_{J'} + \gamma) = \frac{4}{3} \alpha_{e/m} e_Q^2 (2J'+1) \text{max}(\l,\l') k_\gamma^3 \left[I_3^{(0)}(n\l \to n'\l')\right]^2 \left\lbrace \begin{matrix} J & 1 & J' \\ \l' & s & \l \end{matrix} \right\rbrace^2 \,,
\end{equation}
where $\left\lbrace 2 \times 3 \right\rbrace$ is a Wigner $6j$-symbol (see Appendix~\ref{app:subsec:WignerSymbol} on how to compute them).\\
A straight forward computation for the case $n\,^1\!P_1 \to n'\,^1\!S_0 + \gamma$ shows that the leading order decay width is the same also for this transition, since for the initial state we have
\begin{equation}
\l = 1 \,, \quad s = 0 \,, \quad  J = 1 \,,
\end{equation}
the maximal projection with respect to $J$ yields
\begin{equation}
m_J = 1 \quad \Rightarrow \quad m_\l = 1 \,, \quad m_s = 0 \,,
\end{equation}
and the corresponding wave function is thus given by
\begin{equation}
| n\,^1\!P_1 \rangle = | n ; 1 , 1 ; 0 , 0 ; 1 , 1 \rangle = R_{n1}(r) Y_1^1(\Omega_r) \frac{1}{\sqrt{2}} \left( | \uparrow \downarrow \rangle - | \downarrow \uparrow \rangle \right) \,.
\end{equation}
The quantum numbers of the final state are
\begin{align}
\begin{aligned}
& \l' = 0 \,, \quad s' = 0 \,, \quad J' = 0 \,, \\
& m_{\l'} = 0 \,, \quad \Rightarrow \quad m_{s'} = 0 \,, \quad m_{J'} = 0 \,,
\end{aligned}
\end{align}
and thus, the wave function of the final state is given by
\begin{equation}
| n'\,^1\!S_0 \rangle = | n' ; 0 , 0 ; 0 , 0 ; 0 , 0 \rangle = R_{n'0}(r) Y_0^0(\Omega_r) \frac{1}{\sqrt{2}} \left( | \uparrow \downarrow \rangle - | \downarrow \uparrow \rangle \right) \,.
\end{equation}
The matrix element is then given by
\begin{align}
& \M_{\text{E1}}^{(0)}(n\,^1\!P_1 \to n'\,^1\!S_0 + \gamma) = e e_Q \sqrt{\frac{4\pi}{3}} I_3^{(0)}(n1 \to n'0) \\
\nonumber & \quad\quad \times \sum\limits_{\mu = -1}^1 (-1)^{\mu} \sqrt{\frac{9}{4\pi}} \begin{pmatrix} 0 & 1 & 1 \\ 0 & 0 & 0 \end{pmatrix} \begin{pmatrix} 0 & 1 & 1 \\ 0 & 1 & -\mu \end{pmatrix} \times \hat{e}_r \cdot \langle \gamma | \vec{E} | 0 \rangle \\
\nonumber & \quad = e e_Q I_3^{(0)}(n1 \to n'0) \frac{1}{\sqrt{3}} \, \hat{e}_r \cdot \langle \gamma | \vec{E} | 0 \rangle \,,
\end{align}
such that
\begin{equation}
\bar{|\M_{\text{E1}}^{(0)}(n\,^1\!P_1 \to n'\,^1\!S_0 + \gamma)|^2} = e^2 e_Q^2 \left[I_3^{(0)}(n1 \to n'0)\right]^2 \frac{1}{3} \frac{2}{3} k_\gamma^2 \,,
\end{equation}
and we finally arrive at
\begin{align}
\begin{aligned}
\Gamma_{\text{E1}}^{(0)}(n\,^1\!P_1 \to n'\,^1\!S_0 + \gamma) &= \frac{k_\gamma}{2\pi} \bar{|\M_{\text{E1}}^{(0)}(n\,^3\!P_2 \to n'\,^3\!S_1 + \gamma)|^2} \\
&= \frac{4}{9} \alpha_{e/m} e_Q^2 k_\gamma^3 \left[I_3^{(0)}(n1 \to n'0)\right]^2 \,.
\end{aligned}
\end{align}

\section{Higher order operator corrections to the decay width}
\label{sec:HigherOrderOperatorCorrectionsDecayWidth}

Beyond the leading order electric dipole part, Eq.~\eqref{eq:LeadingOrderGammapNRQCDLagrangian}, higher order operators in the Lagrangian~\eqref{eq:GammapNRQCDLagrangian}, higher order corrections to the initial and final state wave functions, and higher order Fock states contribute to the decay width. The contributions of the first type have been derived, e.g., in Refs.~\cite{Pietrulewicz:2011aca,Brambilla:2012be} up to NNLO, $\O(m \alpha_s^6)$. We will refer to them as relativistic corrections to the Lagrangian. The initial and final state wave function corrections will be derived in the following Sections~\ref{sec:FirstOrderWaveFunctionCorrections} and \ref{sec:SecondOrderWaveFunctionCorrections} and corrections due to higher order Fock states will be discussed in Sec.~\ref{sec:NonPerturbativeContributions}.\\
The full decay width for $n\,^3\!P_J \to n'\,^3\!S_1 + \gamma$ is given by
\begin{align}
\label{eq:FullDecayWidth1}
& \Gamma_{n\,^3\!P_J \to n'\,^3\!S_1} = \Gamma^{(0)}_{E_1} \left[ 1 + R^{S=1}(J) - \frac{k_\gamma}{6m} - \frac{k_\gamma^2}{60} \frac{I_5^{(0)}(n1 \to n'0)}{I_3^{(0)}(n1 \to n'0)} \right. \\
\nonumber & \quad + \left. \left( \frac{J(J+1)}{2} - 2 \right) \left( -(1+\kappa_Q^{e/m}) \frac{k_\gamma}{2m} + \frac{1}{m^2} (1 + 2\kappa_Q^{e/m}) \frac{I_2^{(1)}(n1 \to n'0) + 2I_1^{(0)}(n1 \to n'0)}{I_3^{(0)}(n1 \to n'0)} \right) \right] \,,
\end{align}
where $I_N^{(k)}$ is defined in Eq.~\eqref{eq:INk}, the anomalous magnetic moment is defined in Eq.~\eqref{eq:AnomalousMagneticMoment}, and $R^{S=1}(J)$ encodes all the contributions due to initial and final state wave function corrections. The three additional terms encode the relativistic corrections to the Lagrangian.\\
In an analogous manner, we have for $n\,^1\!P_1 \to n'\,^1\!S_0 + \gamma$
\begin{align}
\label{eq:FullDecayWidth2}
\Gamma_{n\,^1\!P_1 \to n'\,^1\!S_0} = \Gamma^{(0)}_{E_1} \left[ 1 + R^{S=0} - \frac{k_\gamma}{6m} - \frac{k_\gamma^2}{60} \frac{I_5^{(0)}(n1 \to n'0)}{I_3^{(0)}(n1 \to n'0)} \right] \,.
\end{align}
We now want to impose the constrains coming from the power counting on the possible terms in the total decay widths, Eqs.~\eqref{eq:FullDecayWidth1} and \eqref{eq:FullDecayWidth2}. This is important, since we are interested in a consistent expression up to NNLO, $\O(m\alpha_s^6)$. This already constrains the leading order decay width, because we need to fix the bottom quark mass $m_b \equiv m$. We do this using the experimental mass of the $b\bar{b}$-ground state, the $\Upsilon(1S)$, and the leading order Coulomb energy\footnote{Note that in Chapter~\ref{chp:QuarkoniumSpectrumPNRQCDWeakCouplingUpToMAlpha4} we took into account higher order corrections to the mass. However, doing so in this case would exceed our goal of NNLO, $\O(m\alpha_s^6)$. Therefore we truncate the series already after the leading order binding energy.}
\begin{equation}
M^{\text{exp.}}(\Upsilon(1S)) = 2 m_b + E_{n=1} \,.
\end{equation}
This fixes the bottom mass to be
\begin{equation}
m_b = \frac{M^{\text{exp.}}(\Upsilon(1S))}{2} \left(1 + \frac{C_F^2 \alpha_s^2}{8} + \O(\alpha_s^3)\right) \,,
\end{equation}
which is the expression for $m_b$ that we will be using for the leading order decay with $\Gamma^{(0)}_{\text{E1}}$. Since the relative accuracy we are interested in is $\alpha_s^2$ suppressed with respect to the leading order, and all the matrix elements that provide such corrections are at least $\alpha_s$ (NLO) or $\alpha_s^2$ (NNLO) suppressed by themselves, we need to truncate the series for $m_b$ after the leading term, hence use
\begin{equation}
m_b = \frac{M^{\text{exp.}}(\Upsilon(1S))}{2}
\end{equation}
for them in order to arrive at a consistent expansion of the decay width up to our desired goal of accuracy.\\
A second issue arises with the matrix elements beyond leading order themselves as it already has been discussed in Sec.~\ref{sec:pNRQCD}. We therefore consider the definition of $R$ as given in Eq.~\eqref{eq:R} and set $\kappa_Q^{e/m}$ to zero. This has to be done, since the anomalous magnetic moment by itself is suppressed by $\alpha_s^2$ with respect to the relativistic corrections and thus would exceed our goal of precision. The relativistic corrections to the Lagrangian, encoded in $2 \bar{\M}_{\text{rel.}}$ have been computed in Refs.~\cite{Pietrulewicz:2011aca,Brambilla:2012be}. They are explicitly not included in the definition of $R$ but enter as additional summands in the Eqs.~\eqref{eq:FullDecayWidth1} and \eqref{eq:FullDecayWidth2}. Thus $R$ as it is given in Eq.~\eqref{eq:R} contains only the corrections to the initial and final states up to relative order $\alpha_s^2$. They are induced by higher orders in the static potential, by higher order potentials steaming from the $\frac{1}{m}$-expansion or by higher order Fock states. In the following we will derive expressions for the pieces entering the function $R$ and discuss the non-perturbative contributions due to higher order Fock states.

\section{First order wave function corrections to the decay width}
\label{sec:FirstOrderWaveFunctionCorrections}

\begin{figure}[t]
\centering
{\includegraphics[clip,trim={5.3cm 24.8cm 11.9cm 3.8cm},width=0.35\textwidth]{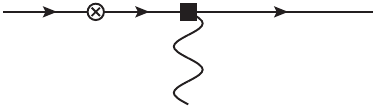}}
\quad\quad\quad
{\includegraphics[
clip,trim={11.1cm 24.4cm 6cm 4.1cm},width=0.35\textwidth]{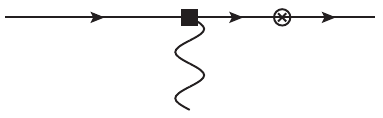}}
\caption[Diagrammatic first order corrections to the initial and final state entering the E1 decay width.]{Diagrammatic first order corrections to the initial and final state entering the E1 decay width, taken from \cite{Pietrulewicz:2011aca}. The first diagram shows a correction to the initial state and the second diagram shows a correction to the final state.}
\label{fig:FirstOrderCorrectionsDecay}
\end{figure}

First order corrections to the decay width are given by matrix elements of the electric dipole operator, where either the initial or the final state receives a correction due to potentials. The corresponding diagrams are of the type depicted in Fig.~\ref{fig:FirstOrderCorrectionsDecay}. There, either the initial (left diagram) or the final state (right diagram) has received a correction according to Eq.~\eqref{eq:FirstOrderWaveFunction}.\\
In order to calculate these matrix elements, we use the method of Coulomb Green functions. However, in contrast to the spectrum, initial and final states are now different and thus we are no longer able to use the orthogonality relations for the associated Laguerre polynomials that allowed us to have analytic expressions.\footnote{Different in this context refers to the fact that the associated Laguerre polynomials do not only appear with different indices but also with different arguments. The latter fact makes it impossible to use the orthogonality relations. This leaves us with two numerically equivalent options: (i) one can perform the sum in $s$ numerically until a desired precision is achieved and thus perform the integrals analytically for each given $s$, which is possible; (ii) one may use the Rodrigues representation~\eqref{eq:RodriguesRepresentationAssociatedLaguerrePolynomials} to trade the integrals for finite sums (interchanging them is allowed, since all the integrals converge absolutely for a given value of $s$ and one may thus interchange integration and performing the finite sum) that explicitly depend on $s$ and subsequently perform the sum in $s$ numerically until a desired precision is achieved.} Furthermore, due to the decomposition of the electric dipole operator~\eqref{eq:DecompositionVecR}, we find a non-trivial angular operator contributing to the matrix element.\\
We first consider the matrix elements for an E1 transition where either the initial or the final state receives a correction due to a central potential $V(\vec{r}\,) = V(r)$. The potentials with a non-trivial angular part shall be dealt with separately below in the Sections~\ref{subsec:FirstOrderMatrixElementsDeltaDistribution} and \ref{subsec:FirstOrderMatrixElementsTensorTerm}. The matrix element for the final state correction then takes the form\footnote{We drop the explicit limits $E \to E_{n'}$ and $\epsilon \to 0$, $\O(\epsilon^0)$, as well as the explicit $E$ dependence of the Coulomb Green function, since the procedure to handle the divergence is known and understood implicitly. For the sake of readability, we also drop the sums in $m \equiv m_{\l}$ and $m' \equiv m_{\l'}$ running from $-\l$ to $\l$ and from $-\l'$ to $\l'$, respectively.}
\begin{align}
& \M_{\text{E1},\text{fin}}^{(1)} = \,^{(1)}\!\langle n' ; \l' ; s' ; J' ; \gamma | \OEone | n ; \l ; s ; J ; 0 \rangle \\
& \quad = \langle n' ; \l' ; s' ; J' ; \gamma | V(r_2) \frac{1}{(E_{n'} - H)'} \OEone(\vec{r}_1) | n ; \l ; s ; J ; 0 \rangle \\
& \quad = -\int \d^3 r_1 \, \d^3 r_2 \, \psi_{n'\l'}^*(\vec{r}_2) V(r_2) G(\vec{r}_2,\vec{r}_1) \, \OEone(\vec{r}_1) \, \psi_{n\l}(\vec{r}_1) \, \chi_{s' J'}^* \chi_{s J} \\
& \quad = - e e_Q \sqrt{\frac{4\pi}{3}} \, N_{n'\l'} N_{n\l} \frac{\mr}{a \lambda \pi} \sum_{\l''=0}^\infty (2\l''+1) \sum\limits_{s=0}^\infty \frac{s!}{(s+\l''+1-\lambda)(s+2\l''+1)!} \\
\nonumber & \quad\quad \times \int\limits_0^\infty \d r_1 \, r_1^3 \r{\lambda}{1}^{\l''} \e{-\frac{1}{2}\r{\lambda}{1}} L_{s}^{2\l''+1}(\r{\lambda}{1}) \r{n}{1}^{\l} \e{-\frac{1}{2}\r{n}{1}} L_{n-\l-1}^{2\l+1}(\r{n}{1}) \\
\nonumber & \quad\quad \times \int\limits_0^\infty \d r_2 \, r_2^2 V(r_2) \r{\lambda}{2}^{\l''} \e{-\frac{1}{2}\r{\lambda}{2}} L_{s}^{2\l''+1}(\r{\lambda}{2}) \rho_{n',2}^{\l'} \e{-\frac{1}{2}\rho_{n',2}} L_{n'-\l'-1}^{2\l'+1}(\rho_{n',2}) \\
\nonumber & \quad\quad \times \frac{4\pi}{2\l''+1} \sum\limits_{m''=-\l''}^{\l''} \sum\limits_{\mu=-1}^1 \int \d\Omega_1 \, Y_{\l''}^{m''*}(\Omega_1) Y_1^{\mu*}(\Omega_1) Y_{\l}^{m}(\Omega_1) \int \d\Omega_2 \, Y_{\l''}^{m''}(\Omega_2) Y_{\l'}^{m'*}(\Omega_2) \\
\nonumber & \quad\quad \times \hat{e}_{r_1} \cdot \langle \gamma | \vec{E} | 0 \rangle \, \chi_{s' J'}^* \chi_{s J} \\
& \quad = - e e_Q \sqrt{\frac{4\pi}{3}} \, N_{n'\l'} N_{n\l} \frac{4\mr}{a \lambda} \sum\limits_{s=0}^\infty \frac{s!}{(s+\l'+1-\lambda)(s+2\l'+1)!} \\
\nonumber & \quad\quad \times \int\limits_0^\infty \d r_1 \, r_1^3 \r{\lambda}{1}^{\l'} \r{n}{1}^{\l} \e{-\frac{1}{2}\r{\lambda}{1}} \e{-\frac{1}{2}\rho_{n,1}} L_{s}^{2\l'+1}(\r{\lambda}{1}) L_{n-\l-1}^{2\l+1}(\rho_{n,1}) \\
\nonumber & \quad\quad \times \int\limits_0^\infty \d r_2 \, r_2^2 V(r_2) \r{\lambda}{2}^{\l'} \rho_{n',2}^{\l'} \e{-\frac{1}{2}\r{\lambda}{2}} \e{-\frac{1}{2}\rho_{n',2}} L_{s}^{2\l'+1}(\r{\lambda}{2}) L_{n'-\l'-1}^{2\l'+1}(\rho_{n',2}) \\
\nonumber & \quad\quad \times \sum\limits_{\mu=-1}^1 (-1)^{m'+\mu} \int \d\Omega_1 \, Y_{\l'}^{-m'}(\Omega_1) Y_1^{-\mu}(\Omega_1) Y_{\l}^{m}(\Omega_1) \times \hat{e}_{r_1} \cdot \langle \gamma | \vec{E} | 0 \rangle \, \chi_{s' J'}^* \chi_{s J} \\
& \quad \equiv e e_Q \widetilde{\M}_{\text{E1},\text{fin}}^{(1)} \times \sum\limits_{\mu=-1}^1 (-1)^{m'+\mu} \int \d\Omega_1 \, Y_{\l'}^{-m'}(\Omega_1) Y_1^{-\mu}(\Omega_1) Y_{\l}^{m}(\Omega_1) \times \hat{e}_{r_1} \cdot \langle \gamma | \vec{E} | 0 \rangle \, \chi_{s' J'}^* \chi_{s J} \,,
\end{align}
where $\widetilde{\M}_{\text{E1},\text{fin}}^{(1)}$ is the radial part of the matrix element containing all the relevant corrections from $V(r)$ and we can identify all the spin, angular momentum, total angular momentum and photon contribution to the matrix element being exactly the same as the one for the leading order matrix element, Eq.~\eqref{eq:SplittingLeadingOrderE1}.\\
In a similar manner we can calculate the matrix element for a E1 transition where the initial state received a correction, hence
\begin{align}
& \M_{\text{E1},\text{ini}}^{(1)} = \langle n' ; \l' ; s' ; J' ; \gamma | \OEone | n ; \l ; s ; J ; 0 \rangle^{(1)} \\
& \quad = \langle n' ; \l' ; s' ; J' ; \gamma | \OEone(\vec{r}_2) \frac{1}{(E_n - H)'} V(r_1) | n ; \l ; s ; J ; 0 \rangle \\
& \quad = -\int \d^3 r_1 \, \d^3 r_2 \, \psi_{n'\l'}^*(\vec{r}_2) \OEone(\vec{r}_2) G(\vec{r}_2,\vec{r}_1) \, V(r_1) \, \psi_{n\l}(\vec{r}_1) \, \chi_{s' J'}^* \chi_{s J} \\
& \quad = - e e_Q \sqrt{\frac{4\pi}{3}} \, N_{n'\l'} N_{n\l} \frac{\mr}{a \lambda \pi} \sum_{\l''=0}^\infty (2\l''+1) \sum\limits_{s=0}^\infty \frac{s!}{(s+\l''+1-\lambda)(s+2\l''+1)!} \\
\nonumber & \quad\quad \times \int\limits_0^\infty \d r_1 \, r_1^2 V(r_1) \r{\lambda}{1}^{\l''} \e{-\frac{1}{2}\r{\lambda}{1}} L_{s}^{2\l''+1}(\r{\lambda}{1}) \rho_{n,1}^{\l} \e{-\frac{1}{2}\rho_{n,1}} L_{n-\l-1}^{2\l+1}(\rho_{n,1}) \\
\nonumber & \quad\quad \times \int\limits_0^\infty \d r_2 \, r_2^3 \r{\lambda}{2}^{\l''} \e{-\frac{1}{2}\r{\lambda}{2}} L_{s}^{2\l''+1}(\r{\lambda}{2}) \rho_{n',2}^{\l'} \e{-\frac{1}{2}\rho_{n',2}} L_{n'-\l'-1}^{2\l'+1}(\rho_{n',2}) \\
\nonumber & \quad\quad \times \frac{4\pi}{2\l''+1} \sum\limits_{m''=-\l''}^{\l''} \sum\limits_{\mu=-1}^1 \int \d\Omega_1 \, Y_{\l''}^{m''*}(\Omega_1) Y_{\l}^{m}(\Omega_1) \int \d\Omega_2 \, Y_{\l'}^{m'*}(\Omega_2) Y_1^{\mu*}(\Omega_2) Y_{\l''}^{m''}(\Omega_2) \\
\nonumber & \quad\quad \times \hat{e}_{r_2} \cdot \langle \gamma | \vec{E} | 0 \rangle \, \chi_{s' J'}^* \chi_{s J} \\
& \quad = - e e_Q \sqrt{\frac{4\pi}{3}} \, N_{n'\l'} N_{n\l} \frac{4\mr}{a \lambda} \sum\limits_{s=0}^\infty \frac{s!}{(s+\l+1-\lambda)(s+2\l+1)!} \\
\nonumber & \quad\quad \times \int\limits_0^\infty \d r_1 \, r_1^2 V(r_1) \r{\lambda}{1}^{\l} \rho_{n,1}^{\l} \e{-\frac{1}{2}\r{\lambda}{1}} \e{-\frac{1}{2}\rho_{n,1}} L_{s}^{2\l+1}(\r{\lambda}{1}) L_{n-\l-1}^{2\l+1}(\rho_{n,1}) \\
\nonumber & \quad\quad \times \int\limits_0^\infty \d r_2 \, r_2^3 \r{\lambda}{2}^{\l} \rho_{n',2}^{\l'}  \e{-\frac{1}{2}\r{\lambda}{2}} \e{-\frac{1}{2}\rho_{n',2}} L_{s}^{2\l+1}(\r{\lambda}{2}) L_{n'-\l'-1}^{2\l'+1}(\rho_{n',2}) \\
\nonumber & \quad\quad \times \sum\limits_{\mu=-1}^1 (-1)^{m'+\mu} \int \d\Omega_2 \, Y_{\l'}^{-m'}(\Omega_2) Y_1^{-\mu}(\Omega_2) Y_{\l}^{m}(\Omega_2) \times \hat{e}_{r_2} \cdot \langle \gamma | \vec{E} | 0 \rangle \, \chi_{s' J'}^* \chi_{s J} \\
& \quad \equiv e e_Q \widetilde{\M}_{\text{E1},\text{ini}}^{(1)} \times \sum\limits_{\mu=-1}^1 (-1)^{m'+\mu} \int \d\Omega_2 \, Y_{\l'}^{-m'}(\Omega_2) Y_1^{-\mu}(\Omega_2) Y_{\l}^{m}(\Omega_2) \times \hat{e}_{r_2} \cdot \langle \gamma | \vec{E} | 0 \rangle \, \chi_{s' J'}^* \chi_{s J} \,,
\end{align}
where, again, $\widetilde{\M}_{\text{E1},\text{ini}}^{(1)}$ encodes all the relevant corrections due to the potential $V(r)$.

\subsection[\texorpdfstring{First order matrix elements containing the $\delta$-distribution}{First order matrix elements containing the delta distribution}]{\texorpdfstring{First order matrix elements containing the {\boldmath$\delta$}-distribution}{First order matrix elements containing the delta distribution}}
\label{subsec:FirstOrderMatrixElementsDeltaDistribution}

The two very generic cases of first order initial and final state corrections both assumed a trivial angular part for the potential. However, this is not the case, once the potential is of the form
\begin{equation}
V(\vec{r}\,) = V \cdot \delta^{(3)}(\vec{r}\,) \,,
\end{equation}
where
\begin{equation}
\delta^{(3)}(\vec{r}\,) \equiv \frac{1}{r^2 \sin\theta} \delta(r) \delta(\theta) \delta(\phi) = \frac{1}{r^2} \delta(r) \delta(\Omega) \,.
\end{equation}
It is important to realize that these potentials can only act as a correction to s-wave states, since these are the only ones whose radial wave functions do not vanish at the origin (also see Appendix~\ref{app:sec:WaveFunctions}). Therefore these corrections can not affect initial states in our case. However, the non-vanishing angular part of the potential no longer allows to find relations among the quantum numbers $\l$, $\l'$ and $\l''$ using the orthonormality of the spherical harmonics. This issue can be overcome, since the angular integral can be computed analytically\footnote{Remember that $\delta(ax)=\frac{1}{|a|}\delta(x)$ for all $a \in \C/\lbrace 0 \rbrace$, such that $\delta(r)=\frac{2}{n a}\delta(\rho_n)$ with $\rho_{n}=\frac{2r}{na}$.}, such that
\begin{align}
& \M_{\text{E1},\text{fin}}^{(1)} = \,^{(1)}\!\langle n' ; 0 ; s' ; J' ; \gamma | \OEone | n ; \l ; s ; J ; 0 \rangle \\
& \quad = \langle n' ; 0 ; s' ; J' ; \gamma | V(\vec{r}_2) \frac{1}{(E_{n'} - H)'} \OEone(\vec{r}_1) | n ; \l ; s ; J ; 0 \rangle \\
& \quad = -\int \d^3 r_1 \, \d^3 r_2 \, \psi_{n'0}^*(\vec{r}_2) V \cdot \delta^{(3)}(\vec{r_2}) G(\vec{r}_2,\vec{r}_1) \, \OEone(\vec{r}_1) \, \psi_{n\l}(\vec{r}_1) \, \chi_{s' J'}^* \chi_{s J} \\
& \quad = - e e_Q \sqrt{\frac{4\pi}{3}} \, N_{n'0} N_{n\l} \frac{\mr}{a \lambda \pi} \sum_{\l''=0}^\infty (2\l''+1) \sum\limits_{s=0}^\infty \frac{s!}{(s+\l''+1-\lambda)(s+2\l''+1)!} \\
\nonumber & \quad\quad \times \int\limits_0^\infty \d r_1 \, r_1^3 \r{\lambda}{1}^{\l''} \e{-\frac{1}{2}\r{\lambda}{1}} L_{s}^{2\l''+1}(\r{\lambda}{1}) \r{n}{1}^{\l} \e{-\frac{1}{2}\r{n}{1}} L_{n-\l-1}^{2\l+1}(\r{n}{1}) \\
\nonumber & \quad\quad \times \int\limits_0^\infty \d r_2 \, r_2^2 V \cdot \frac{1}{r_2^2} \delta(r_2) \r{\lambda}{2}^{\l''} \e{-\frac{1}{2}\r{\lambda}{2}} L_{s}^{2\l''+1}(\r{\lambda}{2}) \e{-\frac{1}{2}\rho_{n',2}} L_{n'-1}^{1}(\rho_{n',2}) \\
\nonumber & \quad\quad \times \frac{4\pi}{2\l''+1} \sum\limits_{m''=-\l''}^{\l''} \sum\limits_{\mu=-1}^1 \int \d\Omega_1 \, Y_{\l''}^{m''*}(\Omega_1) Y_1^{\mu*}(\Omega_1) Y_{\l}^{m}(\Omega_1) \int \d\Omega_2 \, \delta(\Omega_2) Y_{\l''}^{m''}(\Omega_2) Y_{0}^{0*}(\Omega_2) \\
\nonumber & \quad\quad \times \hat{e}_{r_1} \cdot \langle \gamma | \vec{E} | 0 \rangle \, \chi_{s' J'}^* \chi_{s J} \\
& \quad = - e e_Q \sqrt{\frac{4\pi}{3}} \, N_{n'0} N_{n\l} \frac{4 \mr}{a \lambda} \sum_{\l''=0}^\infty \sum\limits_{s=0}^\infty \frac{s!}{(s+\l''+1-\lambda)(s+2\l''+1)!} \\
\nonumber & \quad\quad \times \int\limits_0^\infty \d r_1 \, r_1^3 \r{n}{1}^{\l} \r{\lambda}{1}^{\l''} \e{-\frac{1}{2}\r{\lambda}{1}} \e{-\frac{1}{2}\r{n}{1}} L_{s}^{2\l''+1}(\r{\lambda}{1}) L_{n-\l-1}^{2\l+1}(\r{n}{1}) \\
\nonumber & \quad\quad \times V \cdot \left[\r{\lambda}{2}^{\l''} \e{-\frac{1}{2}\r{\lambda}{2}} \e{-\frac{1}{2}\rho_{n',2}} L_{s}^{2\l''+1}(\r{\lambda}{2}) L_{n'-1}^{1}(\rho_{n',2})\right]_{r_2=0} \\
\nonumber & \quad\quad \times \sum\limits_{m''=-\l''}^{\l''} \sum\limits_{\mu=-1}^1 (-1)^{m''+\mu} \int \d\Omega_1 \, Y_{\l''}^{-m''}(\Omega_1) Y_1^{-\mu}(\Omega_1) Y_{\l}^{m}(\Omega_1) \cdot \left[Y_{\l''}^{m''}(\Omega_2) Y_{0}^{0*}(\Omega_2)\right]_{\Omega_2=0} \\
\nonumber & \quad\quad \times \hat{e}_{r_1} \cdot \langle \gamma | \vec{E} | 0 \rangle \, \chi_{s' J'}^* \chi_{s J} \,.
\end{align}
The condition $r_2 = 0$ directly implies $\l''=0$, and therewith $m''=0$, since otherwise the whole matrix element would vanish, and thus we have
\begin{align}
& \M_{\text{E1},\text{fin}}^{(1)} = - e e_Q \sqrt{\frac{4\pi}{3}} \, N_{n'0} N_{n\l} \frac{4 \mr}{a \lambda} \sum\limits_{s=0}^\infty \frac{s!}{(s+1-\lambda)(s+1)!} \\
\nonumber & \quad\quad \times \int\limits_0^\infty \d r_1 \, r_1^3 \r{n}{1}^{\l} \e{-\frac{1}{2}\r{\lambda}{1}} \e{-\frac{1}{2}\r{n}{1}} L_{s}^{1}(\r{\lambda}{1}) L_{n-\l-1}^{2\l+1}(\r{n}{1}) \\
\nonumber & \quad\quad \times V \cdot \left[\e{-\frac{1}{2}\r{\lambda}{2}} \e{-\frac{1}{2}\rho_{n',2}} L_{s}^{1}(\r{\lambda}{2}) L_{n'-1}^{1}(\rho_{n',2})\right]_{r_2=0} \\
\nonumber & \quad\quad \times \sum\limits_{\mu=-1}^1 (-1)^{\mu} \int \d\Omega_1 \, Y_{0}^{0}(\Omega_1) Y_1^{-\mu}(\Omega_1) Y_{\l}^{m}(\Omega_1) \cdot \left[Y_{0}^{0}(\Omega_2) Y_{0}^{0*}(\Omega_2)\right]_{\Omega_2=0} \\
\nonumber & \quad\quad \times \hat{e}_{r_1} \cdot \langle \gamma | \vec{E} | 0 \rangle \, \chi_{s' J'}^* \chi_{s J} \\
& \quad = - e e_Q \sqrt{\frac{4\pi}{3}} \frac{1}{4\pi} \, N_{n'0} N_{n\l} \frac{4 \mr}{a \lambda} \sum\limits_{s=0}^\infty \frac{s!}{(s+1-\lambda)(s+1)!} \\
\nonumber & \quad\quad \times \int\limits_0^\infty \d r_1 \, r_1^3 \r{n}{1}^{\l} \e{-\frac{1}{2}\r{\lambda}{1}} \e{-\frac{1}{2}\r{n}{1}} L_{s}^{1}(\r{\lambda}{1}) L_{n-\l-1}^{2\l+1}(\r{n}{1}) \\
\nonumber & \quad\quad \times V \cdot L_{s}^{1}(0) L_{n'-1}^{1}(0) \\
\nonumber & \quad\quad \times \sum\limits_{\mu=-1}^1 (-1)^{\mu} \int \d\Omega_1 \, Y_{0}^{0}(\Omega_1) Y_1^{-\mu}(\Omega_1) Y_{\l}^{m}(\Omega_1) \times \hat{e}_{r_1} \cdot \langle \gamma | \vec{E} | 0 \rangle \, \chi_{s' J'}^* \chi_{s J} \\
& \quad \equiv e e_Q \widetilde{\M}_{\text{E1},\text{fin}}^{(1)} \times \sum\limits_{\mu=-1}^1 (-1)^{\mu} \int \d\Omega_1 \, Y_{0}^{0}(\Omega_1) Y_1^{-\mu}(\Omega_1) Y_{\l}^{m}(\Omega_1) \times \hat{e}_{r_1} \cdot \langle \gamma | \vec{E} | 0 \rangle \, \chi_{s' J'}^* \chi_{s J} \,.
\end{align}
Again, $\widetilde{\M}_{\text{E1},\text{fin}}^{(1)}$ encodes the full radial information due to the potential and the rest can be identified with the usual spin/angular part, once one considers explicitly a final state s-wave.

\subsection{First order matrix elements containing the tensor term}
\label{subsec:FirstOrderMatrixElementsTensorTerm}

The potential of the tensor contribution is given by
\begin{equation}
V_T(\vec{r}\,) = \frac{1}{m^2} V_{S_{12}}^{(2)}(r) S_{12}(\hat{r}) = \frac{1}{m^2} \frac{C_F \alpha_s}{4 r^3} S_{12}(\hat{r}) \equiv V_{S_{12}}(r) S_{12}(\hat{r}) \,,
\end{equation}
where $S_{12}(\hat{r}) = 3(\hat{r} \cdot \vec{\sigma}_1)(\hat{r} \cdot \vec{\sigma}_2) - \vec{\sigma}_1 \cdot \vec{\sigma}_2$. The assumption of a trivial angular part of the potentials holds for all of the involved operators, like $\vec{L}^{\,2}$, $\vec{S}^{\,2}$ or $\vec{L} \cdot \vec{S}$, since initial and final states are eigenstates of them. This is not the case for the tensor operator, $S_{12}(\hat{r})$, that explicitly mixes states with $\Delta \l = 2$, e.g. s-waves and d-waves or p-waves and f-waves.\\
The wave function of a given quarkonium state is an eigenfunction of the diagonal term of this admixture with eigenvalue $2 \frac{2\l(\l+1)s(s+1) - 3\chi_{LS} - 6\chi_{LS}^2}{(2\l-1)(2\l+3)}$, as shown, e.g., in Ref.~\cite{Peset:2015vvi}. However, the off diagonal terms contribute to the decay width, since here, in contrast to the spectrum, initial and final state are inherently different. A general framework to incorporate off diagonal operators in quantum mechanical perturbation theory is given in Ref.~\cite{Kiyo:2016jri}. In the following, we derive the matrix elements that are relevant for our discussion of the E1 transition including the mentioned off diagonal contributions:\\
Given four commuting vectors $A_{1}$, $B_{1}$, $C_{1}$ and $D_{1}$, we can, following \cite{Varshalovich:QuantumTheoryAngularMomentum} (Eq.~(19) on page 67 therein), decompose the following irreducible tensor product of rank $0$
\begin{equation}
\Big\lbrace \lbrace A_{1} \otimes B_{1} \rbrace_{2} \otimes \lbrace C_{1} \otimes D_{1} \rbrace_{2} \Big\rbrace_{0} = \frac{1}{\sqrt{5}} \left\lbrace \frac{1}{2} (A \cdot C)(B \cdot D) - \frac{1}{3} (A \cdot B)(C \cdot D) + \frac{1}{2} (A \cdot D)(B \cdot C) \right\rbrace \,,
\end{equation}
where the subscript $2$ indicates the tensor, the subscript $1$ indicates the vector and the subscript $0$ indicates the scalar nature of the objects under consideration. This allows us to decompose
\begin{align}
\Big\lbrace \{\hat{r}\otimes \hat{r}\}_{2} \otimes \{\vec{\sigma}_{1}\otimes \vec{\sigma}_{2}\}_{2} \Big\rbrace_{0} &= \frac{1}{\sqrt{5}} \left\lbrace \frac{1}{2}(\hat{r}\cdot \vec{\sigma}_{1})(\hat{r}\cdot \vec{\sigma}_{2}) - \frac{1}{3}(\hat{r}\cdot \hat{r})(\vec{\sigma}_{1}\cdot \vec{\sigma}_{2}) + \frac{1}{2}(\hat{r}\cdot \vec{\sigma}_{2})(\hat{r}\cdot \vec{\sigma}_{1}) \right\rbrace \\
\nonumber &=\frac{1}{\sqrt{5}} \left[ \frac{1}{2}(\vec{\sigma}_{1}\cdot \hat{r})(\vec{\sigma}_{2}\cdot \hat{r}) - \frac{1}{3}(\vec{\sigma}_{1}\cdot \vec{\sigma}_{2}) + \frac{1}{2}(\vec{\sigma}_{1}\cdot \hat{r})(\vec{\sigma}_{2}\cdot \hat{r}) \right] \\
\nonumber &=\frac{1}{\sqrt{5}} \left[ (\vec{\sigma}_{1}\cdot \hat{r})(\vec{\sigma}_{2}\cdot \hat{r}) - \frac{1}{3}(\vec{\sigma}_{1}\cdot \vec{\sigma}_{2}) \right] \\
\nonumber &=\frac{1}{3\sqrt{5}} \left[ 3(\vec{\sigma}_{1}\cdot \hat{r})(\vec{\sigma}_{2}\cdot \hat{r}) - (\vec{\sigma}_{1}\cdot \vec{\sigma}_{2}) \right] \\
\nonumber &=\frac{1}{3\sqrt{5}} S_{12} \,,
\end{align}
and we can thus write the tensor operator as
\begin{equation}
S_{12} = 3\sqrt{5} \, \Big\lbrace \{\hat{r}\otimes \hat{r}\}_{2} \otimes \{\vec{\sigma}_{1}\otimes \vec{\sigma}_{2}\}_{2} \Big\rbrace_{0} \,.
\end{equation}
Therefore, we have a transition due to the operator $S_{12}$ which has $J=0$ ($M=0$) and thus conserves total angular momentum of the initial and final states. However, this operator is composed of a part which acts as a tensor of rank $2$ in coordinate-space and another part which acts as a tensor of rank $2$ in spin-space.\\
Making use of the Eqs.~(35) on page 65 and (29) on page 479 in \cite{Varshalovich:QuantumTheoryAngularMomentum},
\begin{equation}
\mathcal{M}_{J} \cdot \mathcal{N}_{J} = (-1)^{-J} \sqrt{2J+1} \, \lbrace\mathcal{M}_{J} \otimes \mathcal{N}_{J}\rbrace_{0} \,,
\end{equation}
and
\begin{align}
\begin{aligned}
& \langle j_{1}' j_{2}' j' m' | \hat{P}_{a}(1) \cdot \hat{Q}_{a}(2)| j_{1} j_{2} j m \rangle \\
& \quad\quad\quad = \delta_{j j'} \delta_{m m'} (-1)^{j+j_{1}+j_{2}'} \left\lbrace \begin{matrix} j_{1}' & j_{1} & a \\ j_{2} & j_{2}' & j \end{matrix} \right\rbrace \langle j_{1}' || \hat{P}_{a}(1) || j_{1} \rangle \langle j_{2}' || \hat{Q}_{a}(2) || j_{2} \rangle \,,
\end{aligned}
\end{align}
we have
\begin{align}
& \langle n' ; \l' ; s' ; J' , m' | S_{12} | n ; \l ; s ; J , m \rangle \\
\nonumber & \quad = \langle n' ; \l' ; s' ; J' , m' | 3\sqrt{5} \, \Big\lbrace \{\hat{r} \otimes \hat{r}\}_{2} \otimes \{\vec{\sigma}_{1} \otimes \vec{\sigma}_{2}\}_{2} \Big\rbrace_{0} | n ; \l ; s ; J , m \rangle \\
\nonumber & \quad = 3 \langle n' ; \l' ; s' ; J' , m' | \{\hat{r} \otimes \hat{r}\}_{2} \cdot \{\vec{\sigma}_{1} \otimes \vec{\sigma}_{2}\}_{2} | n ; \l ; s ; J , m \rangle \\
\nonumber & \quad = 3 \delta_{J J'} \delta_{m m'} (-1)^{J+\l+s'} \left\lbrace \begin{matrix} \l' & \l & 2 \\ s & s' & J \end{matrix} \right\rbrace \langle \l' || \{\hat{r} \otimes \hat{r}\}_{2} || \l \rangle \langle s' || \{\vec{\sigma}_{1} \otimes \vec{\sigma}_{2}\}_{2} || s \rangle \,.
\end{align}
Now, we need to compute the two reduced matrix elements of the expression above. This can be done using the Wigner-Eckart theorem (c.f. Eq.~(2) on page 475 of \cite{Varshalovich:QuantumTheoryAngularMomentum}):
\begin{align}
\begin{aligned}
\langle j' m' | \mathcal{M}_{k q} | j m \rangle  &= (-1)^{j'-m'} \begin{pmatrix} j' & k & j \\ -m' & q & m \end{pmatrix} \langle j' || \mathcal{M}_{k} || j \rangle \\
&= (-1)^{2k} \, C_{j m k q}^{j' m'} \frac{\langle j' || \mathcal{M}_{k} || j \rangle}{\sqrt{2j'+1}} \,,
\end{aligned}
\end{align}
where $C_{j m k q}^{j' m'}$ is a Clebsch-Gordan coefficient. We thus have for the coordinate dependent reduced matrix element
\begin{align}
\langle \l' || \{\hat{r} \otimes \hat{r}\}_{2} || \l \rangle &= \langle \l' || \sqrt{\frac{(4\pi) 2!}{(2 \cdot 2 + 1)!!}} \, |\hat{r}|^{2} \, Y_2^q(\Omega) || \l \rangle \\
\nonumber &= \sqrt{\frac{8\pi}{15}} \langle \l' || Y_2^q || \l \rangle \\
\nonumber &= (-1)^{-2 \cdot 2} \sqrt{\frac{8\pi (2 \l' + 1)}{15}} \, \left[C_{\l m2q}^{\l'm_{\l'}}\right]^{-1} \langle \l' m_{\l'} | Y_2^q | \l m_{\l} \rangle \\
\nonumber &= \sqrt{\frac{8\pi (2 \l' + 1)}{15}} \, \left[C_{\ell m2q}^{\l'm_{\l'}}\right]^{-1} \int \d\Omega \, Y_{\l'}^{m_{\l'}*}(\Omega) Y_2^q(\Omega) Y_{\l}^{m_{\l}}(\Omega) \\
\nonumber &= \sqrt{\frac{8\pi (2 \l' + 1)}{15}} \, \left[C_{\ell m2q}^{\l'm_{\l'}}\right]^{-1} \sqrt{\frac{5 (2 \l + 1)}{4\pi (2 \l' + 1)}} C_{\l020}^{\l'0} C_{\l m_{\l}2q}^{\l'm_{\l'}} \\
\nonumber &= \sqrt{\frac{2 (2 \l + 1)}{3}} C_{\l020}^{\l'0} \,,
\end{align}
and for the spin dependent reduced matrix element
\begin{align}
\begin{aligned}
\langle s' || \Big\lbrace \sigma_{1}^{(1)} \otimes \sigma_{1}^{(2)} \Big\rbrace_{2} || s \rangle &= (-1)^{-2 \cdot 2} \sqrt{2 s' + 1} \left[C_{s m_{s} 2 q}^{s' m_{s'}}\right]^{-1} \langle s' m_{s'} | \Big\lbrace \sigma_{1}^{(1)} \otimes \sigma_{1}^{(2)} \Big\rbrace_{2q} | s m_{s} \rangle \\
&= \sqrt{2 s' + 1} \left[C_{s m_{s} 2 q}^{s' m_{s'}}\right]^{-1} \langle s' m_{s'} | \Big\lbrace \sigma_{1}^{(1)} \otimes \sigma_{1}^{(2)} \Big\rbrace_{2q} | s m_{s} \rangle \,,
\end{aligned}
\end{align}
such that
\begin{align}
& \langle s_{1}' m_{s_{1}'} s_{2}' m_{s_{2}'} ; s' m_{s'} | \Big\lbrace \sigma_{1}^{(1)} \otimes \sigma_{1}^{(2)} \Big\rbrace_{2q} | s_{1} m_{s_{1}} s_{2} m_{s_{2}} ; s m_{s} \rangle = \\
\nonumber & \quad = (-1)^{2 \cdot 2} \sqrt{5 (2 s + 1)} C_{s m_{s} 2 q}^{s' m_{s'}} \left\lbrace \begin{matrix} 1 & 1 & 2 \\ s_{1}' & s_{2}' & s' \\ s_{1} & s_{2} & s \end{matrix} \right\rbrace \langle s_{1}' || \sigma_{1}^{(1)} || s_{1} \rangle \langle s_{2}' || \sigma_{1}^{(2)} || s_{2} \rangle \\
\nonumber & \quad = \sqrt{5 (2 s + 1)} C_{s m_{s} 2 q}^{s' m_{s'}} \left\lbrace \begin{matrix} 1 & 1 & 2 \\ s_{1}' & s_{2}' & s' \\ s_{1} & s_{2} & s \end{matrix} \right\rbrace \langle s_{1}' || 2S_{1}^{(1)} || s_{1} \rangle \langle s_{2}' || 2S_{1}^{(2)} || s_{2} \rangle \\
\nonumber & \quad = 4 \sqrt{5 (2 s + 1)} C_{s m_{s} 2 q}^{s' m_{s'}} \left\lbrace \begin{matrix} 1 & 1 & 2 \\ s_{1}' & s_{2}' & s' \\ s_{1} & s_{2} & s \end{matrix} \right\rbrace \delta_{s_{1}s_{1}'} \sqrt{s_{1} (s_{1} + 1) (2 s_{1} + 1)} \delta_{s_{2}s_{2}'} \sqrt{s_{2} (s_{2} + 1) (2 s_{2} + 1)} \\
\nonumber & \quad = 6 \sqrt{5 (2 s + 1)} C_{s m_{s} 2 q}^{s' m_{s'}} \left\lbrace \begin{matrix} 1 & 1 & 2 \\ \frac{1}{2} & \frac{1}{2} & s' \\ \frac{1}{2} & \frac{1}{2} & s \end{matrix} \right\rbrace \,,
\end{align}
where $\left\lbrace 3 \times 3 \right\rbrace$ is a Wigner $9j$-symbol (see Appendix~\ref{app:subsec:WignerSymbol} on how to compute them). The final result is then given by
\begin{align}
& \langle n' ; \l' ; s' ; J' , m' | S_{12} | n ; \l ; s ; J , m \rangle = 3 \delta_{J J'} \delta_{m m'} (-1)^{j+\l+s'} \left\lbrace \begin{matrix} \l' & \l & 2 \\ s & s' & J \end{matrix} \right\rbrace \\
\nonumber & \quad\quad \times (-1)^{\l-2} \sqrt{\frac{2(2\l+1)(2\l'+1)}{3}} \begin{pmatrix} \l & 2 & \l' \\ 0 & 0 & 0 \end{pmatrix} \times 6 \sqrt{5(2s+1)(2s'+1)} \left\lbrace \begin{matrix} 1 & 1 & 2 \\ \frac{1}{2} & \frac{1}{2} & s' \\ \frac{1}{2} & \frac{1}{2} & s \end{matrix} \right\rbrace \\
\nonumber & \quad = \delta_{J J'} \delta_{m m'} (-1)^{3\l+\l'+2+J+s'-2} 6 \sqrt{30} \sqrt{(2\l+1)(2\l'+1)(2s+1)(2s'+1)} \\
\nonumber & \quad\quad \times \left\lbrace \begin{matrix} \l' & \l & 2 \\ s & s' & J \end{matrix} \right\rbrace \begin{pmatrix} \l' & 2 & \l \\ 0 & 0 & 0 \end{pmatrix} \left\lbrace \begin{matrix} 1 & 1 & 2 \\ \frac{1}{2} & \frac{1}{2} & s' \\ \frac{1}{2} & \frac{1}{2} & s \end{matrix} \right\rbrace \\
\nonumber & \quad = \delta_{J J'} \delta_{m m'} (-1)^{\l+\l'+J+s'} 6 \sqrt{30} \sqrt{(2\l+1)(2\l'+1)(2s+1)(2s'+1)} \\
\nonumber & \quad\quad \times \left\lbrace \begin{matrix} \l' & \l & 2 \\ s & s' & J \end{matrix} \right\rbrace \begin{pmatrix} \l' & 2 & \l \\ 0 & 0 & 0 \end{pmatrix} \left\lbrace \begin{matrix} 1 & 1 & 2 \\ \frac{1}{2} & \frac{1}{2} & s' \\ \frac{1}{2} & \frac{1}{2} & s \end{matrix} \right\rbrace \\
\nonumber & \quad = \delta_{J J'} \delta_{m m'} \delta_{s s'} \delta_{s1} (-1)^{J+s'} 2 \sqrt{30} \sqrt{(2\l+1)(2\l'+1)} \left\lbrace \begin{matrix} \l' & \l & 2 \\ 1 & 1 & J \end{matrix} \right\rbrace \begin{pmatrix} \l' & 2 & \l \\ 0 & 0 & 0 \end{pmatrix} \,,
\end{align}
and the last line is because the Wigner $9j$-symbol is only different from zero when $s = s' = 1$. Thus the correction to the initial state matrix element entering the decay width reads
\begin{align}
\M_{\text{E1},\text{ini}}^{(1)} &= \langle n'' ; \l'' ; s'' ; J'' ; \gamma | \OEone | n ; \l ; s ; J ; 0 \rangle^{(1)} \\
& = \langle n'' ; \l'' ; s'' ; J'' | \OEone(\vec{r}_2) \frac{1}{(E_{n} - H)'} V_T(\vec{r}_1) | n ; \l ; s ; J ; 0 \rangle \\
& = -\int \d^3 r_1 \, \d^3 r_2 \, \Phi_{n''\l''s''J''}^*(\vec{r}_2) \OEone(\vec{r}_2) G(\vec{r}_2,\vec{r}_1) V_T(\vec{r}_1) \Phi_{n \l s J}(\vec{r}_1) \\
& = \left(3 \delta_{J J'} \delta_{m m'} (-1)^{J+\l+s'} \left\lbrace\begin{matrix} \l' & \l & 2 \\ s & s' & J \end{matrix}\right\rbrace\right) \\
\nonumber & \quad \times \left(-\sqrt{\frac{2}{3}}\right) \sum\limits_{m_{1},m_{\l'},m_{\l''}} \! (-1)^{-m_{\l''}} \sqrt{\frac{3(2\l'+1)(2\l''+1)}{4\pi}} \begin{pmatrix} \l' & 1 & \l'' \\ 0 & 0 & 0 \end{pmatrix} \begin{pmatrix} \l' & 1 & \l'' \\ m_{\l'} & m_{1} & -m_{\l''} \end{pmatrix} \\
\nonumber & \quad \times (-1)^{\l-2} \, \sqrt{\frac{2(2\l+1)(2\l'+1)}{3}} \begin{pmatrix} \l & 2 & \l' \\ 0 & 0 & 0 \end{pmatrix} \\
\nonumber & \quad \times 6 \sqrt{5(2s+1)(2s'+1)} \left\lbrace\begin{matrix} 1 & 1 & 2 \\ \frac{1}{2} & \frac{1}{2} & s' \\ \frac{1}{2} & \frac{1}{2} & s \end{matrix}\right\rbrace \times \hat{e}_{r_2} \cdot \langle \gamma | \vec{E} | 0 \rangle \\
\nonumber & \quad \times (- e e_Q) \sqrt{\frac{4\pi}{3}} N_{n\l} N_{n''\l''} \frac{4m_{r}}{\lambda a} \sum\limits_{s=0}^{\infty} \frac{s!}{(s+\l'+1-\lambda)(s+2\l'+1)!} \\
\nonumber & \quad \times \int_{0}^{\infty} \d r_{1} \, r_{1}^{2} \, V_{S_{12}}(r_{1}) \left(\frac{2r_{1}}{\lambda a}\right)^{\l'} \e{-\frac{2r_{1}}{2\lambda a}} L_{s}^{2\l'+1}\left(\frac{2r_{1}}{\lambda a}\right) \, \left(\frac{2r_{1}}{n a}\right)^{\l} \e{-\frac{2r_{1}}{2 n a}} L_{n-\l-1}^{2\l+1}\left(\frac{2r_{1}}{n a}\right) \\
\nonumber & \quad \times \int_{0}^{\infty} \d r_{2} \, r_{2}^{3} \, \left(\frac{2r_{2}}{\lambda a}\right)^{\l'} \e{-\frac{2r_{2}}{2\lambda a}} L_{s}^{2\l'+1}\left(\frac{2r_{2}}{\lambda a}\right) \, \left(\frac{2r_{2}}{n''a}\right)^{\l''} \e{-\frac{2r_{2}}{2n''a}} L_{n''-\l''-1}^{2\l''+1}\left(\frac{2r_{2}}{n''a}\right) \,,
\end{align}
where the spin term fixes $s = s' = 1$, the angular integral in $r_2$ fixes $\l' = 1$ and $m_{\l'} = -m_1$ since $\l'' = m_{\l''}=0$. The correction to the final state reads
\begin{align}
\M_{\text{E1},\text{fin}}^{(1)} &= \,^{(1)}\!\langle n'' ; \l'' ; s'' ; J'' ; \gamma | \OEone | n ; \l ; s ; J ; 0 \rangle \\
&= \langle n'' ; \l'' ; s'' ; J'' | V_T(\vec{r}_2) \frac{1}{(E_{n} - H)'} \OEone(\vec{r}_1) | n ; \l ; s ; J ; 0 \rangle \\
&= -\int \d^3 r_1 \, \d^3 r_2 \, \Phi_{n''\l''s''J''}^*(\vec{r}_2) V_T(\vec{r}_2) G(\vec{r}_2,\vec{r}_1) \OEone(\vec{r}_1) \Phi_{n \l s J}(\vec{r}_1) \\
&= \left( 3 \delta_{J' J''} \delta_{m' m''} (-1)^{J'+\l'+s''} \left\lbrace\begin{matrix} \l'' & \l' & 2 \\ s' & s'' & J' \end{matrix}\right\rbrace \right) \\
\nonumber & \quad \times \left(-\sqrt{\frac{2}{3}}\right) \sum\limits_{m_{1},m_{\l},m_{\l'}} (-1)^{-m_{\l'}} \sqrt{\frac{3(2\l+1)(2\l'+1)}{4\pi}} \begin{pmatrix} \l & 1 & \l' \\ 0 & 0 & 0 \end{pmatrix} \begin{pmatrix} \l & 1 & \l' \\ m_{\l} & m_{1} & -m_{\l'} \end{pmatrix} \\
\nonumber & \quad \times (-1)^{\l'-2} \, \sqrt{\frac{2(2\l'+1)(2\l''+1)}{3}} \begin{pmatrix} \l' & 2 & \l'' \\ 0 & 0 & 0 \end{pmatrix} \\
\nonumber & \quad \times 6 \sqrt{5(2s'+1)(2s''+1)} \left\lbrace\begin{matrix} 1 & 1 & 2 \\ \frac{1}{2} & \frac{1}{2} & s'' \\ \frac{1}{2} & \frac{1}{2} & s' \end{matrix}\right\rbrace \times \hat{e}_{r_1} \cdot \langle \gamma | \vec{E} | 0 \rangle \\
\nonumber & \quad \times (- e e_Q) \sqrt{\frac{4\pi}{3}} N_{n\l} N_{n''\l''} \frac{4m_{r}}{\lambda a} \sum\limits_{s=0}^{\infty} \frac{s!}{(s+\l'+1-\lambda)(s+2\l'+1)!} \\
\nonumber & \quad \times \int_{0}^{\infty} \d r_{1} \, r_{1}^{3} \, \left(\frac{2r_{1}}{\lambda a}\right)^{\l'} \e{-\frac{2r_{1}}{2\lambda a}} L_{s}^{2\l'+1}\left(\frac{2r_{1}}{\lambda a}\right) \, \left(\frac{2r_{1}}{n a}\right)^{\l} \e{-\frac{2r_{1}}{2 n a}} L_{n-\l-1}^{2\l+1}\left(\frac{2r_{1}}{n a}\right) \\
\nonumber & \quad \times \int_{0}^{\infty} \d r_{2} \, r_{2}^{2} \, V_{S_{12}}(r_{2}) \left(\frac{2r_{2}}{\lambda a}\right)^{\l'} \e{-\frac{2r_{2}}{2\lambda a}} L_{s}^{2\l'+1}\left(\frac{2r_{2}}{\lambda a}\right) \, \left(\frac{2r_{2}}{n''a}\right)^{\l''} \e{-\frac{2r_{2}}{2n''a}} L_{n''-\l''-1}^{2\l''+1}\left(\frac{2r_{2}}{n''a}\right) \,,
\end{align}
where, again, the spin term fixes $s''=s'=1$, the angular integral in $r_{2}$ fixes $\l'=2$ but, contrary to the initial state correction, there is no constraint on the projections, but the finite sum over them can be performed analytically.\\
One may again define radial matrix elements $\widetilde{\M}_{\text{E1}}^{(1)}$ by simply dividing the above results by the known constant angular/spin factor. This is done in order to be able to compare the numerical results below in Chapter~\ref{chp:NumericalResults}.

\section{Second order wave function corrections to the decay width}
\label{sec:SecondOrderWaveFunctionCorrections}

\begin{figure}[t]
\centering
{\includegraphics[
clip,trim={3.4cm 22.62cm 3.3cm 6.07cm},width=0.7\textwidth]{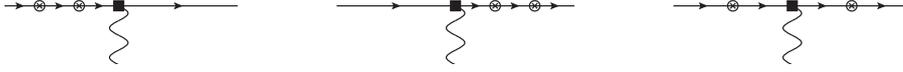}}
\caption[Diagrammatic second order corrections to the initial and final state entering the E1 decay width.]{Diagrammatic second order corrections to the initial and final state entering the E1 decay width, taken from \cite{Pietrulewicz:2011aca}. The first and the second diagram show a second order correction to the initial and final state, respectively and the third diagram shows a first order correction to initial and final state.}
\label{fig:SecondOrderCorrectionsDecay}
\end{figure}

The diagrams contributing to the second order correction of the decay width are depicted in Fig.~\ref{fig:SecondOrderCorrectionsDecay}, where either the initial or the final state receives a second order correction or initial and final state receive a first order correction simultaneously.

The resulting matrix elements are thus given by
\begin{align}
& \M_{\text{E1},\text{ini,ini}}^{(2)} = \langle n' ; \l' ; s' ; J' ; \gamma | \OEone | n ; \l ; s ; J ; 0 \rangle^{(2)} \,, \\
& \M_{\text{E1},\text{fin,fin}}^{(2)} = \,^{(2)}\!\langle n' ; \l' ; s' ; J' ; \gamma | \OEone | n ; \l ; s ; J ; 0 \rangle \,, \\
& \M_{\text{E1},\text{ini,fin}}^{(2)} = \,^{(1)}\!\langle n' ; \l' ; s' ; J' ; \gamma | \OEone | n ; \l ; s ; J ; 0 \rangle^{(1)} \,,
\end{align}
and can be computed, using Eq.~\eqref{eq:SecondOrderMatrixElement}, yielding for the first matrix element
\begin{align}
& \M_{\text{E1},\text{ini,ini}}^{(2)} = \langle n' ; \l' ; s' ; J' ; \gamma | \OEone(\vec{r}_3) \frac{1}{(E_n - H)'} V'(r_2) \frac{1}{(E_n - H)'} V(r_1) | n ; \l ; s ; J ; 0 \rangle \\
\nonumber &\quad\quad - \langle n ; \l ; s ; J | V'(r) | n ; \l ; s ; J \rangle \langle n' ; \l' ; s' ; J' ; \gamma | \OEone(\vec{r}_3) \frac{1}{(E_n - H)'} \1(\vec{r}_2) \frac{1}{(E_n - H)'} V(r_1) | n ; \l ; s ; J ; 0 \rangle \\
\nonumber &\quad\quad - \frac{1}{2} \langle n' ; \l' ; s' ; J' ; \gamma | \OEone(\vec{r}\,) | n ; \l ; s ; J ; 0 \rangle \langle n ; \l ; s ; J | V'(r_3) \frac{1}{(E_n - H)'} \1(\vec{r}_2) \frac{1}{(E_n - H)'} V(r_1) | n ; \l ; s ; J \rangle \\
&\quad = (-1)^2 \int \d^3 r_1 \, \d^3 r_2 \, \d^3 r_3 \, \psi_{n'\l'}^*(\vec{r}_3) \OEone(\vec{r}_3) G(\vec{r}_3,\vec{r}_2) V'(r_2) G(\vec{r}_2,\vec{r}_1) V(r_1) \, \psi_{n\l}(\vec{r}_1) \, \chi_{s' J'}^* \chi_{s J} \\
\nonumber &\quad\quad - \langle n ; \l ; s ; J | V'(r) | n ; \l ; s ; J \rangle \\
\nonumber &\quad\quad\quad \times (-1)^2 \int \d^3 r_1 \, \d^3 r_2 \, \d^3 r_3 \, \psi_{n'\l'}^*(\vec{r}_3) \OEone(\vec{r}_3) G(\vec{r}_3,\vec{r}_2) G(\vec{r}_2,\vec{r}_1) V(r_1) \, \psi_{n\l}(\vec{r}_1) \, \chi_{s' J'}^* \chi_{s J} \\
\nonumber &\quad\quad - \frac{1}{2} \langle n' ; \l' ; s' ; J' ; \gamma | \OEone(\vec{r}\,) | n ; \l ; s ; J ; 0 \rangle \\
\nonumber &\quad\quad\quad \times (-1)^2 \int \d^3 r_1 \, \d^3 r_2 \, \d^3 r_3 \, \psi_{n'\l'}^*(\vec{r}_3) V'(\vec{r}_3) G(\vec{r}_3,\vec{r}_2) G(\vec{r}_2,\vec{r}_1) V(r_1) \, \psi_{n\l}(\vec{r}_1) \, \chi_{s' J'}^* \chi_{s J} \\
&\quad = \int \d^3 r_1 \, \d^3 r_2 \, \d^3 r_3 \, \psi_{n'\l'}^*(\vec{r}_3) \OEone(\vec{r}_3) G(\vec{r}_3,\vec{r}_2) V'(r_2) G(\vec{r}_2,\vec{r}_1) V(r_1) \, \psi_{n\l}(\vec{r}_1) \, \chi_{s' J'}^* \chi_{s J} \\
\nonumber &\quad\quad - \langle n ; \l ; s ; J | V'(r) | n ; \l ; s ; J \rangle \\
\nonumber &\quad\quad\quad \times \int \d^3 r_1 \, \d^3 r_2 \, \d^3 r_3 \, \psi_{n'\l'}^*(\vec{r}_3) \OEone(\vec{r}_3) G(\vec{r}_3,\vec{r}_2) G(\vec{r}_2,\vec{r}_1) V(r_1) \, \psi_{n\l}(\vec{r}_1) \, \chi_{s' J'}^* \chi_{s J} \,.
\end{align}
The last identity holds, because
\begin{align}
& \langle n' ; \l' ; s' ; J' ; \gamma | \OEone(\vec{r}\,) | n ; \l ; s ; J ; 0 \rangle \\
\nonumber &\quad\quad \times \int \d^3 r_1 \, \d^3 r_2 \, \d^3 r_3 \, \psi_{n'\l'}^*(\vec{r}_3) V'(\vec{r}_3) G(\vec{r}_3,\vec{r}_2) G(\vec{r}_2,\vec{r}_1) V(r_1) \, \psi_{n\l}(\vec{r}_1) \, \chi_{s' J'}^* \chi_{s J} \\
&\quad \propto \langle n' ; \l' ; s' ; J' ; \gamma | \OEone(\vec{r}\,) | n ; \l ; s ; J ; 0 \rangle \\
\nonumber &\quad\quad \times \int \d\Omega_1 \, Y_L^{M*}(\Omega_1) Y_\l^m(\Omega_1) \int \d\Omega_2 \, Y_{L'}^{M'*}(\Omega_2) Y_L^M(\Omega_2) \int \d\Omega_3 \, Y_{\l'}^{m'*}(\Omega_3) Y_{L'}^{M'}(\Omega_3) \\
&\quad \propto \langle n' ; \l' ; s' ; J' ; \gamma | \OEone(\vec{r}\,) | n ; \l ; s ; J ; 0 \rangle \delta_{L \l} \delta_{M m} \delta_{L' L} \delta_{M' M} \delta_{\l' L'} \delta_{m' M'} \\
&\quad = \langle n' ; \l ; s' ; J' ; \gamma | \OEone(\vec{r}\,) | n ; \l ; s ; J ; 0 \rangle = 0 \,,
\end{align}
and thus this contribution vanishes exactly, since the electric dipole operator changes the orbital angular momentum by one unit. We therefore have
\begin{align}
& \M_{\text{E1},\text{ini,ini}}^{(2)} = e e_Q \sqrt{\frac{4\pi}{3}} \left(\frac{4\mr}{a}\right)^2 N_{n'\l'} N_{n\l} \sum\limits_{L=0}^\infty \sum\limits_{M=-L}^L \sum\limits_{s=0}^\infty \sum\limits_{L'=0}^\infty \sum\limits_{M'=-L'}^{L'} \sum\limits_{s'=0}^\infty \\
\nonumber &\quad\quad \times \frac{s!}{\lambda (s+L+1-\lambda) (s+2L+1)!} \frac{s'!}{\lambda' (s'+L'+1-\lambda') (s'+2L'+1)!} \\
\nonumber &\quad\quad \times \int \d r_1 \, r_1^2 V(r_1) \r{n}{1}^\l \e{-\frac{1}{2}\r{n}{1}} L_{n-\l-1}^{2\l+1}(\r{n}{1}) \r{\lambda}{1}^L \e{-\frac{1}{2}\r{\lambda}{1}} L_s^{2L+1}(\r{\lambda}{1}) \\
\nonumber &\quad\quad \times \int \d r_2 \, r_2^2 V'(r_2) \r{\lambda}{2}^L \e{-\frac{1}{2}\r{\lambda}{2}} L_s^{2L+1}(\r{\lambda}{2}) \r{\lambda'}{2}^{L'} \e{-\frac{1}{2}\r{\lambda'}{2}} L_{s'}^{2L'+1}(\r{\lambda'}{2}) \\
\nonumber &\quad\quad \times \int \d r_3 \, r_3^3 \r{n'}{3}^{\l'} \e{-\frac{1}{2}\r{n'}{3}} L_{n'-\l'-1}^{2\l'+1}(\r{n'}{3}) \r{\lambda'}{3}^{L'} \e{-\frac{1}{2}\r{\lambda'}{3}} L_{s'}^{2L'+1}(\r{\lambda'}{3}) \\
\nonumber &\quad\quad \times \int \d\Omega_1 \, Y_L^{M*}(\Omega_1) Y_\l^m(\Omega_1) \int \d\Omega_2 \, Y_{L'}^{M'*}(\Omega_2) Y_L^M(\Omega_2) \\
\nonumber &\quad\quad \times \sum\limits_{\mu=-1}^1 \int \d\Omega_3 \, Y_{\l'}^{m'*}(\Omega_3) Y_1^{\mu*}(\Omega_3) Y_{L'}^{M'}(\Omega_3) \\
\nonumber &\quad\quad \times \hat{e}_{r_3} \cdot \langle \gamma | \vec{E} | 0 \rangle \, \chi_{s' J'}^* \chi_{s J} \\
\nonumber &\quad\quad - \langle n ; \l ; s ; J | V'(r) | n ; \l ; s ; J \rangle \, e e_Q \sqrt{\frac{4\pi}{3}} \left(\frac{4\mr}{a}\right)^2 N_{n'\l'} N_{n\l} \sum\limits_{L=0}^\infty \sum\limits_{M=-L}^L \sum\limits_{s=0}^\infty \sum\limits_{L'=0}^\infty \sum\limits_{M'=-L'}^{L'} \sum\limits_{s'=0}^\infty \\
\nonumber &\quad\quad \times \frac{s!}{\lambda (s+L+1-\lambda) (s+2L+1)!} \frac{s'!}{\lambda' (s'+L'+1-\lambda') (s'+2L'+1)!} \\
\nonumber &\quad\quad \times \int \d r_1 \, r_1^2 V(r_1) \r{n}{1}^\l \e{-\frac{1}{2}\r{n}{1}} L_{n-\l-1}^{2\l+1}(\r{n}{1}) \r{\lambda}{1}^L \e{-\frac{1}{2}\r{\lambda}{1}} L_s^{2L+1}(\r{\lambda}{1}) \\
\nonumber &\quad\quad \times \int \d r_2 \, r_2^2 V'(r_2) \r{\lambda}{2}^L \e{-\frac{1}{2}\r{\lambda}{2}} L_s^{2L+1}(\r{\lambda}{2}) \r{\lambda'}{2}^{L'} \e{-\frac{1}{2}\r{\lambda'}{2}} L_{s'}^{2L'+1}(\r{\lambda'}{2}) \\
\nonumber &\quad\quad \times \int \d r_3 \, r_3^3 \r{n'}{3}^{\l'} \e{-\frac{1}{2}\r{n'}{3}} L_{n'-\l'-1}^{2\l'+1}(\r{n'}{3}) \r{\lambda'}{3}^{L'} \e{-\frac{1}{2}\r{\lambda'}{3}} L_{s'}^{2L'+1}(\r{\lambda'}{3}) \\
\nonumber &\quad\quad \times \int \d\Omega_1 \, Y_L^{M*}(\Omega_1) Y_\l^m(\Omega_1) \int \d\Omega_2 \, Y_{L'}^{M'*}(\Omega_2) Y_L^M(\Omega_2) \\
\nonumber &\quad\quad \times \sum\limits_{\mu=-1}^1 \int \d\Omega_3 \, Y_{\l'}^{m'*}(\Omega_3) Y_1^{\mu*}(\Omega_3) Y_{L'}^{M'}(\Omega_3) \\
\nonumber &\quad\quad \times \hat{e}_{r_3} \cdot \langle \gamma | \vec{E} | 0 \rangle \, \chi_{s' J'}^* \chi_{s J} \\
&\quad = e e_Q \sqrt{\frac{4\pi}{3}} \left(\frac{4\mr}{a}\right)^2 N_{n'\l'} N_{n\l} \sum\limits_{s=0}^\infty \sum\limits_{s'=0}^\infty \\
\nonumber &\quad\quad \times \frac{s!}{\lambda (s+\l+1-\lambda) (s+2\l+1)!} \frac{s'!}{\lambda' (s'+\l+1-\lambda') (s'+2\l+1)!} \\
\nonumber &\quad\quad \times \int \d r_1 \, r_1^2 V(r_1) \r{n}{1}^\l \r{\lambda}{1}^{\l} \e{-\frac{1}{2}\r{n}{1}} \e{-\frac{1}{2}\r{\lambda}{1}} L_{n-\l-1}^{2\l+1}(\r{n}{1}) L_s^{2\l+1}(\r{\lambda}{1}) \\
\nonumber &\quad\quad \times \Bigg[ \int \d r_2 \, r_2^2 V'(r_2) \r{\lambda}{2}^{\l} \r{\lambda'}{2}^{\l} \e{-\frac{1}{2}\r{\lambda}{2}} \e{-\frac{1}{2}\r{\lambda'}{2}} L_s^{2\l+1}(\r{\lambda}{2}) L_{s'}^{2\l+1}(\r{\lambda'}{2}) \\
\nonumber &\quad\quad - \langle n , \l , s , j | V'(r) | n , \l , s , j \rangle \int \d r_2 \, r_2^2 \r{\lambda}{2}^{\l} \r{\lambda'}{2}^{\l} \e{-\frac{1}{2}\r{\lambda}{2}} \e{-\frac{1}{2}\r{\lambda'}{2}} L_s^{2\l+1}(\r{\lambda}{2}) L_{s'}^{2\l+1}(\r{\lambda'}{2}) \Bigg] \\
\nonumber &\quad\quad \times \int \d r_3 \, r_3^3 \r{n'}{3}^{\l'} \r{\lambda'}{3}^{\l} \e{-\frac{1}{2}\r{n'}{3}} \e{-\frac{1}{2}\r{\lambda'}{3}} L_{n'-\l'-1}^{2\l'+1}(\r{n'}{3}) L_{s'}^{2\l+1}(\r{\lambda'}{3}) \\
\nonumber &\quad\quad \times \sum\limits_{\mu=-1}^1 (-1)^{m'+\mu} \int \d\Omega_3 \, Y_{\l'}^{-m'}(\Omega_3) Y_1^{-\mu}(\Omega_3) Y_{\l}^{m}(\Omega_3) \\
\nonumber &\quad\quad \times \hat{e}_{r_3} \cdot \langle \gamma | \vec{E} | 0 \rangle \, \chi_{s' J'}^* \chi_{s J} \\
&\quad \equiv e e_Q \widetilde{\M}_{\text{E1},\text{ini,ini}}^{(2)} \times \sum\limits_{\mu=-1}^1 (-1)^{m'+\mu} \int \d\Omega_3 \, Y_{\l'}^{-m'}(\Omega_3) Y_1^{-\mu}(\Omega_3) Y_{\l}^{m}(\Omega_3) \times \hat{e}_{r_3} \cdot \langle \gamma | \vec{E} | 0 \rangle \, \chi_{s' J'}^* \chi_{s J} \,,
\end{align}
where $\widetilde{\M}_{\text{E1},\text{ini,ini}}^{(2)}$ encodes all the relevant corrections due to the potentials $V_1$ and $V_2$. Splitting the sums in $s$ and $s'$ into finite and divergent parts yields four combinatorial possibilities
\begin{equation}
\begin{array}{c|cc}
\times & s' \neq n-\l-1 & s' = n-\l-1 \\
\hline
s \neq n-\l-1 & \lambda=n,\lambda'=n & \lambda=n,\lambda'=\frac{n}{\sqrt{1-\epsilon'}} \\
s = n-\l-1 & \lambda=\frac{n}{\sqrt{1-\epsilon}},\lambda'=n & \lambda=\frac{n}{\sqrt{1-\epsilon}},\lambda'=\frac{n}{\sqrt{1-\epsilon'}}
\end{array}
\end{equation}
that need to be treated such that the divergence cancellation is ensured. We demonstrate the divergence cancellation in Appendix~\ref{app:sec:DemonstrationDivergenceCancellation}. It is important to notice that $\lambda$ and $\lambda'$ are chosen to be proportional to $n$ and not to $n'$, since the correction only affects the initial state. This is also reflected by how the angular momentum quantum numbers are fixed by the orthonormality of the spherical harmonics, setting $L = L' = \l$.\\
The matrix element associated with the second diagram is given by
\begin{align}
& \M_{\text{E1},\text{fin,fin}}^{(2)} = \langle n' ; \l' ; s' ; J' ; \gamma | V'(\vec{r}_3) \frac{1}{(E_{n'} - H)'} V(\vec{r}_2) \frac{1}{(E_{n'} - H)'} \OEone(\vec{r}_1) | n ; \l ; s ; J ; 0 \rangle \\
\nonumber &\quad\quad \!\!\! - \langle n' ; \l' ; s' ; J' | V(r) | n' ; \l' ; s' ; J' \rangle \langle n' ; \l' ; s' ; J' ; \gamma | V'(\vec{r}_3) \frac{1}{(E_n - H)'} \1(\vec{r}_2) \frac{1}{(E_n - H)'} \OEone(r_1) | n ; \l ; s ; J ; 0 \rangle \\
&\quad = (-1)^2 \int \d^3 r_1 \, \d^3 r_2 \, \d^3 r_3 \, \psi_{n'\l'}^*(\vec{r}_3) V'(\vec{r}_3) G(\vec{r}_3,\vec{r}_2) V(\vec{r}_2) G(\vec{r}_2,\vec{r}_1) \OEone(\vec{r}_1) \, \psi_{n\l}(\vec{r}_1) \, \chi_{s' J'}^* \chi_{s J} \\
\nonumber &\quad\quad - \langle n' ; \l' ; s' ; J' | V(r) | n' ; \l' ; s' ; J' \rangle \\
\nonumber &\quad\quad\quad \times (-1)^2 \int \d^3 r_1 \, \d^3 r_2 \, \d^3 r_3 \, \psi_{n'\l'}^*(\vec{r}_3) V'(\vec{r}_3) G(\vec{r}_3,\vec{r}_2) G(\vec{r}_2,\vec{r}_1) \OEone(\vec{r}_1) \, \psi_{n\l}(\vec{r}_1) \, \chi_{s' J'}^* \chi_{s J} \\
&\quad = e e_Q \sqrt{\frac{4\pi}{3}} \left(\frac{4\mr}{a}\right)^2 N_{n'\l'} N_{n\l} \sum\limits_{L=0}^\infty \sum\limits_{M=-L}^L \sum\limits_{s=0}^\infty \sum\limits_{L'=0}^\infty \sum\limits_{M'=-L'}^{L'} \sum\limits_{s'=0}^\infty \\
\nonumber &\quad\quad \times \frac{s!}{\lambda (s+L+1-\lambda) (s+2L+1)!} \frac{s'!}{\lambda' (s'+L'+1-\lambda') (s'+2L'+1)!} \\
\nonumber &\quad\quad \times \int \d r_1 \, r_1^3 \r{n}{1}^\l \e{-\frac{1}{2}\r{n}{1}} L_{n-\l-1}^{2\l+1}(\r{n}{1}) \r{\lambda}{1}^L \e{-\frac{1}{2}\r{\lambda}{1}} L_s^{2L+1}(\r{\lambda}{1}) \\
\nonumber &\quad\quad \times \Bigg[ \int \d r_2 \, r_2^2 V(r_2) \r{\lambda}{2}^L \e{-\frac{1}{2}\r{\lambda}{2}} L_s^{2L+1}(\r{\lambda}{2}) \r{\lambda'}{2}^{L'} \e{-\frac{1}{2}\r{\lambda'}{2}} L_{s'}^{2L'+1}(\r{\lambda'}{2}) \\
\nonumber &\quad\quad - \langle n' , \l' , s' , j' | V(r) | n' , \l' , s' , j' \rangle \int \d r_2 \, r_2^2 \r{\lambda}{2}^L \e{-\frac{1}{2}\r{\lambda}{2}} L_s^{2L+1}(\r{\lambda}{2}) \r{\lambda'}{2}^{L'} \e{-\frac{1}{2}\r{\lambda'}{2}} L_{s'}^{2L'+1}(\r{\lambda'}{2}) \Bigg] \\
\nonumber &\quad\quad \times \int \d r_3 \, r_3^2 V'(r_3) \r{n'}{3}^{\l'} \e{-\frac{1}{2}\r{n'}{3}} L_{n'-\l'-1}^{2\l'+1}(\r{n'}{3}) \r{\lambda'}{3}^{L'} \e{-\frac{1}{2}\r{\lambda'}{3}} L_{s'}^{2L'+1}(\r{\lambda'}{3}) \\
\nonumber &\quad\quad \times \sum\limits_{\mu=-1}^1 \int \d\Omega_1 \, Y_L^{M*}(\Omega_1) Y_1^{\mu*}(\Omega_1) Y_\l^m(\Omega_1) \\
\nonumber &\quad\quad \times \int \d\Omega_2 \, Y_{L'}^{M'*}(\Omega_2) Y_L^M(\Omega_2) \int \d\Omega_3 \, Y_{\l'}^{m'*}(\Omega_3) Y_{L'}^{M'}(\Omega_3) \\
\nonumber &\quad\quad \times \hat{e}_{r_1} \cdot \langle \gamma | \vec{E} | 0 \rangle \, \chi_{s' J'}^* \chi_{s J} \\
&\quad = e e_Q \sqrt{\frac{4\pi}{3}} \left(\frac{4\mr}{a}\right)^2 N_{n'\l'} N_{n\l} \sum\limits_{s=0}^\infty \sum\limits_{s'=0}^\infty \\
\nonumber &\quad\quad \times \frac{s!}{\lambda (s+\l'+1-\lambda) (s+2\l'+1)!} \frac{s'!}{\lambda' (s'+\l'+1-\lambda') (s'+2\l'+1)!} \\
\nonumber &\quad\quad \times \int \d r_1 \, r_1^3 \r{n}{1}^\l \r{\lambda}{1}^{\l'} \e{-\frac{1}{2}\r{n}{1}} \e{-\frac{1}{2}\r{\lambda}{1}} L_{n-\l-1}^{2\l+1}(\r{n}{1}) L_s^{2\l'+1}(\r{\lambda}{1}) \\
\nonumber &\quad\quad \times \Bigg[ \int \d r_2 \, r_2^2 V(r_2) \r{\lambda}{2}^{\l'} \r{\lambda'}{2}^{\l'} \e{-\frac{1}{2}\r{\lambda}{2}} \e{-\frac{1}{2}\r{\lambda'}{2}} L_s^{2\l'+1}(\r{\lambda}{2}) L_{s'}^{2\l'+1}(\r{\lambda'}{2}) \\
\nonumber &\quad\quad - \langle n' , \l' , s' , j' | V(r) | n' , \l' , s' , j' \rangle \int \d r_2 \, r_2^2 \r{\lambda}{2}^{\l'} \r{\lambda'}{2}^{\l'} \e{-\frac{1}{2}\r{\lambda}{2}} \e{-\frac{1}{2}\r{\lambda'}{2}} L_s^{2\l'+1}(\r{\lambda}{2}) L_{s'}^{2\l'+1}(\r{\lambda'}{2}) \Bigg] \\
\nonumber &\quad\quad \times \int \d r_3 \, r_3^2 V'(r_3) \r{n'}{3}^{\l'} \r{\lambda'}{3}^{\l'} \e{-\frac{1}{2}\r{n'}{3}} \e{-\frac{1}{2}\r{\lambda'}{3}} L_{n'-\l'-1}^{2\l'+1}(\r{n'}{3})  L_{s'}^{2\l'+1}(\r{\lambda'}{3}) \\
\nonumber &\quad\quad \times \sum\limits_{\mu=-1}^1 (-1)^{m'+\mu} \int \d\Omega_1 \, Y_{\l'}^{-m'}(\Omega_1) Y_1^{-\mu}(\Omega_1) Y_\l^m(\Omega_1) \\
\nonumber &\quad\quad \times \hat{e}_{r_1} \cdot \langle \gamma | \vec{E} | 0 \rangle \, \chi_{s' J'}^* \chi_{s J} \\
&\quad \equiv e e_Q \widetilde{\M}_{\text{E1},\text{fin,fin}}^{(2)} \times \sum\limits_{\mu=-1}^1 (-1)^{m'+\mu} \int \d\Omega_1 \, Y_{\l'}^{-m'}(\Omega_1) Y_1^{-\mu}(\Omega_1) Y_\l^m(\Omega_1) \times \hat{e}_{r_1} \cdot \langle \gamma | \vec{E} | 0 \rangle \, \chi_{s' J'}^* \chi_{s J} \,,
\end{align}
and the four combinatorial possibilities are
\begin{equation}
\begin{array}{c|cc}
\times & s' \neq n'-\l'-1 & s' = n'-\l'-1 \\
\hline
s \neq n'-\l'-1 & \lambda=n',\lambda'=n' & \lambda=n',\lambda'=\frac{n'}{\sqrt{1-\epsilon'}} \\
s = n'-\l'-1 & \lambda=\frac{n'}{\sqrt{1-\epsilon}},\lambda'=n' & \lambda=\frac{n'}{\sqrt{1-\epsilon}},\lambda'=\frac{n'}{\sqrt{1-\epsilon'}}
\end{array}
\end{equation}
where now $\lambda$ and $\lambda'$ are proportional to $n'$, since the correction affects the final state only.\\
Finally, the third matrix element only involves first order corrections to each of the states and is thus given by
\begin{align}
& \M_{\text{E1},\text{ini,fin}}^{(2)} = \langle n' ; \l' ; s' ; J' ; \gamma | V'(\vec{r}_3) \frac{1}{(E_{n'} - H)'} \OEone(\vec{r}_2) \frac{1}{(E_n - H)'} V(\vec{r}_1) | n ; \l ; s ; J ; 0 \rangle \\
&\quad = (-1)^2 \int \d^3 r_1 \, \d^3 r_2 \, \d^3 r_3 \, \psi_{n'\l'}^*(\vec{r}_3) V'(\vec{r}_3) G(\vec{r}_3,\vec{r}_2) \OEone(\vec{r}_2) G(\vec{r}_2,\vec{r}_1) V(\vec{r}_1) \, \psi_{n\l}(\vec{r}_1) \, \chi_{s' J'}^* \chi_{s J} \\
&\quad = e e_Q \sqrt{\frac{4\pi}{3}} \left(\frac{4\mr}{a}\right)^2 N_{n'\l'} N_{n\l} \sum\limits_{L=0}^\infty \sum\limits_{M=-L}^L \sum\limits_{s=0}^\infty \sum\limits_{L'=0}^\infty \sum\limits_{M'=-L'}^{L'} \sum\limits_{s'=0}^\infty \\
\nonumber &\quad\quad \times \frac{s!}{\lambda (s+L+1-\lambda) (s+2L+1)!} \frac{s'!}{\lambda' (s'+L'+1-\lambda') (s'+2L'+1)!} \\
\nonumber &\quad\quad \times \int \d r_1 \, r_1^2 V(r_1) \r{n}{1}^\l \e{-\frac{1}{2}\r{n}{1}} L_{n-\l-1}^{2\l+1}(\r{n}{1}) \r{\lambda}{1}^L \e{-\frac{1}{2}\r{\lambda}{1}} L_s^{2L+1}(\r{\lambda}{1}) \\
\nonumber &\quad\quad \times \int \d r_2 \, r_2^3 \r{\lambda}{2}^L \e{-\frac{1}{2}\r{\lambda}{2}} L_s^{2L+1}(\r{\lambda}{2}) \r{\lambda'}{2}^{L'} \e{-\frac{1}{2}\r{\lambda'}{2}} L_{s'}^{2L'+1}(\r{\lambda'}{2}) \\
\nonumber &\quad\quad \times \int \d r_3 \, r_3^2 V'(r_3) \r{n'}{3}^{\l'} \e{-\frac{1}{2}\r{n'}{3}} L_{n'-\l'-1}^{2\l'+1}(\r{n'}{3}) \r{\lambda'}{3}^{L'} \e{-\frac{1}{2}\r{\lambda'}{3}} L_{s'}^{2L'+1}(\r{\lambda'}{3}) \\
\nonumber &\quad\quad \times \int \d\Omega_1 \, Y_L^{M*}(\Omega_1) Y_\l^m(\Omega_1) \\
\nonumber &\quad\quad \times \sum\limits_{\mu=-1}^1 \int \d\Omega_2 \, Y_{L'}^{M'*}(\Omega_2) Y_1^{\mu*}(\Omega_2) Y_L^M(\Omega_2) \\
\nonumber &\quad\quad \times \int \d\Omega_3 \, Y_{\l'}^{m'*}(\Omega_3) Y_{L'}^{M'}(\Omega_3) \\
\nonumber &\quad\quad \times \hat{e}_{r_2} \cdot \langle \gamma | \vec{E} | 0 \rangle \, \chi_{s' J'}^* \chi_{s J} \\
&\quad = e e_Q \sqrt{\frac{4\pi}{3}} \left(\frac{4\mr}{a}\right)^2 N_{n'\l'} N_{n\l} \sum\limits_{s=0}^\infty \sum\limits_{s'=0}^\infty \\
\nonumber &\quad\quad \times \frac{s!}{\lambda (s+\l+1-\lambda) (s+2\l+1)!} \frac{s'!}{\lambda' (s'+\l'+1-\lambda') (s'+2\l'+1)!} \\
\nonumber &\quad\quad \times \int \d r_1 \, r_1^2 V(r_1) \r{n}{1}^\l \r{\lambda}{1}^{\l} \e{-\frac{1}{2}\r{n}{1}} \e{-\frac{1}{2}\r{\lambda}{1}} L_{n-\l-1}^{2\l+1}(\r{n}{1}) L_s^{2\l+1}(\r{\lambda}{1}) \\
\nonumber &\quad\quad \times \int \d r_2 \, r_2^3 \r{\lambda}{2}^{\l} \r{\lambda'}{2}^{\l'} \e{-\frac{1}{2}\r{\lambda}{2}} \e{-\frac{1}{2}\r{\lambda'}{2}} L_s^{2\l+1}(\r{\lambda}{2}) L_{s'}^{2\l'+1}(\r{\lambda'}{2}) \\
\nonumber &\quad\quad \times \int \d r_3 \, r_3^2 V'(r_3) \r{n'}{3}^{\l'} \r{\lambda'}{3}^{\l'} \e{-\frac{1}{2}\r{n'}{3}} \e{-\frac{1}{2}\r{\lambda'}{3}} L_{n'-\l'-1}^{2\l'+1}(\r{n'}{3}) L_{s'}^{2\l'+1}(\r{\lambda'}{3}) \\
\nonumber &\quad\quad \times \sum\limits_{\mu=-1}^1 (-1)^{m'+\mu} \int \d\Omega_2 \, Y_{\l'}^{-m'}(\Omega_2) Y_1^{-\mu}(\Omega_2) Y_{\l}^m(\Omega_2) \\
\nonumber &\quad\quad \times \hat{e}_{r_2} \cdot \langle \gamma | \vec{E} | 0 \rangle \, \chi_{s' J'}^* \chi_{s J} \\
&\quad \equiv e e_Q \widetilde{\M}_{\text{E1},\text{ini,fin}}^{(2)} \times \sum\limits_{\mu=-1}^1 (-1)^{m'+\mu} \int \d\Omega_2 \, Y_{\l'}^{-m'}(\Omega_2) Y_1^{-\mu}(\Omega_2) Y_{\l}^m(\Omega_2) \times \hat{e}_{r_2} \cdot \langle \gamma | \vec{E} | 0 \rangle \, \chi_{s' J'}^* \chi_{s J} \,,
\end{align}
and the four combinatorial possibilities are
\begin{equation}
\begin{array}{c|cc}
\times & s' \neq n'-\l'-1 & s' = n'-\l'-1 \\
\hline
s \neq n-\l-1 & \lambda=n,\lambda'=n' & \lambda=n,\lambda'=\frac{n'}{\sqrt{1-\epsilon'}} \\
s = n-\l-1 & \lambda=\frac{n}{\sqrt{1-\epsilon}},\lambda'=n' & \lambda=\frac{n}{\sqrt{1-\epsilon}},\lambda'=\frac{n'}{\sqrt{1-\epsilon'}}
\end{array}
\end{equation}
where now $\lambda$ and $\lambda'$ are proportional to $n$ and $n'$, respectively, since each state is affected by one correction independently.\\
This final result gives us all the formulae needed in order to compute the E1 decay width up to NNLO, $\O(m\alpha_s^6)$, in pNRQCD at weak coupling for a p-wave to s-wave transition.

\section{Non-perturbative color octet contributions to the decay width}
\label{sec:NonPerturbativeContributions}

\begin{figure}[t]
\centering
{\includegraphics[
clip,trim={2.5cm 14.3cm 2.4cm 3.5cm},width=0.7\textwidth]{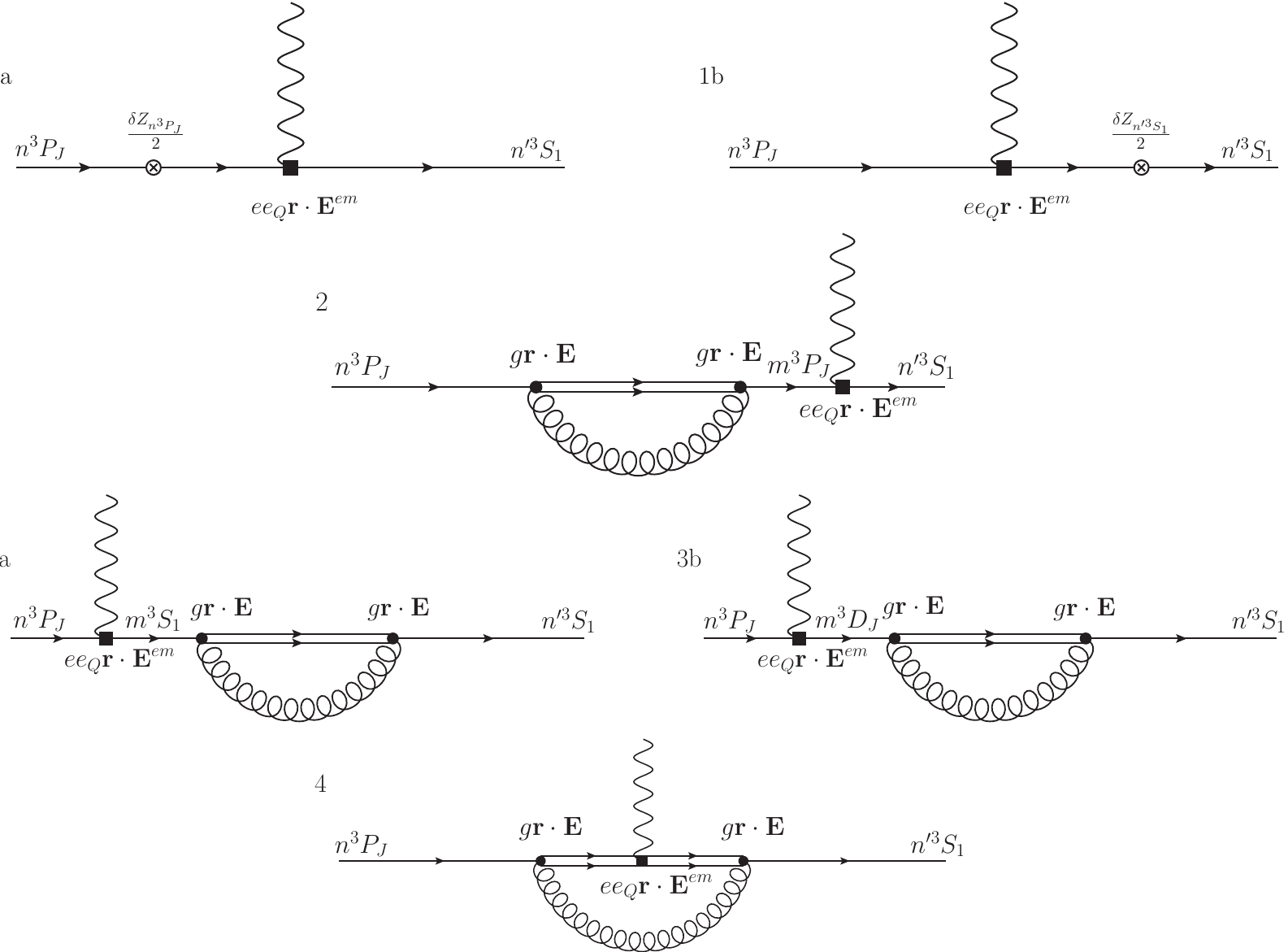}}
\caption[Color-octet contributions to the E1 transition $n\,^3\!P_J \to n'\,^3\!S_1 + \gamma$.]{Color-octet contributions to the E1 transition $n\,^3\!P_J \to n'\,^3\!S_1 + \gamma$ for weakly-coupled states. The double lines represent intermediate octet states and the curly lines represent ultra soft gluons. Taken from\cite{Brambilla:2012be}.}
\label{fig:NonPerturbativeDiagrams}
\end{figure}

Apart from the perturbative analysis performed until now, there are also additional non-perturbative contributions to the electric dipole decay width.\footnote{An overview, including the formulas we will use below, may be found in Ref.~\cite{Brambilla:2012be}.} A diagrammatic representation of those contributions is presented in Fig.~\ref{fig:NonPerturbativeDiagrams}. In the case of magnet dipole transitions all of these contributions vanish, since the magnetic dipole operator is independent of $r$ and commutes with the kinetic energy \cite{Brambilla:2005zw,Brambilla:2012be}.
They are corrections due to higher order Fock states and steam from the coupling of either the initial or the final quarkonium states to low-energy degrees of freedom. The lowest higher order Fock state consists of an ultra-soft gluon and a quarkonium in a color-octet configuration. Thus, the electric dipole operator may now additionally couple to the quark-antiquark octet state in the Lagrangian~\eqref{eq:pNRQCDLagrangian}, explicitly through terms like
\begin{equation}
\label{eq:OctetLagrangian} \L_{\text{pNRQCD}} \ni V_A(r) \, \text{tr}\lbrace O^\dagger g \vec{r} \cdot \vec{E} S + S^\dagger g \vec{r} \cdot \vec{E} O \rbrace \,.
\end{equation}
The first two diagrams (1a and 1b) of Fig.~\ref{fig:NonPerturbativeDiagrams} correspond to the renormalization $\frac{1}{2} (\delta Z_{n\,^3\!P_J} + \delta Z_{n'\,^3\!S_1})$ of the initial and final states. Diagram 2 depicts the coupling of an intermediate octet state to the initial p-wave quarkonium state and diagram 3a depicts the same situation for the final s-wave quarkonium state. In diagram 3b the intermediate octet state couples an intermediate d-wave color singlet to the final state s-wave quarkonium. Note further that in all of the diagrams, 2, 3a, and 3b, the intermediate color singlet quarkonium state has the principal quantum number $m$, which may different from the initial $n$ or final $n'$ principal quantum number. In this three cases the photon is radiated from the color singlet initial or final state quarkonium, whereas in diagram 4 the intermediate octet state radiates the photon according to Eq.~\eqref{eq:OctetLagrangian}.\\
The matrix elements \cite{Brambilla:2012be}, corresponding to Fig.~\ref{fig:NonPerturbativeDiagrams} are given by
\begin{align}
\label{eq:MatrixElementNormalizationState} & 
\begin{aligned}
& \M_{n\,^3\!P_J \to n'\,^3\!S_1 + \gamma}^{\text{fig. \ref{fig:NonPerturbativeDiagrams}, 1}} = \frac{\M_{n\,^3\!P_J \to n'\,^3\!S_1 + \gamma}^{(0)}}{12} \int\limits_{0}^{\infty} \d t \, t \, \langle \text{vac} | g E^{a\,i}(\vec{R},t) \phi(t,0)^{\text{adj}}_{a b} g E^{b\,i}(\vec{R},0) | \text{vac} \rangle \\
& \quad \times \left[\,^{(0)}\!\langle n\,^3\!P_J | r^j \e{-\i(H_{\text{o}}^{(0)}-E_{n1}^{(0)}) t} r^j | n\,^3\!P_J \rangle^{(0)} + \,^{(0)}\!\langle n'\,^3\!S_1 | r^j \e{-\i(H_{\text{o}}^{(0)}-E_{n'0}^{(0)}) t} r^j | n'\,^3\!S_1 \rangle^{(0)}\right] \,,
\end{aligned} \\
&
\begin{aligned}
& \M_{n\,^3\!P_J \to n'\,^3\!S_1 + \gamma}^{\text{fig. \ref{fig:NonPerturbativeDiagrams}, 2}} = - \frac{\i}{6} \sum\limits_{m \neq n} \frac{\M_{m\,^3\!P_J \to n'\,^3\!S_1 + \gamma}^{(0)}}{E_{n1}^{(0)} - E_{m1}^{(0)}} \int\limits_{0}^{\infty} \d t \, \langle \text{vac} | g E^{a\,i}(\vec{R},t) \phi(t,0)^{\text{adj}}_{a b} g E^{b\,i}(\vec{R},0) | \text{vac} \rangle \\
& \quad \times \,^{(0)}\!\langle m\,^3\!P_J | r^j \e{-\i(H_{\text{o}}^{(0)}-E_{n1}^{(0)}) t} r^j | n\,^3\!P_J \rangle^{(0)} \,,
\end{aligned} \\
&
\begin{aligned}
& \M_{n\,^3\!P_J \to n'\,^3\!S_1 + \gamma}^{\text{fig. \ref{fig:NonPerturbativeDiagrams}, 3a}} = - \frac{\i}{6} \sum\limits_{m \neq n'} \frac{\M_{n\,^3\!P_J \to m\,^3\!S_1 + \gamma}^{(0)}}{E_{n'0}^{(0)} - E_{m0}^{(0)}} \int\limits_{0}^{\infty} \d t \, \langle \text{vac} | g E^{a\,i}(\vec{R},t) \phi(t,0)^{\text{adj}}_{a b} g E^{b\,i}(\vec{R},0) | \text{vac} \rangle \\
& \quad \times \,^{(0)}\!\langle n'\,^3\!S_1 | r^j \e{-\i(H_{\text{o}}^{(0)}-E_{n'0}^{(0)}) t} r^j | m\,^3\!S_1 \rangle^{(0)} \,,
\end{aligned} \\
& \M_{n\,^3\!P_J \to n'\,^3\!S_1 + \gamma}^{\text{fig. \ref{fig:NonPerturbativeDiagrams}, 3b}} = 0 \,, \\
&
\begin{aligned}
& \M_{n\,^3\!P_J \to n'\,^3\!S_1 + \gamma}^{\text{fig. \ref{fig:NonPerturbativeDiagrams}, 4}} = - \frac{\i}{6} \int\limits_{0}^{\infty} \d t \, \int\limits_{0}^{t} \d t' \, \langle \text{vac} | g E^{a\,i}(\vec{R},t) \phi(t,0)^{\text{adj}}_{a b} g E^{b\,i}(\vec{R},0) | \text{vac} \rangle \\
& \quad \times \,^{(0)}\!\langle n'\,^3\!S_1 | r^{j} \e{-\i(H_{\text{o}}^{(0)}-E_{n'0}^{(0)}) (t-t')} \, e e_Q k_\gamma \, \vec{\epsilon}^{\,*}(\sigma) \cdot \vec{r} \, \e{-\i(H_{\text{o}}^{(0)}-E_{n1}^{(0)}) t'} r^{j} | n\,^3\!P_J \rangle^{(0)} \,,
\end{aligned}
\end{align}
where $\M_{n\,^3P_J \to n'\,^3S_1 + \gamma}^{\text{fig. \ref{fig:NonPerturbativeDiagrams}, 3b}}$ vanishes, since the scalar $r^j \, \e{-\i(H_{\text{o}}^{(0)}-E_{n'0}^{(0)}) t} \, r^j$ cannot change the orbital angular momentum. The vacuum state $| \text{vac} \rangle$ contains no heavy quarks and only ultra soft gluons. The Wilson line in the adjoint representation is given by
\begin{equation}
\label{eq:WilsonLine} \phi(t,0)_{a b}^{\text{adj}} = \left[\exp{-\i g \int\limits_{0}^{t} \d t' \, A_0(\vec{R},t')^{\text{adj}}}\right]_{a b} \,,
\end{equation}
and the leading order octet Hamiltonian is $H_{\text{o}}^{(0)} = -\frac{\nabla_r^2}{m} + V_{\text{o}}^{(0)}$ with $V_{\text{o}}^{(0)} = \frac{1}{2 N_c} \frac{\alpha_s}{r}$ \cite{Brambilla:2004jw}.\\
We consider $E \gg \Lambda_{\text{QCD}}$ which implies that the chromo-electric correlator reduces to the two gluon condensate and all of the above terms contribute. However, we just consider the renormalization to the states here. The renormalization is related to the self-energy of a generic state $H$, which was first derived in \cite{Brambilla:1999xf}. The self-energy is - up to a factor of $\frac{1}{N_C}$ that has been reabsorbed into the normalization of the vacuum - given by
\begin{align}
\label{eq:SelfEnergy}
\begin{aligned}
\delta E_{H(\lambda)} &= -\frac{\i}{6} \int\limits_{0}^{\infty} \d t \, \langle \text{vac} | g E^{a\,i}(\vec{R},t) \phi(t,0)_{a b}^{\text{adj}} g E^{b\,i}(\vec{R},0) | \text{vac} \rangle \\
& \times \,^{(0)}\!\langle H(\vec{0},\lambda) | r^{j} \e{-\i(H_{\text{o}}^{(0)} - E_{H}^{(0)})t} r^{j} | H(\vec{0},\lambda) \rangle^{(0)} \,.
\end{aligned}
\end{align}
The normalization of the state is given by the derivative of the self-energy with respect to the energy, hence by
\begin{align}
\label{eq:NormalizationState}
\begin{aligned}
\delta Z_{H(\lambda)} &= \frac{\partial \delta E_{H(\lambda)}}{\partial E_{H}^{(0)}} = \frac{1}{6} \int\limits_{0}^{\infty} \d t \, t \, \langle \text{vac} | g E^{a\,i}(\vec{R},t) \phi(t,0)_{a b}^{\text{adj}} g E^{b\,i}(\vec{R},0) | \text{vac} \rangle \\
& \times \,^{(0)}\!\langle H(\vec{0},\lambda) | r^{j} \e{-\i(H_{\text{o}}^{(0)} - E_{H}^{(0)})t} r^{j} | H(\vec{0},\lambda) \rangle^{(0)} \,.
\end{aligned}
\end{align}
In order to incorporate the contribution to the normalization of the states, we use that\footnote{In this context $A = \lim\limits_{\epsilon \to 0} (H_{\text{o}}^{(0)} - E_{H}^{(0)} - \i\epsilon)$, such that Im[$A$]$<0$ and the integral converges.}
\begin{equation}
\int\limits_{0}^{\infty} \d t \, \e{-\i A t} = \frac{-\i}{A} \,, \quad \text{for Im}[A] < 0 \,.
\end{equation}
We furthermore assume that the chromo-electric condensate is given by a local condensate, hence
\begin{equation}
\langle \text{vac} | g E^{a\,i}(\vec{R},t) \phi(t,0)_{a b}^{\text{adj}} g E^{b\,i}(\vec{R},0) | \text{vac} \rangle \simeq \langle \text{vac} | g^2 E^{a\,i} \delta_{a b} E^{b\,i} | \text{vac} \rangle = \langle \text{vac} | g^2 E^2 | \text{vac} \rangle \,.
\end{equation}
This is equivalent to taking the limit $t \to 0$ in Eq.~\eqref{eq:WilsonLine}, which then evaluates to $\phi(t,0)_{a b}^{\text{adj}} \overset{t \to 0}{\longrightarrow} \delta_{a b}$, and the assumption is valid as long as $E \gtrsim \Lambda_{\text{QCD}}$. We thus may approximate the self-energy~\eqref{eq:SelfEnergy} by
\begin{align}
\begin{aligned}
\delta E_{H(\lambda)} &\simeq -\frac{\i}{6} \langle \text{vac} | g^2 E^{a\,i} \delta_{a b} E^{b\,i} | \text{vac} \rangle \int\limits_{0}^{\infty} \d t \, ^{(0)}\!\langle H(\vec{0},\lambda) | r \e{-\i(H_{\text{o}}^{(0)} - E_{H}^{(0)})t} r | H(\vec{0},\lambda) \rangle^{(0)} \\
&= -\frac{1}{6} \langle \text{vac} | g^2 E^{a\,i} \delta_{a b} E^{b\,i} | \text{vac} \rangle \, ^{(0)}\!\langle H(\vec{0},\lambda) | r \frac{1}{(H_{\text{o}}^{(0)} - E_{H}^{(0)})} r | H(\vec{0},\lambda) \rangle^{(0)} \,.
\end{aligned}
\end{align}
The normalization of the state, Eq.~\eqref{eq:NormalizationState}, is then given by
\begin{align}
\delta Z_{H(\lambda)} &\simeq -\frac{1}{6} \langle \text{vac} | g^2 E^{a\,i} \delta_{a b} E^{b\,i} | \text{vac} \rangle \frac{\partial}{\partial E_{H}^{(0)}} \, ^{(0)}\!\langle H(\vec{0},\lambda) | r \frac{1}{\big(H_{\text{o}}^{(0)} - E_{H}^{(0)}\big)} r | H(\vec{0},\lambda) \rangle^{(0)} \\
\nonumber &= -\frac{1}{6} \langle \text{vac} | g^2 E^{a\,i} \delta_{a b} E^{b\,i} | \text{vac} \rangle \, ^{(0)}\!\langle H(\vec{0},\lambda) | r \frac{1}{\big(H_{\text{o}}^{(0)} - E_{H}^{(0)}\big)^2} r | H(\vec{0},\lambda) \rangle^{(0)} \\
\nonumber &= -\frac{1}{6} \langle \text{vac} | g^2 E^{a\,i} \delta_{a b} E^{b\,i} | \text{vac} \rangle \, ^{(0)}\!\langle H(\vec{0},\lambda) | r \frac{1}{\big(H_{\text{o}}^{(0)} - E_{H}^{(0)}\big)} \1 \frac{1}{\big(H_{\text{o}}^{(0)} - E_{H}^{(0)}\big)} r | H(\vec{0},\lambda) \rangle^{(0)} \,,
\end{align}
and thus the contributions of diagram 1a and 1b of Fig.~\ref{fig:NonPerturbativeDiagrams} to the matrix element~\eqref{eq:MatrixElementNormalizationState} read
\begin{align}
\label{eq:MatrixElementNormalizationStateFinal} & \M_{n\,^3\!P_J \to n'\,^3\!S_1 + \gamma}^{\text{fig. 8, 1}} \simeq \M_{n\,^3\!P_J \to n'\,^3\!S_1 + \gamma}^{(0)} \frac{1}{2} \left(-\frac{1}{6}\right) \langle \text{vac} | g^2 E^{a\,i} \delta_{a b} E^{b\,i} | \text{vac} \rangle \\
\nonumber & \quad \times \left[\,^{(0)}\!\langle n\,^3\!P_J | r \frac{1}{(H_{\text{o}}^{(0)} - E_{H}^{(0)})} \1 \frac{1}{(H_{\text{o}}^{(0)} - E_{H}^{(0)})} r | n\,^3\!P_J \rangle^{(0)} \right. \\
\nonumber & \quad\quad \left. + \,^{(0)}\!\langle n'\,^3\!S_1 | r \frac{1}{(H_{\text{o}}^{(0)} - E_{H}^{(0)})} \1 \frac{1}{(H_{\text{o}}^{(0)} - E_{H}^{(0)})} r | n'\,^3\!S_1 \rangle^{(0)} \right] \,,
\end{align}
where the additional factor of $\frac{1}{2}$ is due to the definition of $\frac{\delta Z_{H(\lambda)}}{2}$ in the respective diagrams.\\
The object $\frac{1}{(H_{\text{o}}^{(0)} - E_{H}^{(0)})}$ is the octet Coulomb Green function which is given by \cite{Voloshin:1979uv}
\begin{equation}
G_{\text{o}}(\vec{r}_1,\vec{r}_2;E) = \sum\limits_{\l'=0}^\infty (2\l'+1) P_{\l'}(\hat{r}_1 \cdot \hat{r}_2) G_{\l',o}(r_1,r_2;E) \,,
\end{equation}
where
\begin{equation}
G_{\l',{\text{o}}}(r_1,r_2;E) = \frac{m k}{2\pi} (2 k r_1)^{\l'} (2 k r_2)^{\l'} \e{-k(r_1+r_2)} \sum\limits_{s=0}^\infty \frac{L_{s}^{2\l'+1}(2 k r_1)L_{s}^{2\l'+1}(2 k r_2) \, s!}{(s+\l'+1+\frac{m \alpha_s}{2 \cdot 2N_{\text{c}} k})[(s+2\l'+1)!]} \,,
\end{equation}
and the energy is defined as
\begin{equation}
E = - \frac{k^2}{m} = - \frac{\mr C_F^2 \alpha_s^2}{2 n^2} \quad \Rightarrow \quad k = \frac{\mr C_F \alpha_s}{n} = \frac{1}{n a} \,.
\end{equation}
We thus have
\begin{align}
& G_{\text{o}}(\vec{r}_1,\vec{r}_2;E) = \frac{4\pi m k}{2\pi} \sum\limits_{\l'=0}^\infty \sum\limits_{s=0}^\infty \frac{s!}{(s+\l'+1+\frac{m \alpha_s}{2 \cdot 2N_{\text{c}} k})[(s+2\l'+1)!]} \\
\nonumber &\quad\quad \times (2 k r_1)^{\l'} \e{-k r_1} L_{s}^{2\l'+1}(2 k r_1) (2 k r_2)^{\l'} \e{-k r_2} L_{s}^{2\l'+1}(2 k r_2) \\
\nonumber &\quad\quad \times \sum\limits_{m'=-\l'}^{\l'} Y_{\l'}^{m'*}(\Omega_1) Y_{\l'}^{m'}(\Omega_2) \\
&\quad = \frac{4\mr}{n a} \sum\limits_{\l'=0}^\infty \sum\limits_{s=0}^\infty \frac{s!}{(s+\l'+1+\frac{n}{2N_{\text{c}} C_F})[(s+2\l'+1)!]} \\
\nonumber &\quad\quad \times \left(\frac{2 r_1}{n a}\right)^{\l'} \e{-\frac{1}{n a} r_1} L_{s}^{2\l'+1}\left(\frac{2 r_1}{n a}\right) \, \left(\frac{2 r_2}{n a}\right)^{\l'} \e{-\frac{1}{n a} r_2} L_{s}^{2\l'+1}\left(\frac{2 r_2}{n a}\right) \\
\nonumber &\quad\quad \times \sum\limits_{m'=-\l'}^{\l'} Y_{\l'}^{m'*}(\Omega_1) Y_{\l'}^{m'}(\Omega_2) \,,
\end{align}
which is exactly the singlet Coulomb Green function, except for the denominator that reads in the singlet case $(s+\l'+1-\lambda)$ and got replaced by $(s+\l'+1+\frac{n}{2N_{\text{c}} C_F})$ in the octet case. This amounts to the change of the variable $\lambda \to -\frac{n}{2N_{\text{c}} C_F}$, that explicitly takes care of the change in the potential of the Hamiltonian, going from a attractive one $\propto -C_F$ in the singlet case to a repulsive one $\propto +\frac{1}{2N_{\text{c}}}$ in the octet case. This change in sign also cures the divergence for one particular value of $s$ that is present in the singlet case, such that we do not need the prescription using $\lambda$ but can directly use $n$.\\
The matrix element we need to compute then takes the following form, where we again use the dimensionless parameter $\rho_i = \frac{2 r_i}{n a}$,
\begin{align}
& ^{(0)}\!\langle n ; \l | r \frac{1}{(H_{\text{o}}^{(0)} - E_{H}^{(0)})} \1 \frac{1}{(H_{\text{o}}^{(0)} - E_{H}^{(0)})} r | n ; \l \rangle^{(0)} \\
& \quad = \int \d^3 r_1 \d^3 r_2 \d^3 r_3 \, \psi_{n\l}^*(\vec{r}_3) r_3 G_{\text{o}}(\vec{r}_3,\vec{r}_2;E) G_{\text{o}}(\vec{r}_2,\vec{r}_1;E) r_1 \psi_{n\l}(\vec{r}_1) \\
& \quad = \left(\frac{4\mr}{n a}\right)^2 N_{n\l}^2 \sum\limits_{\l'=0}^\infty \sum\limits_{\l''=0}^\infty \sum\limits_{s=0}^\infty \sum\limits_{s'=0}^\infty \frac{s!}{(s+\l'+1+\frac{n}{2N_{\text{c}} C_F})[(s+2\l'+1)!]} \\
\nonumber & \quad\quad \times \frac{s'!}{(s'+\l''+1+\frac{n}{2N_{\text{c}} C_F})[(s'+2\l''+1)!]} \\
\nonumber & \quad\quad \times \int \d r_1 \, r_1^3 \rho_1^{\l} \e{-\frac{1}{2} \rho_1} L_{n-\l-1}^{2\l+1}\left(\rho_1\right) \rho_1^{\l'} \e{-\frac{1}{2} \rho_1} L_{s}^{2\l'+1}\left(\rho_1\right) \\
\nonumber & \quad\quad \times \int \d r_2 \, r_2^2 \rho_2^{\l'} \e{-\frac{1}{2} \rho_2} L_{s}^{2\l'+1}\left(\rho_2\right) \rho_2^{\l''} \e{-\frac{1}{2} \rho_2} L_{s'}^{2\l''+1}\left(\rho_2\right) \\
\nonumber & \quad\quad \times \int \d r_3 \, r_3^3 \rho_3^{\l''} \e{-\frac{1}{2} \rho_3} L_{s'}^{2\l''+1}\left(\rho_3\right) \rho_3^{\l} \e{-\frac{1}{2} \rho_3} L_{n-\l-1}^{2\l+1}\left(\rho_3\right) \\
\nonumber & \quad\quad \times \sum\limits_{m'=-\l'}^{\l'} \sum\limits_{m''=-\l''}^{\l''} \int \d\Omega_1 Y_{\l'}^{m'*}(\Omega_1) Y_{\l}^{m}(\Omega_1) \int \d\Omega_2 Y_{\l''}^{m''*}(\Omega_2) Y_{\l'}^{m'}(\Omega_2) \int \d\Omega_3 Y_{\l}^{m*}(\Omega_3) Y_{\l''}^{m''}(\Omega_3) \\
\label{eq:BeginNonPerturbative} & \quad = \left(\frac{4\mr}{n a}\right)^2 N_{n\l}^2 \left(\frac{n a}{2}\right)^{11} \sum\limits_{s=0}^\infty \sum\limits_{s'=0}^\infty \frac{s!}{(s+\l+1+\frac{n}{2N_{\text{c}} C_F})[(s+2\l+1)!]} \\
\nonumber & \quad\quad \times \frac{s'!}{(s'+\l+1+\frac{n}{2N_{\text{c}} C_F})[(s'+2\l+1)!]} \\
\nonumber & \quad\quad \times \int \d \rho_1 \, \rho_1^{3+2\l} \e{-\rho_1} L_{n-\l-1}^{2\l+1}\left(\rho_1\right) L_{s}^{2\l+1}\left(\rho_1\right) \\
\nonumber & \quad\quad \times \int \d \rho_2 \, \rho_2^{2+2\l} \e{-\rho_2} L_{s}^{2\l+1}\left(\rho_2\right) L_{s'}^{2\l+1}\left(\rho_2\right) \\
\nonumber & \quad\quad \times \int \d \rho_3 \, \rho_3^{3+2\l} \e{-\rho_3} L_{s'}^{2\l+1}\left(\rho_3\right) L_{n-\l-1}^{2\l+1}\left(\rho_3\right) \\
\label{eq:EndNonPerturbative} & \quad = \left(\frac{4\mr}{n a}\right)^2 N_{n\l}^2 \left(\frac{n a}{2}\right)^{11} \\
\nonumber & \quad \times \Bigg\lbrace \Big[ \sum\limits_{s=0}^2 \frac{s!}{(s+\l+1+\frac{n}{2N_{\text{c}} C_F})[(s+2\l)!]} \frac{1}{(s+\l+1+\frac{n}{2N_{\text{c}} C_F})} \\
\nonumber & \quad\quad \times \Gamma(3+\alpha) \frac{\Gamma(3)}{\Gamma(4-n+\l)} \frac{\Gamma(3)}{\Gamma(3-s) [(s)!]} \times \Gamma(3+\alpha) \frac{\Gamma(3)}{\Gamma(3-s) [(s)!]} \frac{\Gamma(3)}{\Gamma(4-n+\l)} \Big] \\
\nonumber & \quad - \Big[ \sum\limits_{s=1}^2 \frac{s!}{(s+\l+1+\frac{n}{2N_{\text{c}} C_F})[(s+2\l)!]} \frac{1}{(s+\l+\frac{n}{2N_{\text{c}} C_F})} \\
\nonumber & \quad\quad \times \Gamma(3+\alpha) \frac{\Gamma(3)}{\Gamma(4-n+\l)} \frac{\Gamma(3)}{\Gamma(3-s) [(s)!]} \times \Gamma(3+\alpha) \frac{\Gamma(3)}{\Gamma(4-s) [(s-1)!]} \frac{\Gamma(3)}{\Gamma(4-n+\l)} \Big] \\
\nonumber & \quad - \Big[ \sum\limits_{s=0}^1 \frac{s!}{(s+\l+1+\frac{n}{2N_{\text{c}} C_F})[(s+2\l+1)!]} \frac{(s+1)}{(s+\l+2+\frac{n}{2N_{\text{c}} C_F})} \\
\nonumber & \quad\quad \times \Gamma(3+\alpha) \frac{\Gamma(3)}{\Gamma(4-n+\l)} \frac{\Gamma(3)}{\Gamma(3-s) [(s)!]} \times \Gamma(3+\alpha) \frac{\Gamma(3)}{\Gamma(2-s) [(s+1)!]} \frac{\Gamma(3)}{\Gamma(4-n+\l)} \Big] \\
\nonumber & \quad + \Big[ \sum\limits_{s=0}^2 \frac{s!}{(s+\l+1+\frac{n}{2N_{\text{c}} C_F})[(s+2\l+1)!]} \frac{(s+1)}{(s+\l+1+\frac{n}{2N_{\text{c}} C_F})} \\
\nonumber & \quad\quad \times \Gamma(3+\alpha) \frac{\Gamma(3)}{\Gamma(4-n+\l)} \frac{\Gamma(3)}{\Gamma(3-s) [(s)!]} \times \Gamma(3+\alpha) \frac{\Gamma(3)}{\Gamma(3-s) [(s)!]} \frac{\Gamma(3)}{\Gamma(4-n+\l)} \Big] \Bigg\rbrace \,.
\end{align}
The derivation of the final form, Eq.~\eqref{eq:EndNonPerturbative}, from Eq.~\eqref{eq:BeginNonPerturbative} is quite lengthy and we present it in Appendix~\ref{app:sec:NonPerturbativeContribution}. We, somewhat surprisingly, find that the two infinite sums in $s$ and $s'$ can be reduced to only a finite number of contributions. Since the result is proportional to $a^6 \sim \alpha_s^{-6}$, we expect a highly scale dependent result. Furthermore, the result for the matrix element scales like $n^6$ which basically restricts the analysis to the lowest lying states, since otherwise the corrections easily exceeds the leading order effect. This scaling behavior with $n$ and $\alpha_{s}$ has already been found, e.g., in Refs.~\cite{Voloshin:1978hc,Leutwyler:1980tn,Voloshin:1979uv,Titard:1993nn}. Taking it into account, we expect the $2\,^3\!P_J$ contribution in Eq.~\eqref{eq:MatrixElementNormalizationStateFinal} to be $\sim 2$ orders of magnitude bigger than the $1\,^3S_1$ contribution by a simple prefactor comparison. The numerical results confirm this prediction, since we find
\begin{align}
& ^{(0)}\!\langle 2 ; 1 | r \frac{1}{(H_{\text{o}}^{(0)} - E_{H}^{(0)})} \1 \frac{1}{(H_{\text{o}}^{(0)} - E_{H}^{(0)})} r | 2 ; 1 \rangle^{(0)} = \frac{85670.6}{m^7 \alpha_s^6} \,, \\
& ^{(0)}\!\langle 1 ; 0 | r \frac{1}{(H_{\text{o}}^{(0)} - E_{H}^{(0)})} \1 \frac{1}{(H_{\text{o}}^{(0)} - E_{H}^{(0)})} r | 1 ; 0 \rangle^{(0)} = \frac{434.364}{m^7 \alpha_s^6} \,.
\end{align}
Having the results for the matrix elements, we now need to estimate the chromo-electric condensate $\langle \text{vac} | g^2 E^2 | \text{vac} \rangle$. It is related to the two gluon condensate $\langle \text{vac} | \alpha_s G^2 | \text{vac} \rangle$ that may be extracted from lattice computations, where $G_{\mu\nu}^a$ is the field strength tensor. Lorentz invariance of the vacuum requires that the electric and magnetic pair densities are of equal magnitude and opposite sign \cite{Leutwyler:1980tn,Voloshin:1979uv,Titard:1993nn}, and are given by
\begin{align}
& \langle \text{vac} | E^{a,i} E^{b,j} | \text{vac} \rangle = \frac{1}{24} \delta^{a b} \delta^{i j} \langle \text{vac} | E^2 | \text{vac} \rangle = -\frac{1}{96} \delta^{a b} \delta^{i j} \langle \text{vac} | G^2 | \text{vac} \rangle \,, \\
& \langle \text{vac} | B^{a,i} B^{b,j} | \text{vac} \rangle = \frac{1}{24} \delta^{a b} \delta^{i j} \langle \text{vac} | B^2 | \text{vac} \rangle = +\frac{1}{96} \delta^{a b} \delta^{i j} \langle \text{vac} | G^2 | \text{vac} \rangle \,,
\end{align}
such that
\begin{align}
& \langle \text{vac} | g^2 E^2 | \text{vac} \rangle = -\frac{1}{4} \langle \text{vac} | g^2 G^2 | \text{vac} \rangle \,, \\
& \langle \text{vac} | g^2 B^2 | \text{vac} \rangle = +\frac{1}{4} \langle \text{vac} | g^2 G^2 | \text{vac} \rangle \,,
\end{align}
and the factor $\frac{1}{24} = \frac{1}{3} \frac{1}{N_{\text{c}}^2-1}$ is due to averaging in position and color space (note that $\delta_{a b} \delta_{a b} = N_{\text{c}}^2-1$ and $\delta_{i j} \delta_{i j} = 3$). We thus finally have
\begin{equation}
\langle \text{vac} | g^2 E^2 | \text{vac} \rangle = - \pi \langle \text{vac} | \alpha_s G^2 | \text{vac} \rangle \,,
\end{equation}
and some values for $\langle \text{vac} | \alpha_s G^2 | \text{vac} \rangle$ [GeV$^4$] can be found in the literature:
\begin{equation}
\langle \text{vac} | \alpha_s G^2 | \text{vac} \rangle = \left\lbrace
\begin{matrix}
(4.2 \pm 2.0) \cdot 10^{-2} & \text{ from \cite{Titard:1993nn}} \\
(7.5 \pm 2.5) \cdot 10^{-2} & \text{ from \cite{Narison:1995tw}} \\
(7.7 \pm 8.7) \cdot 10^{-2} & \text{ from \cite{Bali:2014sja}}
\end{matrix}
\right.
\end{equation}

\clearpage
\thispagestyle{empty}
\clearpage

\chapter{Numerical results}
\label{chp:NumericalResults}

In this chapter we give a detailed numerical analysis of the processes $\chi_{b J}(1P) \to \Upsilon(1S) + \gamma$ and $h_b(1P) \to \eta_{b}(1S) + \gamma$ and give theoretical predictions for the partial and total widths of the $\chi_{b J}(1P)$, for $J=0,1,2$, and of the $h_b(1P)$.

\section{Parameters}

\begin{figure}[t]
\centering
{\includegraphics[clip,trim={0.0cm 1.2cm 0.0cm 0.0cm},width=\textwidth]{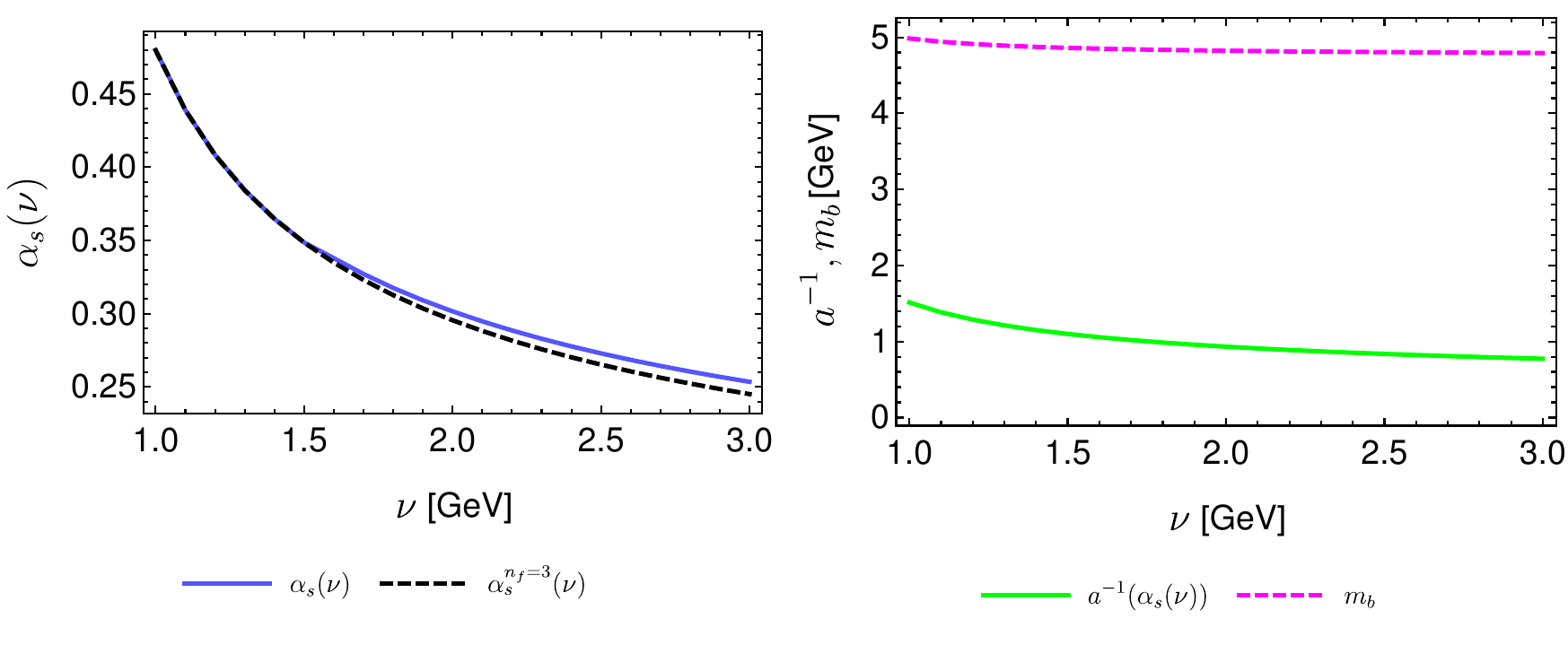}}
\caption[Scale dependence of several parameters.]{Scale dependence of several parameters. \textit{Left panel:} Scale dependence of $\alpha_s(\nu)$ (solid blue curve) and of $\alpha_s^{n_f=3}(\nu)$ (dashed black curve). \textit{Right panel:} Scale dependence of the inverse Bohr radius $a^{-1}$ (solid green curve) and residual scale dependence of $m_b$ entering the leading order decay width (dashed magenta curve).}
\label{fig:ParameterScaleDependence}
\end{figure}

\begin{table}[H]
\centering
\caption[Experimental quarkonia masses.]{Experimental quarkonia masses in [GeV] 
taken from \cite{PDG:2014}.}
\label{tab:QuarkoniaMasses}
\begin{tabular}{r|c|c|c|c|c|c}
physical state & $\Upsilon(1S)$ & $\chi_{b0}(1P)$ & $\chi_{b1}(1P)$ & $\chi_{b2}(1P)$ & $\eta_b(1S)$ & $h_b(1P)$ \\
$n\,^{2s+1}\!\l_J$ & $1\,^3\!S_1$ & $2\,^3\!P_0$ & $2\,^3\!P_1$ & $2\,^3\!P_2$ & $1\,^1\!S_0$ & $2\,^1\!P_1$ \\
\hline
exp. mass & 9.460 & 9.859 & 9.893 & 9.912 & 9.398 & 9.899
\end{tabular} 
\end{table}
The parameters that enter the calculation are
\begin{equation}
n_f = 3 \,, \quad\quad e_Q = -\frac{1}{3} \,, \quad\quad \alpha_{e/m} = \frac{e^2}{4\pi} = \frac{1}{137.0359991} \,,
\end{equation}
where $n_f$ is the number of massless flavors, $e_Q$ is the electric charge of the bottom quark in units of the electron charge $e$ and $\alpha_{e/m}$ is the electromagnetic fine structure constant. We furthermore assume that the c-quark decouples. The masses of the initial and final quarkonium states are chosen to be the experimental ones, listed in Table~\ref{tab:QuarkoniaMasses}.\\
The corresponding photon energies are determined by the kinematics of the two body decay (see Fig.~\ref{fig:kinematics}) and are given by
\begin{equation}
k_\gamma = \frac{m_i^2 - m_f^2}{2 m_i} = \left\lbrace \begin{array}{ll}
391.1~\text{MeV} \,, & \chi_{b0}(1P) \to \Upsilon(1S) + \gamma \\
423.0~\text{MeV} \,, & \chi_{b1}(1P) \to \Upsilon(1S) + \gamma \\
441.6~\text{MeV} \,, & \chi_{b2}(1P) \to \Upsilon(1S) + \gamma \\
488.3~\text{MeV} \,, & h_b(1P) \to \eta_b(1S) + \gamma
\end{array} \right.
\end{equation}
The running of the strong fine structure constant $\alpha_s(\nu)$ as a function of the scale $\nu$ is taken into account at 4-loop accuracy with $n_f$ massless flavors. This is done using the MATHEMATICA package RunDec \cite{Chetyrkin:2000yt}.\footnote{As mentioned before, sample code including a short explanation can be found in Appendix~\ref{app:sec:RunDec}.} The resulting scale dependence is shown in Fig.~\ref{fig:ParameterScaleDependence}. There, the solid blue curve takes into account the automatic decoupling at the respective thresholds and the dashed black curve is the one we are going to follow, which does not take into account the c-quark as a dynamical degree of freedom.\\
As already argued in Sec.~\ref{sec:HigherOrderOperatorCorrectionsDecayWidth}, the bottom quark mass $m_b$ that should enter the leading order decay width includes the static Coulomb energy that is a function of $\alpha_s(\nu)$. Its residual dependence on the scale $\nu$ is shown in Fig.~\ref{fig:ParameterScaleDependence}. The quark mass entering the corrections to the decay width and, e.g., the inverse Bohr radius $a^{-1} = C_F \mr \alpha_s$, however, is fixed to the $\Upsilon(1S)$ mass only. Therefore the scale dependence of $a(\alpha_s(\nu))^{-1}$, also shown in Fig.~\ref{fig:ParameterScaleDependence}, is only due to the running of $\alpha_s(\nu)$. Some values are given in Table~\ref{tab:Parameters}.
\begin{table}[t]
\centering
\caption{Some parameters and their scale dependence.}
\label{tab:Parameters}
\begin{tabular}{r|c|c|c|c|c}
$\nu$ [GeV] & $1.0$ & $1.5$ & $2.0$ & $2.5$ & $3.0$ \\
\hline
$\alpha_s^{4\text{ loops, } n_f=3}(\nu)$ & $0.479778$ & $0.345836$ & $0.295478$ & $0.265205$ & $0.245092$ \\
\hline
$m_b(\nu)$ [GeV] & $4.98515$ & $4.86138$ & $4.82374$ & $4.80525$ & $4.79415$ \\
\hline
$a(\nu)^{-1}$ [GeV] & $1.51295$ & $1.09909$ & $0.931769$ & $0.836306$ & $0.77288$ \\
\hline
$a(\nu)$ [GeV$^{-1}$] & $0.660961$ & $0.909847$ & $1.07323$ & $1.19573$ & $1.29386$ \\
\hline
$a(\nu)$ [fm] & $0.130209$ & $0.17924$ & $0.211426$ & $0.23556$ & $0.254891$
\end{tabular} 
\end{table}

\section[\texorpdfstring{The $2\,^3\!P_{J=0,1,2} \to 1\,^3\!S_1 + \gamma$ processes}{The 2 3PJ012 to 1 3S1 + gamma processes}]{\texorpdfstring{The {\boldmath$2\,^3\!P_{J=0,1,2} \to 1\,^3\!S_1 + \gamma$} processes}{The 2 3PJ012 to 1 3S1 + gamma processes}}

\begin{figure}[H]
	\centering
	{\includegraphics[clip,trim={0.0cm 2.8cm 0.0cm 1.1cm},width=\textwidth]{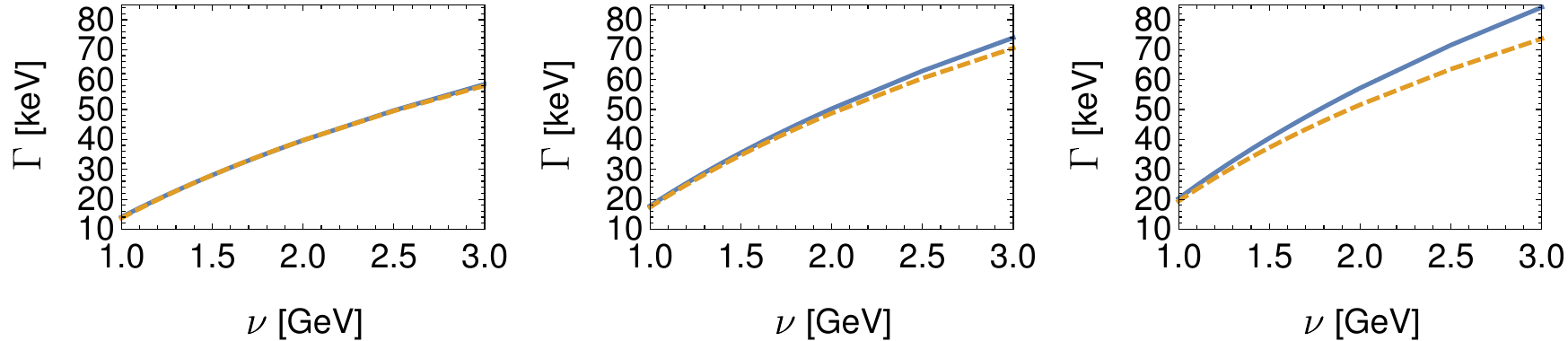}}
	\\[0.2cm]
	{\includegraphics[clip,trim={0.0cm 3.3cm 0.0cm 1.2cm},width=\textwidth]{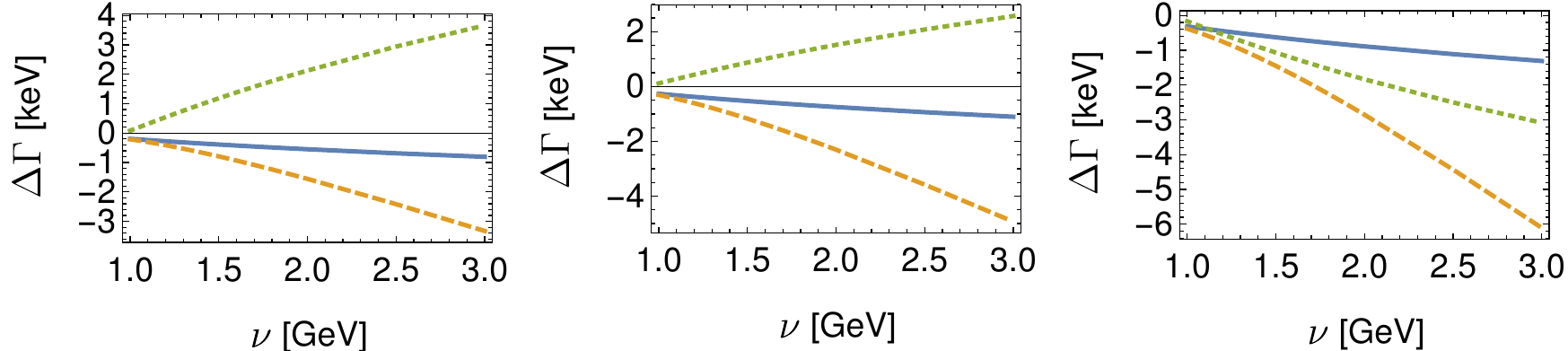}}
	\caption[Relativistic corrections to the decays $\chi_{b_{0,1,2}}(1P) \to \Upsilon(1S) + \gamma$.]{\textit{First row:} Leading order decay widths (solid blue curves) and the included effect of the relativistic corrections on the decay widths of $\chi_{b_{0,1,2}}(1P) \to \Upsilon(1S) + \gamma$ ($2\,^3\!P_{J=0,1,2} \to 1\,^3\!S_1 + \gamma$) (dashed orange curves). \textit{Second row:} Relative contribution of the three relativistic corrections to the decay width $\chi_{b_{0,1,2}}(1P) \to \Upsilon(1S) + \gamma$ according to Eq.~\eqref{eq:FullDecayWidth1}. The solid blue curve corresponds to the first contribution ($\propto -k_{\gamma}/m$), the dashed orange to the second ($\propto -k_{\gamma}^{2} I_{5}^{(0)}/I_{3}^{(0)}$) and the dotted green to the third one ($\propto J(J+1)/2 - 2$), respectively. The three panels in each series refer to the three cases $J=0,1,2$.}
	\label{fig:23PJ13S1-relativistic}
\end{figure}

We start the numerical analysis of the $\chi_{bJ}(1P) \to \Upsilon(1S) + \gamma$ ($2\,^3\!P_J \to 1\,^3\!S_1 + \gamma$), for $J=0,1,2$, with the relativistic contributions of Eq.~\eqref{eq:FullDecayWidth1}, steaming from higher order operators. Therefore we set $R^{S=1}(J)$ and, in order to stay within our goal of accuracy, also $\kappa_Q^{e/m}$ to be zero. The resulting decay widths and the relative corrections induced by the three contributions are shown in Fig.~\ref{fig:23PJ13S1-relativistic}. As one can clearly see, already the leading order decay width comes with a pronounced dependence on the scale $\nu$. This is due to the residual $\alpha_s(\nu)$ dependence of the Bohr radius. This scale dependence could possibly be reduced in a renormalization group improved approach, where the full static potential is put into the Schrödinger equation and the running of $\alpha_s$ is taken into account in a different manner. The relativistic effects are small as expected. For the $\chi_{b0}$ they are negligible and they are of the order of $\sim 2\%$ or $\sim 5\%$ for the $\chi_{b1}$ and for the $\chi_{b2}$, respectively. This can be seen from the second row in Fig.~\ref{fig:23PJ13S1-relativistic}, where for $J=0$ the relativistic contributions cancel, which is not the case for $J=2$, yielding
\begin{figure}[H]
	\centering
	{\includegraphics[clip,trim={0.0cm 1.3cm 0.0cm 1.0cm},width=\textwidth]{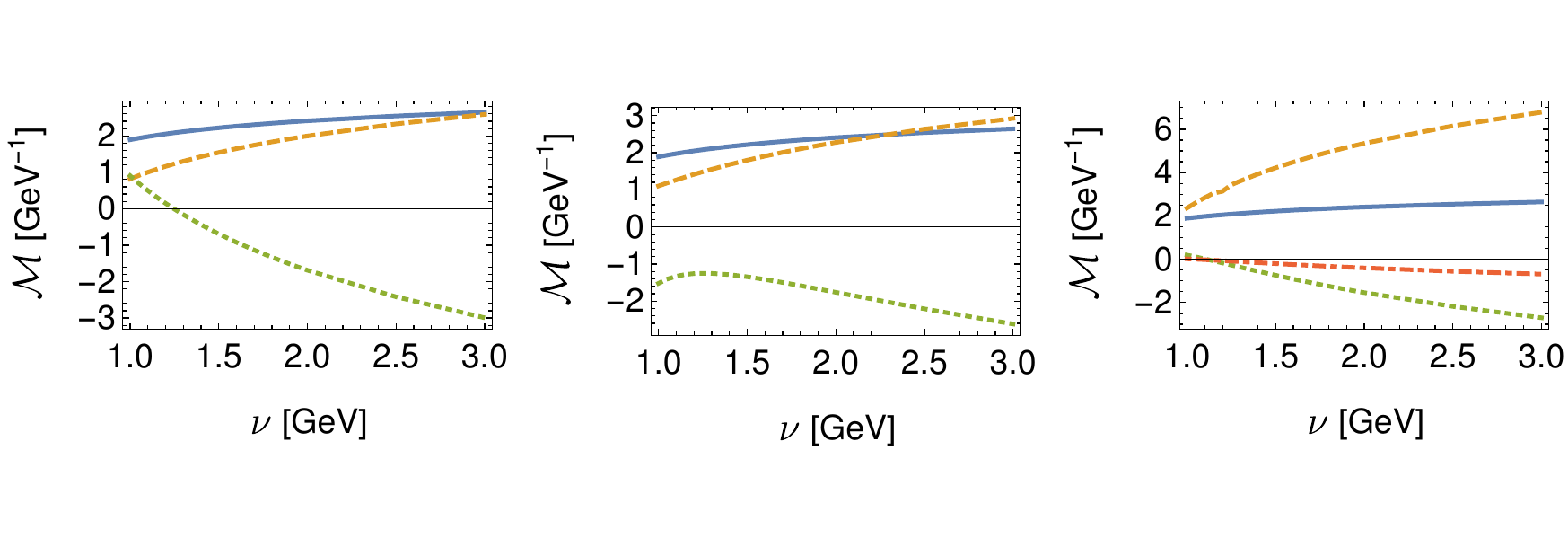}}
	\caption[Radiative corrections to the decays $\chi_{b_{0,1,2}}(1P) \to \Upsilon(1S) + \gamma$.]{Matrix elements contributing to the decay width $\chi_{b_{0,1,2}}(1P) \to \Upsilon(1S) + \gamma$ ($2\,^3\!P_{J=0,1,2} \to 1\,^3\!S_1 + \gamma$), induced by higher order corrections to the static potential. The solid blue lines indicate the leading order matrix element, the dashed orange lines indicate initial and the dotted green lines indicate final state corrections. \textit{Left panel:} NLO corrections, induced by a single insertion of the NLO static potential $V_{a1}$. \textit{Middle panel:} NNLO corrections, induced by a single insertion of the NNLO static potential $V_{a2}$. \textit{Right panel:} NNLO corrections, induced either by a double insertion or by two single insertions of the NLO static potential $V_{a1}$. The latter of those corresponds to the dot-dashed red line, where a first order correction to both, initial and final, states has been considered.}
	\label{fig:23PJ13S1-static}
\end{figure}
the $\sim 5\%$ effect.\\
The corrections due to hard and soft gluons, encoded in the corrections to the static potential, give rise to several initial and final state corrections. They also involve second order corrections to the states. They explicitly depend on the scale $\nu$, due to the logs, and one therefore expects a significant scale dependence of the resulting matrix elements. This is indeed the case as visible in Fig.~\ref{fig:23PJ13S1-static}. It is to be noted, that, since none of these potentials comes with an explicit or implicit dependence on either $\l$, $s$
\begin{figure}[H]
	\centering
	{\includegraphics[clip,trim={0.0cm 4.3cm 0.0cm 1.1cm},width=\textwidth]{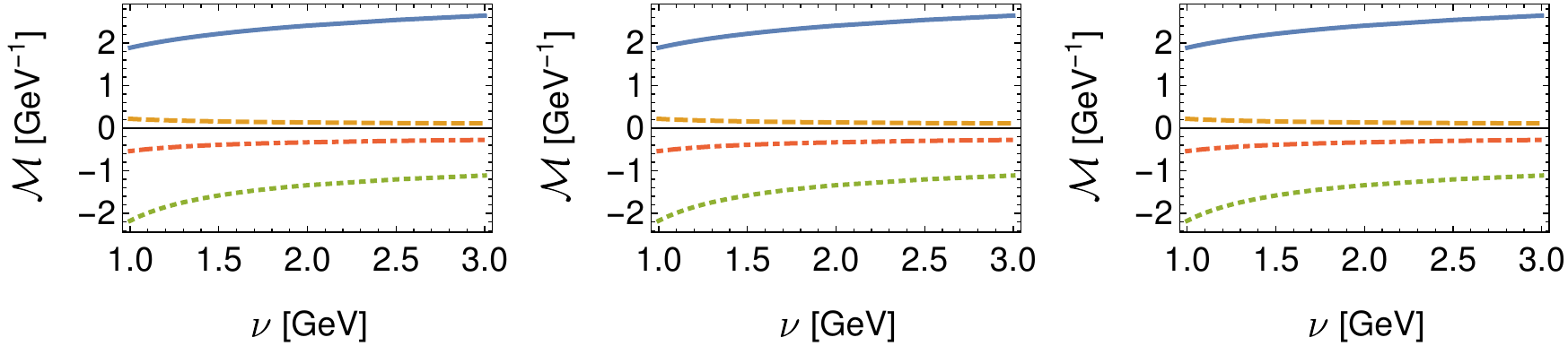}}
	\\[0.2cm]
	{\includegraphics[clip,trim={0.0cm 5.7cm 0.0cm 1.1cm},width=\textwidth]{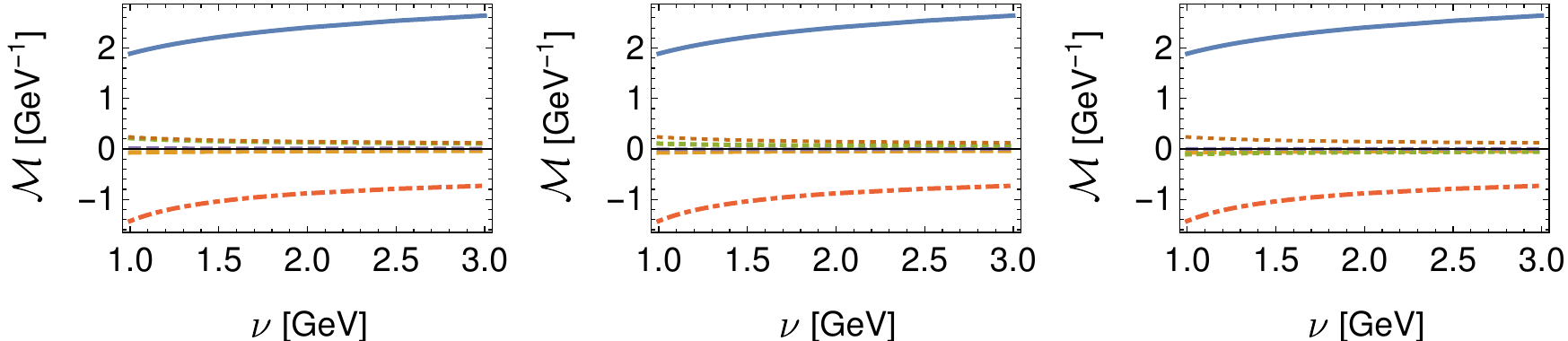}}
	\\[0.2cm]
	{\includegraphics[clip,trim={0.0cm 5.0cm 0.0cm 1.1cm},width=\textwidth]{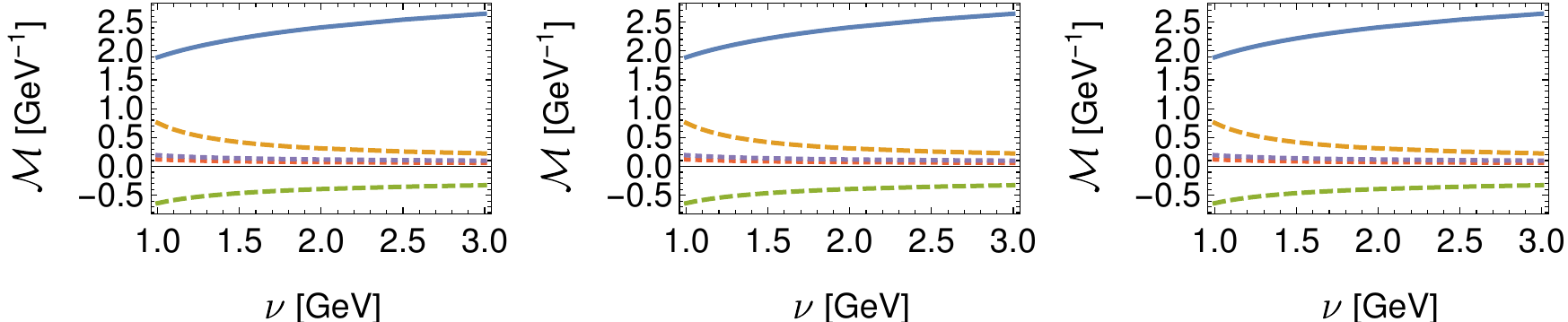}}
	\caption[Relativistic $\frac{1}{m}$ and $\frac{1}{m^2}$ corrections to the decays $\chi_{b_{0,1,2}}(1P) \to \Upsilon(1S) + \gamma$.]{Matrix elements contributing to the decay width $\chi_{b_{0,1,2}}(1P) \to \Upsilon(1S) + \gamma$ ($2\,^3\!P_{J=0,1,2} \to 1\,^3\!S_1 + \gamma$), induced by the relativistic $\frac{1}{m}$ and $\frac{1}{m^2}$ potentials. The three panels in each series refer to the three cases $J=0,1,2$. \textit{First row:} Matrix elements of the leading order (solid blue) and NNLO $V_r^{(1)}$ (dashed orange and dotted green lines for initial and final state corrections, respectively) and $V_r^{(2)}$ (dot-dashed red line for final state correction) corrections. \textit{Second row:} Matrix elements of the leading order (solid blue) and NNLO orbital angular momentum (dashed orange for initial state correction), spin-orbit (dotted green for initial state correction), spin (dot-dashed red for final state correction) and tensor corrections (dashed violet and dotted brown for initial and final state corrections, respectively). \textit{Third row:} Matrix elements of the leading order (solid blue), $V_{p^{2}}^{(2)}$ (dashed orange and green for initial and final state corrections, respectively) and quartic kinetic (dotted red and violet for initial and final state corrections, respectively) corrections.}
	\label{fig:23PJ13S1-others}
\end{figure}
or $J$, the result is universal for all p-wave to s-wave transitions. The left and middle panels refer to the first order initial and final wave function corrections coming from $a_1(\nu,r)$ and $a_2(\nu,r)$, respectively. The right panel refers to the second order correction due to the $a_1(\nu,r)$ term of the static potential. Among the features shown by the panels, the following are of particular interest: (i) the matrix elements clearly exceed the value of the LO one. To some extend this is also due to the involved factors steaming from the $\beta$-function that are large. (ii) The matrix elements depend quite strongly on the scale $\nu$, especially for small $\nu$; in some sense, we expected such behavior from the numerical analysis of the M1 transitions in Refs.~\cite{Brambilla:2005zw,Pineda:2013lta}. This issue in principle always exists, once a perturbative series containing logs is truncated. A possible solution might be using the renormalization group improved Lagrangian \cite{Pineda:2001ra}, which is fundamentally different from the analysis we perform herein, and therefore we will not consider it in this work. (iii) The zero crossing in some of the matrix elements comes from the logarithms in the Eqs.~\eqref{eq:a1} and \eqref{eq:a2}. This depends on the power (odd in the first two and even in the last cases, respectively) of the logarithms that enter the potential. The scale where this effect occurs is the same in all of them, being $\nu \sim 1.2$~GeV. (iv) Initial and final state corrections partially cancel each other order by order.\\
In contrast to the higher-order corrections of the static potential, the remaining potentials due to the relativistic $\frac{1}{m}$-expansion behave very well in perturbation theory. Their respective matrix elements are shown in Fig.~\ref{fig:23PJ13S1-others}. As one can see, most of the contributions are small, except for the initial state correction induced by $V_r^{(1)}$ and the correction due to $V_{S^2}$. This is because the $V_r^{(1)}$ potential is only suppressed by one power of $m$ and the p-wave has a stronger dependence on $\nu$ entering via the Bohr radius $a$. The spin correction is relatively large because of the factor in the potential. The moderate corrections induced by the quadratic correction, $-\frac{\nabla_{r}^{4}}{4m^{3}}$, to the kinetic energy (dashed lines in the third row) almost cancel between initial and final state corrections. The overall dependence on the scale $\nu$ is smooth in all of the cases but a slight trend towards larger values with respect to decreasing scale can be observed. This is due to the scale dependence of the Bohr radius, which has a strong scale dependence coming from the running of $\alpha_s$.\\
In the next step we need to sum all the above matrix elements order by order. The result is shown in the first row of Fig.~\ref{fig:23PJ13S1-orderBYorder}. The most important features are the strong scale dependence and the numerically big values, especially of the total (NLO+NNLO) matrix element. For the purpose of estimating the impact of each order, we also show the corresponding decay widths at each given order in the second row of Fig.~\ref{fig:23PJ13S1-orderBYorder}.\footnote{Note that each of the given orders (the one of the matrix elements as well as the one of the decay width) refers to the relative order with respect to the respective leading order. As we elaborated in Sec.~\ref{sec:pNRQCD}: $\Gamma_{NLO} \not\propto \M_{NLO}^2$, etc.}
\begin{figure}[t]
\centering
{\includegraphics[clip,trim={0.0cm 4.2cm 0.0cm 1.1cm},width=\textwidth]{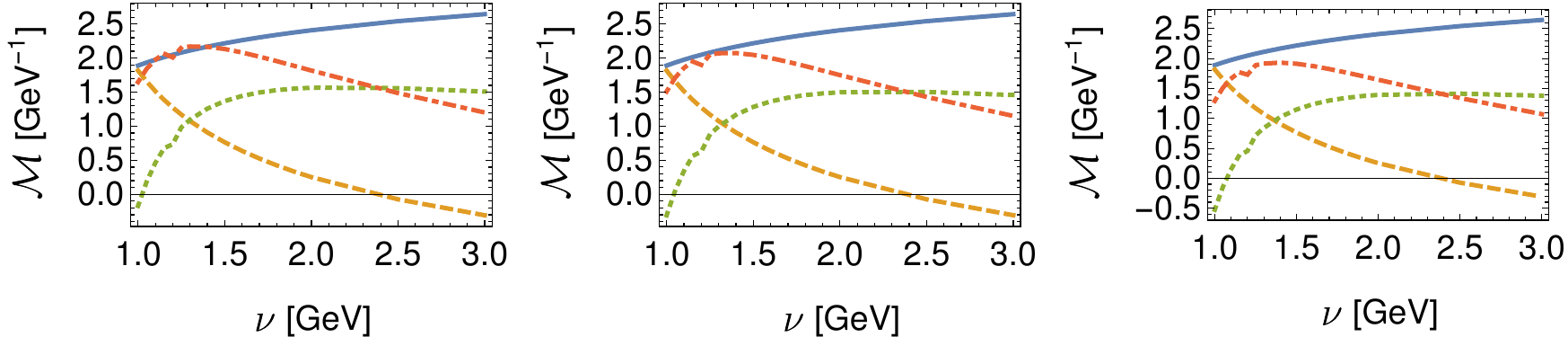}}
\\[0.2cm]
{\includegraphics[clip,trim={0.0cm 4.2cm 0.0cm 1.1cm},width=\textwidth]{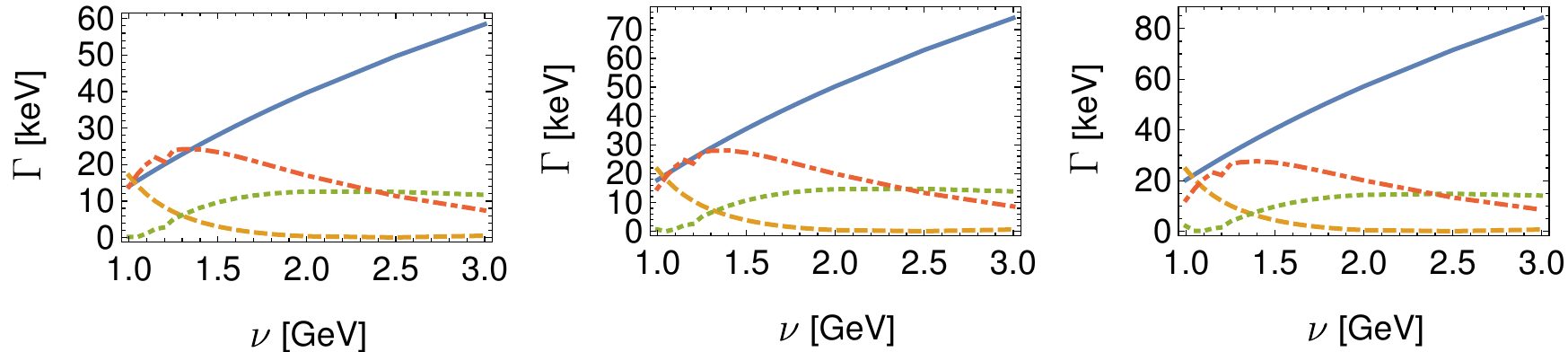}}
\caption[Order by order matrix elements and decay widths of the decays $\chi_{b_{0,1,2}}(1P) \to \Upsilon(1S) + \gamma$.]{Total matrix elements and decay widths for the transitions $\chi_{b_{0,1,2}}(1P) \to \Upsilon(1S) + \gamma$ ($2\,^3\!P_{J=0,1,2} \to 1\,^3\!S_1 + \gamma$) at different orders. The three panels in each series refer to the three cases $J=0,1,2$. \textit{First row:} Total matrix elements at leading order (solid blue), NLO (dashed orange), NNLO (dotted green) and NLO+NNLO (dot-dashed red). \textit{Second row:} Decay widths at (not up to) leading order (solid blue), NLO (dashed orange), NNLO (dotted green) and NLO+NNLO (dot-dashed red).}
\label{fig:23PJ13S1-orderBYorder}
\end{figure}
From both plot sequences one can see that the overall impact of corrections decreases with increasing total angular momentum $J$. For $J=0$ the total matrix element and the total corrections to the decay width exceed the respective leading order curves, whereas for $J=1$ they just touch each other and finally for $J=2$ they do not touch each other at all. The kink, visible in the NNLO (and subsequently also in NLO+NNLO) matrix element at $\sim 1.2$~GeV can be traced back to either the zero crossing or to the maximum in the matrix elements induced by the static potential. The total NLO and NNLO matrix elements show again a zero crossing and the combined NLO+NNLO matrix element has a clear maximum. The zero crossings yield vanishing contribution in the respective decay widths, as visible in the second row.\\
The final result~\eqref{eq:FullDecayWidth1} is shown in Fig.~\ref{fig:23PJ13S1-FinalResult}.
\begin{figure}[t]
\centering
{\includegraphics[clip,trim={0.0cm 0.0cm 0.0cm 0.0cm},width=\textwidth]{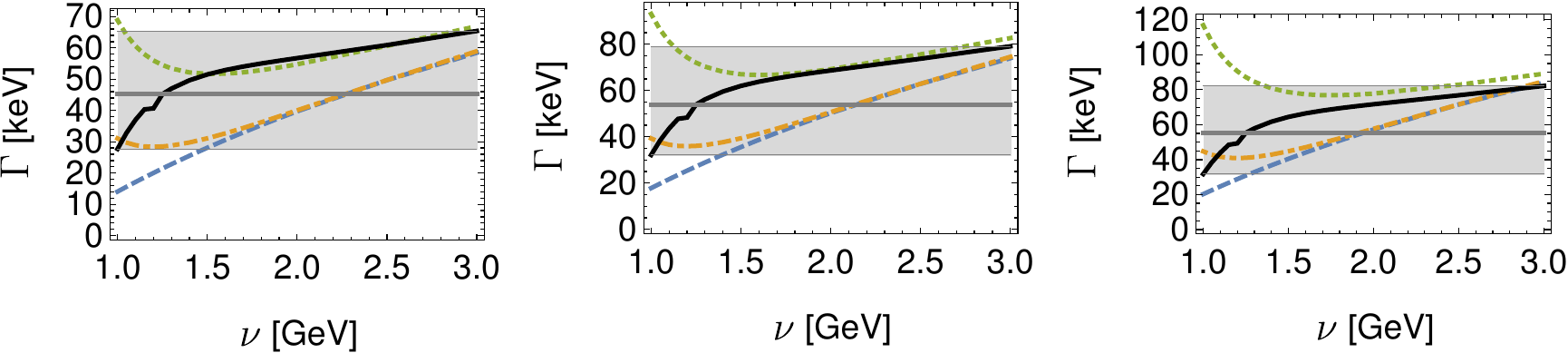}}
\caption[Final decay widths of the decays $\chi_{b_{0,1,2}}(1P) \to \Upsilon(1S) + \gamma$.]{Final decay widths according to Eq.~\eqref{eq:FullDecayWidth1} for the transitions $\chi_{b_{0,1,2}}(1P) \to \Upsilon(1S) + \gamma$ ($2\,^3\!P_{J=0,1,2} \to 1\,^3\!S_1 + \gamma$). The three panels refer to the three cases $J=0,1,2$. The dashed blue curves are the leading order decay widths, the dashed orange ones incorporate NLO corrections and the solid black curves are our final result, incorporating NLO+NNLO and relativistic corrections. The dotted green curves are equivalent to the black ones but omit all corrections to the static potential. We take our final values at $\nu = 1.25$~GeV and the gray bands indicate the associated uncertainties.}
\label{fig:23PJ13S1-FinalResult}
\end{figure}
It is clearly visible that incorporating NLO and especially NNLO corrections diminishes the scale dependence of the decay width. For instance, for the $J=1$ case, the leading order covers the range of $\sim (17-74)$~keV, incorporating the NLO contribution shrinks the range to $\sim (35-75)$~keV, and finally also incorporating NNLO corrections results in a range of $\sim (32-79)$~keV. Although a slight shift towards higher upper bounds is noticeable, the whole range and thus the overall scale dependence decreases. On the other hand omitting the corrections to the static potential, hence setting $a_{1}(\nu,r)$ and $a_{2}(\nu,r)$ to zero, results in a curve that exhibits a completely different behavior. It is not only shifted to values significantly above the leading order, but also the shape, especially in the low energy regime, completely changes. We may therefore conclude that although the logs give rise to non-negligible contributions, they must not be omitted but should be incorporated in a renormalization group improved way.\\
We choose our final result to be the one at the scale that self-consistently solves the Bohr radius
\begin{equation}
\frac{1}{a} = \frac{m C_F \alpha_s(\frac{1}{a})}{2} \,,
\end{equation}
yielding $\nu = \frac{1}{a} = 1.25$~GeV.\\
Other theoretical estimates of physical observables within non-relativistic EFTs follow the prescription of minimal sensitivity with respect to the scale $\nu$. In our case, see Fig.~\ref{fig:23PJ13S1-FinalResult}, it is difficult to establish and thus we take our values at the scale established above. Another option would have been to take the typical momentum transfer $p \sim m v \sim 1.5$~GeV which can be obtained noting that the b-quark mass is given by $m \sim 4.8$~GeV and that $v^2 \sim 0.1$ in bottomonium systems. We want to stress that this estimation excludes parts of the leading order decay width, but completely includes the NLO, and by construction, also the NNLO widths. Therefore, we overall obtain a better result than pure leading order.\\
We estimate the associated uncertainty with the one coming from the scaling behavior, since this is the dominant effect. We furthermore tried to estimate the generic effect of a NNNLO contribution. For that, we modified Eq.~\eqref{eq:FullDecayWidth1} to incorporate an additional summand $\text{const.} \times \alpha_s^3$ with a constant of order one. However, this only generates very small variations around our final (solid black) curve that are completely covered by our error estimation due to the scale dependence.\\
The overall convergence of the perturbative series is not as good as we hoped for. This becomes apparent, once one considers the difference between the LO and NLO, and between the NLO and NNLO results. Having a convergent series the latter two should be as close or even closer together than the first-mentioned. This is not the case here but is very likely to be cured in a renormalization improved approach as demonstrated for M1 transitions in \cite{Pineda:2013lta}. However, the uncertainty associated with this issue is totally covered by the uncertainty due to the scale variation.\\
Overall, we investigated the three most common sources of uncertainties as mentioned above. We furthermore want to note that additional sources of uncertainties are given by the input parameters, being the masses of initial and final states, the resulting associated uncertainty in $k_\gamma$ and the running of $\alpha_s$. If one assumes that these quantities are accurate within $\lesssim (1-3)\%$, their impact is again covered by the uncertainty due to the scale variation. Our final results at NNLO read
\begin{align}
& \Gamma_{\text{E1}}(\chi_{b0}(1P) \to \Upsilon(1S) + \gamma) = 45^{+20}_{-18}~\text{keV} \,, \\
& \Gamma_{\text{E1}}(\chi_{b1}(1P) \to \Upsilon(1S) + \gamma) = 54^{+25}_{-22}~\text{keV} \,, \\
& \Gamma_{\text{E1}}(\chi_{b2}(1P) \to \Upsilon(1S) + \gamma) = 55^{+27}_{-24}~\text{keV} \,.
\end{align}
We compare our results with total and leading order values obtained in several other theoretical approaches. Among them are a non-relativistic constituent quark model (CQM) \cite{Segovia:2016xqb}, a relativistic quark model (R) \cite{Ebert:2002pp}, the Godfrey-Isgur potential model (GI) \cite{Godfrey:2015dia} and a Buchmüller-Tye potential-model (BT) \cite{Grotch:1984gf}. The results are presented in Table~\ref{tab:ResultsComparisonChiB}.
\begin{table}[t]
\centering
\caption[Comparison of $\chi_{b J}(1P) \to \Upsilon(1S) + \gamma$ decay widths.]{Comparison of our decay widths of $\chi_{b J}(1P) \to \Upsilon(1S) + \gamma$ at LO, NLO and NNLO with a non-relativistic constituent quark model (CQM) \cite{Segovia:2016xqb}, a relativistic quark model (R) \cite{Ebert:2002pp}, the Godfrey-Isgur potential model (GI) \cite{Godfrey:2015dia} and with the LO and total decay widths obtained in a Buchmüller-Tye potential-model (BT) \cite{Grotch:1984gf}. All decay widths are given in units of keV.}
\label{tab:ResultsComparisonChiB}
\begin{tabular}{c|Sc|Sc|Sc|Sc|Sc|Sc|Sc|Sc}
Mode & LO & NLO & NNLO & CQM & R & GI & BT (LO) & BT \\
\hline
$\chi_{b0}(1P) \to \Upsilon(1S) + \gamma$ & 21.45 & 28.50 & 45.22 & 28.07 & 29.9 & 23.8 & 30.0 & 25.7 \\
$\chi_{b1}(1P) \to \Upsilon(1S) + \gamma$ & 27.15 & 36.07 & 53.94 & 35.66 & 36.6 & 29.5 & 35.9 & 29.8 \\
$\chi_{b2}(1P) \to \Upsilon(1S) + \gamma$ & 30.89 & 41.03 & 55.34 & 39.15 & 40.2 & 32.8 & 41.0 & 33.0
\end{tabular}
\end{table}
Our final results are bigger by $\sim 50\%$ compared with the non-relativistic and relativistic potential model values. Roughly the same holds for the Godfrey-Isgur potential model. However, they are still compatible within our uncertainties. On the other hand, the agreement among our leading order result and the one obtained in a Buchmüller-Tye potential-model is remarkable, especially since the used approaches differ considerably. The computations done in the Buchmüller-Tye potential-model only take the first excited state into account when calculating corrections to the initial and final states. They are justifying this by stating that: (i) coefficients of intermediate states are suppressed due to large energy denominators; (ii) a $\Delta n > 1$ leads to negligibly small radial overlap integrals with respect to the relevant operator $r$. This argumentation in principle is right, however, we find that numerically it does not hold that strictly during our analysis. We therefore incorporated at least the first 10 excited states in order to reach a numerically stable result. In the case of the radiative corrections to the static potential we did actually include at least the first 15 excited states.\\
Overall we can conclude that in general our results are bigger than those obtained in different theoretical approaches. However, due to our rather big uncertainty, associated with the sensitivity to the scale, they remain compatible. The fact that our leading order values agrees quite well with the quoted results is notable, since a clear distinction between LO, NLO and NNLO is not easy or even not possible in different approaches. A reduction of the scale dependence and an improvement of the convergence of the perturbative series with respect to our results in an renormalization improved approach is likely to further improve compatibility in the future.\\
From the experimental side, apart from the branching fraction, little is known 
about the partial and total decay widths under study. Nevertheless, we use the 
branching fractions given by the PDG \cite{PDG:2016} and our theoretical 
results for the partial widths, in order to predict the total widths of the 
$\chi_{b J}$-family. The results are given in Table~\ref{tab:ResultsChiB}.
\begin{table}[t]
\centering
\caption[$\chi_{b J}(1P) \to \Upsilon(1S) + \gamma$ results and 
predictions.]{Results and predictions for the partial and total decay widths of 
$\chi_{b J}(1P) \to \Upsilon(1S) + \gamma$. The branching fractions are given 
by the PDG \cite{PDG:2016}, the partial widths correspond to our final results 
(black lines in Fig.~\ref{fig:23PJ13S1-FinalResult}) including the theoretical 
uncertainties (gray bands in Fig.~\ref{fig:23PJ13S1-FinalResult}), and the 
total widths are predictions.}
\label{tab:ResultsChiB}
\begin{tabular}{c|Sc|Sc|Sc}
Mode & Fraction $\mathcal{B} = \frac{\Gamma_i}{\Gamma}$ [PDG] & Partial width $\Gamma_i$ & Total width $\Gamma$ \\
\hline
$\chi_{b0}(1P) \to \Upsilon(1S) + \gamma$ & $(1.76 \pm 0.35)\%$ & $45^{+20}_{-18}$~keV & $2.6^{+1.3}_{-1.1}$~MeV \\
$\chi_{b1}(1P) \to \Upsilon(1S) + \gamma$ & $(33.9 \pm 2.2)\%$ & $54^{+25}_{-22}$~keV & $159^{+75}_{-65}$~keV \\
$\chi_{b2}(1P) \to \Upsilon(1S) + \gamma$ & $(19.1 \pm 1.2)\%$ & $55^{+27}_{-24}$~keV & $290^{+142}_{-124}$~keV
\end{tabular}
\end{table}
The Belle collaboration reports a upper limit on the total decay width of the $\chi_{b0}(1P)$ at 90\% confidence level \cite{Abdesselam:2016xbr:MB}, stating that $\Gamma(\chi_{b0}(1P)) < 2.4$~MeV which is well within the uncertainty band of our prediction. These uncertainties are obtained via standard Gaussian uncertainty propagation, where the total width
\begin{equation}
\Gamma = \frac{\Gamma_i}{\mathcal{B}} \,,
\end{equation}
is the quotient of the branching fraction $\mathcal{B}$ and the partial width $\Gamma_i$, and thus the uncertainty is given by
\begin{equation}
\Delta \Gamma = \Gamma \cdot \sqrt{\left(\frac{\Delta \mathcal{B}}{\mathcal{B}}\right)^2 + \left(\frac{\Delta \Gamma_i}{\Gamma_i}\right)^2 + 2 \cdot \frac{\sigma_{\mathcal{B}\Gamma_i}}{\mathcal{B} \cdot \Gamma_i}} \,,
\end{equation}
where $\sigma_{\mathcal{B}\Gamma_i}$ encodes the cross-correlation that, in this case, is zero.

\section[\texorpdfstring{The $2\,^1\!P_1 \to 1\,^1\!S_0 + \gamma$ process}{The 2 1P1 to 1 1S0 + gamma process}]{\texorpdfstring{The {\boldmath$2\,^1\!P_1 \to 1\,^1\!S_0 + \gamma$} process}{The 2 1P1 to 1 1S0 + gamma process}}

\begin{figure}[H]
	\centering
	{\includegraphics[clip,trim={0.0cm 1.9cm 0.0cm 0.2cm},width=\textwidth]{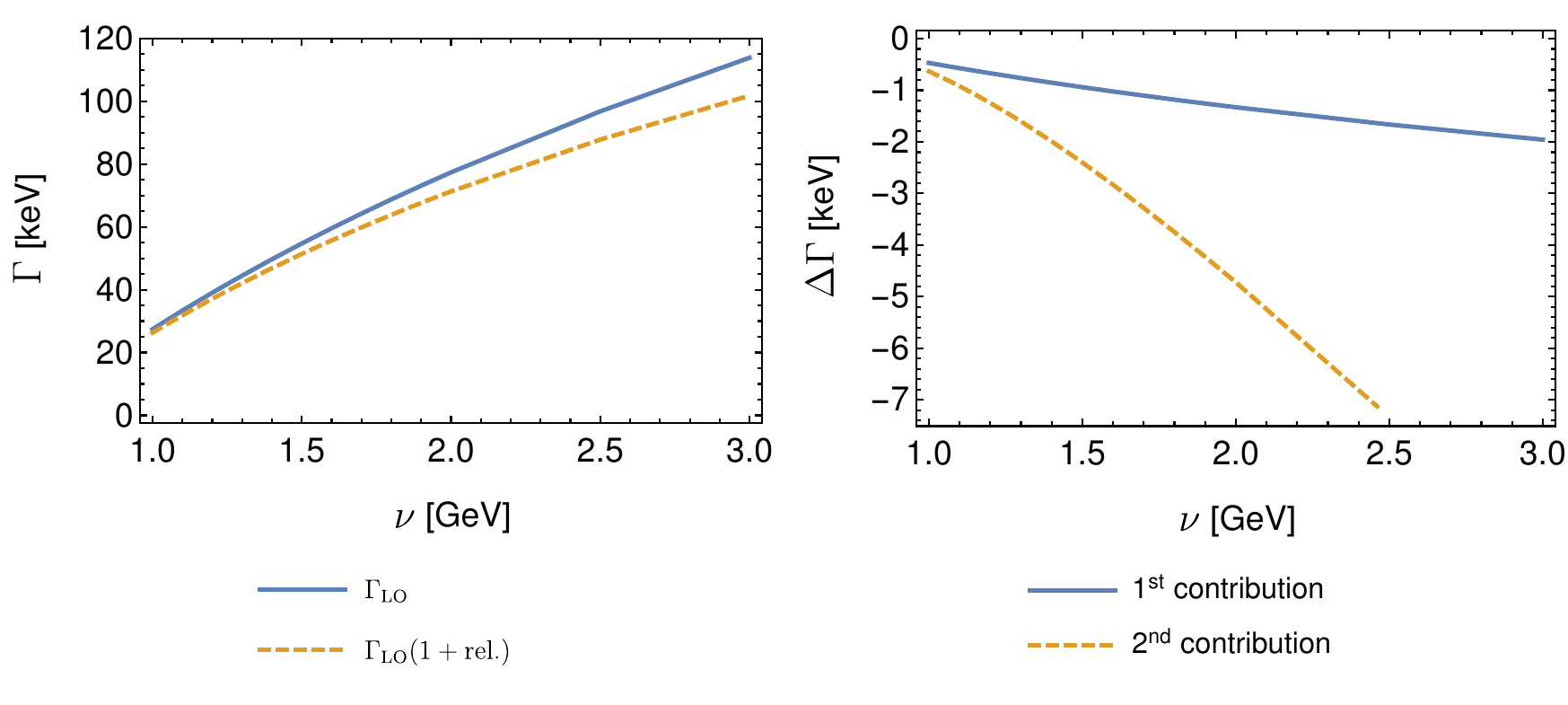}}
	\caption[Relativistic corrections to the decay $h_b(1P) \to \eta_b(1S) + \gamma$.]{\textit{Left panel:} Leading order decay width (solid blue curve) and the included effect of the relativistic corrections on the decay width of $h_b(1P) \to \eta_b(1S) + \gamma$ ($2\,^1\!P_1 \to 1\,^1\!S_0 + \gamma$) (dashed orange curve). \textit{Right panel:} Relative contribution of the two relativistic corrections to the decay width $h_b(1P) \to \eta_b(1S) + \gamma$ according to Eq.~\eqref{eq:FullDecayWidth2}. The solid blue curve corresponds to the first contribution ($\propto -k_{\gamma}/m$) and the dashed orange to the second one ($\propto -k_{\gamma}^{2} I_{5}^{(0)}/I_{3}^{(0)}$), respectively.}
	\label{fig:21P111S0-relativistic}
\end{figure}

\begin{figure}[H]
	\centering
	{\includegraphics[clip,trim={0.0cm 4.4cm 0.0cm 1.1cm},width=\textwidth]{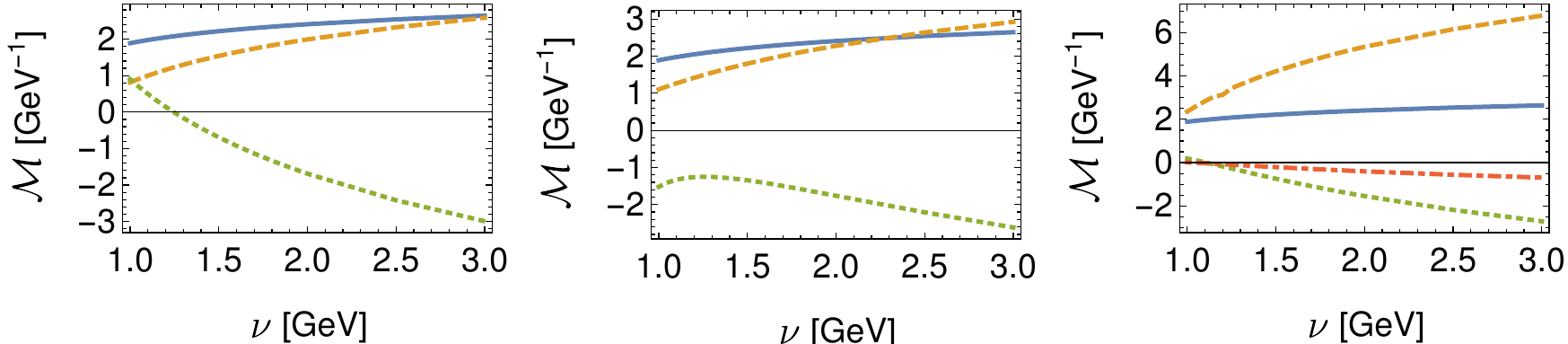}}
	\caption[Radiative corrections to the decay $h_b(1P) \to \eta_b(1S) + \gamma$.]{Matrix elements contributing to the decay width $h_b(1P) \to \eta_b(1S) + \gamma$ ($2\,^1\!P_1 \to 1\,^1\!S_0 + \gamma$), induced by higher order corrections to the static potential. The solid blue lines indicate the leading order matrix element, the dashed orange lines indicate initial and the dotted green lines indicate final state corrections. \textit{Left panel:} NLO corrections, induced by a single insertion of the NLO static potential $V_{a1}$. \textit{Middle panel:} NNLO corrections, induced by a single insertion of the NNLO static potential $V_{a2}$. \textit{Right panel:} NNLO corrections, induced either by a double insertion or by two single insertions of the NLO static potential $V_{a1}$. The latter of those corresponds to the dot-dashed red line, where a first order correction to both, initial and final, states has been considered.}
	\label{fig:21P111S0-static}
\end{figure}

\begin{figure}[H]
	\centering
	{\includegraphics[clip,trim={0.0cm 4.9cm 0.0cm 1.1cm},width=\textwidth]{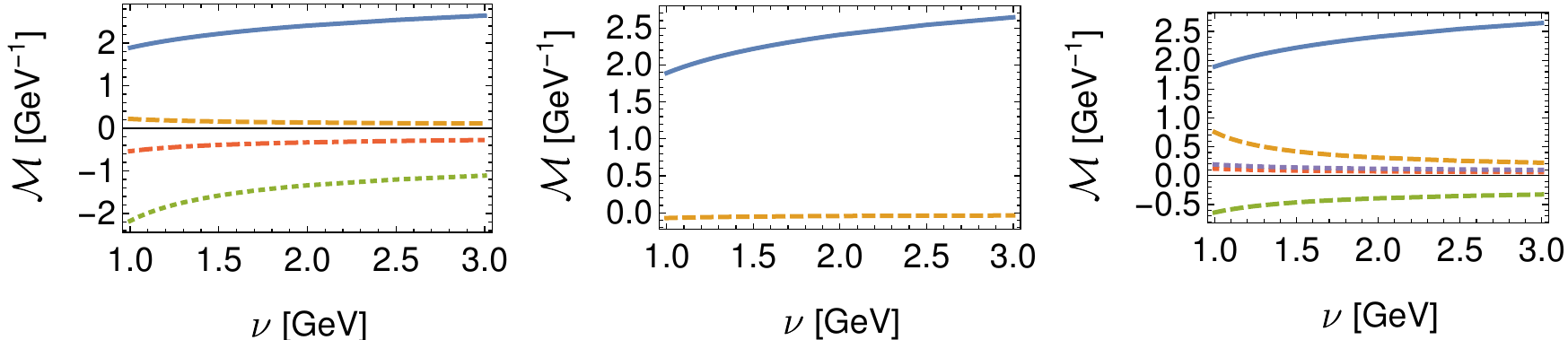}}
	\caption[Relativistic $\frac{1}{m}$ and $\frac{1}{m^2}$ corrections to the decay $h_b(1P) \to \eta_b(1S) + \gamma$.]{Matrix elements contributing to the decay width $h_b(1P) \to \eta_b(1S) + \gamma$ ($2\,^1\!P_1 \to 1\,^1\!S_0 + \gamma$), induced by the relativistic $\frac{1}{m}$ and $\frac{1}{m^2}$ potentials. \textit{Left panel:} Matrix elements of the leading order (solid blue) and NNLO $V_r^{(1)}$ (dashed orange and dot-dashed green lines for initial and final state corrections, respectively) and $V_r^{(2)}$ (dotted red line for final state correction) corrections. \textit{Middle Panel:} Matrix elements of the leading order (solid blue) and NNLO orbital angular momentum (dashed orange for initial state correction). Note that all other corrections due to spin-orbit, spin and tensor vanish exactly due to the quantum numbers of initial and final state, respectively. \textit{Right panel:} Matrix element of the leading order (solid blue), $V_{p^{2}}^{(2)}$ (dashed orange and green for initial and final state corrections, respectively) and quartic kinetic (dotted red and violet for initial and final state corrections, respectively) corrections.}
	\label{fig:21P111S0-others}
\end{figure}

\begin{figure}[H]
	\centering
	{\includegraphics[clip,trim={0.0cm 4.2cm 0.0cm 1.1cm},width=\textwidth]{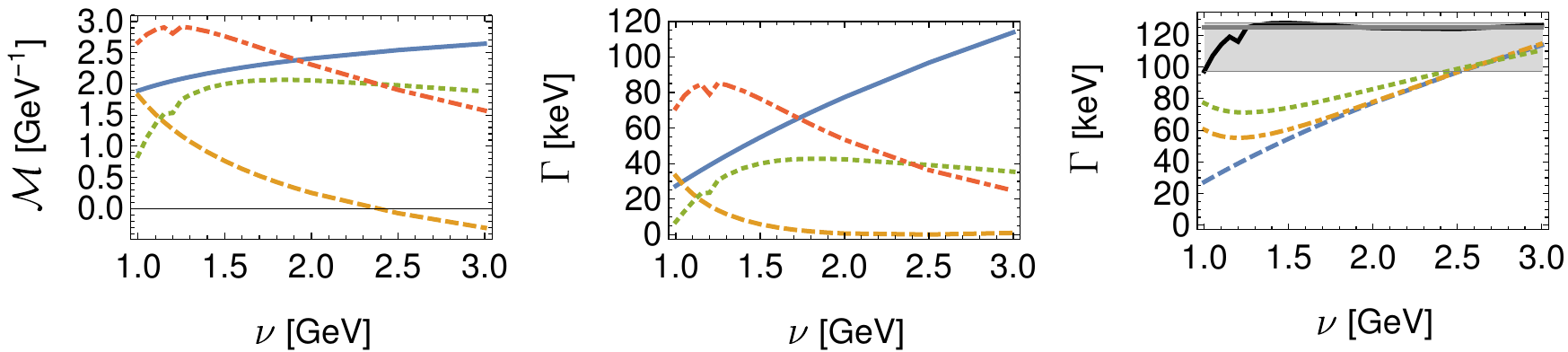}}
	\caption[Order by order matrix elements and decay width and final decay width of the decay $h_b(1P) \to \eta_b(1S) + \gamma$.]{Total matrix elements and decay widths at different orders, and final decay widths according to Eq.~\eqref{eq:FullDecayWidth2} for the transition $h_b(1P) \to \eta_b(1S) + \gamma$ ($2\,^1\!P_1 \to 1\,^1\!S_0 + \gamma$). \textit{Left panel:} Total matrix elements at leading order (solid blue), NLO (dashed orange), NNLO (dotted green) and NLO+NNLO (dot-dashed red). \textit{Middle panel:} Decay widths at (not up to) leading order (solid blue), NLO (dashed orange), NNLO (dotted green) and NLO+NNLO (dot-dashed red). \textit{Right panel:} The dashed blue curve is the leading order decay width, the dashed orange one incorporates NLO corrections and the solid black curve is our final result incorporating NLO+NNLO and relativistic corrections. The dotted green curve is equivalent to the black one but omits all corrections to the static potential. We take our final value at $\nu = 1.25$~GeV and the gray band indicates the associated uncertainty.}
	\label{fig:21P111S0-orderBYorderFinalResult}
\end{figure}

We start the numerical analysis of the $h_b(1P) \to \eta_b(1S) + \gamma$ ($2\,^1\!P_1 \to 1\,^1\!S_0 + \gamma$), with the relativistic contributions of Eq.~\eqref{eq:FullDecayWidth2}, steaming from higher order operators. Therefore we set $R^{S=0}$ to be zero. The resulting decay width and the relative corrections induced by the two contributions are shown in Fig.~\ref{fig:21P111S0-relativistic}. In comparison with the same analysis for the $\chi_{bJ}$ decays one can see that the leading order decay width and the relativistic effects are larger here. This, however, is an effect due to $k_\gamma$ only. The photon energy $k_\gamma$ increases for increasing $J$ in the $\chi_{bJ}$ cases and the energy of the emitted photon is even bigger in the case of the $h_b$ decay. The fact that this photon energy enters in third power explains the overall increasing effect that is still rather small, being of the order of $\sim 10\%$, as expected.\\
The corrections due to hard and soft gluons, encoded in the corrections to the static potential, give rise to the exact same contribution as in the case of $\chi_{bJ}(1P) \to \Upsilon(1S) + \gamma$, as visible in Fig.~\ref{fig:21P111S0-static}. This is, as already mentioned, because none of these potentials comes with an explicit or implicit dependence on either $\l$, $s$ or $J$. The important features and conclusions have already been discussed in the previous
section.\\
The respective matrix elements due to potentials steaming from the relativistic $\frac{1}{m}$-expansion are shown in Fig.~\ref{fig:21P111S0-others}. The only difference with respect to the $\chi_{bJ}(1P) \to \Upsilon(1S) + \gamma$ transition is the absence of the spin-orbit, spin, and tensor contributions. This has a major impact on the total NNLO matrix element, since especially the $S^2$ spin induced correction comes with a sizable negative contribution that is now absent, especially in the low energy region.\\
The order by order sum over all the above matrix elements is shown in Fig.~\ref{fig:21P111S0-orderBYorderFinalResult}. Again, the kink, visible in the NNLO and NLO+NNLO matrix elements at $\sim 1.2$~GeV can be traced back to the zero crossing or maximum in the matrix elements induced by the static potential, see Fig.~\ref{fig:21P111S0-static}. In this case it appears more pronounced. The non-existence of several negative contributions at NNLO yields a much more pronounced dependence on the scale $\nu$ for values $\nu \lesssim 1.5$~GeV. The resulting total matrix element in this regime, and subsequently the resulting decay width, clearly exceeds the leading order.
This behavior, however, leads to a noticeable overall mitigation of the scale dependence of our final result~\eqref{eq:FullDecayWidth2} that is shown in Fig.~\ref{fig:21P111S0-orderBYorderFinalResult}. The effect of non-existent contributions heavily amplifies the diminishing of the scale dependence, once NLO and especially NNLO corrections are incorporated to the decay width. The leading order decay width, for instance, covers the range of $\sim (27-114)$~keV. Incorporating the NLO contribution shrinks the range to $\sim (64-115)$~keV, and finally also incorporating NNLO corrections results in a range of $\sim (97-127)$~keV. Again, we observe a slight shift towards higher upper bounds, but the whole range and thus the overall scale dependence decreases drastically. On the other hand omitting the corrections to the static potential results in a curve that is quite close to the leading order one at higher scales and would also significantly diminish the scale dependence for lower values of $\nu$. This is a noticeable contrast with respect to the $\chi_{bJ}(1P) \to \Upsilon(1S) + \gamma$ case, but is understandable, since there we observe several additional contributions that are not present here. We may conclude that although the, here absent, contributions behave well in a perturbative manner, they yield a significant contribution, since their absolute impact is of order $75\%$ for $\nu = 1.0$~GeV. Again, the whole analysis is very likely to benefit from a renormalization group improved treatment.\\
Our final result is taken at $\nu = 1.25$~GeV following the prescription of self-consistently solving the Bohr radius. Scales of minimal sensitivity exist at $\nu \sim 1.5$~GeV and $\nu \sim 2.3$~GeV. Although the resulting value for the partial decay width would not change drastically, these $\nu$-values are not compatible with the typical scale of momentum transfer and are therefore disregarded. The main difference with respect to the final results from the $\chi_{bJ}$ decays is the overall much smaller scale dependence and the difference in the shape of the curve for larger values of $\nu$. Here we observe almost no scale dependence and the overall result with respect to scale dependence is significantly better than the pure leading order. We also want to point out that omitting the corrections to the static potential (dotted green curve) gives rise to a significantly different result than in the $\chi_{bJ}$ case. There the final curve omitting the radiative corrections lies above the included final value, whereas in the $h_b$ case the curve omitting the radiative corrections is below the included final result.\\
As in the previous section the estimation of NNNLO effects with a prefactor of order one are covered by the uncertainty of the scale dependence. In contrast thereto, the convergence of the perturbative series is worse and we might underestimate the overall uncertainty of our result by simply using the uncertainty assigned to the scale dependence. It therefore is very likely that the $h_b$ decay might benefit even more from a renormalization improved approach. Our final result at NNLO reads
\begin{equation}
\Gamma_{\text{E1}}(h_b(1P) \to \eta_b(1S) + \gamma) = 125^{+3}_{-27}~\text{keV} \,.
\end{equation}
A comparison of our results with the same theoretical approaches than for the $\chi_{bJ}$-family is given in Table~\ref{tab:ResultsComparisonHB1}. There are no predictions using a Buchmüller-Tye potential-model but there are predictions coming from the light-front quark model (LFQM) \cite{Shi:2016cef:MB} and screened potential models with zeroth-order wave functions (SNR0) and first-order relativistically corrected wave functions (SNR1) \cite{Li:2009nr}.
\begin{table}[t]
\centering
\caption[Comparison of $h_b(1P) \to \eta_b(1S) + \gamma$ decay widths.]{Comparison of our decay widths of $h_b(1P) \to \eta_b(1S) + \gamma$ at LO, NLO and NNLO with a non-relativistic constituent quark model (CQM) \cite{Segovia:2016xqb}, a relativistic quark model (R) \cite{Ebert:2002pp}, the Godfrey-Isgur potential model (GI) \cite{Godfrey:2015dia}, the light-front quark model (LFQM) \cite{Shi:2016cef:MB} and screened potential models with zeroth-order wave functions (SNR0) and first-order relativistically corrected wave functions (SNR1) \cite{Li:2009nr}. All decay widths are given in units of keV.}
\label{tab:ResultsComparisonHB1}
\begin{tabular}{c|Sc|Sc|Sc|Sc|Sc|Sc|Sc|Sc}
Mode & LO & NLO & NNLO & CQM & R & GI & LFQM & SNR$_{0/1}$ \\
\hline
$h_b(1P) \to \eta_b(1S) + \gamma$ & 41.77 & 55.48 & 124.83 & 43.7 & 52.6 & 35.7 & 37.5 & 55.8 / 36.3
\end{tabular}
\end{table}
Due to the very dominant contribution of the radiative corrections to the static potential that overall increase to total matrix element and due to the non-existence of several angular contributions that tend to decrease the total matrix element in the case of the $\chi_{b J}(1P) \to \Upsilon(1S) + \gamma$, none of these results lies within the error band of our final result. Our leading order and partially also our NLO results are compatible with the different approaches. As already mentioned before, we expect the renormalization improved approach to milden the impact of the radiative corrections and to improve the convergence of the perturbative series, such that one should expect a more comparable result in the future.\\
Using the branching fraction given by the PDG \cite{PDG:2016} and our 
theoretical result for the partial width, we can predict the total width of the 
$h_b$. The result is given in Table~\ref{tab:ResultsHB} and the uncertainties 
are again obtained via standard Gaussian uncertainty propagation.
\begin{table}[t]
\centering
\caption[$h_b(1P) \to \eta_b(1S) + \gamma$ result and prediction.]{Result and 
prediction for the partial and total decay width of $h_b(1P) \to \eta_b(1S) + 
\gamma$. The branching fraction is given by the PDG \cite{PDG:2016}, the 
partial width corresponds to our final result (black line in 
Fig.~\ref{fig:21P111S0-orderBYorderFinalResult}) including the theoretical 
uncertainty (gray band in Fig.~\ref{fig:21P111S0-orderBYorderFinalResult}), and 
the total width is a prediction.}
\label{tab:ResultsHB}
\begin{tabular}{c|Sc|Sc|Sc}
Mode & Fraction $\mathcal{B} = \frac{\Gamma_i}{\Gamma}$ [PDG] & Partial width $\Gamma_i$ & Total width $\Gamma$ \\
\hline
$h_b(1P) \to \eta_b(1S) + \gamma$ & $(52^{+6}_{-5})\%$ & $125^{+3}_{-27}$~keV & $240^{+28}_{-57}$~keV
\end{tabular}
\end{table}

\section{Analysis of the non-perturbative contribution to the wave function renormalization}

\begin{figure}[t!]
\centering
{\includegraphics[clip,trim={0.0cm 3.3cm 0.0cm 0.0cm},width=0.5\textwidth]{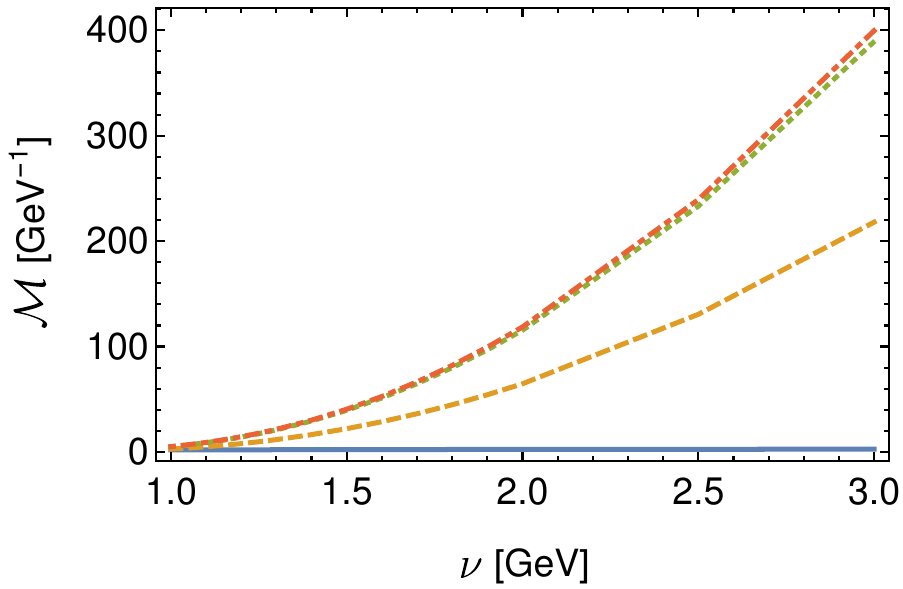}}
\caption[Scaling behavior of the non-perturbative correction to the wave function renormalization.]{Scaling behavior of the non-perturbative correction to the wave function renormalization contributing to the transition $2\,^3\!P_J \to 1\,^3\!S_1 + \gamma$. The solid blue curve is the leading order and the dashed orange, the dotted green and the dash-dotted red ones represent the three different values $\langle \alpha_s G^2 \rangle = 0.042~\text{GeV}^{-4}$, $\langle \alpha_s G^2 \rangle = 0.075~\text{GeV}^{-4}$ and $\langle \alpha_s G^2 \rangle = 0.077~\text{GeV}^{-4}$, respectively.}
\label{fig:NonPerturbative}
\end{figure}
Finally, we want to analyze the impact of the non-perturbative contribution to the wave function renormalization, derived in Sec.~\ref{sec:NonPerturbativeContributions}. Using the final result obtained there, together with the three literature values for the two gluon condensate, we obtain
\begin{align}
\M_{n\,^{3}P_J \to n'\,^{3}S_1 + \gamma}^{\text{fig. 8, 1}} \simeq \M_{n\,^{3}P_J \to n'\,^{3}S_1 + \gamma}^{(0)} \times \left\lbrace
\begin{matrix}
\frac{0.0178699}{\alpha_s^6} \\
\frac{0.0319105}{\alpha_s^6} \\
\frac{0.0327615}{\alpha_s^6}
\end{matrix}
\right. \,.
\end{align}
The scaling behavior is shown in Fig.~\ref{fig:NonPerturbative}. As already anticipated, the results exceed the leading order matrix element by several orders of magnitude with increasing $\nu$. However, in the very low energy regime, where $\nu \lesssim 1.1$~GeV, the contribution is smaller than the leading order one but the overall scaling behavior makes this a contribution that is completely out of proportion. This behavior is also the reason why we did explicitly not incorporate the contribution into our final analysis of the decay width.\\
A consistent procedure to incorporate this contribution and the question of its interpretation is subject to discussion and active research at this time. A state of the art discussion about the two gluon condensate and its role in non-perturbative computations may be found in Refs.~\cite{Pineda:1996nw,Bali:2014sja}. In the non-perturbative regime of pNRQCD, where the potentials themselves have to be computed on, e.g., the lattice, also the approximation as a local condensate breaks down.

\clearpage
\thispagestyle{empty}
\clearpage

\chapter{Summary and outlook}
\label{chp:SummaryOutlook}

In this work we have derived detailed analytic expressions for the partial decay widths of the processes $n\,^3\!P_J \to n'\,^3\!S_1 + \gamma$, for $J=0,1,2$ and $n\,^1\!P_1 \to n'\,^1\!S_0 + \gamma$ in pNRQCD up to NNLO, $\O(m\alpha_s^6)$, using quantum mechanical perturbation theory. The leading order decay width was obtained by solving the associated Schrödinger equation with a Coulomb type potential which is the leading order static potential. Radiative corrections to the static potential at order $v$ and $v^2$, and higher order potentials in the $\frac{1}{m}$-expansion at order $v^2$ have been included in a systematic way. Both of them contribute to the decay width via corrections to the initial and final state quarkonia. Relativistic corrections to the pNRQCD Lagrangian at order $v^2$ provide corrections to the leading order electric dipole operator and have been included accordingly. Furthermore we discussed the impact of non-perturbative color octet contributions entering the decay width at relative order $v^2$ via higher order Fock states. We explicitly derived an analytic expression for the contribution to the wave function renormalization under the assumption of a local two gluon condensate and discussed its scaling behavior.\\
The numerical analysis of the perturbative corrections to the wave functions reveals a severe dependence on the scale $\nu$ which is dominated by the logarithmic contributions steaming from the radiative corrections to the static potential and from the running of $\alpha_s(\nu)$ that affects primarily the Bohr radius $a = \frac{1}{\mr C_F \alpha_s}$. This severe scale dependence is taken into account by the large uncertainties that accompanies our final results. Most of the other contributions, especially corrections to the kinetic energy and most of the spin/angular corrections, behave well in perturbation theory. A bigger contribution, however, is obtained from the potentials that are proportional to the $\delta$-distribution, especially for low values of the scale $\nu$. Nevertheless, we are still able to give predictions for the total widths of the $\chi_{b J}$-family and of the $h_b$ that, within errors lie within known upper bounds from experiment and are of the order of similar decay widths in the heavy meson sector. Furthermore, our results for the $\chi_{b J}$-family comprises the predictions from non-relativistic constituent quark models and the leading order results match the ones from a Buchmüller-Tye potential-model within a $\sim 7\%$ deviation. In the case of the $h_b$ the results are not comparable with other theoretical approaches, although there are several available. The error again is quite big but far smaller than in the $\chi_{b J}$-family. Due to the absence of several negative matrix elements that vanish due to the spin 0 nature of initial and final state quarkonia, the result for the partial decay width is significantly bigger. Apart from the branching fraction there is no experimental data available for this transition, yet the predicted total width seems reasonable.\\
We are expecting quite some change with respect to the scaling behavior once the full static potential is included in a renormalization improved way. This is work in progress for a future publication. The static potential should be included into the Schrödinger equation which then needs to be solved numerically. Furthermore, making the substitution $\nu \to \frac{1}{r}$ in the appropriate low distance region and also changing the running of $\alpha_s$ by incorporating $\alpha_{V_{\text{s}}} = \alpha_s(1/r) + a_1(1/r,r) \alpha_s^2(1/r) + a_2(1/r,r) \alpha_s^3(1/r)$ accordingly is assumed to give a numerically more stable result and should diminish the scale dependence. As already mention this approach delivered reasonable and stable results in M1 transitions, however one should keep in mind that there the leading order is a constant, since the matrix element does not depend on $r$ but evaluate to $\delta_{n n'}$, since the magnetic dipole operator acts as identity in spacial and angular momentum space. In the case of E1 transitions this is not the case and thus already at leading order we have a scale dependence entering via the Bohr radius.

\clearpage
\thispagestyle{empty}
\clearpage

%
%
%
%

\appendix

\chapter{Notations and conventions}
\label{app:chp:NotationsAndConvention}

During this work the following notations and conventions are used:

\begin{itemize}
\item $c = \hbar = 1$
\item The metric tensor is given by $g_{\mu\nu} = g^{\mu\nu} = \mathrm{diag}\{1,-1,-1,-1\}$.
\item Einstein's summation convention is used.
\item Greek Lorentz indices run from 0 to 3, Latin ones from 1 to 3.
\item $x^{\mu} = (x_0,\vec{x})^T$ and $x_{\mu} = g_{\mu\nu}x^{\nu} = (x_0,-\vec{x})^T$ with the product $a \cdot b = a_{\mu}b^{\mu} = a_0b_0 - \vec{a} \cdot \vec{b}$.
\item The commutator is defined via $[A,B] = AB - BA$ and the anticommutator via $\lbrace A,B \rbrace = AB + BA$.
\item The four gamma matrices are defined such that they generate a Clifford algebra via the anticommutation relation
\begin{equation}
\{\gamma^{\mu},\gamma^{\nu}\} = 2g^{\mu\nu} \,.
\end{equation}
They can be viewed as a basis of a representation space of the 4-vector representation of the Lorentz-group sitting inside the Clifford algebra. The elements
\begin{equation}
\sigma^{\mu\nu} = \frac{\i}{2} \left[\gamma^{\mu},\gamma^{\nu}\right]
\end{equation}
form a representation of the Lie-algebra of the Lorentz-group.\\
A fifth gamma matrix can be defined as
\begin{equation}
\gamma^5 = \i\gamma^0\gamma^1\gamma^2\gamma^3 = \frac{\i}{4!} \epsilon_{\mu\nu\rho\sigma} \gamma^\mu \gamma^\nu \gamma^\rho \gamma^\sigma \,.
\end{equation}
It is useful in the context of chirality, since it allows to define the left- and right-handed projector as
\begin{equation}
\mathcal{P}_{\text{L,R}} = \frac{\1_{4 \times 4} \mp \gamma^5}{2} \,.
\end{equation}
The fifth gamma matrix is hermitian, has eigenvalues $\pm 1$ and anticommutes with the four other gamma matrices, hence fulfills
\begin{align}
(\gamma_5)^{\dagger} = \gamma_5 \,, \quad\quad (\gamma_5)^2 = \1_{4 \times 4} \,, \quad\quad \{\gamma^5,\gamma^{\mu}\} = 0 \,.
\end{align}
The set $\lbrace \gamma^0,\gamma^1,\gamma^2,\gamma^3,\i\gamma^5 \rbrace$ therefore forms the basis of the Clifford algebra in 5 spacetime dimensions for the metric signature (1,4).\\
In general, the gamma matrices fulfill the following identities
\begin{align}
\begin{aligned}
& \gamma^{\mu}\gamma_{\mu}=4\1_{4\times4} \,, && \gamma^{\mu}\gamma^{\nu} = g^{\mu\nu} - 2\sigma^{\mu\nu} \,, \\
& \gamma^{\mu}\gamma^{\nu}\gamma_{\mu} = -2\gamma^{\nu} \,, && \gamma^\mu\gamma^\nu\gamma^\rho\gamma_\mu = 4g^{\nu\rho} \,, \\
& \gamma^\mu\gamma^\nu\gamma^\rho\gamma^\sigma\gamma_\mu = -2\gamma^\sigma\gamma^\rho\gamma^\nu \,, \quad\quad && \gamma^\mu\gamma^\nu\gamma^\rho = g^{\mu\nu}\gamma^\rho + g^{\nu\rho}\gamma^\mu - g^{\mu\rho}\gamma^\nu - \i\epsilon^{\sigma\mu\nu\rho}\gamma_\sigma\gamma^5 \,,
\end{aligned}
\end{align}
the following trace identities, where the trace over any odd number of gamma matrices vanishes,
\begin{align}
\begin{aligned}
& \mathrm{tr}\{\gamma^{\mu}\gamma^{\nu}\} = 4g^{\mu\nu} \,, && \mathrm{tr}\{\gamma^{\mu}\gamma^{\nu}\gamma^{\rho}\gamma^{\sigma}\} = 4\(g^{\mu\nu}g^{\rho\sigma} - g^{\mu\rho}g^{\nu\sigma} + g^{\mu\sigma}g^{\nu\rho}\) \,, \\
& \mathrm{tr}\{\gamma^5\} = \mathrm{tr}\{\gamma^\mu\gamma^\nu\gamma^5\} = 0 \,, && \mathrm{tr}\{\gamma^\mu\gamma^\nu\gamma^\rho\gamma^\sigma\gamma^5\} = 4\i\epsilon^{\mu\nu\rho\sigma} \,, \\
& \mathrm{tr}\{\gamma^{\mu 1}\dots\gamma^{\mu n}\} = \mathrm{tr}\{\gamma^{\mu n}\dots\gamma^{\mu 1}\} \,,
\end{aligned}
\end{align}
and are normalized, since
\begin{align}
\begin{aligned}
& (\gamma^0)^{\dagger} = \gamma^0 \quad \Leftrightarrow \quad (\gamma^0)^2 = \1_{4 \times 4} \,, \\
& (\gamma^i)^{\dagger} = -\gamma^i \quad \Leftrightarrow \quad (\gamma^i)^2 = -\1_{4 \times 4} \,, \\
\Rightarrow & (\gamma^\mu)^{\dagger} = \gamma^0 \gamma^\mu \gamma^0 \,.
\end{aligned}
\end{align}
Finally, in Dirac representation, the gamma matrices are given by
\begin{align}
\begin{aligned}
& \gamma^0 = \begin{pmatrix} 1 & 0 & 0 & 0 \\ 0 & 1 & 0 & 0 \\ 0 & 0 & -1 & 0 \\ 0 & 0 & 0 & -1 \end{pmatrix} \,, \quad\quad \gamma^1 = \begin{pmatrix} 0 & 0 & 0 & 1 \\ 0 & 0 & 1 & 0 \\ 0 & -1 & 0 & 0 \\ -1 & 0 & 0 & 0 \end{pmatrix} \,, \quad\quad \gamma^2 = \begin{pmatrix} 0 & 0 & 0 & -\i \\ 0 & 0 & \i & 0 \\ 0 & \i & 0 & 0 \\ -\i & 0 & 0 & 0 \end{pmatrix} \,, \\
& \gamma^3 = \begin{pmatrix} 0 & 0 & 1 & 0 \\ 0 & 0 & 0 & -1 \\ -1 & 0 & 0 & 0 \\ 0 & 1 & 0 & 0 \end{pmatrix} \,, \quad\quad\quad \gamma^5 = \begin{pmatrix} 0 & 0 & 1 & 0 \\ 0 & 0 & 0 & 1 \\ 1 & 0 & 0 & 0 \\ 0 & 1 & 0 & 0 \end{pmatrix} \,.
\end{aligned}
\end{align}
\item Feynman slashed quantities are defined as $\slashed{a} = \gamma^{\mu}a_{\mu}$ with $\slashed{a}^2 = a^2$ and $\slashed{a}\slashed{b} = ab - \i a_{\mu}\sigma^{\mu\nu}b_{\nu}$.
\item The three hermitian and traceless Pauli matrices are given by
\begin{equation}
\sigma_1 = \begin{pmatrix} 0 & 1 \\ 1 & 0 \end{pmatrix} \,, \quad\quad \sigma_2 = \begin{pmatrix} 0 & -\i \\ \i & 0 \end{pmatrix} \,, \quad\quad \sigma_3 = \begin{pmatrix} 1 & 0 \\ 0 & -1 \end{pmatrix} \,,
\end{equation}
and thus fulfill $\sigma_i^{\dagger} = \sigma_i$ and $\mathrm{tr}\{\sigma_i\} = 0$. Furthermore they are normalized with respect to the trace, since $\mathrm{tr}\{\sigma_i\sigma_j\} = 2\delta_{i j}$.\\
The generators, $\tau_i = \frac{1}{2} \sigma_i$, of the Lie-group SU(2) fulfill
\begin{align}
\begin{aligned}
& \left[\tau_i,\tau_j\right] = \i \epsilon_{i j k} \tau_k \,, \\
& \left\lbrace \tau_i,\tau_j \right\rbrace = \frac{1}{2} \delta_{i j} \,,
\end{aligned}
\end{align}
where the real and totally antisymmetric structure constants of the corresponding Lie-algebra are given by the Levi-Civita-symbol using the convention $\epsilon_{123}=1$.\\
From the above relations one may derive
\begin{align}
\begin{aligned}
& \sigma_i\sigma_j = \i\epsilon_{i j k}\sigma_k + \delta_{i j} \1_{2 \times 2} \,, \\
& \epsilon_{i k l} \epsilon_{j k l} = 2 \delta_{i j} \,.
\end{aligned}
\end{align}
Using the Pauli matrices, one can write the gamma matrices in the compact form
\begin{equation}
\gamma^0 = \begin{pmatrix} \1_{2 \times 2} & 0 \\ 0 & -\1_{2 \times 2} \end{pmatrix} \,, \quad\quad \gamma^i = \begin{pmatrix} 0 & \sigma^i \\ -\sigma^i & 0 \end{pmatrix} \,, \quad\quad \gamma^5 = \begin{pmatrix} 0 & \1_{2 \times 2} \\ \1_{2 \times 2} & 0 \end{pmatrix} \,.
\end{equation}
Defining $\sigma^\mu = (\1_{2 \times 2},\vec{\sigma})^T$ and $\bar{\sigma}^\mu = (\1_{2 \times 2},-\vec{\sigma})^T$, one may in short hand notation cast the gamma matrices in the Weyl (chiral) representation as
\begin{equation}
\gamma^\mu = \begin{pmatrix} 0 & \sigma^\mu \\ \bar{\sigma}^\mu & 0 \end{pmatrix} \,, \quad\quad \gamma^5 = \begin{pmatrix} -\1_{2 \times 2} & 0 \\ 0 & \1_{2 \times 2} \end{pmatrix} \,.
\end{equation}
The Weyl (chiral) representation can be obtained from the Dirac representation via an unitary transformation:
\begin{equation}
\gamma^\mu_{\text{W}} = U \gamma^\mu_{\text{D}} U^{-1} \,, \quad \text{with  } U = \frac{1}{\sqrt{2}} \begin{pmatrix} 1 & 1 \\ -1 & 1 \end{pmatrix} \,, \quad U^{-1} = U^\dagger = \frac{1}{\sqrt{2}} \begin{pmatrix} 1 & -1 \\ 1 & 1 \end{pmatrix} \,.
\end{equation}
\item The eight hermitian and traceless Gell-Mann matrices are given by
\begin{align}
\begin{aligned}
& \lambda^1 = \begin{pmatrix} 0 & 1 & 0 \\ 1 & 0 & 0 \\ 	0 & 0 & 0 \end{pmatrix} \,, \quad \lambda^2 = \begin{pmatrix} 0 & -\i & 0 \\ \i & 0 & 0 \\ 0 & 0 & 0 \end{pmatrix} \,, \quad \lambda^3 = \begin{pmatrix} 1 & 0 & 0 \\ 	0 & -1 & 0 \\ 0 & 0 & 0 \end{pmatrix} \,, \\
& \lambda^4 = \begin{pmatrix} 0 & 0 & 1 \\ 0 & 0 & 0 \\ 1 & 0 & 0 \end{pmatrix} \,, \quad \lambda^5 = \begin{pmatrix} 0 & 0 & -\i \\ 0 & 0 & 0 \\ \i & 0 & 0 \end{pmatrix} \,, \quad \lambda^6 = \begin{pmatrix} 0 & 0 & 0 \\ 0 & 0 & 1 \\ 0 & 1 & 0 \end{pmatrix} \,, \\
& \lambda^7 = \begin{pmatrix} 0 & 0 & 0 \\ 0 & 0 & -\i \\ 0 & \i & 0 \end{pmatrix} \,, \quad \lambda^8 = \frac{1}{\sqrt{3}}\begin{pmatrix} 1 & 0 & 0 \\ 0 & 1 & 0 \\ 	0 & 0 & -2 \end{pmatrix} \,,
\end{aligned}
\end{align}
and thus fulfill $\lambda_a^{\dagger} = \lambda_a$ and $\mathrm{tr}\{\lambda_a\} = 0$. Furthermore they are normalized with respect to the trace, since $\mathrm{tr}\{\lambda_a\lambda_b\} = 2\delta_{a b}$.\\
The generators, $t_a = \frac{1}{2} \lambda_a$, of the Lie-group SU(3) fulfill
\begin{align}
\begin{aligned}
& \left[t_a,t_b\right] = \i f_{a b c} t_c \,, \\
& \left\lbrace t_a,t_b \right\rbrace = \frac{1}{3} \delta_{a b} \1_{3 \times 3} + d_{a b c} t_c \,,
\end{aligned}
\end{align}
where the real, totally antisymmetric and, up to permutations, non-vanishing structure constants $f$ of the corresponding Lie-algebra are given by
\begin{align}
\begin{aligned}
& f_{123} = 1 \,, \quad\quad f_{458} = f_{678} = \frac{\sqrt{3}}{2} \,, \\
& f_{147} = -f_{156} = f_{246} = f_{257} = f_{345} = -f_{367} = \frac{1}{2} \,,
\end{aligned}
\end{align}
and the symmetric coefficients $d$ are given by
\begin{align}
\begin{aligned}
& d_{118} = d_{228} = d_{338} = -d_{888} = \frac{1}{\sqrt{3}} \,, \quad\quad d_{448} = d_{558} = d_{668} = d_{778} = -\frac{1}{2\sqrt{3}} \,,\\
& d_{146} = d_{157} = -d_{247} = d_{256} = d_{344} = d_{355} = -d_{366} = -d_{377} = \frac{1}{2} \,.
\end{aligned}
\end{align}
From the above relations one may derive the relations for the Casimir operators $C_F$ and $C_A$ of the fundamental and the adjoint representations, respectively
\begin{align}
\begin{aligned}
& t_{i j}^a t_{j k}^a = C_F \delta_{i k} \,, \\
& f^{a c d} f^{b c d} = C_A \delta^{a b} \,.
\end{aligned}
\end{align}
\end{itemize}

\section{Group theoretical and static potential and QCD related constants}
\label{app:sec:Constants}

The group theoretical constants entering the computations in this work are given by
\begin{equation}
T_F = \frac{1}{2} \,, \quad\quad C_A = N_{\text{c}} = 3 \,, \quad\quad C_F = \frac{N_{\text{c}}^2 - 1}{2 N_{\text{c}}} = \frac{4}{3} \,.
\end{equation}
The expansion parameters $a$ of the static potential, see e.g. \cite{Pineda:2011dg}, are given by
\begin{align}
\begin{aligned}
& a_1 = \frac{31 C_A - 20 n_f T_F}{9} \,, \\	
& a_2 = \frac{400 T_F^2 n_f^2}{81} - C_F T_F n_f \left(\frac{55}{3} - 16 \zeta(3)\right) + C_A^2 \left(\frac{4343}{162} + \frac{16\pi^2 - \pi^4}{4} +	\frac{22 \zeta(3)}{3}\right) \\
&\quad - C_A T_F n_f \left(\frac{1798}{81} + \frac{56 \zeta(3)}{3}\right) \,,
\end{aligned}
\end{align}
where $\zeta(x) = \sum\limits_{n=1}^\infty \frac{1}{n^x}$ denotes the Riemann Zeta function.\\
The coefficients of the $\beta$-function, first computed in Refs.~\cite{Gross:1973ju,Politzer:1973fx,PhysRevLett.30.1343,Gross:1974cs}, listed, e.g., in Ref.~\cite{Pineda:2011dg}, are given by
\begin{equation}
\beta_0 = \frac{11}{3} C_A - \frac{4}{3} T_F n_f \,, \quad\quad \beta_1 = \frac{34}{3} C_A^2 - \frac{20}{3}  C_A T_F n_f - 4 C_F T_F n_f \,.
\end{equation}
The pole mass expansion coefficients $d$, see Ref.~\cite{Brambilla:2001qk}, are given by
\begin{align}
\begin{aligned}
& d_1 = \frac{307}{32} + \frac{\pi^2}{3} + \frac{\pi^2 \ln{2}}{9} - \frac{\zeta(3)}{6} + n_f \left(-\frac{71}{144} - \frac{\pi^2}{18}\right) \\
& \quad \simeq 13.4434 - 1.04137 n_f \,, \\
& d_2 = \frac{8462917}{93312} + \frac{652841 \pi^2}{38880} - \frac{695 \pi^4}{7776} - \frac{575 \pi^2 \ln{2}}{162} \\
& \quad - \frac{22 \pi^2 \lnSquare{2}}{81} - \frac{55 \mathrm{ln}^4(2)}{162} - \frac{220 \text{Li}_4\left(\frac{1}{2}\right)}{27} + \frac{58 \zeta(3)}{27} - \frac{1439 \pi^2 \zeta(3)}{432} + \frac{1975 \zeta(5)}{216} \\
& \quad + n_f \Big(-\frac{231847}{23328} - \frac{991 \pi^2}{648} + \frac{61 \pi^4}{1944} - \frac{11 \pi^2 \ln{2}}{81} + \frac{2 \pi^2 \lnSquare{2}}{81} + \frac{\mathrm{ln}^4(2)}{81} \\
& \quad + \frac{8 \text{Li}_4\left(\frac{1}{2}\right)}{27} - \frac{241 \zeta(3)}{72}\Big) + n_f^2 \Big(\frac{2353}{23328} + \frac{13 \pi^2}{324} + \frac{7 \zeta(3)}{54}\Big) \\
& \quad \simeq 190.391 - 26.6551 n_f + 0.652691 n_f^2 \,,
\end{aligned}
\end{align}
where $\text{Li}_n(z) = \sum\limits_{k=1}^\infty \frac{z^k}{k^n}$ is the polylogarithm function.

\section{Explicit expressions of the Coulomb wave functions}
\label{app:sec:WaveFunctions}

We explicitly display the first few radial Coulomb wave functions $R_{n\l}(r)$ with respect to Eq.~\eqref{eq:ZerothOrderWaveFunction}. They are consistent with the ones given in \cite{BransdenJoachain:QuantumMechanics} (Eq.~(7.140) therein) and read
\begin{align}
\begin{aligned}
R_{10}(r) &= \frac{2}{\sqrt{a^3}} \, \e{-\frac{r}{a}} \,, \\
R_{20}(r) &= \frac{-\frac{r}{a} + 2}{2 \sqrt{2 a^3}} \, \e{-\frac{r}{2 a}} \,, \\
R_{21}(r) &= \frac{\frac{r}{a}}{2 \sqrt{6 a^3}} \, \e{-\frac{r}{2 a}} \,, \\
R_{30}(r) &= 2 \frac{2 \frac{r^2}{a^2} - 18 \frac{r}{a} + 27}{81 \sqrt{3 a^3}} \, \e{-\frac{r}{3 a}} \,, \\
R_{31}(r) &= 2 \frac{-2 \frac{r^2}{a^2} + 12 \frac{r}{a}}{81 \sqrt{6 a^3}} \e{-\frac{r}{3 a}} \,, \\
R_{32}(r) &= 2 \frac{2 \frac{r^2}{a^2}}{81 \sqrt{30 a^3}} \, \e{-\frac{r}{3 a}} \,.
\end{aligned}
\end{align}
The corresponding spherical harmonics $Y_{\l m}(\theta,\phi)$ read
\begin{align}
\begin{aligned}
Y_{00}(\theta,\phi) &= \frac{1}{\sqrt{4\pi}} \\
Y_{1m}(\theta,\phi) &= \left\lbrace \begin{array}{ll}
\frac{1}{2}\sqrt{\frac{3}{2\pi}} \cdot \e{-\i\phi} \cdot \sin(\theta) & ,m = -1 \\
\frac{1}{2}\sqrt{\frac{3}{\pi}} \cdot \cos(\theta) & ,m = 0 \\
-\frac{1}{2}\sqrt{\frac{3}{2\pi}} \cdot \e{\i\phi} \cdot \sin(\theta) & ,m = 1
\end{array} \right. \\
Y_{2m}(\theta,\phi) &= \left\lbrace \begin{array}{ll}
\frac{1}{4}\sqrt{\frac{15}{2\pi}} \cdot \e{-2\i\phi} \cdot \sin^2(\theta) & ,m = -2 \\
\frac{1}{2}\sqrt{\frac{15}{2\pi}} \cdot \e{-\i\phi} \cdot \sin(\theta) \cdot \cos(\theta) & ,m = -1 \\
\frac{1}{4}\sqrt{\frac{5}{\pi}} \cdot (3\cos^2(\theta) - 1) & ,m = 0 \\
\frac{-1}{2}\sqrt{\frac{15}{2\pi}} \cdot \e{\i\phi} \cdot \sin(\theta) \cdot \cos(\theta) & ,m = 1 \\
\frac{1}{4}\sqrt{\frac{15}{2\pi}} \cdot \e{2\i\phi} \cdot \sin^2(\theta) & ,m = 2
\end{array} \right.
\end{aligned}
\end{align}
Setting $r = 0$ in the above expressions makes explicit that only s-waves do not vanish at the origin.

\clearpage

\section{How to use the MATHEMATICA package RunDec}
\label{app:sec:RunDec}

We give the general procedure and two examples on how to determine $\alpha_s(\nu)$ for $n_f=3$ and $n_f=4$ massless flavors using the MATHEMATICA package RunDec. For further instructions see \cite{Chetyrkin:2000yt}. Our starting point is $\alpha_s^{(5)}(M_Z) = 0.118$ and $M_Z=91.18$~GeV, being the value of $\alpha_s$ at the $Z$-boson scale with 5 massless flavors:
\begin{itemize}
\item General procedure:\hfill
\begin{Mathematica}
	In[1]:= Import["RunDec.m"];
\end{Mathematica}
\begin{Mathematica}
	In[2]:= NumDef
	Out[2]= {asMz -> 0.118,Mz -> 91.18,Mt -> 175,Mb -> 4.7,
		 Mc -> 1.6,muc -> 1.2,mub -> 3.97,Mtau -> 1.777}
\end{Mathematica}
The routine AsRunDec takes as input $\alpha_s(\mu_0)$, $\mu_0$, $\mu$ and the number of loops and returns $\alpha_s(\mu)$.\\
The routine AlphasExact takes as input $\alpha_s(\mu)$, $\mu$, $\nu$, $n_f$ and the number of loops and returns $\alpha_s(\nu)$.
\item $n_f=3$:\hfill\\
Remind that in the RunDec the mass of the c-quark is $M_{\text{c}} = 1.6$~GeV and so we have to go to values below this to obtain $\alpha_s^{(3)}(\mu)$. We choose $\mu=1.0$~GeV and obtain with 4-loop accuracy:
\begin{Mathematica}
	In[3]:= mu1GeV=1.0;
		loops4=4;
		alsmu1GeV=AsRunDec[asMz/.NumDef,Mz/.NumDef,mu1GeV,
			loops4]
	Out[3]= 0.479778
\end{Mathematica}
Running this up to $\nu=3.0$~GeV, where we have 4 active flavors, yields at 4-loop accuracy:
\begin{Mathematica}
	In[4]:= nu3GeV=3.0;
		flavor4=4;
		AlphasExact[alsmu1GeV,mu1GeV,nu3GeV,flavor4,loops4]
	Out[4]= 0.258185
\end{Mathematica}
\item $n_f=4$:\hfill\\
Remind that in the RunDec the mass of the b-quark is $M_{\text{b}} = 4.7$~GeV and so we have to go to values below this to obtain $\alpha_s^{(4)}(\mu)$. We choose $\mu=4.0$~GeV and obtain with 4-loop accuracy:
\begin{Mathematica}
	In[5]:= mu4GeV=4.0;
		alsmu4GeV=AsRunDec[asMz/.NumDef,Mz/.NumDef,mu4GeV,
			loops4]
	Out[5]= 0.228270
\end{Mathematica}
Running this up to $\nu=6.0$~GeV, where we have 5 active flavors, yields at 4-loop accuracy:
\begin{Mathematica}
	In[6]:= nu6GeV=6.0;
		flavor5=5;
		AlphasExact[alsmu4GeV,mu4GeV,nu6GeV,flavor5,loops4]
	Out[6]= 0.188513
\end{Mathematica}
\end{itemize}

\clearpage
\thispagestyle{empty}
\clearpage

\chapter{Functions and Expectation values}
\label{app:chp:FunctionsExpectationValues}

We list a set of functions and definitions including some finite sums and expectation values of single and double potential insertions that are used throughout this work. They may partly be found e.g. in the appendices of Refs.~\cite{Pineda:1997hz,Peset:2015zga,Peset:2015vvi}; if not, they will be derived explicitly. We furthermore show the divergence cancellation associated with the Coulomb Green function and derive the non-perturbative contribution to the wave function renormalization.

\section{Functions}
\label{app:sec:Functions}

The following finite sums appear throughout this work:
\begin{align}
& S_p(N) = \sum\limits_{i=1}^N \frac{1}{i^p} \,, \\
& \Delta S_{1a} = S_1(n+\l) - S_1(n-\l-1) \,, \\
& \Sigma_a(n,\l) = \Sigma_3^{(m)}(n,\l) + \Sigma_3^{(k)}(n,\l) + \frac{2}{n} \Sigma_2^{(k)}(n,\l) \,, \\
& \Sigma_p^{(m)}(n,\l) = \frac{(n+\l)!}{(n-\l-1)!} \sum\limits_{m=-\l}^\l \frac{R(l,m)}{(n+m)^p} S_1(n+m) \,, \\
& \Sigma_p^{(k)}(n,\l) = \frac{(n-\l-1)!}{(n+\l)!} \sum\limits_{k=1}^{n-\l-1} \frac{(k+2\l)!}{(k-1)!(k+\l-n)^p} \,, \\
& R(\l,m) = \frac{(-1)^{\l-m}}{(\l+m)!(\l-m)!} \,.
\end{align}

\subsection{The Gamma, Digamma and Polygamma functions}
\label{app:subsec:GammaDigammaPolygamma}

The (complete) Gamma function\footnote{We refer to it as the complete Gamma function, since it may also be generalized to the upper incomplete Gamma function $\Gamma(a,x)$ and the lower incomplete Gamma function $\gamma(a,x)$, such that $\Gamma(a,0) = \Gamma(a)$ and $\Gamma(a,x) + \gamma(a,x) = \Gamma(a)$.} \cite{mathworld:GammaFunction} $\Gamma(n)$ is defined to be an extension of the factorial to complex and real number arguments. It is related to the factorial by
\begin{equation}
\Gamma(n) = (n-1)! \,.
\end{equation}
It is analytic everywhere in the complex $z$-plane, except at $z=0,-1,-2,\dots$, and the residue at $z=-k$ is given by $\frac{(-1)^k}{k!}$. The Gamma function can be defined as an integral for Re$(z) > 0$ via
\begin{equation}
\Gamma(z) = \int\limits_0^\infty \d t \, t^{z-1} \e{-t} = 2 \int\limits_0^\infty \d t \, t^{2z-1} \e{-t^2} \,.
\end{equation}
A beautiful relationship between $\Gamma(z)$ and the Riemann zeta function $\zeta(z)$ is given for Re$(z) > 0$ by
\begin{equation}
\Gamma(z) \zeta(z) = \int\limits_0^\infty \d u \, \frac{u^{z-1}}{\e{u} - 1} \,.
\end{equation}
The gamma function satisfies the functional equations
\begin{equation}
\Gamma(1+z) = z \Gamma(z) \quad\quad \text{and} \quad\quad \Gamma(1-z) = -z \Gamma(-z) \,,
\end{equation}
and differentiating yields
\begin{equation}
\Gamma'(z) = \frac{\d}{\d z} \Gamma(z) = \Gamma(z) \psi(z) = \Gamma(z) \psi_0(z) \,,
\end{equation}
where $\psi(z)$ is the Digamma function \cite{mathworld:DigammaFunction} and $\psi_0(z)$ is the zeroth Polygamma function \cite{mathworld:PolygammaFunction}.\\
The Digamma function is defined as
\begin{equation}
\psi(z) = \frac{\d}{\d z} \ln{\Gamma(z)} = \frac{\Gamma'(z)}{\Gamma(z)} \,,
\end{equation}
and satisfies
\begin{equation}
\psi(z) = \int\limits_0^\infty \d t \, \left(\frac{\e{-t}}{t} - \frac{\e{-z t}}{1 - \e{-t}}\right) = (-1) \sum\limits_{k=0}^\infty \frac{1}{(z+k)} \,.
\end{equation}
The Polygamma function is defined as
\begin{equation}
\psi_n(z) = \frac{\d^{n+1}}{\d z^{n+1}} \ln{\Gamma(z)} = \frac{\d^n}{\d z^n} \psi(z) \,,
\end{equation}
and satisfies
\begin{equation}
\psi_n(z) = (-1)^{n+1} n! \sum\limits_{k=0}^\infty \frac{1}{(z+k)^{n+1}} \,,
\end{equation}
and one has the identity
\begin{equation}
\psi(z) = \psi_0(z) \,.
\end{equation}
They are implemented into the MATHEMATICA language as \MathematicaInline{Gamma[z]} and \MathematicaInline{PolyGamma[n,z]}, respectively.

\subsection{The Laguerre polynomials}

The Laguerre polynomials \cite{mathworld:LaguerrePolynomial} are the solution of the Laguerre differential equation \cite{mathworld:LaguerreDifferentialEquation}
\begin{equation}
x y'' + (1-x)y' + n y = 0 \,,
\end{equation}
where $n \in \N_0$. They are given by the sum
\begin{equation}
L_n(x) = \sum\limits_{k=0}^n \frac{(-1)^k}{k!} \begin{pmatrix} n \\ k \end{pmatrix} x^k \,,
\end{equation}
where $\begin{pmatrix} n \\ k \end{pmatrix}$ is the binomial coefficient. Their Rodrigues representation is given by
\begin{equation}
L_n(x) = \frac{\e{x}}{n!} \frac{\d^n}{\d x^n} (x^n \e{-x})
\end{equation}
and the generating functional for the Laguerre polynomials is given by
\begin{equation}
g(x,z) = \frac{\exp{-\frac{x z}{1-z}}}{1-z} \,.
\end{equation}
They are implemented into the MATHEMATICA language as \MathematicaInline{LaguerreL[n,x]}.

\subsection{The associated Laguerre polynomials}
\label{app:subsec:AssociatedLaguerrePolynomials}

The associated Laguerre polynomials \cite{mathworld:AssociatedLaguerrePolynomial} are the solution of the associated Laguerre differential equation \cite{mathworld:LaguerreDifferentialEquation}
\begin{equation}
x y'' + (k+1-x)y' + n y = 0 \,,
\end{equation}
where $n \in \N_0$ and $k \leq n$. They are given by the sum
\begin{equation}
L_n^k(x) = \frac{1}{n!} \sum\limits_{i=0}^n \frac{n!}{i!} \begin{pmatrix} k+n \\ n-i \end{pmatrix} (-x)^i \,,
\end{equation}
where $\begin{pmatrix} k+n \\ n-i \end{pmatrix}$ is the binomial coefficient. They are connected to the Laguerre polynomials via
\begin{equation}
L_n^k(x) = (-1)^k \frac{\d^k}{\d x^k} L_{n+k}(x)
\end{equation}
and their Rodrigues representation is given by
\begin{align}
\begin{aligned}
\label{eq:RodriguesRepresentationAssociatedLaguerrePolynomials} L_n^k(x) & = \frac{\e{x} x^{-k}}{n!} \frac{\d^n}{\d x^n} (\e{-x} x^{n+k}) \\
& = \sum\limits_{m=0}^n (-1)^m \frac{(n+k)!}{(n-m)!(k+m)!m!} x^m
\end{aligned}
\end{align}
and the generating functional for the associated Laguerre polynomials is given by
\begin{equation}
g(x,z) = \frac{\exp{-\frac{x z}{1-z}}}{(1-z)^{k+1}} \,.
\end{equation}
The associated Laguerre polynomials are orthogonal over $[0,\infty)$ with respect to the weighting function $x^k \e{-x}$, hence
\begin{equation}
\label{eq:OrthogonalityAssociatedLaguerreI} \int\limits_0^\infty \d x \, \e{-x} x^k L_n^k(x) L_m^k(x) = \frac{(n+k)!}{n!} \delta_{n m} \,.
\end{equation}
They also satisfy
\begin{equation}
\label{eq:OrthogonalityAssociatedLaguerreII} \int\limits_0^\infty \d x \, \e{-x} x^{k+1} \left[L_n^k(x)\right]^2 = \frac{(n+k)!}{n!} (2n+k+1) \,,
\end{equation}
and are related to the Hermite polynomials via
\begin{align}
\begin{aligned}
& H_{2n}(x) = (-1)^n 2^{2n} n! L_n^{-\frac{1}{2}}(x^2) \,, \\
& H_{2n+1}(x) = (-1)^n 2^{2n+1} n! x L_n^{\frac{1}{2}}(x^2) \,.
\end{aligned}
\end{align}
The relations~\eqref{eq:OrthogonalityAssociatedLaguerreI} and \eqref{eq:OrthogonalityAssociatedLaguerreII} can be generalized \cite{Titard:1993nn,Kiyo:2014uca} for $\beta < 0$ to
\begin{align}
\label{eq:OrthogonalityAssociatedLaguerreIII}
\begin{aligned}
& \int\limits_0^\infty \d z \, z^{\alpha+\beta} \e{-z} \big[L_m^\alpha(z)\big]^2 = \sum_{k=0}^m \frac{\Gamma(1+\alpha+\beta+k)}{k!} \left[\frac{\Gamma(m-k-\beta)}{\Gamma(-\beta) [(m-k)!]}\right]^2 \,, \\
& \int\limits_0^\infty \d z \, z^{\alpha+\beta} \e{-z} L_n^\alpha(z) L_m^\alpha(z) = \sum_{k=0}^{\text{min}(n,m)} \frac{\Gamma(1+\alpha+\beta+k)}{k!} \frac{\Gamma(n-k-\beta)}{\Gamma(-\beta) [(n-k)!]} \frac{\Gamma(m-k-\beta)}{\Gamma(-\beta) [(m-k)!]} \,,
\end{aligned}
\end{align}
and even for $\beta \ge 0$ \cite{Titard:1993nn} to
\begin{align}
\label{eq:OrthogonalityAssociatedLaguerreIV}
\begin{aligned}
& \int\limits_0^\infty \d z \, z^{\alpha+\beta} \e{-z} \big[L_m^\alpha(z)\big]^2 = \sum_{k=0}^m \frac{\Gamma(1+\alpha+\beta+k)}{k!} \left[\frac{\Gamma(1+\beta)}{\Gamma(1+\beta-m+k) [(m-k)!]}\right]^2 \,, \\
& \int\limits_0^\infty \d z \, z^{\alpha+\beta} \e{-z} L_n^\alpha(z) L_m^\alpha(z) \\
& \quad\quad\quad = \sum_{k=0}^{\text{min}(n,m)} \frac{\Gamma(1+\alpha+\beta+k)}{k!} \frac{\Gamma(1+\beta)}{\Gamma(1+\beta-n+k) [(n-k)!]} \frac{\Gamma(1+\beta)}{\Gamma(1+\beta-m+k) [(m-k)!]} \,.
\end{aligned}
\end{align}
Finally, \cite{GalindoPascual:QuantumMechanicsI} (Eqs.~(A75) and (A76) therein), provides the very useful recursion relations
\begin{align}
\begin{aligned}
\label{eq:RecursionAssociatedLaguerre}
&  L_n^{k-1}(z) = L_n^k(z) - L_{n-1}^k(z) \\
& z L_n^{k+1}(z) = (n+k+1) L_n^k(z) - (n+1) L_{n+1}^k(z) \,.
\end{aligned}
\end{align}
They are implemented into the MATHEMATICA language as \MathematicaInline{LaguerreL[n,k,x]}.

\subsection{The Legendre polynomials}
\label{app:subsec:Legendre}

The Legendre polynomials \cite{mathworld:LegendrePolynomial} are the solution of the Legendre differential equation \cite{mathworld:LegendreDifferentialEquation}
\begin{equation}
(1-x^2)y'' - 2 x y' + n(n+1) y = 0 \,,
\end{equation}
where $n \in \N_0$. They are given by the contour integral
\begin{equation}
P_n(z) = \frac{1}{2\pi\i} \oint \d t \, (1 - 2 t z + t^2)^{-\frac{1}{2}} t^{-n-1} \,,
\end{equation}
where the contour encloses the origin and is traversed in a counterclockwise direction. Their Rodrigues representation is given by
\begin{equation}
P_n(z) = \frac{1}{2^n n!} \frac{\d^n}{\d x^n} (x^2-1)^n
\end{equation}
and the generating functional for the Legendre polynomials is given by
\begin{equation}
g(t,z) = (1 - 2 z t + t^2)^{-\frac{1}{2}} = \sum\limits_{n=0}^\infty P_n(z) t^n \,.
\end{equation}
The Legendre polynomials are orthogonal over $(-1,1)$ with respect to the weighting function $1$, hence
\begin{equation}
\int\limits_{-1}^1 \d z \, P_n(z) P_m(z) = \frac{2}{2n+1} \delta_{n m} \,.
\end{equation}
They also satisfy
\begin{equation}
\sum\limits_{n=0}^\infty \frac{2n+1}{2} P_n(z') P_n(z) = \delta(z'-z) \,,
\end{equation}
where $\delta(z'-z)$ is the delta distribution. Following \cite{GalindoPascual:QuantumMechanicsI} (Eq.~(A41) therein), we may cast the Legendre polynomials as
\begin{equation}
P_\l(\cos\theta) = \frac{4\pi}{2\l+1} \sum\limits_{m=-\l}^\l Y_\l^{m*}(\theta_1,\phi_1)Y_\l^m(\theta_2,\phi_2) \,,
\end{equation}
once we have two directions in space, characterized by the spherical coordinates $(\theta_1,\phi_1)$ and $(\theta_2,\phi_2)$, where $\theta$ is the angle between them and the $Y_{\l m}(\theta,\phi)$ are the spherical harmonics.\\
They are implemented into the MATHEMATICA language as \MathematicaInline{LegendreP[n,x]}.

\subsection{The associated Legendre polynomials}

The associated Legendre polynomials \cite{mathworld:AssociatedLegendrePolynomial} are the solution of the associated Legendre differential equation \cite{mathworld:AssociatedLegendreDifferentialEquation}
\begin{equation}
\left[(1-x^2)y'\right]' + \left[n(n+1)-\frac{\l^2}{1-x^2}\right]y = 0 \,,
\end{equation}
where $n,\l \in \N_0$ and $0 \leq \l \leq n$. They are connected to the Legendre polynomials via
\begin{equation}
P_n^\l(z) = (-1)^\l (1-z^2)^{\frac{\l}{2}} \frac{\d^\l}{\d z^\l} P_n(z)
\end{equation}
and their Rodrigues representation is given by
\begin{align}
P_n^\l(z)  = \frac{(-1)^\l}{2^\l \l!} (1-z^2)^{\frac{\l}{2}} \frac{\d^{n+\l}}{\d z^{n+\l}} (x^2-1)^n \,,
\end{align}
allowing for $-n \leq \l \leq n$.\\
The associated Legendre polynomials are orthogonal over $[-1,1]$ with respect to the weighting function $1$, hence
\begin{equation}
\int\limits_{-1}^1 \d z \, P_n^\l(z) P_m^\l(z) = \frac{2}{2n+1} \frac{(n+\l)!}{(n-\l)!} \delta_{n m} \,,
\end{equation}
and orthogonal over $[-1,1]$ with respect to $\l$ with the weighting function $(1-z^2)^{-1}$, hence
\begin{equation}
\int\limits_{-1}^1 \d z \, \frac{1}{1-z^2} P_n^\l(z) P_n^{\l'}(z) = \frac{(n+\l)!}{\l(\l-n)!} \delta_{\l\l'} \,.
\end{equation}
They also satisfy
\begin{equation}
P_n^{-\l}(z) = (-1)^\l \frac{(n-\l)!}{(n+\l)!} P_n^\l(z) \quad\quad \text{and} \quad\quad P_n^\l(-z) = (-1)^{n+\l} P_n^\l(z) \,.
\end{equation}
They are implemented into the MATHEMATICA language as \MathematicaInline{LegendreP[n,l,x]}.

\subsection{The spherical harmonics}
\label{app:subsec:SphericalHarmonics}

In spherical coordinates the Laplace operator takes the form
\begin{equation}
\Delta = \frac{\partial^2}{\partial r^2} + \frac{2}{r} \frac{\partial}{\partial r} + \frac{1}{r^2} \left(\frac{\partial^2}{\partial \theta^2} + \frac{\cos\theta}{\sin\theta} \frac{\partial}{\partial \theta} + \frac{1}{\sin^2\theta} \frac{\partial^2}{\partial \phi^2}\right) = \Delta_r + \frac{1}{r^2} \Delta_\Omega \,.
\end{equation}
The spherical harmonics \cite{mathworld:SphericalHarmonic} are the angular part of the solution of the Laplace equation in spherical coordinates
\begin{equation}
\Delta f(r,\Omega) = 0 \,.
\end{equation}
Thus the defining differential equation is given by the spherical harmonic differential equation \cite{mathworld:SphericalHarmonicDifferentialEquation}
\begin{equation}
\left(\frac{\partial^2}{\partial \theta^2} + \frac{\cos\theta}{\sin\theta} \frac{\partial}{\partial \theta} + \frac{1}{\sin^2\theta} \frac{\partial^2}{\partial \phi^2} + \l(\l+1)\right) u(\Omega) = 0 \,,
\end{equation}
where $\l \in \N_0$. The spherical harmonics are given by
\begin{equation}
Y_{\l m}(\Omega) = \frac{1}{\sqrt{2\pi}} \sqrt{\frac{2\l+1}{2} \frac{(\l-m)!}{(\l+m)!}} P_\l^m(\cos\theta) \e{\i m \phi} \,,
\end{equation}
where $-\l \leq m \leq \l$ and $P_\l^m(\cos\theta)$ is the associated Legendre polynomial. They are orthonormal on the unit sphere with respect to the weighting function $1$, hence
\begin{equation}
\int \d\Omega \, Y_{\l m}^*(\Omega) Y_{\l' m'}(\Omega) = \delta_{\l\l'} \delta_{mm'}
\end{equation}
and form a complete set, hence
\begin{equation}
\sum\limits_{\l=0}^\infty \sum\limits_{m=-\l}^\l Y_{\l m}^*(\Omega') Y_{\l m}(\Omega) = \delta(\phi - \phi') \delta(\cos\theta - \cos\theta') \,.
\end{equation}
The complex conjugate is given by
\begin{equation}
Y_{\l -m}(\Omega) = (-1)^m Y_{\l m}^*(\Omega) \quad\quad \text{and} \quad\quad Y_{\l m}^*(\Omega) = (-1)^m Y_{\l -m}(\Omega) \,.
\end{equation}
Furthermore the spherical harmonics satisfy
\begin{align}
\begin{aligned}
& \sum\limits_{m=-\l}^\l Y_{\l m}^*(\Omega) Y_{\l m}(\Omega) = \frac{2\l+1}{4\pi} \,, \\
& \int \d\Omega \, Y_{\l m}(\Omega) Y_{\l' m'}(\Omega) Y_{\l'' m''}(\Omega) = \sqrt{\frac{(2\l+1)(2\l'+1)(2\l''+1)}{4\pi}} \begin{pmatrix} \l & \l' & \l'' \\ 0 & 0 & 0 \end{pmatrix} \begin{pmatrix} \l & \l' & \l'' \\ m & m' & m'' \end{pmatrix} \,,
\end{aligned}
\end{align}
where $\left( 2 \times 3 \right)$ is a Wigner $3j$-symbol (see Appendix~\ref{app:subsec:WignerSymbol} on how to compute them) that can be related to Clebsch-Gordan coefficients.\footnote{Remember that the space $H_\l$ of spherical harmonics of degree $\l$ is a representation of the symmetry group SO(3) and its double-cover SU(2). To be precise, $H_\l$ in fact is an irreducible representation of SU(3).}\\
They are implemented into the MATHEMATICA language as \MathematicaInline{SphericalHarmonicY[l,m,theta,phi]}.

\subsection{The Hermite polynomials}

The Hermite polynomials \cite{mathworld:HermitePolynomial} are the solution of the Hermite differential equation \cite{mathworld:HermiteDifferentialEquation}
\begin{equation}
y'' - 2 x y' + 2 n y = 0 \,,
\end{equation}
where $n \in \N_0$. They are given by the contour integral
\begin{equation}
H_n(z) = \frac{n!}{2\pi\i} \oint \d t \, \e{-t^2 + 2 t z} t^{-n-1} \,,
\end{equation}
where the contour encloses the origin and is traversed in a counterclockwise direction. Their Rodrigues representation is given by
\begin{equation}
H_n(x) = (-1)^n \e{x^2} \frac{\d^n}{\d x^n} \e{-x^2} \,,
\end{equation}
and they can be represented by the sum
\begin{equation}
H_n(x) = \left\lbrace \begin{array}{cl}
n! \sum\limits_{\l=0}^{\frac{n}{2}} \frac{(-1)^{\frac{n}{2}-\l}}{(2\l)! \left(\frac{n}{2}-\l\right)!} (2x)^{2\l} & \text{ for even } n \\
n! \sum\limits_{\l=0}^{\frac{n-1}{2}} \frac{(-1)^{\frac{n-1}{2}-\l}}{(2\l+1)! \left(\frac{n-1}{2}-\l\right)!} (2x)^{2\l+1} & \text{ for odd } n
\end{array} \right.
\end{equation}
The Hermite polynomials are orthogonal over $(-\infty,\infty)$ with respect to the weighting function $\e{-x^2}$, hence
\begin{equation}
\int\limits_{-\infty}^\infty \d x \, \e{-x^2} H_n(x) H_m(x) = 2^n n! \sqrt{\pi} \delta_{n m} \,.
\end{equation}
They are related to the derivative of the error-function via
\begin{equation}
H_n(z) = \frac{(-1)^n}{2} \sqrt{\pi} \e{z^2} \frac{\d^{n+1}}{\d z^{n+1}} \text{erf}(z) \,,
\end{equation}
allowing for negative $n$. They also satisfy
\begin{align}
& H_n(-x) = (-1)^n H_n(x) \,,
\end{align}
and are related to the associated Laguerre polynomials via
\begin{equation}
H_{2n}(x) = (-1)^n 2^{2n} n! L_n^{-\frac{1}{2}}(x^2) \quad\quad \text{and} \quad\quad H_{2n+1}(x) = (-1)^n 2^{2n+1} n! x L_n^{\frac{1}{2}}(x^2) \,.
\end{equation}
They are implemented into the MATHEMATICA language as \MathematicaInline{HermiteH[n,x]}.

\subsection{The Wigner symbols}
\label{app:subsec:WignerSymbol}

In order to define the Wigner symbols we first need the triangle coefficient \cite{mathworld:TriangleCoefficient}, given by
\begin{equation}
\Delta(a,b,c) = \frac{(a+b-c)! (a-b+c)! (-a+b+c)!}{(a+b+c+1)!} \,.
\end{equation}
The generic Wigner symbols are related to the Clebsch-Gordan coefficients.\\
The specific $3j$-symbols \cite{mathworld:Wigner3j-Symbol} are given, using the Racah formula, by
\begin{align}
\begin{aligned}
\begin{pmatrix} a & b & c \\ \alpha & \beta & \gamma \end{pmatrix} &= (-1)^{a-b-\gamma} \sqrt{\Delta(a,b,c)} \, || \sum\limits_t || \frac{(-1)^t}{x} \\
& \times \sqrt{(a+\alpha)! (a-\alpha)! (b+\beta)! (b-\beta)! (c+\gamma)! (c-\gamma)!} \,,
\end{aligned}
\end{align}
where
\begin{equation}
x = t! (c-b+t+\alpha)! (c-a+t-\beta)! (a+b-c-t)! (a-t-\alpha)! (b-t+\beta)! \,,
\end{equation}
and the reduced sum $|| \sum\limits_t ||$ only cover these integers $t$, for witch all the factorials have non-negative arguments.\\
They are implemented into the MATHEMATICA language as\\
\MathematicaInline{ThreeJSymbol[\{a,alpha\},\{b,beta\},\{c,gamma\}]}.\\
The specific $6j$-symbols \cite{mathworld:Wigner6j-Symbol} are given, using the Racah formula, by
\begin{align}
\begin{aligned}
\left\lbrace\begin{matrix} j_1 & j_2 & j_3 \\ J_1 & J_2 & J_3 \end{matrix}\right\rbrace &= || \sum\limits_t || \frac{(-1)^t (t+1)!}{y} \\
& \times \sqrt{\Delta(j_1,j_2,j_3) \Delta(j_1,J_2,J_3) \Delta(J_1,j_2,J_3) \Delta(J_1,J_2,j_3)} \,,
\end{aligned}
\end{align}
where
\begin{align}
\begin{aligned}
y &= (t-j_1-j_2-j_3)! (t-j_1-J_2-J_3)! (t-J_1-j_2-J_3)! (t-J_1-J_2-j_3)! \\
&\times (j_1+j_2+J_1+J_2-t)! (j_2+j_3+J_2+J_3-t)! (j_3+j_1+J_3+J_1-t)! \,,
\end{aligned}
\end{align}
and the reduced sum $|| \sum\limits_t ||$ only cover these integers $t$, for witch all the factorials have non-negative arguments.\\
They are implemented into the MATHEMATICA language as \MathematicaInline{SixJSymbol[\{j1,j2,j3\},\{J1,J2,J3\}]}.\\
The specific $9j$-symbols \cite{mathworld:Wigner9j-Symbol} are given in terms of $6j$-symbols as
\begin{align}
\begin{aligned}
\left\lbrace\begin{matrix} j_1 & j_2 & j_3 \\ j_4 & j_5 & j_6 \\ j_7 & j_8 & j_9 \end{matrix}\right\rbrace &= \sum\limits_g (-1)^{2g} (2g+1) \left\lbrace\begin{matrix} j_1 & j_4 & j_7 \\ j_8 & j_9 & g \end{matrix}\right\rbrace \left\lbrace\begin{matrix} j_2 & j_5 & j_8 \\ j_4 & g & j_6 \end{matrix}\right\rbrace \left\lbrace\begin{matrix} j_3 & j_6 & j_9 \\ g & j_1 & j_2 \end{matrix}\right\rbrace \,,
\end{aligned}
\end{align}
where $g$ runs over all integers allowed by the $6j$-symbols.\\
They can be implemented into the MATHEMATICA language as
\begin{Mathematica}
	nineJSymbol[{j1_,j2_,j3_},{j4_,j5_,j6_},{j7_,j8_,j9_}]:=
	 Module[{gmin,gmax},
	  gmin=Max[{Abs[j1-j9],Abs[j4-j8],Abs[j2-j6]}];
	  gmax=Min[{Abs[j1+j9],Abs[j4+j8],Abs[j2+j6]}];
	  Sum[(-1)^(2g)*(2g+1)*SixJSymbol[{j1,j4,j7},{j8,j9,g}]*
	   SixJSymbol[{j2,j5,j8},{j4,g,j6}]*SixJSymbol[{j3,j6,j9},
	    {g,j1,j2}],{g,gmin,gmax}]
	];
\end{Mathematica}

\section{Expectation values for single potential insertions}
\label{app:sec:ExpectationValuesSinglePotentialInsertions}

We list a set of single potential insertion expectation values, using the master integrals from Refs.~\cite{Pineda:1997hz,Peset:2015zga,Peset:2015vvi}.

\begin{align}
\label{eq:ExpectationValueP**4} & \langle n \l | \nabla_r^4 | n \l \rangle = \frac{1}{(n a)^4} \left( \frac{8n}{2\l+1} - 3 \right) \,, \\
\label{eq:ExpectationValueAntiCom1Over2RandP**2} & \langle n \l | \left\lbrace \frac{1}{2r},-\nabla_r^2 \right\rbrace | n \l \rangle = -a^{-3} \left( \frac{1}{n^4} - \frac{4}{(2\l + 1) n^3} \right) \,, \\
\label{eq:ExpectationValue1OverR} & \langle n \l| \frac{1}{r} | n \l \rangle = \frac{1}{n^2 a} \,, \\
\label{eq:ExpectationValue1OverR**2} & \langle n \l| \frac{1}{r^2} | n \l \rangle = \frac{2}{(2\l+1) n^3 a^2} \,, \\
\label{eq:ExpectationValue1OverR**3} & \langle n \l| \frac{1}{r^3} | n \l \rangle = \frac{2 (1 - \delta_{\l0})}{\l(\l+1)(2\l+1) (n a)^3} \,, \\
\label{eq:ExpectationValueR-3} & \langle n \l | \text{reg } \frac{1}{r^3} | n \l \rangle = \frac{2}{(n a)^3} \left[\left( \ln{\frac{n a}{2}} - S_1(n) - \frac{n-1}{2n} \right) 2\delta_{\l0} + \frac{1 - \delta_{\l0}}{\l(\l+1)(2\l+1)} \right] \,, \\
\label{eq:ExpectationValueDelta3} & \langle n \l | \delta^{(3)}(\vec{r}\,) | n \l \rangle = \frac{1}{\pi (n a)^3} \delta_{\l0} \,, \\
\label{eq:ExpectationValueLnMuROverR} & \langle n \l| \frac{\ln{\mr r}}{r} | n \l \rangle = \frac{1}{n^2 a} \left(\ln{\frac{n a \mr}{2}} - \gamma_E + S_1(n+\l)\right) \,, \\
\label{eq:ExpectationValueLn**2MuROverR} & \begin{aligned}
& \langle n \l| \frac{\ln{\mr r}^2}{r} | n \l \rangle = \frac{1}{n^2 a} \Bigg( \ln{\frac{n a \mr}{2}}^2 + 2 \psi(n+\l+1) \ln{\frac{n a \mr}{2}} + \psi(n+\l+1)^2 \\
& \quad + \psi'(n+\l+1) + \theta(n-\l-2) \frac{2\Gamma(n-\l)}{\Gamma(n+\l+1)} \sum\limits_{j=0}^{n-\l-2} \frac{\Gamma(2\l+2+j)}{j!(n-\l-1-j)^2} \Bigg) \,.
\end{aligned}
\end{align}

\subsection{Derivation of the expectation values containing derivatives}

Since $\vec{p} = -\i\vec{\nabla}_r$ is a hermitian operator, we can let it act on the initial or final state and can furthermore make use of the equation of motion~\eqref{eq:ZerothOrderSchroedingerEquation}. Thus we have
\begin{align}
^{(0)}\!\langle n\l | \vec{\nabla}_r^{4} | n\l \rangle^{(0)} &= 4\mr^2 \, ^{(0)}\!\langle n\l | \frac{\vec{\nabla}_r^{2}}{2\mr} \frac{\vec{\nabla}_r^{2}}{2\mr} | n\ell \rangle^{(0)} \\
\nonumber & = 4\mr^2 \, ^{(0)}\!\langle n\l | \left[E_{n\l} + \frac{C_F \alpha_s}{r}\right] \left[E_{n\l} + \frac{C_F \alpha_s}{r}\right] | n\l \rangle^{(0)} \\
\nonumber & = 4\mr^2 \left\lbrace E_{n\l}^2 + 2E_{n\l} C_{F} \alpha_s \, ^{(0)}\!\langle n\l | \frac{1}{r} | n\l \rangle^{(0)} + C_F^2 \alpha_s^2 \, ^{(0)}\!\langle n\l| \frac{1}{r^2} | n\l \rangle^{(0)} \right\rbrace \\
\nonumber & = 4\mr^2 \left\lbrace \frac{\mr^2 C_F^4 \alpha_s^4}{4n^4} - 2 \frac{\mr C_F^2 \alpha_s^2}{2n^2} C_F \alpha_s \frac{1}{a n^2} + C_F^2 \alpha_s^2 \frac{2}{a^2 n^3 (2\l+1)} \right\rbrace \\
\nonumber & = \frac{1}{a^4 n^4} \left(1 - 4 + \frac{8n}{(2\l+1)}\right) \\
\nonumber & = \frac{1}{a^4 n^4} \left(\frac{8n}{(2\l+1)} - 3\right) \,,
\end{align}
and
\begin{align}
^{(0)}\!\langle n\l | \left\lbrace \frac{1}{2r} , -\vec{\nabla}_r^{2} \right\rbrace | n\l \rangle^{(0)} &= \mr \, ^{(0)}\!\langle n\l | \left[\frac{1}{r} \frac{-\vec{\nabla}_r^2}{2\mr} + \frac{-\vec{\nabla}_r^2}{2\mr} \frac{1}{r} \right] | n\l \rangle^{(0)} \\
\nonumber & = \mr \, ^{(0)}\!\langle n\l | \left[ \frac{1}{r} \left(E_{n\l} + \frac{C_F \alpha_s}{r} \right) + \left( E_{n\l} + \frac{C_F \alpha_s}{r} \right) \frac{1}{r} \right] | n\l \rangle^{(0)} \\
\nonumber & = 2\mr \left[ E_{n\l} \, ^{(0)}\!\langle n\l | \frac{1}{r} | n\l \rangle^{(0)} + C_F \alpha_s \, ^{(0)}\!\langle n\l | \frac{1}{r^2} | n\l \rangle^{(0)} \right] \\
\nonumber & = 2\mr \left[ -\frac{\mr C_F^2 \alpha_s^2}{2n^2} \frac{1}{a n^2} + C_F \alpha_s \frac{2}{a^2 n^3 (2\l+1)} \right] \\
\nonumber & = -\frac{1}{a^3} \left[ \frac{1}{n^4} - \frac{4}{n^3 (2\l+1)} \right] \,.
\end{align}

\subsection[\texorpdfstring{Derivation of the expectation values containing the $\delta$-distribution}{Derivation of the expectation values containing the delta-distribution}]{\texorpdfstring{Derivation of the expectation values containing the {\boldmath $\delta$}-distribution}{Derivation of the expectation values containing the delta-distribution}}

The $\delta$-distribution in three dimensions has the representation
\begin{equation}
\delta^{(3)}(\vec{r}\,) = \frac{1}{r \sin(\theta)} \delta(r) \delta(\theta) \delta(\phi) \,,
\end{equation}
and thus
\begin{align}
^{(0)}\!\langle n\l | \delta^{(3)}(\vec{r}\,) | n\l \rangle^{(0)} &= \int\limits_{0}^{\infty} \d r \, R_{n\l}^*(r) R_{n\l}(r) \delta(r) \int\limits_{0}^{\pi} \int\limits_{0}^{2\pi} \d\theta \, \d\phi \, Y_{\l m}^*(\theta,\phi) Y_{\l m}(\theta,\phi) \delta(\theta) \delta(\phi) \\
\nonumber & = \frac{\delta_{\l 0}}{4\pi} |R_{n\l}(0)|^2 \\
\nonumber & = \frac{\delta_{\l 0}}{4\pi} \left( \frac{2}{n a} \right)^3 \frac{(n-\l-1)!}{2n[(n+\l)!]} \left[ L_{n-\l-1}^{2\l+1}(0) \right]^{2} \\
\nonumber & = \frac{\delta_{\l 0}}{4\pi} \left( \frac{2}{n a} \right)^3 \frac{(n-\l-1)!}{2n[(n+\l)!]} \left[ \frac{(n+\l)!}{(n-\l-1)!(2\l+1)!} \right]^2 \\
\nonumber & = \frac{\delta_{\l 0}}{4\pi} \left( \frac{2}{n a} \right)^3 \frac{(n+\l)!}{2n(n-\l-1)![(2\l+1)!]^2} \\
\nonumber & = \frac{\delta_{\l 0}}{4\pi} \left( \frac{2}{n a} \right)^3 \frac{n!}{2n(n-1)!} \\
\nonumber & = \frac{1}{\pi n^3 a^3} \delta_{\l 0} \,,
\end{align}
where we explicitly set $\l=0$ in the step before the last.

\subsection[\texorpdfstring{Derivation of the expectation values for $\frac{1}{r}$, $\frac{\ln{r}}{r}$ and $\frac{\lnSquare{r}}{r}$}{Derivation of the expectation values for 1/r, ln r/r and ln**2 r/r}]{\texorpdfstring{Derivation of the expectation values for {\boldmath$\frac{1}{r}$}, {\boldmath$\frac{\ln{r}}{r}$} and {\boldmath$\frac{\lnSquare{r}}{r}$}}{Derivation of the expectation values for 1/r, ln r/r and ln**2 r/r}}

For small $\epsilon$ we have the expansion
\begin{equation}
r^\epsilon = 1 + \epsilon \ln{r} + \frac{1}{2} \epsilon^2 \lnSquare{r} + \O(\epsilon^3) \,,
\end{equation}
and we expand the corresponding expectation value in the same way
\begin{equation}
\label{eq:ExpectationValueEpsilonExpansion} \langle n\l | \frac{r^\epsilon}{r} | n\l \rangle = \langle n\l | 
\frac{1}{r} | n\l \rangle + \epsilon \langle n\l | \frac{\ln{r}}{r} | n\l \rangle + \frac{1}{2} \epsilon^2 \langle n\l | \frac{\lnSquare{r}}{r} | n\l \rangle + \O(\epsilon^3) \,.
\end{equation}
The left hand side of Eq.~\eqref{eq:ExpectationValueEpsilonExpansion} can be calculated explicitly
\begin{align}
\langle n\l | \frac{r^\epsilon}{r} | n\l \rangle &= \int \d \Omega_r \, \int\limits_0^\infty \d r \, r^2 R_{n\l}^*(r) Y_{\l m}^*(\Omega_r) r^{\epsilon-1} R_{n\l}(r) Y_{\l m}(\Omega_r) \\
\nonumber & = \int\limits_0^\infty \d r \, r^{\epsilon+1} R_{n\l}^2(r) \\
\nonumber & = N_{n\l}^2 \left(\frac{n \alpha_s}{2}\right)^{\epsilon+2} \int\limits_0^\infty \d\rho \, \rho^{2\l+1+\epsilon} \, \e{-\rho} \left[L_{n-\l-1}^{2\l+1}(\rho)\right]^2 \\
\nonumber & = N_{n\l}^2 \left(\frac{n \alpha_s}{2}\right)^{\epsilon+2} \int\limits_0^\infty \d\rho \, \rho^{a+\epsilon} \, \e{-\rho} \left[L_{m}^{a}(\rho)\right]^2 \\
\nonumber & = \frac{(n-\l-1)!}{2n(n+\l)!} \left(\frac{n \alpha_s}{2}\right)^{\epsilon-1} \sum\limits_{k=0}^m \frac{\Gamma(1+a+\epsilon+k)}{k!} \left\lbrace \frac{\Gamma(m-\epsilon-k)}{\Gamma(-\epsilon) (m-k)!} \right\rbrace^2 \\
\nonumber & = \frac{(n-\l-1)!}{2n(n+\l)!} \left(\frac{n \alpha_s}{2}\right)^{\epsilon-1} \sum_{k=0}^m \frac{\Gamma(1+a+\epsilon+k)}{k!} \\
\nonumber & \quad \times \left(-\frac{\epsilon}{m-k} + \frac{\epsilon^2(\psi(m-k)+\gamma_E)}{m-k} + \O(\epsilon^3) \right) \left(\delta_{m k} - \frac{\epsilon \delta_{m\geq k+1}}{m-k} + \O(\epsilon^2)\right) \\
\nonumber & = \frac{(n-\l-1)!}{2n(n+\l)!} \left(\frac{n \alpha_s}{2}\right)^{\epsilon-1} \sum_{k=0}^m \frac{\Gamma(1+a+\epsilon+k)}{k!} \\
\nonumber & \quad \times \left\lbrace \left(-\frac{\epsilon}{m-k} \delta_{m k} + \frac{\epsilon^2(\psi(m-k)+\gamma_E)}{(m-k)^2} \delta_{m k} + \frac{\epsilon^2 \delta_{m\geq k+1}}{m-k} + \O(\epsilon^3)\right) \right\rbrace \\
\nonumber & = \frac{1}{\alpha_s n^2} + \epsilon \frac{1}{\alpha_s n^2} \left[ \ln{\frac{n \alpha_s}{2}} + \psi(n+\l+1) \right] \\
\nonumber & \quad + \frac{1}{2} \epsilon^2 \frac{1}{\alpha_s n^2} \Bigg\lbrace \left[ \ln{\frac{n \alpha_s}{2}} + \psi(n+\l+1) \right]^2 + \psi'(n+\l+1) \\
\nonumber & \quad + \theta(n-\l-2) \frac{2\Gamma(n-\l)}{\Gamma(n+\l+1)} \sum_{k=0}^{n-\l-2} \frac{\Gamma(2\l+2+k)}{\Gamma(k+1)(n-\l-k-1)^2} \Bigg\rbrace + \O(\epsilon^3) \,,
\end{align}
and by comparing order-by-order in $\epsilon$, we obtain the desired results for the different matrix elements.

\section{Expectation values for double potential insertions}
\label{app:sec:ExpectationValuesDoublePotentialInsertions}

We list a set of double potential insertion expectation values, using the master integrals from Refs.~\cite{Pineda:1997hz,Peset:2015zga,Peset:2015vvi}.

\begin{align}
\label{eq:ExpValDoublePot1OverR1OverR} & \langle n \l | \frac{1}{r} \frac{1}{(E_n - H)'} \frac{1}{r} | n \l \rangle = -\frac{\mr}{2n^2} \,, \\
& \langle n \l | \frac{1}{r} \frac{1}{(E_n - H)'} \frac{1}{r^2} | n \l \rangle = -\frac{2\mr}{(2\l+1) n^3 a} \,, \\
& \langle n \l | \frac{1}{r} \frac{1}{(E_n - H)'} \frac{1}{r^3} | n \l \rangle = -\frac{3 \mr (1-\delta_{\l0})}{\l(\l+1)(2\l+1)n^3 a^2} \,, \\
& \langle n \l | \frac{1}{r} \frac{1}{(E_n - H)'} \delta^{(3)}(\vec{r}\,) | n \l \rangle = -\frac{3 \mr}{2\pi n^3 a^2}\delta_{\l0} \,, \\
& \langle n \l | \frac{\ln{r \e{\gamma_E}}}{r} \frac{1}{(E_n - H)'} \frac{1}{r} | n \l \rangle = -\frac{\mr}{2n^2}\left( \ln{\frac{n a}{2}} + S_1(n+\l) - 1 \right) \,, \\
& \langle n \l | \frac{\ln{r \e{\gamma_E}}}{r} \frac{1}{(E_n - H)'} \frac{1}{r^2} | n \l \rangle \\
\nonumber & \quad = \frac{2 \mr}{(2\l+1)n^3 a} \left[\frac{1}{2} + n \left(\frac{\pi^2}{6} - \Sigma_2^{(k)}(n,\l) - \Sigma_2^{(m)}(n,\l)\right) - \ln{\frac{n a}{2}} - S_1(n+\l) \right] \,, \\
& \langle n \l | \frac{\ln{r \e{\gamma_E}}}{r} \frac{1}{(E_n - H)'} \frac{1}{r^3} | n \l \rangle \\
\nonumber & \quad = \frac{2 \mr (1-\delta_{\l0})}{\l(\l+1)(2\l+1)n^3 a^2} \left[\frac{1}{2} - \frac{3}{2} \ln{\frac{n a}{2}} - \frac{3}{2} S_1(n+\l) + \Sigma_1^{(m)}(n,\l) + \l \left(\Sigma_1^{(m)}(n,\l) + \Sigma_1^{(k)}(n,\l)\right) \right. \\
\nonumber & \quad\quad \left. + \frac{n\pi^2}{6} - n \left(\Sigma_2^{(m)}(n,\l) + \Sigma_2^{(k)}(n,\l)\right) \right] \,, \\
& \langle n \l | \frac{\ln{r \e{\gamma_E}}}{r} \frac{1}{(E_n - H)'} \delta^{(3)}(\vec{r}\,) | n \l \rangle = \frac{\mr \delta_{\l0}}{\pi n^3 a^2} \left[\frac{1}{2} + \frac{n\pi^2}{6} - n \Sigma_2^{(k)}(n,0) - \frac{3}{2} \ln{\frac{n a}{2}} - \frac{3}{2} S_1(n) \right] \,.
\end{align}

\subsection{Derivation of some expectation values for double potential insertions}
\label{app:subsec:DerivationExpectationValuesDoublePotentialInsertions}

We now want to demonstrate the decomposition of a generic expectation value with a double potential insertion into the two master integrals $I_1$ and $I_2$. We have
\begin{align}
& \langle n'\l' | V \frac{1}{(E - H)'} V' | n\l \rangle = -\int \d^3 r_1 \, \d^3 r_2 \, \psi_{n'\l'}^*(\vec{r}_2) V(\vec{r}_2) G(\vec{r}_2,\vec{r}_1) V'(\vec{r}_1) \psi_{n\l}(\vec{r}_1) \\
& \quad = -\int \d^3 r_1 \, \d^3 r_2 \, \psi_{n'\l'}^*(\vec{r}_2) V(\vec{r}_{2}) \sum_{\l''=0}^\infty \frac{\mr (2\l''+1)}{a \lambda \pi} P_{\l''}(\hat{r}_1 \cdot \hat{r}_2) \r{\lambda}{1}^{\l''} \r{\lambda}{2}^{\l''} \e{-\frac{1}{2}(\r{\lambda}{1}+\r{\lambda}{2})} \\
\nonumber & \quad\quad \times \sum\limits_{s=0}^\infty \frac{L_{s}^{2\l''+1}(\r{\lambda}{1}) L_{s}^{2\l''+1}(\r{\lambda}{2}) s!}{(s+\l''+1-\lambda)(s+2\l''+1)!} V'(\vec{r}_1) \psi_{n\l}(\vec{r}_1) \\
& \quad = -N_{n'\l'} N_{n\l} \frac{\mr}{a \lambda \pi} \sum_{\l''=0}^\infty (2\l''+1) \sum\limits_{s=0}^\infty \frac{s!}{(s+\l''+1-\lambda)(s+2\l''+1)!} \\
\nonumber & \quad\quad \times \int\limits_0^\infty \d r_1 \, r_1^2 V'(r_1) \r{\lambda}{1}^{\l''} \e{-\frac{1}{2}\r{\lambda}{1}} L_{s}^{2\l''+1}(\r{\lambda}{1}) \rho_{n,1}^\l \e{-\frac{1}{2}\rho_{n,1}} L_{n-\l-1}^{2\l+1}(\rho_{n,1}) \\
\nonumber & \quad\quad \times \int\limits_0^\infty \d r_2 \, r_2^2 V(r_2) \r{\lambda}{2}^{\l''} \e{-\frac{1}{2}\r{\lambda}{2}} L_{s}^{2\l''+1}(\r{\lambda}{2}) \rho_{n',2}^{\l'} \e{-\frac{1}{2}\rho_{n',2}} L_{n'-\l'-1}^{2\l'+1}(\rho_{n',2}) \\
\nonumber & \quad\quad \times \int \d\Omega_1 \, \d\Omega_2 \, V'(\Omega_1) Y_{\l}^{m}(\Omega_1) V(\Omega_2) Y_{\l'}^{m'*}(\Omega_2) P_{\l''}(\hat{r}_1 \cdot \hat{r}_2) \\
& \quad = -N_{n'\l'} N_{n\l} \frac{4\mr}{a \lambda} \sum_{\l''=0}^\infty \sum\limits_{s=0}^\infty \frac{s!}{(s+\l''+1-\lambda)(s+2\l''+1)!} \\
\nonumber & \quad\quad \times \int\limits_0^\infty \d r_1 \, r_1^2 V'(r_1) \r{\lambda}{1}^{\l''} \rho_{n,1}^\l \e{-\frac{1}{2}\r{\lambda}{1}} \e{-\frac{1}{2}\rho_{n,1}} L_{s}^{2\l''+1}(\r{\lambda}{1}) L_{n-\l-1}^{2\l+1}(\rho_{n,1}) \\
\nonumber & \quad\quad \times \int\limits_0^\infty \d r_2 \, r_2^2 V(r_2) \r{\lambda}{2}^{\l''} \rho_{n',2}^{\l'} \e{-\frac{1}{2}\r{\lambda}{2}} \e{-\frac{1}{2}\rho_{n',2}} L_{s}^{2\l''+1}(\r{\lambda}{2}) L_{n'-\l'-1}^{2\l'+1}(\rho_{n',2}) \\
\nonumber & \quad\quad \times \sum\limits_{m''=-\l''}^{\l''} \int \d\Omega_1 \, Y_{\l''}^{m''*}(\Omega_1) V'(\Omega_1) Y_{\l}^{m}(\Omega_1) \int \d\Omega_2 \, Y_{\l'}^{m'*}(\Omega_2) V(\Omega_2) Y_{\l''}^{m''}(\Omega_2) \,.
\end{align}
From here we can clearly see that for trivial angular parts of the potentials, $V'(\Omega_1) = V(\Omega_2) = \1$, the orthonormality relation of the spherical harmonics enforces $l = l' = l''$ and $m = m' = m''$, as it is the case upon calculating corrections to the energy. We then have
\begin{align}
& \langle n\l | V \frac{1}{(E_n - H)'} V' | n\l \rangle = -N_{n\l}^2 \frac{4\mr}{a \lambda} \sum\limits_{s=0}^\infty \frac{s!}{(s+\l+1-\lambda)(s+2\l+1)!} \\
\nonumber & \quad\quad \times \int\limits_0^\infty \d r_1 \, r_1^2 V'(r_1) \r{\lambda}{1}^{\l} \rho_{n,1}^\l \e{-\frac{1}{2}\r{\lambda}{1}} \e{-\frac{1}{2}\rho_{n,1}} L_{s}^{2\l+1}(\r{\lambda}{1}) L_{n-\l-1}^{2\l+1}(\rho_{n,1}) \\
\nonumber & \quad\quad \times \int\limits_0^\infty \d r_2 \, r_2^2 V(r_2) \r{\lambda}{2}^{\l} \rho_{n,2}^{\l} \e{-\frac{1}{2}\r{\lambda}{2}} \e{-\frac{1}{2}\rho_{n,2}} L_{s}^{2\l+1}(\r{\lambda}{2}) L_{n-\l-1}^{2\l+1}(\rho_{n,2}) \,.
\end{align}
We now consider the case $\lambda = n$ for which the sum over $s$ becomes singular once $s = n - \l - 1$ and we thus have, writing $r_i = \frac{n a}{2} \rho_i$ and only considering the finite part,
\begin{align}
& \left.\langle n\l | V \frac{1}{(E_n - H)'} V' | n\l \rangle\right|_{\begin{array}{l} \text{finite} \\ \lambda = n \end{array}} = -N_{n\l}^2 \frac{4\mr}{n a} \sum\limits_{s=0,s \neq n-\l-1}^\infty \frac{s!}{(s+\l+1-n)(s+2\l+1)!} \\
\nonumber & \quad\quad \times \left(\frac{n a}{2}\right)^3 \int\limits_0^\infty \d \rho_1 \, \rho_1^2 V'(\rho_1) \rho_1^{\l} \rho_1^\l \e{-\frac{1}{2}\rho_1} \e{-\frac{1}{2}\rho_1} L_{s}^{2\l+1}(\rho_1) L_{n-\l-1}^{2\l+1}(\rho_1) \\
\nonumber & \quad\quad \times \left(\frac{n a}{2}\right)^3 \int\limits_0^\infty \d \rho_2 \, \rho_2^2 V(\rho_2) \rho_2^{\l} \rho_2^{\l} \e{-\frac{1}{2}\rho_2} \e{-\frac{1}{2}\rho_2} L_{s}^{2\l+1}(\rho_2) L_{n-\l-1}^{2\l+1}(\rho_2) \\
& \quad = -\frac{(n-\l-1)!}{(n+\l)!} \frac{\mr}{n} \sum\limits_{s=0,s \neq n-\l-1}^\infty \frac{s!}{(s+\l+1-n)(s+2\l+1)!} \\
\nonumber & \quad\quad \times \frac{n a}{2} \int\limits_0^\infty \d \rho_1 \, \rho_1^{2+2\l} V'(\rho_1) \e{-\rho_1} L_{s}^{2\l+1}(\rho_1) L_{n-\l-1}^{2\l+1}(\rho_1) \\
\nonumber & \quad\quad \times \frac{n a}{2} \int\limits_0^\infty \d \rho_2 \, \rho_2^{2+2\l} V(\rho_2) \e{-\rho_2} L_{s}^{2\l+1}(\rho_2) L_{n-\l-1}^{2\l+1}(\rho_2)
\end{align}
and for the divergent part, where we have $\r{\lambda}{i} = \rho_{n,i} \sqrt{1-\epsilon}$,\footnote{Once all ambiguities are eliminated, we will drop the explicit subscript $n$ for the sake of readability.}
\begin{align}
& \left.\langle n\l | V \frac{1}{(E_n - H)'} V' | n\l \rangle\right|_{\begin{array}{l} s = n-\l-1 \\ \lambda = \frac{n}{\sqrt{1-\epsilon}} \end{array}} \\
& \quad = -N_{n\l}^2 \frac{4\mr}{a \lambda} \frac{(n-\l-1)!}{(n-\l-1+\l+1-\lambda)(n-\l-1+2\l+1)!} \\
\nonumber & \quad\quad \times \left(\frac{n a}{2}\right)^3 \int\limits_0^\infty \d \rho_{n,1} \, \rho_{n,1}^2 V'(\rho_{n,1}) \r{\lambda}{1}^{\l} \rho_{n,1}^\l \e{-\frac{1}{2}\r{\lambda}{1}} \e{-\frac{1}{2}\rho_{n,1}} L_{n-\l-1}^{2\l+1}(\r{\lambda}{1}) L_{n-\l-1}^{2\l+1}(\rho_{n,1}) \\
\nonumber & \quad\quad \times \left(\frac{n a}{2}\right)^3 \int\limits_0^\infty \d \rho_{n,2} \, \rho_{n,2}^2 V(\rho_{n,2}) \r{\lambda}{2}^{\l} \rho_{n,2}^{\l} \e{-\frac{1}{2}\r{\lambda}{2}} \e{-\frac{1}{2}\rho_{n,2}} L_{n-\l-1}^{2\l+1}(\r{\lambda}{2}) L_{n-\l-1}^{2\l+1}(\rho_{n,2}) \\
& \quad = -\frac{(n-\l-1)!}{(n+\l)!} \frac{\mr}{\frac{n}{\sqrt{1-\epsilon}}} \frac{(n-\l-1)!}{(n-\frac{n}{\sqrt{1-\epsilon}})(n+\l)!} \\
\nonumber & \quad\quad \times \frac{n a}{2} \int\limits_0^\infty \d \rho_{n,1} \, \rho_{n,1}^2 V'(\rho_{n,1}) \left(\rho_{n,1}\sqrt{1-\epsilon}\right)^{\l} \rho_{n,1}^\l \e{-\frac{1}{2}\rho_{n,1}\sqrt{1-\epsilon}} \e{-\frac{1}{2}\rho_{n,1}} L_{n-\l-1}^{2\l+1}(\rho_{n,1}\sqrt{1-\epsilon}) L_{n-\l-1}^{2\l+1}(\rho_{n,1}) \\
\nonumber & \quad\quad \times \frac{n a}{2} \int\limits_0^\infty \d \rho_{n,2} \, \rho_{n,2}^2 V(\rho_{n,2}) \left(\rho_{n,2}\sqrt{1-\epsilon}\right)^{\l} \rho_{n,2}^{\l} \e{-\frac{1}{2}\rho_{n,2}\sqrt{1-\epsilon}} \e{-\frac{1}{2}\rho_{n,2}} L_{n-\l-1}^{2\l+1}(\rho_{n,2}\sqrt{1-\epsilon}) L_{n-\l-1}^{2\l+1}(\rho_{n,2}) \\
& \quad = -\left(\frac{(n-\l-1)!}{(n+\l)!}\right)^2 \frac{\mr}{n^2} \frac{1-\epsilon}{\sqrt{1-\epsilon}-1} \\
\nonumber & \quad\quad \times \frac{n a}{2} \int\limits_0^\infty \d \rho_1 \, \rho_1^{2+2\l} V'(\rho_1) \left(\sqrt{1-\epsilon}\right)^{\l} \e{-\frac{1}{2}\rho_1\sqrt{1-\epsilon}} \e{-\frac{1}{2}\rho_1} L_{n-\l-1}^{2\l+1}(\rho_1\sqrt{1-\epsilon}) L_{n-\l-1}^{2\l+1}(\rho_1) \\
\nonumber & \quad\quad \times \frac{n a}{2} \int\limits_0^\infty \d \rho_2 \, \rho_2^{2+2\l} V(\rho_2) \left(\sqrt{1-\epsilon}\right)^{\l} \e{-\frac{1}{2}\rho_2\sqrt{1-\epsilon}} \e{-\frac{1}{2}\rho_2} L_{n-\l-1}^{2\l+1}(\rho_2\sqrt{1-\epsilon}) L_{n-\l-1}^{2\l+1}(\rho_2)
\end{align}
We can therefore conclude
\begin{equation}
\langle n\l | V \frac{1}{(E_n - H)'} V' | n\l \rangle = I_1(V,V') + I_2(V,V') \,,
\end{equation}
with
\begin{align}
& I_1(V,V') = -\frac{(n-\l-1)!}{(n+\l)!} \frac{\mr}{n} \sum\limits_{s=0,s \neq n-\l-1}^\infty \frac{s!}{(s+\l+1-n)(s+2\l+1)!} \\
\nonumber & \quad \times \frac{n a}{2} \int\limits_0^\infty \d \rho_1 \, \rho_1^{2+2\l} V'(\rho_1) \e{-\rho_1} L_{s}^{2\l+1}(\rho_1) L_{n-\l-1}^{2\l+1}(\rho_1) \\
\nonumber & \quad \times \frac{n a}{2} \int\limits_0^\infty \d \rho_2 \, \rho_2^{2+2\l} V(\rho_2) \e{-\rho_2} L_{s}^{2\l+1}(\rho_2) L_{n-\l-1}^{2\l+1}(\rho_2) \,, \\
& I_2(V,V') = -\left(\frac{(n-\l-1)!}{(n+\l)!}\right)^2 \frac{\mr}{n^2} \frac{1-\epsilon}{\sqrt{1-\epsilon}-1} \\
\nonumber & \quad \times \frac{n a}{2} \int\limits_0^\infty \d \rho_1 \, \rho_1^{2+2\l} V'(\rho_1) \left(\sqrt{1-\epsilon}\right)^{\l} \e{-\frac{1}{2}\rho_1\sqrt{1-\epsilon}} \e{-\frac{1}{2}\rho_1} L_{n-\l-1}^{2\l+1}(\rho_1\sqrt{1-\epsilon}) L_{n-\l-1}^{2\l+1}(\rho_1) \\
\nonumber & \quad \left. \times \frac{n a}{2} \int\limits_0^\infty \d \rho_2 \, \rho_2^{2+2\l} V(\rho_2) \left(\sqrt{1-\epsilon}\right)^{\l} \e{-\frac{1}{2}\rho_2\sqrt{1-\epsilon}} \e{-\frac{1}{2}\rho_2} L_{n-\l-1}^{2\l+1}(\rho_2\sqrt{1-\epsilon}) L_{n-\l-1}^{2\l+1}(\rho_2) \right|_{\epsilon \to 0 \,, \, \O(\epsilon^0)} \,,
\end{align}
where it is very important to realize that one should take the limit $\epsilon\to 0$ and take the $\O(\epsilon^0)$ term as we will show below in Appendix~\ref{app:sec:DemonstrationDivergenceCancellation}. As an example we reproduce the expectation value for a double potential insertion of the form $\frac{1}{r}$, Eq.~\eqref{eq:ExpValDoublePot1OverR1OverR}:
\begin{align}
& I_1\left(\frac{1}{r},\frac{1}{r}\right) = -\frac{\mr}{n} \frac{(n-\l-1)!}{(n+\l)!} \sum\limits_{s=0,s \neq n-\l-1}^\infty \frac{s!}{(s+\l+1-n)(s+2\l+1)!} \\
\nonumber & \quad\quad \times \int\limits_0^\infty \d \rho_1 \, \rho_1^{2\l+1} \e{-\rho_1} L_s^{2\l+1}(\rho_1) L_{n-\l-1}^{2\l+1}(\rho_1) \\
\nonumber & \quad\quad \times \int\limits_0^\infty \d \rho_2 \, \rho_2^{2\l+1} \e{-\rho_2} L_s^{2\l+1}(\rho_2) L_{n-\l-1}^{2\l+1}(\rho_2) \\
\nonumber & \quad = -\frac{\mr}{n} \frac{(n-\l-1)!}{(n+\l)!} \sum\limits_{s=0,s \neq n-\l-1}^\infty \frac{s!}{(s+\l+1-n)(s+2\l+1)!} \\
\nonumber & \quad\quad \times \frac{\Gamma(1+2\l+n-\l-1)}{(n-\l-1)!} \delta_{s,n-\l-1} \frac{\Gamma(1+2\l+n-\l-1)}{(n-\l-1)!} \delta_{s,n-\l-1} \\
\nonumber & \quad = 0 \,, \\
& I_2\left(\frac{1}{r},\frac{1}{r}\right) = - \frac{\mr}{n^2} \frac{1-\epsilon}{\sqrt{1-\epsilon}-1} \left[\frac{(n-\l-1)!}{(n+\l)!}\right]^2 \\
\nonumber & \quad\quad \times (1-\epsilon)^{\frac{\l}{2}} \int\limits_0^\infty \d \rho_1 \, \rho_1^{2\l+1} \e{-\frac{\rho_1}{2}(1+\sqrt{1-\epsilon})} L_{n-\l-1}^{2\l+1}(\rho_1\sqrt{1-\epsilon}) L_{n-\l-1}^{2\l+1}(\rho_1) \\
\nonumber & \quad\quad \times (1-\epsilon)^{\frac{\l}{2}} \int\limits_0^\infty \d \rho_2 \, \rho_2^{2\l+1} \e{-\frac{\rho_2}{2}(1+\sqrt{1-\epsilon})} L_{n-\l-1}^{2\l+1}(\rho_2\sqrt{1-\epsilon}) L_{n-\l-1}^{2\l+1}(\rho_2) \\
\nonumber & \quad = - \frac{\mr}{n^2} \frac{1-\epsilon}{\sqrt{1-\epsilon}-1} \left[\frac{(n-\l-1)!}{(n+\l)!}\right]^2 \\
\nonumber & \quad\quad \times \Bigg\lbrace\int\limits_0^\infty \d \rho_1 \, \rho_1^{2\l+1} \e{-\rho_1} [L_{n-\l-1}^{2\l+1}(\rho_1)]^2 \\
\nonumber & \quad\quad + \epsilon \Bigg[\frac{1}{2} \int\limits_0^\infty \d \rho_1 \, \rho_1^{2\l+1} \e{-\rho_1} L_{n-\l-1}^{2\l+1}(\rho_1) \left[(n+\l) L_{n-\l-2}^{2\l+1}(\rho_1) - (n-\l-1) L_{n-\l-1}^{2\l+1}(\rho_1)\right] \\
\nonumber & \quad\quad + \frac{\l}{2} \int\limits_0^\infty \d \rho_1 \, \rho_1^{2\l+1} \e{-\rho_1} [L_{n-\l-1}^{2\l+1}(\rho_1)]^2 + \frac{1}{4} \int\limits_0^\infty \d \rho_1 \, \rho_1^{2\l+2} \e{-\rho_1} [L_{n-\l-1}^{2\l+1}(\rho_1)]^2\Bigg]\Bigg\rbrace \\
\nonumber & \quad\quad \times \Bigg\lbrace\int\limits_0^\infty \d \rho_2 \, \rho_2^{2\l+1} \e{-\rho_2} [L_{n-\l-1}^{2\l+1}(\rho_2)]^2 \\
\nonumber & \quad\quad + \epsilon \Bigg[\frac{1}{2} \int\limits_0^\infty \d \rho_2 \, \rho_2^{2\l+1} \e{-\rho_2} L_{n-\l-1}^{2\l+1}(\rho_2) \left[(n+\l) L_{n-\l-2}^{2\l+1}(\rho_2) - (n-\l-1) L_{n-\l-1}^{2\l+1}(\rho_2)\right] \\
\nonumber & \quad\quad + \frac{\l}{2} \int\limits_0^\infty \d \rho_2 \, \rho_2^{2\l+1} \e{-\rho_2} [L_{n-\l-1}^{2\l+1}(\rho_2)]^2 + \frac{1}{4} \int\limits_0^\infty \d \rho_2 \, \rho_2^{2\l+2} \e{-\rho_2} [L_{n-\l-1}^{2\l+1}(\rho_2)]^2\Bigg]\Bigg\rbrace \\
\nonumber & \quad = - \frac{\mr}{n^2} \frac{1-\epsilon}{\sqrt{1-\epsilon}-1} \left[\frac{(n-\l-1)!}{(n+\l)!}\right]^2 \\
\nonumber & \quad\quad \times \Bigg\lbrace\int\limits_0^\infty \d \rho_1 \, \rho_1^{2\l+1} \e{-\rho_1} [L_{n-\l-1}^{2\l+1}(\rho_1)]^2 + \epsilon \Bigg[\frac{n+\l}{2} \int\limits_0^\infty \d \rho_1 \, \rho_1^{2\l+1} \e{-\rho_1} L_{n-\l-1}^{2\l+1}(\rho_1) L_{n-\l-2}^{2\l+1}(\rho_1) \\
\nonumber & \quad\quad + \frac{1-n}{2} \int\limits_0^\infty \d \rho_1 \, \rho_1^{2\l+1} \e{-\rho_1} [L_{n-\l-1}^{2\l+1}(\rho_1)]^2 + \frac{1}{4} \int\limits_0^\infty \d \rho_1 \, \rho_1^{2\l+2} \e{-\rho_1} [L_{n-\l-1}^{2\l+1}(\rho_1)]^2\Bigg]\Bigg\rbrace \\
\nonumber & \quad\quad \times \Bigg\lbrace\int\limits_0^\infty \d \rho_2 \, \rho_2^{2\l+1} \e{-\rho_2} [L_{n-\l-1}^{2\l+1}(\rho_2)]^2 + \epsilon \Bigg[\frac{n+\l}{2} \int\limits_0^\infty \d \rho_2 \, \rho_2^{2\l+1} \e{-\rho_2} L_{n-\l-1}^{2\l+1}(\rho_2) L_{n-\l-2}^{2\l+1}(\rho_2) \\
\nonumber & \quad\quad + \frac{1-n}{2} \int\limits_0^\infty \d \rho_2 \, \rho_2^{2\l+1} \e{-\rho_2} [L_{n-\l-1}^{2\l+1}(\rho_2)]^2 + \frac{1}{4} \int\limits_0^\infty \d \rho_2 \, \rho_2^{2\l+2} \e{-\rho_2} [L_{n-\l-1}^{2\l+1}(\rho_2)]^2\Bigg]\Bigg\rbrace \\
\nonumber & \quad = - \frac{\mr}{n^2} \frac{1-\epsilon}{\sqrt{1-\epsilon}-1} \left[\frac{(n-\l-1)!}{(n+\l)!}\right]^2 \\
\nonumber & \quad\quad \times \frac{\Gamma(n+\l+1)}{(n-\l-1)!} + \epsilon \left[\frac{n+\l}{2} 0 + \frac{1-n}{2} \frac{\Gamma(n+\l+1)}{(n-\l-1)!} + \frac{1}{4} \frac{2n\Gamma(n+\l+1)}{(n-\l-1)!}\right] \\
\nonumber & \quad\quad \times \frac{\Gamma(n+\l+1)}{(n-\l-1)!} + \epsilon \left[\frac{n+\l}{2} 0 + \frac{1-n}{2} \frac{\Gamma(n+\l+1)}{(n-\l-1)!} + \frac{1}{4} \frac{2n\Gamma(n+\l+1)}{(n-\l-1)!}\right] \\
\nonumber & \quad = - \frac{\mr}{n^2} \frac{1-\epsilon}{\sqrt{1-\epsilon}-1} \left[\frac{(n-\l-1)!}{(n+\l)!}\right]^2 \left[\frac{\Gamma(n+\l+1)}{(n-\l-1)!}\right]^2 \left(1+\frac{\epsilon}{2}\right)^2 \\
\nonumber & \quad = - \frac{\mr}{n^2} \frac{1-\epsilon}{\sqrt{1-\epsilon}-1} \left(1+\frac{\epsilon}{2}\right)^2 \\
\nonumber & \quad = - \frac{\mr}{n^2} \left(-\frac{2}{\epsilon} + \frac{1}{2} + \frac{13\epsilon}{8} + \O(\epsilon^2)\right) \,.
\end{align}
Taking the limit $\epsilon \to 0$ and picking up the term of $\O(\epsilon^{0})$, we obtain $I_2\left(\frac{1}{r},\frac{1}{r}\right) = - \frac{\mr}{2n^2}$ which is the desired result.

\section{Demonstration of the divergence cancellation}
\label{app:sec:DemonstrationDivergenceCancellation}

We start again from the matrix element~\eqref{eq:Connection1} and first restrict ourselves to the types of matrix elements appearing in the spectrum, hence the ones with same initial and final state:
\begin{align}
& \langle n\l | V \frac{1}{(E_n - H)'} V' | n\l \rangle = \int \d^3 r_1 \, \d^3 r_2 \, \psi_{n\l}^*(\vec{r}_2) \, V(\vec{r}_2) \, G'(\vec{r}_2,\vec{r}_1) \, V'(\vec{r}_1) \, \psi_{n\l}(\vec{r}_1) \\
& = - \lim_{E \to E_n} \int \d^3 r_1 \, \d^3 r_2 \, \psi_{n\l}^*(\vec{r}_2) \, V(\vec{r}_2) \, \left(G(\vec{r}_1,\vec{r}_2,E) - \frac{|\psi_{n\l}|^2}{E - E_n}\right) \, V'(\vec{r}_1) \, \psi_{n\l}(\vec{r}_1) \\
& = - \lim_{E \to E_n} \Bigg\lbrace \int \d^3 r_1 \, \d^3 r_2 \, \psi_{n\l}^*(\vec{r}_2) \, V(\vec{r}_2) \, G(\vec{r}_1,\vec{r}_2,E) \, V'(\vec{r}_1) \, \psi_{n\l}(\vec{r}_1) \\
\nonumber & \quad + \int \d^3 r_1 \, \d^3 r_2 \, \psi_{n\l}^*(\vec{r}_2) \, V(\vec{r}_2) \, \frac{\psi_{n\l}^*(\vec{r}_1) \psi_{n\l}(\vec{r}_2)}{E - E_n} \, V'(\vec{r}_1) \, \psi_{n\l}(\vec{r}_1) \Bigg\rbrace \\
& = - \lim_{E \to E_n} \Bigg\lbrace\Bigg[ N_{n\l} N_{n\l} \frac{4\mr}{a \lambda} \sum_{\l''=0}^\infty \sum\limits_{s=0}^\infty \frac{s!}{(s+\l''+1-\lambda)(s+2\l''+1)!} \\
\nonumber & \quad\quad \times \int\limits_0^\infty \d r_1 \, r_1^2 V'(r_1) \r{\lambda}{1}^{\l''} \rho_{n,1}^\l \e{-\frac{1}{2}\r{\lambda}{1}} \e{-\frac{1}{2}\rho_{n,1}} L_{s}^{2\l''+1}(\r{\lambda}{1}) L_{n-\l-1}^{2\l+1}(\rho_{n,1}) \\
\nonumber & \quad\quad \times \int\limits_0^\infty \d r_2 \, r_2^2 V(r_2) \r{\lambda}{2}^{\l''} \rho_{n,2}^{\l} \e{-\frac{1}{2}\r{\lambda}{2}} \e{-\frac{1}{2}\rho_{n,2}} L_{s}^{2\l''+1}(\r{\lambda}{2}) L_{n-\l-1}^{2\l+1}(\rho_{n,2}) \\
\nonumber & \quad\quad \times \sum\limits_{m''=-\l''}^{\l''} \int \d\Omega_1 \, Y_{\l''}^{m''*}(\Omega_1) V'(\Omega_1) Y_{\l}^{m}(\Omega_1) \int \d\Omega_2 \, Y_{\l}^{m*}(\Omega_2) V(\Omega_2) Y_{\l''}^{m''}(\Omega_2) \Bigg] \\
\nonumber & \quad + \Bigg[ \frac{1}{E - E_n} N_{n\l}^2 N_{n\l}^2 \\
\nonumber & \quad\quad \times \int \d r_1 \, r_1^2 V'(\vec{r}_1) \r{n}{1}^\l \r{n}{1}^\l \e{-\frac{1}{2}\r{n}{1}} \e{-\frac{1}{2}\r{n}{1}} L_{n-\l-1}^{2\l+1}(\rho_{n,1}) L_{n-\l-1}^{2\l+1}(\rho_{n,1}) \\
\nonumber & \quad\quad \times \int \d r_2 \, r_1^2 V(\vec{r}_2) \r{n}{2}^\l \r{n}{2}^\l \e{-\frac{1}{2}\r{n}{2}} \e{-\frac{1}{2}\r{n}{2}} L_{n-\l-1}^{2\l+1}(\rho_{n,2}) L_{n-\l-1}^{2\l+1}(\rho_{n,2}) \\
\nonumber & \quad\quad \int \d\Omega_1 \, Y_{\l}^{m*}(\Omega_1) V'(\Omega_1) Y_{\l}^{m}(\Omega_1) \int \d\Omega_2 \, Y_{\l}^{m*}(\Omega_2) V(\Omega_2) Y_{\l}^{m}(\Omega_2) \Bigg]\Bigg\rbrace
\end{align}
One could now assume that the angular parts of the potentials, $V'(\Omega_1)$ and $V(\Omega_2)$ are equal to identity, which would impose $\l = \l''$ and thus $m = m''$. Or one could argue that the limit $E \to E_n$ directly hits the pole in each expression in square brackets individually and this divergence can only be cured if $\l = \l''$ holds.\footnote{This limit is equivalent to the limit $\lambda \to n$ and thus it is equivalent to the limits $\epsilon \to 0$, while $E \to E_n (1 - \epsilon)$, which corresponds to $\epsilon \to 0$, while $\lambda \to \frac{n}{\sqrt{1 - \epsilon}}$.} And it has to be cured, since the matrix element is a physical observable and thus has to be finite. Either way, we arrive at
\begin{align}
& \langle n\l | V \frac{1}{(E_n - H)'} V' | n\l \rangle = \int \d^3 r_1 \, \d^3 r_2 \, \psi_{n\l}^*(\vec{r}_2) \, V(\vec{r}_2) \, G_{n\l}'(\vec{r}_2,\vec{r}_1) \, V'(\vec{r}_1) \, \psi_{n\l}(\vec{r}_1) \\
& = - \lim_{E \to E_n} \Bigg\lbrace\Bigg[ N_{n\l} N_{n\l} \frac{4\mr}{a \lambda} \sum\limits_{s=0}^\infty \frac{s!}{(s+\l+1-\lambda)(s+2\l+1)!} \\
\nonumber & \quad\quad \times \int\limits_0^\infty \d r_1 \, r_1^2 V'(r_1) \r{\lambda}{1}^{\l} \rho_{n,1}^\l \e{-\frac{1}{2}\r{\lambda}{1}} \e{-\frac{1}{2}\rho_{n,1}} L_{s}^{2\l+1}(\r{\lambda}{1}) L_{n-\l-1}^{2\l+1}(\rho_{n,1}) \\
\nonumber & \quad\quad \times \int\limits_0^\infty \d r_2 \, r_2^2 V(r_2) \r{\lambda}{2}^{\l} \rho_{n,2}^{\l} \e{-\frac{1}{2}\r{\lambda}{2}} \e{-\frac{1}{2}\rho_{n,2}} L_{s}^{2\l+1}(\r{\lambda}{2}) L_{n-\l-1}^{2\l+1}(\rho_{n,2}) \\
\nonumber & \quad\quad \times \int \d\Omega_1 \, Y_{\l}^{m*}(\Omega_1) V'(\Omega_1) Y_{\l}^{m}(\Omega_1) \int \d\Omega_2 \, Y_{\l}^{m*}(\Omega_2) V(\Omega_2) Y_{\l}^{m}(\Omega_2) \Bigg] \\
\nonumber & \quad + \Bigg[ \frac{1}{E - E_n} N_{n\l}^2 N_{n\l}^2 \\
\nonumber & \quad\quad \times \int \d r_1 \, r_1^2 V'(\vec{r}_1) \r{n}{1}^\l \r{n}{1}^\l \e{-\frac{1}{2}\r{n}{1}} \e{-\frac{1}{2}\r{n}{1}} L_{n-\l-1}^{2\l+1}(\rho_{n,1}) L_{n-\l-1}^{2\l+1}(\rho_{n,1}) \\
\nonumber & \quad\quad \times \int \d r_2 \, r_1^2 V(\vec{r}_2) \r{n}{2}^\l \r{n}{2}^\l \e{-\frac{1}{2}\r{n}{2}} \e{-\frac{1}{2}\r{n}{2}} L_{n-\l-1}^{2\l+1}(\rho_{n,2}) L_{n-\l-1}^{2\l+1}(\rho_{n,2}) \\
\nonumber & \quad\quad \int \d\Omega_1 \, Y_{\l}^{m*}(\Omega_1) V'(\Omega_1) Y_{\l}^{m}(\Omega_1) \int \d\Omega_2 \, Y_{\l}^{m*}(\Omega_2) V(\Omega_2) Y_{\l}^{m}(\Omega_2) \Bigg]\Bigg\rbrace
\end{align}
We rewrite the limit, exhibiting the crucial value of the sum to be $s = n-\l-1$. However, this is exactly the value appearing in the second square bracket. We thus now disregard the finite terms, including the angular integrals and arrive at\footnote{Remember that $\r{\frac{n}{\sqrt{1-\epsilon}}}{i} = \r{n}{i} \sqrt{1-\epsilon}$.}
\begin{align}
& \left.\langle n\l | V \frac{1}{(E_n - H)'} V' | n\l \rangle\right|_{\text{divergent}} \\
& = - N_{n\l}^2 \lim_{\epsilon \to 0} \Bigg\lbrace\Bigg[ \frac{4\mr}{a \frac{n}{\sqrt{1-\epsilon}}} \frac{(n-\l-1)!}{(n-\l-1+\l+1-\frac{n}{\sqrt{1-\epsilon}})(n-\l-1+2\l+1)!} \\
\nonumber & \quad\quad \times \int\limits_0^\infty \d r_1 \, r_1^2 V'(r_1) \r{n}{1}^{\l} (\sqrt{1-\epsilon})^\l \rho_{n,1}^\l \e{-\frac{1}{2}\r{n}{1}\sqrt{1-\epsilon}} \e{-\frac{1}{2}\rho_{n,1}} L_{n-\l-1}^{2\l+1}(\r{n}{1}\sqrt{1-\epsilon}) L_{n-\l-1}^{2\l+1}(\rho_{n,1}) \\
\nonumber & \quad\quad \times \int\limits_0^\infty \d r_2 \, r_2^2 V(r_2) \r{n}{2}^{\l} (\sqrt{1-\epsilon})^\l \rho_{n,2}^{\l} \e{-\frac{1}{2}\r{n}{2}\sqrt{1-\epsilon}} \e{-\frac{1}{2}\rho_{n,2}} L_{n-\l-1}^{2\l+1}(\r{n}{2}\sqrt{1-\epsilon}) L_{n-\l-1}^{2\l+1}(\rho_{n,2}) \Bigg] \\
\nonumber & \quad + \Bigg[ \frac{1}{-\epsilon E_n} N_{n\l}^2 \\
\nonumber & \quad\quad \times \int \d r_1 \, r_1^2 V'(\vec{r}_1) \r{n}{1}^\l \r{n}{1}^\l \e{-\frac{1}{2}\r{n}{1}} \e{-\frac{1}{2}\r{n}{1}} L_{n-\l-1}^{2\l+1}(\rho_{n,1}) L_{n-\l-1}^{2\l+1}(\rho_{n,1}) \\
\nonumber & \quad\quad \times \int \d r_2 \, r_1^2 V(\vec{r}_2) \r{n}{2}^\l \r{n}{2}^\l \e{-\frac{1}{2}\r{n}{2}} \e{-\frac{1}{2}\r{n}{2}} L_{n-\l-1}^{2\l+1}(\rho_{n,2}) L_{n-\l-1}^{2\l+1}(\rho_{n,2}) \Bigg]\Bigg\rbrace \\
& = - N_{n\l}^2 \lim_{\epsilon \to 0} \Bigg\lbrace\Bigg[ \frac{1}{-2E_n} N_{n\l}^2 \frac{1-\epsilon}{\sqrt{1-\epsilon}-1} \\
\nonumber & \quad\quad \times \int\limits_0^\infty \d r_1 \, r_1^2 V'(r_1) \r{n}{1}^{\l} (\sqrt{1-\epsilon})^\l \rho_{n,1}^\l \e{-\frac{1}{2}\r{n}{1}\sqrt{1-\epsilon}} \e{-\frac{1}{2}\rho_{n,1}} L_{n-\l-1}^{2\l+1}(\r{n}{1}\sqrt{1-\epsilon}) L_{n-\l-1}^{2\l+1}(\rho_{n,1}) \\
\nonumber & \quad\quad \times \int\limits_0^\infty \d r_2 \, r_2^2 V(r_2) \r{n}{2}^{\l} (\sqrt{1-\epsilon})^\l \rho_{n,2}^{\l} \e{-\frac{1}{2}\r{n}{2}\sqrt{1-\epsilon}} \e{-\frac{1}{2}\rho_{n,2}} L_{n-\l-1}^{2\l+1}(\r{n}{2}\sqrt{1-\epsilon}) L_{n-\l-1}^{2\l+1}(\rho_{n,2}) \Bigg] \\
\nonumber & \quad + \Bigg[ \frac{1}{-\epsilon E_n} N_{n\l}^2 \\
\nonumber & \quad\quad \times \int \d r_1 \, r_1^2 V'(\vec{r}_1) \r{n}{1}^\l \r{n}{1}^\l \e{-\frac{1}{2}\r{n}{1}} \e{-\frac{1}{2}\r{n}{1}} L_{n-\l-1}^{2\l+1}(\rho_{n,1}) L_{n-\l-1}^{2\l+1}(\rho_{n,1}) \\
\nonumber & \quad\quad \times \int \d r_2 \, r_1^2 V(\vec{r}_2) \r{n}{2}^\l \r{n}{2}^\l \e{-\frac{1}{2}\r{n}{2}} \e{-\frac{1}{2}\r{n}{2}} L_{n-\l-1}^{2\l+1}(\rho_{n,2}) L_{n-\l-1}^{2\l+1}(\rho_{n,2}) \Bigg]\Bigg\rbrace \\
& = - N_{n\l}^4 \lim_{\epsilon \to 0} \Bigg\lbrace\Bigg[ \frac{1}{-2E_n} \frac{1-\epsilon}{\sqrt{1-\epsilon}-1} \\
\nonumber & \quad\quad \times \int\limits_0^\infty \d r_1 \, r_1^2 V'(r_1) \r{n}{1}^{\l} (\sqrt{1-\epsilon})^\l \rho_{n,1}^\l \e{-\frac{1}{2}\r{n}{1}\sqrt{1-\epsilon}} \e{-\frac{1}{2}\rho_{n,1}} L_{n-\l-1}^{2\l+1}(\r{n}{1}\sqrt{1-\epsilon}) L_{n-\l-1}^{2\l+1}(\rho_{n,1}) \\
\nonumber & \quad\quad \times \int\limits_0^\infty \d r_2 \, r_2^2 V(r_2) \r{n}{2}^{\l} (\sqrt{1-\epsilon})^\l \rho_{n,2}^{\l} \e{-\frac{1}{2}\r{n}{2}\sqrt{1-\epsilon}} \e{-\frac{1}{2}\rho_{n,2}} L_{n-\l-1}^{2\l+1}(\r{n}{2}\sqrt{1-\epsilon}) L_{n-\l-1}^{2\l+1}(\rho_{n,2}) \Bigg] \\
\nonumber & \quad + \Bigg[ \frac{1}{-\epsilon E_n} \\
\nonumber & \quad\quad \times \int \d r_1 \, r_1^2 V'(\vec{r}_1) \r{n}{1}^\l \r{n}{1}^\l \e{-\frac{1}{2}\r{n}{1}} \e{-\frac{1}{2}\r{n}{1}} L_{n-\l-1}^{2\l+1}(\rho_{n,1}) L_{n-\l-1}^{2\l+1}(\rho_{n,1}) \\
\nonumber & \quad\quad \times \int \d r_2 \, r_1^2 V(\vec{r}_2) \r{n}{2}^\l \r{n}{2}^\l \e{-\frac{1}{2}\r{n}{2}} \e{-\frac{1}{2}\r{n}{2}} L_{n-\l-1}^{2\l+1}(\rho_{n,2}) L_{n-\l-1}^{2\l+1}(\rho_{n,2}) \Bigg]\Bigg\rbrace
\end{align}
Expanding the first square brackets yields
\begin{align}
& \frac{1}{-2E_n} \frac{1-\epsilon}{\sqrt{1-\epsilon}-1} \\
\nonumber & \quad \times \r{n}{1}^{\l} (\sqrt{1-\epsilon})^\l \rho_{n,1}^\l \e{-\frac{1}{2}\r{n}{1}\sqrt{1-\epsilon}} \e{-\frac{1}{2}\rho_{n,1}} L_{n-\l-1}^{2\l+1}(\r{n}{1}\sqrt{1-\epsilon}) L_{n-\l-1}^{2\l+1}(\rho_{n,1}) \\
\nonumber & \quad \times \r{n}{2}^{\l} (\sqrt{1-\epsilon})^\l \rho_{n,2}^{\l} \e{-\frac{1}{2}\r{n}{2}\sqrt{1-\epsilon}} \e{-\frac{1}{2}\rho_{n,2}} L_{n-\l-1}^{2\l+1}(\r{n}{2}\sqrt{1-\epsilon}) L_{n-\l-1}^{2\l+1}(\rho_{n,2}) \\
& = \frac{1}{\epsilon E_n} \\
\nonumber & \quad\quad \times \rho_{n,1}^{\l} \rho_{n,1}^{\l} \e{-\frac{1}{2}\rho_{n,1}} \e{-\frac{1}{2}\rho_{n,1}} L_{n-\l-1}^{2\l+1}(\rho_{n,1}) L_{n-\l-1}^{2\l+1}(\rho_{n,1}) \\
\nonumber & \quad\quad \times \rho_{n,2}^{\l} \rho_{n,2}^{\l} \e{-\frac{1}{2}\rho_{n,2}} \e{-\frac{1}{2}\rho_{n,2}} L_{n-\l-1}^{2\l+1}(\rho_{n,2}) L_{n-\l-1}^{2\l+1}(\rho_{n,2}) \\
\nonumber & \quad + \#_0 \, \epsilon^0 + \#_1 \, \epsilon + \O(\epsilon^2)
\end{align}
and all the terms linear or of higher order in $\epsilon$ vanish in the limit $\epsilon \to 0$. The finite part $\#_0$ is the one we extract and the divergent part cancels exactly.\\
For the matrix elements appearing in the calculation of the E1 transition we find for the correction to the initial state
\begin{align}
& \,^{(0)}\!\langle n'\l' | \OEone | n\l \rangle^{(1)} = \,^{(0)}\!\langle n'\l' | \OEone \left( \sum\limits_{k \neq n} \frac{\,^{(0)}\!\langle k\l | V | n\l \rangle^{(0)}}{E_n^{(0)} - E_{k}^{(0)}} | k\l \rangle^{(0)} \right) \\
\nonumber & \quad = \langle n'\l' | \OEone \left( \sum\limits_{k \neq n} \frac{\langle k\l | V | n\l \rangle}{E_{n} - E_{k}} | k\l \rangle \right) = \langle n'\l' | \OEone \left( \sum\limits_{k \neq n} \frac{| k\l \rangle \langle k\l |}{E_{n} - E_{k}} V | n\l \rangle \right) \\
\nonumber & \quad = \lim_{E \to E_{n}} \left[ \langle n'\l' | \OEone \frac{\sum_{k} | k\l \rangle \langle k\l |}{H - E} V | n\l \rangle - \langle n'\l' | \OEone \frac{| n\l \rangle \langle n\l |}{E_{n} - E} V | n\l \rangle \right] \\
\nonumber & \quad = - \lim_{E \to E_{n}} \left[ \langle n'\l' | \OEone \frac{\sum_{k} | k\l \rangle \langle k\l |}{E - H} V | n\l \rangle - \langle n'\l' | \OEone \frac{| n\l \rangle \langle n\l |}{E - E_{n}} V | n\l \rangle \right] \\
\nonumber & \quad = - \lim_{E \to E_{n}} \left[ \langle n'\l' | \OEone \frac{\1}{E - H} V | n\ell \rangle - \langle n'\l' | \OEone \frac{\mathcal{P}(n)}{E - E_{n}} V | n\l \rangle \right] \\
\nonumber & \quad = - \langle n'\l' | \OEone \lim_{E \to E_{n}} \left[ \frac{\1}{E - H} - \frac{\mathcal{P}(n)}{E - E_{n}} \right] V | n\l \rangle \\
\nonumber & \quad = - \int \d^{3} r_{1} \, \d^{3} r_{2} \, \psi_{n'\l'}^{*}(\vec{r}_{2}) \OEone(\vec{r}_{2}) \left[ \lim_{E \to E_{n}} \left( G(\vec{r}_{1},\vec{r}_{2},E) - \frac{\psi_{n\l}^{*}(\vec{r_{1}})\psi_{n\l}(\vec{r_{2}})}{E - E_{n}} \right) \right] V(\vec{r}_{1}) \psi_{n\l}(\vec{r}_{1}) \\
\nonumber & \quad = - \int \d^{3} r_{1} \, \d^{3} r_{2} \, \psi_{n'\l'}^{*}(\vec{r}_{2}) \OEone(\vec{r}_{2}) G'(\vec{r}_{1},\vec{r}_{2}) V(\vec{r}_{1}) \psi_{n\l}(\vec{r}_{1}) \\
\nonumber & \quad = - \langle n'\l' | \OEone \frac{\1}{(E_n - H)'} V | n\l \rangle \,,
\end{align}
and therefore the divergence is particular of the initial state because the Green function inherits its quantum numbers via the term
\begin{equation}
\lim_{E \to E_{n}} \frac{1}{E - E_{n}} = \lim_{\epsilon \to 0} \frac{1}{E_{n}(1 - \epsilon) - E_{n}} = \lim_{\epsilon \to 0} \frac{1}{\epsilon E_{n}} \,.
\end{equation}
From the construction above, it is clear that the corrections to the final state must give a divergence that is a particular property of the final state only, because the Green function inherits its quantum numbers via the term
\begin{equation}
\lim_{E \to E_{n'}} \frac{1}{E - E_{n'}} = \lim_{\epsilon' \to 0} \frac{1}{E_{n'}(1 - \epsilon') - E_{n'}} = \lim_{\epsilon' \to 0} \frac{1}{\epsilon' E_{n'}} \,,
\end{equation}
or more explicitly:
\begin{align}
& \, ^{(1)}\!\langle n'\l' | \OEone | n\l \rangle^{(0)} = \left( \sum\limits_{k \neq n'} \, ^{(0)}\!\langle k\l' | \frac{\, ^{(0)}\!\langle n'\l' | V | k\l' \rangle^{(0)}}{E_{n'}^{(0)} - E_{k}^{(0)}} \right) \OEone | n\l \rangle^{(0)} \\
\nonumber & \quad = \left( \sum\limits_{k \neq n'} \langle k\l' | \frac{\langle n'\l' | V | k\l' \rangle}{E_{n'} - E_{k}} \right) \OEone | n\l \rangle = \left( \langle n'\l' | V \sum\limits_{k \neq n'} \frac{| k\l' \rangle \langle k\l' |}{E_{n'} - E_{k}} \right) \OEone | n\l \rangle \\
\nonumber & \quad = \lim_{E \to E_{n'}} \left[ \langle n'\l' | V \frac{\sum_{k} | k\l' \rangle \langle k\l' |}{H - E} \OEone | n\l \rangle - \langle n'\l' | V \frac{| n'\l' \rangle \langle n'\l' |}{E_{n'} - E} \OEone | n\l \rangle \right] \\
\nonumber & \quad = - \lim_{E \to E_{n'}} \left[ \langle n'\l' | V \frac{\sum_{k} | k\l' \rangle \langle k\l' |}{E - H} \OEone | n\l \rangle - \langle n'\l' | V \frac{| n'\l' \rangle \langle n'\l' |}{E - E_{n'}} \OEone | n\l \rangle \right] \\
\nonumber & \quad = - \lim_{E \to E_{n'}} \left[ \langle n'\l' | V \frac{\1}{E - H} \OEone | n\l \rangle - \langle n'\l' | V \frac{\mathcal{P}(n)}{E - E_{n'}} \OEone | n\l \rangle \right] \\
\nonumber & \quad = - \langle n'\l' | V \lim_{E \to E_{n'}} \left[ \frac{\1}{E - H} - \frac{\mathcal{P}(n)}{E - E_{n'}} \right] \OEone | n\l \rangle \\
\nonumber & \quad = - \int \d^{3} r_{1} \, \d^{3} r_{2} \, \psi_{n'\l'}^{*}(\vec{r}_{2}) V(\vec{r}_{2}) \left[ \lim_{E \to E_{n'}} \left( G(\vec{r}_{1},\vec{r}_{2},E) - \frac{\psi_{n'\l'}^{*}(\vec{r_{1}})\psi_{n'\l'}(\vec{r_{2}})}{E - E_{n'}} \right) \right] \OEone(\vec{r}_{1}) \psi_{n\l}(\vec{r}_{1}) \\
\nonumber & \quad = - \int \d^{3} r_{1} \, \d^{3} r_{2} \, \psi_{n'\l'}^{*}(\vec{r}_{2}) V(\vec{r}_{2}) G'(\vec{r}_{1},\vec{r}_{2}) \OEone(\vec{r}_{1}) \psi_{n\l}(\vec{r}_{1}) \\
\nonumber & \quad = - \langle n'\l' | V \frac{\1}{(E_{n'} - H)'} \OEone | n\l \rangle \,.
\end{align}
An intermediate conclusion we can draw already is that expectation values with double potential insertions and matrix elements with one correction to either the initial or final state are free of divergences. This is due to the fact that all of these objects contain only one Green function that inherits its quantum numbers from the state the correction is applied to. Therefore, during the actual computation, we can take advantage of this, by simply neglecting the $\mathcal{P}(n)$ part, but therefore as well neglecting the $\frac{1}{\epsilon}$ divergence after the expanding and thus simply taking the limit $\epsilon \to 0$ while keeping the term $\O(\epsilon^0)$ only. This simple picture changes slightly, however not conceptually, in the case where we have two Green functions involved, as we will see now.\\
For the second order correction there are three different terms\footnote{This whole argumentation would also hold conceptually for the third order correction to the energy, which would be of order $m \alpha_s^5$ and thus beyond our scope. Therefore the only objects involving two Green functions are the matrix elements contributing to the decay width at order $m \alpha_s^6$.}:
\begin{equation}
^{(1)}\!\langle n'\l' | \OEone | n\l \rangle^{(1)} \,, \quad\quad ^{(2)}\!\langle n'\l' | \OEone | n\l \rangle^{(0)} \quad\quad \text{and} \quad\quad ^{(0)}\!\langle n'\l' | \OEone | n\l \rangle^{(2)} \,.
\end{equation}
Following the argumentation above, it is clear that the first one has two simple divergences, one which goes as $\frac{1}{\epsilon}$, associated with the initial state, and the other one that behaves as $\frac{1}{\epsilon'}$, associated with the final state. For the latter two, each individual Green Function has a simple divergence as well. However, each of the individual Green functions is a finite object and all the arising divergences are canceled. Thus, in the latter two cases, one can no longer neglect the $\mathcal{P}(n)$ parts and drop the divergences because this would lead to admixture, and thus to $\frac{1}{\epsilon^2}$ or $\frac{1}{\epsilon'^2}$ divergences that do not exist. In order to overcome this in the actual computation, we treat each Green function with a different regulator ($\epsilon$ and $\epsilon'$) even if both of them act on the same state (initial or final). By doing so we avoid admixture and guarantee an exact cancellation of all arising divergences while still keeping the computational effort as low as possible. Consider the example of the Green functions
\begin{equation}
G = \frac{G_{-1}}{\epsilon} + G_{0} + G_{1} \epsilon + \dots \,, \quad \text{and} \quad G' = \frac{G'_{-1}}{\epsilon'} + G'_{0} + G'_{1} \epsilon' + \dots \,.
\end{equation}
If we explicitly subtract the divergences individually and take limits $\epsilon \to 0$ and $\epsilon' \to 0$ the product of them would be
\begin{equation}
\left(G - \frac{G_{-1}}{\epsilon}\right) \left(G' - \frac{G'_{-1}}{\epsilon'}\right) = G_{0} G'_{0} \,,
\end{equation}
as it should be. However, if we would use the same regulator for both of them and simply drop the divergence we would end up with the wrong result
\begin{equation}
G G' = \frac{G_{-1} G'_{-1}}{\epsilon^2} + \frac{G_{-1} G'_{0} + G'_{-1} G_{0}}{\epsilon} + G_{0} G'_{0} + 2 G_{-1} G'_{-1} + \dots
\end{equation}
which after dropping the divergent terms and taking the limit $\epsilon \to 0$ gives the additional term $2 G_{-1} G'_{-1}$ that yields a wrong result.\\
Finally we can conclude that the following holds:
\begin{align}
& \langle n\l | V \frac{1}{(E_n - H)'} V' | n\l \rangle = - \int \d^3 r_1 \, \d^3 r_2 \, \psi_{n\l}^*(\vec{r}_2) \, V(\vec{r}_2) \, G'(\vec{r}_2,\vec{r}_1) \, V'(\vec{r}_1) \, \psi_{n\l}(\vec{r}_1) \\
\nonumber & = \left. - \int \d^3 r_1 \, \d^3 r_2 \, \psi_{n\l}^*(\vec{r}_2) \, V(\vec{r}_2) \, G(\vec{r}_2,\vec{r}_1,E) \, V'(\vec{r}_1) \, \psi_{n\l}(\vec{r}_1)  \right|_{\begin{array}{l} E \to E_n \\ \epsilon \to 0 , \O(\epsilon^0) \end{array}} \\
& \, ^{(1)}\!\langle n'\l' | \OEone | n\l \rangle = \langle n'\l' | V \frac{1}{(E_{n'} - H)'} \OEone | n\l \rangle \\
\nonumber & = - \int \d^3 r_1 \, \d^3 r_2 \, \psi_{n'\l'}^*(\vec{r}_2) \, V(\vec{r}_2) \, G'(\vec{r}_2,\vec{r}_1) \, \OEone(\vec{r}_1) \, \psi_{n\l}(\vec{r}_1) \\
\nonumber & = \left. - \int \d^3 r_1 \, \d^3 r_2 \, \psi_{n'\l'}^*(\vec{r}_2) \, V(\vec{r}_2) \, G(\vec{r}_2,\vec{r}_1,E) \, \OEone(\vec{r}_1) \, \psi_{n\l}(\vec{r}_1) \right|_{\begin{array}{l} E \to E_{n'} \\ \epsilon' \to 0 , \O(\epsilon'^0) \end{array}} \\
& \langle n'\l' | \OEone | n\l \rangle^{(1)} = \langle n'\l' | \OEone \frac{1}{(E_n - H)'} V | n\l \rangle \\
\nonumber & = - \int \d^3 r_1 \, \d^3 r_2 \, \psi_{n'\l'}^*(\vec{r}_2) \, \OEone(\vec{r}_2) \, G'(\vec{r}_2,\vec{r}_1) \, V(\vec{r}_1) \, \psi_{n\l}(\vec{r}_1) \\
\nonumber & = \left. - \int \d^3 r_1 \, \d^3 r_2 \, \psi_{n'\l'}^*(\vec{r}_2) \, \OEone(\vec{r}_2) \, G(\vec{r}_2,\vec{r}_1,E) \, V(\vec{r}_1) \, \psi_{n\l}(\vec{r}_1) \right|_{\begin{array}{l} E \to E_n \\ \epsilon \to 0 , \O(\epsilon^0)	\end{array}} \\
& \, ^{(1)}\!\langle n'\l' | \OEone | n\l \rangle^{(1)} = \langle n'\l' | V' \frac{1}{(E_{n'} - H)'} \OEone \frac{1}{(E_n - H)'} V | n\l \rangle \\
\nonumber & = \int \d^3 r_1 \, \d^3 r_2 \, \d^3 r_3 \, \psi_{n'\l'}^*(\vec{r}_3) \, V'(\vec{r}_3) \, G'(\vec{r}_3,\vec{r}_2) \, \OEone(\vec{r}_2) \, G'(\vec{r}_2,\vec{r}_1) \, V(\vec{r}_1) \, \psi_{n\l}(\vec{r}_1) \\
\nonumber & = \left. \int \d^3 r_1 \, \d^3 r_2 \, \d^3 r_3 \, \psi_{n'\l'}^*(\vec{r}_3) \, V'(\vec{r}_3) \, G(\vec{r}_3,\vec{r}_2,E') \, \OEone(\vec{r}_2) \, G(\vec{r}_2,\vec{r}_1,E) \, V(\vec{r}_1) \, \psi_{n\l}(\vec{r}_1) \right|_{\!\!\begin{array}{l} E \to E_n (1-\epsilon) \\ E' \to E_{n'} (1-\epsilon') \\ \epsilon \to 0 , \O(\epsilon^0) \\ \epsilon' \to 0 , \O(\epsilon'^0) \end{array}} \\
& \, ^{(2)}\!\langle n'\l' | \OEone | n\l \rangle = \langle n'\l' | V' \frac{1}{(E_{n'} - H)'} V \frac{1}{(E_{n'} - H)'} \OEone | n\l \rangle \\
\nonumber & = - \int \d^3 r_1 \, \d^3 r_2 \, \d^3 r_3 \, \psi_{n'\l'}^*(\vec{r}_3) \, V'(\vec{r}_3) \, G'(\vec{r}_3,\vec{r}_2) \, V(\vec{r}_2) \, G'(\vec{r}_2,\vec{r}_1) \, \OEone(\vec{r}_1) \, \psi_{n\l}(\vec{r}_1) \\
\nonumber & = \left. \! - \! \int \d^3 r_1 \, \d^3 r_2 \, \d^3 r_3 \, \psi_{n'\l'}^*(\vec{r}_3) \, V'(\vec{r}_3) \, G(\vec{r}_3,\vec{r}_2,E') \, V(\vec{r}_2) \, G(\vec{r}_2,\vec{r}_1,E) \, \OEone(\vec{r}_1) \, \psi_{n\l}(\vec{r}_1) \right|_{\!\!\begin{array}{l} E \to E_{n'} (1-\epsilon) \\ E' \to E_{n'} (1-\epsilon') \\ \epsilon \to 0 , \O(\epsilon^0) \\ \epsilon' \to 0 , \O(\epsilon'^0) \end{array}} \\
& \langle n'\l' | \OEone | n\l \rangle^{(2)} = \langle n'\l' | \OEone \frac{1}{(E_n - H)'} V' \frac{1}{(E_n - H)'} V | n\l \rangle \\
\nonumber & = - \int \d^3 r_1 \, \d^3 r_2 \, \d^3 r_3 \, \psi_{n'\l'}^*(\vec{r}_3) \, \OEone(\vec{r}_3) \, G'(\vec{r}_3,\vec{r}_2) \, V'(\vec{r}_2) \, G'(\vec{r}_2,\vec{r}_1) \, V(\vec{r}_1) \, \psi_{n\l}(\vec{r}_1) \\
\nonumber & = \left. \! - \! \int \d^3 r_1 \, \d^3 r_2 \, \d^3 r_3 \, \psi_{n'\l'}^*(\vec{r}_3) \, \OEone(\vec{r}_3) \, G(\vec{r}_3,\vec{r}_2,E') \, V'(\vec{r}_2) \, G(\vec{r}_2,\vec{r}_1,E) \, V(\vec{r}_1) \, \psi_{n\l}(\vec{r}_1) \right|_{\!\!\begin{array}{l} E \to E_n (1-\epsilon) \\ E' \to E_n (1-\epsilon') \\ \epsilon \to 0 , \O(\epsilon^0) \\ \epsilon' \to 0 , \O(\epsilon'^0) \end{array}}
\end{align}
Using this method significantly simplifies the computations, since one does not need to explicitly calculate the divergent $\mathcal{P}(n)$ part. During all of this work we suppress the additional $E$ in the Coulomb Green function as well as the limits, since both these notations are understood implicitly.

\section{Derivation of the non-perturbative contribution to the wave function renormalization}
\label{app:sec:NonPerturbativeContribution}

We derive the final form of Eq.~\eqref{eq:BeginNonPerturbative}, that is given by
\begin{align}
& ^{(0)}\!\langle n ; \l | r \frac{1}{(H_{\text{o}}^{(0)} - E_{H}^{(0)})} \1 \frac{1}{(H_{\text{o}}^{(0)} - E_{H}^{(0)})} r | n ; \l \rangle^{(0)} \\
& \quad = \left(\frac{4\mr}{n a}\right)^2 N_{n\l}^2 \left(\frac{n a}{2}\right)^{11} \sum\limits_{s=0}^\infty \sum\limits_{s'=0}^\infty \frac{s!}{(s+\l+1+\frac{n}{2N_{\text{c}} C_F})[(s+2\l+1)!]} \\
\nonumber & \quad\quad \times \frac{s'!}{(s'+\l+1+\frac{n}{2N_{\text{c}} C_F})[(s'+2\l+1)!]} \\
\nonumber & \quad\quad \times \int \d \rho_1 \, \rho_1^{3+2\l} \e{-\rho_1} L_{n-\l-1}^{2\l+1}\left(\rho_1\right) L_{s}^{2\l+1}\left(\rho_1\right) \\
\nonumber & \quad\quad \times \int \d \rho_2 \, \rho_2^{2+2\l} \e{-\rho_2} L_{s}^{2\l+1}\left(\rho_2\right) L_{s'}^{2\l+1}\left(\rho_2\right) \\
\nonumber & \quad\quad \times \int \d \rho_3 \, \rho_3^{3+2\l} \e{-\rho_3} L_{s'}^{2\l+1}\left(\rho_3\right) L_{n-\l-1}^{2\l+1}\left(\rho_3\right) \,.
\end{align}
We may now exploit the completeness and recursion relations for the associated Laguerre polynomials, given in Appendix~\ref{app:subsec:AssociatedLaguerrePolynomials}. The intuitive but naive way would be to use Eqs.~\eqref{eq:OrthogonalityAssociatedLaguerreIII} and \eqref{eq:OrthogonalityAssociatedLaguerreIV} to trade all the three integrals for finite sums depending on $s$ and $s'$. However, exploiting the recursion relations~\eqref{eq:RecursionAssociatedLaguerre} for the $\rho_2$ integral first, and then using Eq.~\eqref{eq:OrthogonalityAssociatedLaguerreI} allows us to avoid the numerically non-trivial double series in $s$ and $s'$, since we may rewrite
\begin{align}
& \int \d \rho_2 \, \rho_2^{2+2\l} \e{-\rho_2} L_{s}^{2\l+1}\left(\rho_2\right) L_{s'}^{2\l+1}\left(\rho_2\right) = \int \d \rho_2 \, \rho_2^{2\l} \e{-\rho_2} \left[\rho_2 L_{s}^{2\l+1}\left(\rho_2\right)\right] \left[\rho_2 L_{s'}^{2\l+1}\left(\rho_2\right)\right] \\
& \quad = \int \d \rho_2 \, \rho_2^{2\l} \e{-\rho_2} \left[(s+2\l+1) L_{s}^{2\l}\left(\rho_2\right) - (s+1) L_{s+1}^{2\l}\left(\rho_2\right)\right] \\
\nonumber & \quad\quad \times \left[(s'+2\l+1) L_{s'}^{2\l}\left(\rho_2\right) - (s'+1) L_{s'+1}^{2\l}\left(\rho_2\right)\right] \\
& \quad = (s+2\l+1) (s'+2\l+1) \int \d \rho_2 \, \rho_2^{2\l} \e{-\rho_2} L_{s}^{2\l}\left(\rho_2\right) L_{s'}^{2\l}\left(\rho_2\right) \\
\nonumber & \quad\quad - (s+2\l+1) (s'+1) \int \d \rho_2 \, \rho_2^{2\l} \e{-\rho_2} L_{s}^{2\l}\left(\rho_2\right) L_{s'+1}^{2\l}\left(\rho_2\right) \\
\nonumber & \quad\quad - (s+1) (s'+2\l+1) \int \d \rho_2 \, \rho_2^{2\l} \e{-\rho_2} L_{s+1}^{2\l}\left(\rho_2\right) L_{s'}^{2\l}\left(\rho_2\right) \\
\nonumber & \quad\quad + (s+1) (s'+1) \int \d \rho_2 \, \rho_2^{2\l} \e{-\rho_2} L_{s+1}^{2\l}\left(\rho_2\right) L_{s'+1}^{2\l}\left(\rho_2\right) \\
& \quad = (s+2\l+1) (s'+2\l+1) \frac{(s+2\l)!}{s!} \delta_{s s'} - (s+2\l+1) (s'+1) \frac{(s+2\l)!}{s!}  \delta_{s s'+1} \\
\nonumber & \quad\quad - (s+1) (s'+2\l+1) \frac{(s+1+2\l)!}{(s+1)!}  \delta_{s+1s'} + (s+1) (s'+1) \frac{(s+1+2\l)!}{(s+1)!}  \delta_{s+1s'+1} \\
& \quad = \frac{(s+2\l+1)^2 [(s+2\l)!]}{s!} \delta_{s s'} - \frac{(s+2\l+1) [(s+2\l)!]}{(s-1)!}  \delta_{s s'+1} \\
\nonumber & \quad\quad - \frac{(s+2\l+2) [(s+2\l+1)!]}{s!}  \delta_{s+1s'} + \frac{(s+1) [(s+2\l+1)!]}{s!}  \delta_{s+1s'+1} \,.
\end{align}
This allows us to eliminate the series in $s'$ due to the Kronecker deltas and subsequently perform the integrals in $\rho_1$ and $\rho_3$, using the relations~\eqref{eq:OrthogonalityAssociatedLaguerreIII} and \eqref{eq:OrthogonalityAssociatedLaguerreIV}, which finally leads to one series in $s$ and several finite sums depending on $s$ only:
\begin{align}
& ^{(0)}\!\langle n ; \l | r \frac{1}{(H_{\text{o}}^{(0)} - E_{H}^{(0)})} \1 \frac{1}{(H_{\text{o}}^{(0)} - E_{H}^{(0)})} r | n ; \l \rangle^{(0)} \\
& \quad = \left(\frac{4\mr}{n a}\right)^2 N_{n\l}^2 \left(\frac{n a}{2}\right)^{11} \\
\nonumber & \quad \times \Bigg\lbrace \Big[ \sum\limits_{s=0}^\infty \sum\limits_{s'=0}^\infty \frac{s!}{(s+\l+1+\frac{n}{2N_{\text{c}} C_F})[(s+2\l+1)!]} \frac{s'!}{(s'+\l+1+\frac{n}{2N_{\text{c}} C_F})[(s'+2\l+1)!]} \\
\nonumber & \quad\quad \times \int \d \rho_1 \, \rho_1^{3+2\l} \e{-\rho_1} L_{n-\l-1}^{2\l+1}\left(\rho_1\right) L_{s}^{2\l+1}\left(\rho_1\right) \\
\nonumber & \quad\quad \times \int \d \rho_3 \, \rho_3^{3+2\l} \e{-\rho_3} L_{s'}^{2\l+1}\left(\rho_3\right) L_{n-\l-1}^{2\l+1}\left(\rho_3\right) \\
\nonumber & \quad\quad \times \frac{(s+2\l+1)^2 [(s+2\l)!]}{s!} \delta_{s s'} \Big] \\
\nonumber & \quad - \Big[ \sum\limits_{s=0}^\infty \sum\limits_{s'=0}^\infty \frac{s!}{(s+\l+1+\frac{n}{2N_{\text{c}} C_F})[(s+2\l+1)!]} \frac{s'!}{(s'+\l+1+\frac{n}{2N_{\text{c}} C_F})[(s'+2\l+1)!]} \\
\nonumber & \quad\quad \times \int \d \rho_1 \, \rho_1^{3+2\l} \e{-\rho_1} L_{n-\l-1}^{2\l+1}\left(\rho_1\right) L_{s}^{2\l+1}\left(\rho_1\right) \\
\nonumber & \quad\quad \times \int \d \rho_3 \, \rho_3^{3+2\l} \e{-\rho_3} L_{s'}^{2\l+1}\left(\rho_3\right) L_{n-\l-1}^{2\l+1}\left(\rho_3\right) \\
\nonumber & \quad\quad \times \frac{(s+2\l+1) [(s+2\l)!]}{(s-1)!} \delta_{s s'+1} \Big] \\
\nonumber & \quad - \Big[ \sum\limits_{s=0}^\infty \sum\limits_{s'=0}^\infty \frac{s!}{(s+\l+1+\frac{n}{2N_{\text{c}} C_F})[(s+2\l+1)!]} \frac{s'!}{(s'+\l+1+\frac{n}{2N_{\text{c}} C_F})[(s'+2\l+1)!]} \\
\nonumber & \quad\quad \times \int \d \rho_1 \, \rho_1^{3+2\l} \e{-\rho_1} L_{n-\l-1}^{2\l+1}\left(\rho_1\right) L_{s}^{2\l+1}\left(\rho_1\right) \\
\nonumber & \quad\quad \times \int \d \rho_3 \, \rho_3^{3+2\l} \e{-\rho_3} L_{s'}^{2\l+1}\left(\rho_3\right) L_{n-\l-1}^{2\l+1}\left(\rho_3\right) \\
\nonumber & \quad\quad \times \frac{(s+2\l+2) [(s+2\l+1)!]}{s!} \delta_{s+1s'} \Big] \\
\nonumber & \quad + \Big[ \sum\limits_{s=0}^\infty \sum\limits_{s'=0}^\infty \frac{s!}{(s+\l+1+\frac{n}{2N_{\text{c}} C_F})[(s+2\l+1)!]} \frac{s'!}{(s'+\l+1+\frac{n}{2N_{\text{c}} C_F})[(s'+2\l+1)!]} \\
\nonumber & \quad\quad \times \int \d \rho_1 \, \rho_1^{3+2\l} \e{-\rho_1} L_{n-\l-1}^{2\l+1}\left(\rho_1\right) L_{s}^{2\l+1}\left(\rho_1\right) \\
\nonumber & \quad\quad \times \int \d \rho_3 \, \rho_3^{3+2\l} \e{-\rho_3} L_{s'}^{2\l+1}\left(\rho_3\right) L_{n-\l-1}^{2\l+1}\left(\rho_3\right) \\
\nonumber & \quad\quad \times \frac{(s+1) [(s+2\l+1)!]}{s!} \delta_{s+1s'+1} \Big] \Bigg\rbrace \\
& \quad = \left(\frac{4\mr}{n a}\right)^2 N_{n\l}^2 \left(\frac{n a}{2}\right)^{11} \\
\nonumber & \quad \times \Bigg\lbrace \Big[ \sum\limits_{s=0}^\infty \frac{s!}{(s+\l+1+\frac{n}{2N_{\text{c}} C_F})[(s+2\l)!]} \frac{1}{(s+\l+1+\frac{n}{2N_{\text{c}} C_F})} \\
\nonumber & \quad\quad \times \int \d \rho_1 \, \rho_1^{3+2\l} \e{-\rho_1} L_{n-\l-1}^{2\l+1}\left(\rho_1\right) L_{s}^{2\l+1}\left(\rho_1\right) \\
\nonumber & \quad\quad \times \int \d \rho_3 \, \rho_3^{3+2\l} \e{-\rho_3} L_{s}^{2\l+1}\left(\rho_3\right) L_{n-\l-1}^{2\l+1}\left(\rho_3\right) \Big] \\
\nonumber & \quad - \Big[ \sum\limits_{s=0}^\infty \frac{s!}{(s+\l+1+\frac{n}{2N_{\text{c}} C_F})[(s+2\l)!]} \frac{1}{(s+\l+\frac{n}{2N_{\text{c}} C_F})} \\
\nonumber & \quad\quad \times \int \d \rho_1 \, \rho_1^{3+2\l} \e{-\rho_1} L_{n-\l-1}^{2\l+1}\left(\rho_1\right) L_{s}^{2\l+1}\left(\rho_1\right) \\
\nonumber & \quad\quad \times \int \d \rho_3 \, \rho_3^{3+2\l} \e{-\rho_3} L_{s-1}^{2\l+1}\left(\rho_3\right) L_{n-\l-1}^{2\l+1}\left(\rho_3\right) \Big] \\
\nonumber & \quad - \Big[ \sum\limits_{s=0}^\infty \frac{s!}{(s+\l+1+\frac{n}{2N_{\text{c}} C_F})[(s+2\l+1)!]} \frac{(s+1)}{(s+\l+2+\frac{n}{2N_{\text{c}} C_F})} \\
\nonumber & \quad\quad \times \int \d \rho_1 \, \rho_1^{3+2\l} \e{-\rho_1} L_{n-\l-1}^{2\l+1}\left(\rho_1\right) L_{s}^{2\l+1}\left(\rho_1\right) \\
\nonumber & \quad\quad \times \int \d \rho_3 \, \rho_3^{3+2\l} \e{-\rho_3} L_{s+1}^{2\l+1}\left(\rho_3\right) L_{n-\l-1}^{2\l+1}\left(\rho_3\right) \Big] \\
\nonumber & \quad + \Big[ \sum\limits_{s=0}^\infty \frac{s!}{(s+\l+1+\frac{n}{2N_{\text{c}} C_F})[(s+2\l+1)!]} \frac{(s+1)}{(s+\l+1+\frac{n}{2N_{\text{c}} C_F})} \\
\nonumber & \quad\quad \times \int \d \rho_1 \, \rho_1^{3+2\l} \e{-\rho_1} L_{n-\l-1}^{2\l+1}\left(\rho_1\right) L_{s}^{2\l+1}\left(\rho_1\right) \\
\nonumber & \quad\quad \times \int \d \rho_3 \, \rho_3^{3+2\l} \e{-\rho_3} L_{s}^{2\l+1}\left(\rho_3\right) L_{n-\l-1}^{2\l+1}\left(\rho_3\right) \Big] \Bigg\rbrace \\
& = \quad \left(\frac{4\mr}{n a}\right)^2 N_{n\l}^2 \left(\frac{n a}{2}\right)^{11} \\
\nonumber & \quad \times \Bigg\lbrace \Big[ \sum\limits_{s=0}^\infty \frac{s!}{(s+\l+1+\frac{n}{2N_{\text{c}} C_F})[(s+2\l)!]} \frac{1}{(s+\l+1+\frac{n}{2N_{\text{c}} C_F})} \\
\nonumber & \quad\quad \times \sum_{k=0}^{\text{min}(n-\l-1,s)} \frac{\Gamma(3+\alpha+k)}{k!} \frac{\Gamma(3)}{\Gamma(4-n+\l+k) [(n-\l-1-k)!]} \frac{\Gamma(3)}{\Gamma(3-s+k) [(s-k)!]} \\
\nonumber & \quad\quad \times \sum_{k'=0}^{\text{min}(s,n-\l-1)} \frac{\Gamma(3+\alpha+k')}{k'!} \frac{\Gamma(3)}{\Gamma(3-s+k') [(s-k')!]} \frac{\Gamma(3)}{\Gamma(4-n+\l+k') [(n-\l-1-k')!]} \Big] \\
\nonumber & \quad - \Big[ \sum\limits_{s=0}^\infty \frac{s!}{(s+\l+1+\frac{n}{2N_{\text{c}} C_F})[(s+2\l)!]} \frac{1}{(s+\l+\frac{n}{2N_{\text{c}} C_F})} \\
\nonumber & \quad\quad \times \sum_{k=0}^{\text{min}(n-\l-1,s)} \frac{\Gamma(3+\alpha+k)}{k!} \frac{\Gamma(3)}{\Gamma(4-n+\l+k) [(n-\l-1-k)!]} \frac{\Gamma(3)}{\Gamma(3-s+k) [(s-k)!]} \\
\nonumber & \quad\quad \times \sum_{k'=0}^{\text{min}(s-1,n-\l-1)} \frac{\Gamma(3+\alpha+k')}{k'!} \frac{\Gamma(3)}{\Gamma(4-s+k') [(s-1-k')!]} \frac{\Gamma(3)}{\Gamma(4-n+\l+k') [(n-\l-1-k')!]} \Big] \\
\nonumber & \quad - \Big[ \sum\limits_{s=0}^\infty \frac{s!}{(s+\l+1+\frac{n}{2N_{\text{c}} C_F})[(s+2\l+1)!]} \frac{(s+1)}{(s+\l+2+\frac{n}{2N_{\text{c}} C_F})} \\
\nonumber & \quad\quad \times \sum_{k=0}^{\text{min}(n-\l-1,s)} \frac{\Gamma(3+\alpha+k)}{k!} \frac{\Gamma(3)}{\Gamma(4-n+\l+k) [(n-\l-1-k)!]} \frac{\Gamma(3)}{\Gamma(3-s+k) [(s-k)!]} \\
\nonumber & \quad\quad \times \sum_{k'=0}^{\text{min}(s+1,n-\l-1)} \frac{\Gamma(3+\alpha+k')}{k'!} \frac{\Gamma(3)}{\Gamma(2-s+k') [(s+1-k')!]} \frac{\Gamma(3)}{\Gamma(4-n+\l+k') [(n-\l-1-k')!]} \Big] \\
\nonumber & \quad + \Big[ \sum\limits_{s=0}^\infty \frac{s!}{(s+\l+1+\frac{n}{2N_{\text{c}} C_F})[(s+2\l+1)!]} \frac{(s+1)}{(s+\l+1+\frac{n}{2N_{\text{c}} C_F})} \\
\nonumber & \quad\quad \times \sum_{k=0}^{\text{min}(n-\l-1,s)} \frac{\Gamma(3+\alpha+k)}{k!} \frac{\Gamma(3)}{\Gamma(4-n+\l+k) [(n-\l-1-k)!]} \frac{\Gamma(3)}{\Gamma(3-s+k) [(s-k)!]} \\
\nonumber & \quad\quad \times \sum_{k'=0}^{\text{min}(s,n-\l-1)} \frac{\Gamma(3+\alpha+k')}{k'!} \frac{\Gamma(3)}{\Gamma(3-s+k') [(s-k')!]} \frac{\Gamma(3)}{\Gamma(4-n+\l+k') [(n-\l-1-k')!]} \Big] \Bigg\rbrace \,,
\end{align}
where we introduced the abbreviation $\alpha \equiv 2\l+1$. Because the states we are interested in are the $2\,^3\!P_J$ and the $1\,^3S_1$, it is immediately clear that, since for them $n-\l-1 = 0$, in each of the finite sums in $k$ and $k'$, only $k=0$ and $k'=0$ terms contribute, respectively. This is due to the respective denominators $(n-\l-1-k)!$ and $(n-\l-1-k')!$ that are present in all of them.\\
For higher radial excitations, hence for hypothetical cases were $n-\l-1 = \xi > 0$, the finite sums are restricted to $k \leq \xi$ and $k' \leq \xi$, respectively.\footnote{We note, however, that due to denominators of the form $\Gamma(\xi-n+\l)$, $\xi \in \N$, arbitrarily high excitations are not possible.}\\
Explicitly implementing the condition $n-\l-1 = 0$ and therewith performing the resulting trivial finite sums in $k$ and $k'$, respectively, yields
\begin{align}
& \left.^{(0)}\!\langle n ; \l | r \frac{1}{(H_{\text{o}}^{(0)} - E_{H}^{(0)})} \1 \frac{1}{(H_{\text{o}}^{(0)} - E_{H}^{(0)})} r | n ; \l \rangle^{(0)}\right|_{n-\l-1=0} \\
& \quad = \left(\frac{4\mr}{n a}\right)^2 N_{n\l}^2 \left(\frac{n a}{2}\right)^{11} \\
\nonumber & \quad \times \Bigg\lbrace \Big[ \sum\limits_{s=0}^\infty \frac{s!}{(s+\l+1+\frac{n}{2N_{\text{c}} C_F})[(s+2\l)!]} \frac{1}{(s+\l+1+\frac{n}{2N_{\text{c}} C_F})} \\
\nonumber & \quad\quad \times \Gamma(3+\alpha) \frac{\Gamma(3)}{\Gamma(4-n+\l)} \frac{\Gamma(3)}{\Gamma(3-s) [(s)!]} \times \Gamma(3+\alpha) \frac{\Gamma(3)}{\Gamma(3-s) [(s)!]} \frac{\Gamma(3)}{\Gamma(4-n+\l)} \Big] \\
\nonumber & \quad - \Big[ \sum\limits_{s=0}^\infty \frac{s!}{(s+\l+1+\frac{n}{2N_{\text{c}} C_F})[(s+2\l)!]} \frac{1}{(s+\l+\frac{n}{2N_{\text{c}} C_F})} \\
\nonumber & \quad\quad \times \Gamma(3+\alpha) \frac{\Gamma(3)}{\Gamma(4-n+\l)} \frac{\Gamma(3)}{\Gamma(3-s) [(s)!]} \times \Gamma(3+\alpha) \frac{\Gamma(3)}{\Gamma(4-s) [(s-1)!]} \frac{\Gamma(3)}{\Gamma(4-n+\l)} \Big] \\
\nonumber & \quad - \Big[ \sum\limits_{s=0}^\infty \frac{s!}{(s+\l+1+\frac{n}{2N_{\text{c}} C_F})[(s+2\l+1)!]} \frac{(s+1)}{(s+\l+2+\frac{n}{2N_{\text{c}} C_F})} \\
\nonumber & \quad\quad \times \Gamma(3+\alpha) \frac{\Gamma(3)}{\Gamma(4-n+\l)} \frac{\Gamma(3)}{\Gamma(3-s) [(s)!]} \times \Gamma(3+\alpha) \frac{\Gamma(3)}{\Gamma(2-s) [(s+1)!]} \frac{\Gamma(3)}{\Gamma(4-n+\l)} \Big] \\
\nonumber & \quad + \Big[ \sum\limits_{s=0}^\infty \frac{s!}{(s+\l+1+\frac{n}{2N_{\text{c}} C_F})[(s+2\l+1)!]} \frac{(s+1)}{(s+\l+1+\frac{n}{2N_{\text{c}} C_F})} \\
\nonumber & \quad\quad \times \Gamma(3+\alpha) \frac{\Gamma(3)}{\Gamma(4-n+\l)} \frac{\Gamma(3)}{\Gamma(3-s) [(s)!]} \times \Gamma(3+\alpha) \frac{\Gamma(3)}{\Gamma(3-s) [(s)!]} \frac{\Gamma(3)}{\Gamma(4-n+\l)} \Big] \Bigg\rbrace \,.
\end{align}
At this point one may realize that now additionally each individual sum in $s$ is constrained to a finite number of summands, due to denominators of the form $\Gamma(\xi-s)$, $\xi \in \N$, and the matrix element can thus be computed exactly. This actually also holds for the hypothetical cases were $n-\l-1 > 0$, although there the constrains will be functions of $n-\l-1$. Including the constrains in $s$, the matrix element takes the form
\begin{align}
& \left.^{(0)}\!\langle n ; \l | r \frac{1}{(H_{\text{o}}^{(0)} - E_{H}^{(0)})} \1 \frac{1}{(H_{\text{o}}^{(0)} - E_{H}^{(0)})} r | n ; \l \rangle^{(0)}\right|_{n-\l-1=0} \\
& \quad = \left(\frac{4\mr}{n a}\right)^2 N_{n\l}^2 \left(\frac{n a}{2}\right)^{11} \\
\nonumber & \quad \times \Bigg\lbrace \Big[ \sum\limits_{s=0}^2 \frac{s!}{(s+\l+1+\frac{n}{2N_{\text{c}} C_F})[(s+2\l)!]} \frac{1}{(s+\l+1+\frac{n}{2N_{\text{c}} C_F})} \\
\nonumber & \quad\quad \times \Gamma(3+\alpha) \frac{\Gamma(3)}{\Gamma(4-n+\l)} \frac{\Gamma(3)}{\Gamma(3-s) [(s)!]} \times \Gamma(3+\alpha) \frac{\Gamma(3)}{\Gamma(3-s) [(s)!]} \frac{\Gamma(3)}{\Gamma(4-n+\l)} \Big] \\
\nonumber & \quad - \Big[ \sum\limits_{s=1}^2 \frac{s!}{(s+\l+1+\frac{n}{2N_{\text{c}} C_F})[(s+2\l)!]} \frac{1}{(s+\l+\frac{n}{2N_{\text{c}} C_F})} \\
\nonumber & \quad\quad \times \Gamma(3+\alpha) \frac{\Gamma(3)}{\Gamma(4-n+\l)} \frac{\Gamma(3)}{\Gamma(3-s) [(s)!]} \times \Gamma(3+\alpha) \frac{\Gamma(3)}{\Gamma(4-s) [(s-1)!]} \frac{\Gamma(3)}{\Gamma(4-n+\l)} \Big] \\
\nonumber & \quad - \Big[ \sum\limits_{s=0}^1 \frac{s!}{(s+\l+1+\frac{n}{2N_{\text{c}} C_F})[(s+2\l+1)!]} \frac{(s+1)}{(s+\l+2+\frac{n}{2N_{\text{c}} C_F})} \\
\nonumber & \quad\quad \times \Gamma(3+\alpha) \frac{\Gamma(3)}{\Gamma(4-n+\l)} \frac{\Gamma(3)}{\Gamma(3-s) [(s)!]} \times \Gamma(3+\alpha) \frac{\Gamma(3)}{\Gamma(2-s) [(s+1)!]} \frac{\Gamma(3)}{\Gamma(4-n+\l)} \Big] \\
\nonumber & \quad + \Big[ \sum\limits_{s=0}^2 \frac{s!}{(s+\l+1+\frac{n}{2N_{\text{c}} C_F})[(s+2\l+1)!]} \frac{(s+1)}{(s+\l+1+\frac{n}{2N_{\text{c}} C_F})} \\
\nonumber & \quad\quad \times \Gamma(3+\alpha) \frac{\Gamma(3)}{\Gamma(4-n+\l)} \frac{\Gamma(3)}{\Gamma(3-s) [(s)!]} \times \Gamma(3+\alpha) \frac{\Gamma(3)}{\Gamma(3-s) [(s)!]} \frac{\Gamma(3)}{\Gamma(4-n+\l)} \Big] \Bigg\rbrace \,.
\end{align}

\clearpage
\thispagestyle{empty}
\clearpage

\chapter{Alternative method to calculate the first order correction to the wave function}
\label{app:chp:AlternativeFirstOrderWaveFunction}

The method of Coulomb Green functions is not the only one to compute corrections to the wave function. Although it has proven to be very successful, it suffers from the need of summing over infinitely many intermediate states. Operators that can be written in the form of ladder operators, like in the case of the quantum mechanical harmonic oscillator, lead to a finite amount of intermediate states only. However, in the case of Hydrogen like systems such a formalism is not known and one eventually has to perform the sum over intermediate states numerically. We now want to present an alternative method that circumvents this drawback.\\
In order to calculate the first order correction to the wave function, $\psi_1$, we make use of the following ansatz\footnote{It is due to Prof. Norbert Kaiser that we became aware of this alternative method. Nevertheless, Chapter~8.1 in \cite{BransdenJoachain:QuantumMechanics} uses a similar method on very general ground but for a discrete case. We extend and specify the method mentioned there to the continuous case.}, assuming that the unperturbed Hamiltonian is of the form
\begin{equation}
H_0 = \left( \frac{-1}{2\mr} \nabla^2 + V^{(0)} \right) \,,
\end{equation}
being the leading order Schrödinger Hamiltonian~\eqref{eq:ZerothOrderSchroedingerEquation}. The leading order static potential~\eqref{eq:LeadingOrderStaticPotential} is given by $V^{(0)} = -C_F \frac{\alpha_s}{r}$ and the perturbation is a function of the radial component only, hence
\begin{equation}
H_1 = V(r) \,.
\end{equation}
The first order correction to the energy, Eq.~\eqref{eq:FirstOrderEnergy}, is given by $E_1 = \langle \psi_0 | H_1 | \psi_0 \rangle$ and we have the condition
\begin{equation}
(E_0 - H_0) \psi_1 = (H_1 - E_1) \psi_0 \,,
\end{equation}
that can be verified easily be multiplying both sides by $\psi_0^*$ from the left and integrating, yielding
\begin{align}
& \langle \psi_0 | (E_0 - H_0) | \psi_1 \rangle = \langle \psi_0 | \hspace{-0.4cm} \underbrace{(H_1 - E_1)}_{= 0 \text{ by construction}} \hspace{-0.4cm} | \psi_0 \rangle \\
& \langle \psi_0 | E_0 | \psi_1 \rangle - \hspace{-0.9cm} \underbrace{\langle \psi_0 | H_0}_{= E_0 \langle \psi_0| \text{ since } H_0 = H_0^*} \hspace{-0.9cm} | \psi_1 \rangle = 0 \\
& E_0 \underbrace{\langle \psi_0 | \psi_1 \rangle}_{= 0} - E_0 \underbrace{\langle \psi_0 | \psi_1 \rangle}_{= 0} = 0 \,,
\end{align}
where we assumed orthogonality between the unperturbed and the perturbed wave function in the last step.\\
We now assume that we can construct $\psi_1$ from $\psi_0$ by means of an operator $\O = F(r)$ that is a function of the radial component only, hence
\begin{equation}
\psi_1 = \O \psi_0 = F(r) \psi_0 \,.
\end{equation}
This implies $\psi_0$ to be an eigenfunction of $\O$ and thus the commutator of $H_0$ and $\O$ vanishes and we have the condition
\begin{equation}
[\O,H_0] \psi_0 = (H_1 - E_1) \psi_0 \,.
\end{equation}
This is the central equation from which we can derive a differential equation for $\O$, whose solution will give us the function $F(r)$ that we need to construct the first order correction to the wave function.\\
Using spherical coordinates we have
\begin{equation}
\nabla^2 = \frac{\partial^2}{\partial r^2} + \frac{2}{r} \frac{\partial}{\partial r}
\end{equation}
and can thus derive
\begin{align}
[\O,H_0] \psi_0 &= \O \left( \frac{-1}{2\mr} \nabla^2 - C_F \frac{\alpha_s}{r} \right) \psi_0 - \left( \frac{-1}{2\mr} \nabla^2 - C_F \frac{\alpha_s}{r} \right) \O \psi_0 \\
&= \O \left( \frac{-1}{2\mr} \nabla^2 \right) \psi_0 - \left( \frac{-1}{2\mr} \nabla^2 \right) \O \psi_0 \\
&= \frac{-1}{2\mr} \left[ \O (\nabla^2 \psi_0) - \nabla^2 (\O \psi_0) \right] \\
&= \frac{-1}{2\mr} \left[ \O (\nabla^2 \psi_0) - \O (\nabla^2 \psi_0) - (\nabla^2 \O) \psi_0 - 2 (\nabla \O)(\nabla \psi_0) \right] \\
&= \frac{-1}{2\mr} \left[ - (\nabla^2 \O) \psi_0 - 2 (\nabla \O)(\nabla \psi_0) \right] \\
&= \frac{-1}{2\mr} \left[ - \left( \O'' + \frac{2}{r} \O' \right) - 2 \O' \frac{\psi_0'}{\psi_0} \right] \psi_0 = (H_1 - E_1) \psi_0 \,.
\end{align}
We have now arrived at the desired differential equation for the operator $\O$, namely
\begin{equation}
\frac{1}{2\mr} \left[\O'' + \left( \frac{2}{r} + 2 \frac{\psi_0'}{\psi_0} \right) \O' \right] = H_1 - E_1 \,,
\end{equation}
that has the solution, using $\O = F(r)$,
\begin{equation}
F'(r) = c_1 \frac{\psi_0(r)^{-2}}{r^2} + \frac{2 \mr \psi_0(r)^{-2}}{r^2} \int\limits_0^r \d R \, R^2 \psi_0(R)^2 (V(R) - E_1) \,.
\end{equation}
For physical reasons we demand $F'$ not to be singular at the origin and therefore we set $c_1 = 0$. $F(r)$ and therewith the operator $\O$ can then be obtained by
\begin{equation}
F(r) = \int\limits_0^r \d R \, F'(R) + c_2 \,,
\end{equation}
where $c_2$ is another integration constant that can be fixed by demanding orthogonality with respect to the zeroth order wave function, hence
\begin{equation}
\langle \psi_1 | \psi_0 \rangle = \int\limits_0^\infty \d r \, r^2 (F(r) \psi_0)^* \psi_0 = 0 \,.
\end{equation}
Having fixed $c_2$ we may now cast the final result as
\begin{equation}
\psi_1(r) = F(r) \psi_0(r) \,.
\end{equation}

\begin{description}
\item[Example:] We drop the $\frac{1}{2\mr}$-factor and take the zeroth order wave function to be $\psi_0 = 2 \e{-r}$ and assume a linear perturbation $V(r) = r$. We then find the first order energy correction to be
\begin{equation}
E_1 = \int\limits_0^\infty \d r \, r^3 4 \e{-2r} = \frac{3}{2} \,.
\end{equation}
The differential equation then takes the form
\begin{equation}
\left[F'' + \left( \frac{2}{r} - 2 \right) F' \right] = r - \frac{3}{2}
\end{equation}
having the solution
\begin{equation}
F'(r) = \frac{\e{2r}}{4 r^2} \int\limits_0^r \d R \, R^2 4 \e{-2R} (R - \frac{3}{2}) = -\frac{r}{2} \,,
\end{equation}
where we already set $c_1 = 0$. This amounts to
\begin{equation}
F(r) = - \int\limits_0^r \d R \, \frac{R}{2} = -\frac{r^2}{4} + c_2 \,,
\end{equation}
where orthogonality implies
\begin{equation}
\int\limits_0^\infty \d r \, r^2 \left[\left(-\frac{r^2}{4} + c_2\right) 2\e{-r}\right]^* 2\e{-r} = c_2 - \frac{3}{4} = 0 \quad \Rightarrow c_2 = \frac{3}{4}
\end{equation}
and we finally arrive at
\begin{align}
\begin{aligned}
& F(r) = -\frac{r^2}{4} + \frac{3}{4} = \frac{1}{4} (-r^2 + 3) \\
\Rightarrow & \psi_1 = \frac{1}{2} (-r^2 + 3) \e{-r} \,,
\end{aligned}
\end{align}
\end{description}
which in deed is the correct solution. However, it turns out that for potentials proportional to $\ln{r}$ or $\frac{1}{r^3}$ this differential equation ansatz is no longer a practicable way to obtain the first order wave function and therefore we shall be using the method of Coulomb Green functions.

\clearpage
\thispagestyle{empty}
\clearpage

\backmatter

\listoffigures

\clearpage
\thispagestyle{empty}
\clearpage

\listoftables

\clearpage
\thispagestyle{empty}
\clearpage

\bibliography{mybib,../../Dokumente/Mendeley_Desktop/library}
\bibliographystyle{alpha}

\clearpage
\thispagestyle{empty}
\clearpage

\chapter{Acknowledgments}
\label{app:chp:Acknowledgments}

First, I want to thank PD~Dr.~Antonio Vairo and Prof.~Dr.~Nora Brambilla for the opportunity to write this Master thesis and for their continuous supervision, encouragement and support. I am grateful for the opportunities to present my work at the \textit{XIIth Quark Confinement and the Hadron Spectrum} and during the \textit{10th Science Week} of the \textit{Excellence Cluster Universe} and for the opportunity to attend the \textit{Summer School on Symmetries, Fundamental Interactions and Cosmology}. I acknowledge and appreciative the financial support of the Physik Department of the Technische Universität München that allowed me to attend the XIIth Quark Confinement and the Hadron Spectrum conference.\\
I am deeply indebted to my supervisor Dr.~Jorge Segovia for mentoring and helping me during the whole process of learning, understanding, calculating and finally writing. Thank you for spending hours with me; discussing and explaining the obvious and the hidden.\\
In addition, I want to thank all of the people working at T30f and T39 for the great time I had with you. Thanks also for the hours of inspiring talks, not only on the subjects of this thesis but on physics and politics and various other topics as well. Special thanks go to Vladyslav Shtabovenko for saving me much time by explaining various aspects of MATHEMATICA and to Prof.~Dr.~Norbert Kaiser for always taking time in order to discuss issues on analytical computations and for providing the hint that lead to the ansatz described in Appendix~\ref{app:chp:AlternativeFirstOrderWaveFunction}.\\
Finally, I would like to thank my family and my friends for supporting me in any imaginable way. I owe you everything.

\end{document}